\documentclass[a4paper,12pt]{amsart}
\calclayout

\usepackage{enumerate}
\usepackage{enumitem}
\usepackage{amsmath}    
\usepackage{amssymb}    
\usepackage{bm}         
\usepackage{graphicx}   
\usepackage{color}      

\usepackage{natbib}    
\usepackage{epsf}
\usepackage{lscape}
\bibpunct{(}{)}{;}{a}{,}{,}

\newcommand{\by}{{\bf y}}

\newcommand{\bz}{{\bf z}}

\newcommand{\bw}{{\bf w}}

\newcommand{\bzeta}{\bm{\zeta}}

\newcommand{\bmu}{\bm{\mu}}

\newcommand{\btheta}{\bm{\theta}}
\newcommand{\tbtheta}{\widetilde{\boldsymbol{\theta}}}
\newcommand{\bvartheta}{\boldsymbol{\vartheta}}
\newcommand{\tbvartheta}{\widetilde{\boldsymbol{\vartheta}}}

\newcommand{\Mk}{\mathcal{M}_k}

\newtheorem{thm}{Theorem}

\newtheorem{prop}[thm]{Proposition}

\newtheorem{defin}{Proposition}
\newtheorem{define}{Definition}

\theoremstyle{definition}

 \theoremstyle{remark}

\numberwithin{equation}{section}
\usepackage{array}
\usepackage{layout}
\newtheorem{defin1}{\textbf{Corollary}}
\newtheorem{defin2}{\textbf{Theorem}}
\usepackage[norelsize]{algorithm2e}
\usepackage{chngcntr}
\usepackage{apptools}

\usepackage{hyperref}
\usepackage[all]{hypcap}
\usepackage{bookmark}

\title{On choosing mixture components via non-local priors}

\author{Jairo F\'uquene, Mark Steel, David Rossell}
\thanks{Corresponding author: David Rossell (rosselldavid@gmail.com)}

\begin{document}

\date{}

\maketitle

\begin{center}
\textbf{Abstract}

\end{center}

Choosing the number of mixture components remains an elusive challenge. Model selection criteria can be either overly liberal or conservative and return poorly-separated components of limited practical use. We formalize non-local priors (NLPs) for mixtures and show how they lead to well-separated components with non-negligible weight, interpretable as distinct subpopulations. We also propose an estimator for posterior model probabilities under local and non-local priors, showing that Bayes factors are ratios of posterior to prior empty-cluster probabilities. The estimator is widely applicable and helps set thresholds to drop unoccupied components in overfitted mixtures. We suggest default prior parameters based on multi-modality for Normal/T mixtures and minimal informativeness for categorical outcomes. We characterise theoretically the NLP-induced sparsity, derive tractable expressions and algorithms. We fully develop Normal, Binomial and product Binomial mixtures but the theory, computation and principles hold more generally. We observed a serious lack of sensitivity of the Bayesian information criterion (BIC), insufficient parsimony of the AIC and a local prior, and a mixed behavior of the singular BIC. We also considered overfitted mixtures, their performance was competitive but depended on tuning parameters. Under our default prior elicitation NLPs offered a good compromise between sparsity and power to detect meaningfully-separated components.

\vspace*{.3in}


\section{Introduction}

Mixture models have many applications,
 {\it e.g.} in human genetics \citep{Shork}, false discovery rate
control \citep{Efron}, signal deconvolution
\citep{West}, density estimation \citep{escobar} and
cluster analysis \citep{raftery,baudry2012combining}.
See \cite{book_silvia} and \cite{robert4} for an extensive treatment.
Despite having such a fundamental role,
their irregular nature (multi-modal and unbounded likelihood, non-identifiability)
creates difficulties in choosing the number of components both in the
Bayesian and frequentist paradigms.
As discussed below, although existing formal criteria
may achieve model selection consistency as the sample size grows to infinity \citep{gassiat2013consistent},
in practice they often lead to too many or too few components
and require the data analyst to perform some ad-hoc post-processing.
Our main contributions are proposing the use of non-local priors (NLPs) to select the number of components,
characterizing the properties of the associated inference (improved sparsity) and proposing computationally tractable algorithms.
This includes the ECP algorithm, a new strategy to obtain posterior model probabilities applicable
both to local and non-local priors.
We also emphasise prior elicitation to obtain default prior parameters
and illustrate the framework in popular families
that include Normal, Student-$t$
(T), Binomial and product Binomial mixtures.
Our formulation, theory and computational algorithms hold more generally, however.

Consider a sample $\by=(\by_{1},...,\by_{n})$
of independent observations from a finite mixture
where $\by_{i} \in \mathbb{R}^{p}$ arises from the density
\begin{align}\label{likelihood1}
p(\by_{i}\mid\bvartheta_{k},\Mk)=\sum_{j=1}^{k}\eta_{j}p(\by_{i}\mid\btheta_{j}).
\end{align}
The component densities $p(\by\mid\btheta_{j})$ are indexed by a parameter
$\btheta_{j}\in\Theta$,
$\boldsymbol{\eta}=(\eta_{1},...,\eta_{k}) \in \mathcal{E}_{k}$ denotes the weights,
$\mathcal{E}_{k}$ the unit simplex
and $\Mk$ the model with $k$ components. Our main goal is to infer $k$.
For simplicity we assume that there is an upper bound $K$
such that $k \in \{1,\ldots,K\}$,
{\it e.g.} given by subject-matter considerations,
but our framework remains valid for a prior
distribution on $k$ with support on the natural numbers.
The whole parameter is
$\bvartheta_{k}=(\btheta,\boldsymbol{\eta}) \in \Theta_k= \Theta^k \times \mathcal{E}_k$
where  $\btheta=(\btheta_{1},...,\btheta_{k})$.
As an example, in Normal mixtures
$p(\by \mid \btheta_j)= \text{N}(\by; \bmu_j, \Sigma_j)$ and $\btheta_j=(\bmu_j, \Sigma_j)$ where $\bmu_{j} \in \mathbb{R}^p$ is the mean
and $\Sigma_{j}$ the covariance matrix of component $j$.
One may also consider heavy-tailed alternatives such as T densities $p(\by \mid \btheta_j)= \text{T}(\by; \bmu_j, \Sigma_j,\upsilon_{j})$,
where $\btheta_j=(\bmu_j, \Sigma_j, \upsilon_{j})$ and $\upsilon_{j}$ is the degrees of freedom parameter.
Another class illustrated here are product Binomial mixtures with mass function
$p(\by_i \mid \btheta_j)=\prod_{f=1}^{p}{L_{if} \choose y_{if}}\theta_{jf}^{y_{if}}(1-\theta_{jf})^{L_{if}-y_{if}}$,
where $\by_i=(y_{i1},\ldots,y_{ip})$
are the number of successes for individual $i$ across $p$ outcomes,
$L_{if}$ the number of trials and $\theta_{jf}$ the success probability for outcome $f$ under component $j$,
and $\btheta_j=(\theta_{j1},\ldots,\theta_{jp})$.
The case $p=1$ corresponds to a Binomial mixture.
Throughout, we assume that $\by$ are generated by
$p(\by \mid \bvartheta_{k^*}^*,\mathcal{M}_{k^*})$
for some $k^*\in \{1,\ldots,K\}$, $\bvartheta_{k^*}^* \in \Theta_{k^*}$.

Mixtures suffer from a lack of identifiability that plays a
fundamental role both in estimation and model selection.
This issue can be caused by the invariance of the likelihood
to relabeling the components or by posing overfitted models
that could be equivalently defined with $k'<k$ components,
{\it e.g.} setting $\eta_j=0$ or $\btheta_i=\btheta_j$ for some $i \neq j$.
Relabeling (also known as label switching) is due to there being
$k!$ ways of rearranging the components
that give the same $p(\by \mid \bvartheta_k,\Mk)$.
Although relabelling creates some technical difficulties, it does not seriously
hamper inference. For instance, if $k=k^*$ then the maximum likelihood estimator (MLE)
is consistent and asymptotically Normal as $n \rightarrow \infty$ in the quotient topology \citep{redner},
and from a Bayesian perspective the integrated likelihood behaves asymptotically as in regular models \citep{crawford:1994}.
Non-identifiability due to overfitting has more serious consequences,
{\it e.g.} estimates for $p(\by \mid \bvartheta_k,\Mk)$ are consistent under mild conditions
\citep{van}
but the MLE and posterior mode of $\bvartheta_k$ can behave erratically
\citep{leroux,judith,ho:2016}.
In addition, as we now discuss, frequentist and Bayesian methods to choose $\Mk$ can behave unsatisfactorily.

The literature on criteria to choose $k$ is too large to cover here,
the reader is referred to \cite{Richardson}, \cite{raftery}, \cite{baudry2012combining} and \cite{gassiat2013consistent}.
We review a few model-based criteria, as these are most closely related to our proposal
and can be applied to any probability model.
From a frequentist perspective
the likelihood ratio test between $\Mk$ and $\mathcal{M}_{k+1}$ may diverge as $n \rightarrow \infty$
when data truly arise from $\Mk$
unless restrictions on the parameters or likelihood penalties are imposed (\cite{Gosh,Liu,chen}).
As an alternative one may consider criteria such as the
Bayesian information criterion (BIC), Akaike's information criterion (AIC),
the integrated complete likelihood \citep{biernacki:2000} or the singular BIC (\cite{drton2017bayesian}, sBIC).
Although the BIC justification as an approximation to the Bayesian evidence
\citep{Schwarz} is not valid for overfitted mixtures,
it is often adopted as a useful criterion \citep{raftery}.
One issue is that the BIC ignores that $p(\by \mid \bvartheta_k,\Mk)$
has $k!$ maxima, causing a loss of sensitivity to detect truly present components.
More importantly, the dimensionality penalty $p_k=\mbox{dim}(\Theta_k)$ used by the BIC
is too large for overfitted mixtures \citep{watanabe:2013}, again decreasing power.
These theoretical observations align with the empirical results we present here.
The sBIC builds on \cite{watanabe:2009,watanabe:2013}
to improve the BIC's asymptotic approximation of the integrated likelihood.
In our results the sBIC over-penalized model complexity in some examples (albeit less so than the BIC)
but under-penalized in others, where it gave similar results to the AIC.

From a Bayesian perspective, model selection is often based on
the posterior probability
$P(\Mk\mid\by)=
p(\by\mid\Mk)P(\Mk)/p(\by)$,
where $P(\Mk)$ is the prior probability,
\begin{equation}\label{Integred}
p(\by\mid\Mk)=\int_{\Theta_{k}}p(\by\mid\bvartheta_{k},\Mk)p(\bvartheta_{k}\mid\Mk)
d\bvartheta_{k}
\end{equation}
the integrated (or marginal) likelihood and
$p(\bvartheta_{k} \mid\Mk)$ a prior distribution under $\Mk$.
One may also use Bayes factors $B_{k',k}(\by)=p(\by\mid\mathcal{M}_{k'})/p(\by\mid\Mk)$ to compare any pair $\mathcal{M}_{k'}, \Mk$.
A common argument for \eqref{Integred}
is that it automatically penalizes overly complex models,
however this parsimony is not as strong as one would ideally wish.
To gain intuition, for regular models with fixed $p_k$ one obtains
\begin{align}
\log p(\by \mid \Mk)= \log p(\by \mid \hat{\bvartheta}_k,\Mk) - \frac{p_k}{2} \log(O_p(n)) + O_p(1)
\label{eq:asymp_marglhood_regular}
\end{align}
as $n \rightarrow \infty$ \citep{dawid:1999}.
This implies that
$B_{k^*,k}(\by)$ grows exponentially as $n \rightarrow \infty$ when $\mathcal{M}_{k^*} \not\subset \Mk$
but is only $O_p(n^{-(p_k-p_{k^*})/2})$ when $\mathcal{M}_{k^*} \subset \Mk$.
That is, overfitted models are only penalized at a slow polynomial rate.
Key to the current manuscript, \cite{david1} showed that either faster polynomial or quasi-exponential rates
are obtained by letting $p(\bvartheta_k \mid \Mk)$ be a NLP (defined below).
Expression (\ref{eq:asymp_marglhood_regular}) remains valid for many mixtures with $k \leq k^*$
(including Normal mixtures, \cite{crawford:1994}),
however this is no longer the case for $k>k^*$.
Using algebraic statistics, \cite{watanabe:2009,watanabe:2013}
gave expressions analogous to (\ref{eq:asymp_marglhood_regular}) for $k>k^*$
where $p_k/2$ is replaced by a rational number $\lambda \in [p_{k^*}/2,p_k/2]$ called the {\it real canonical threshold}
and the remainder term is $O_p(\log \log n)$ instead of $O_p(1)$.
The exact value of $\lambda$ is complicated but the implication is that $p_k$ in (\ref{eq:asymp_marglhood_regular})
imposes an overly stringent penalty that can decrease the sensitivity of the BIC,
and also that the Bayes factor to penalize overfitted $k>k^*$ mixtures is $B_{k,k^*}(\by)= O_p(n^{-(\lambda - p_{k^*}/2)})$.
That is, akin to regular models, $k>k^*$ is penalized only at a slow polynomial rate.
These results align with those in \cite{chambaz:2008}.
Denoting by $\hat{k}=\arg\max_k P(\Mk \mid \by)$,
these authors found that the frequentist probability $P_{\vartheta_{k^*}^*}(\hat{k} < k^*)=O(e^{-an})$ but in
contrast $P_{\vartheta^*}(\hat{k} > k^*)=O((\mbox{log}n)^b/\sqrt{n})$ for some constants $a,b>0$,
again implying that spurious components are not sufficiently penalized.  We emphasize that these
results apply to a wide class of priors but not to the
NLP class proposed in this paper, for which faster rates are attained.
Note also that the BIC and related likelihood penalties
attain consistency as $n \to \infty$ for fairly general mixtures \citep{gassiat2013consistent},
as long as $\log(n)$ is replaced by a rate strictly between $\log\log(n)$ and $n$,
but as illustrated here for finite (potentially quite large) $n$ the BIC can lack sensitivity.

An interesting alternative to considering $k \in \{1,\ldots,K\}$ is to set a single large $k$
and subsequently discard unoccupied components, a strategy often referred to as {\it overfitted mixtures}.
\cite{judith} showed that the prior on the weights $p(\bm{\eta}\mid\Mk)$
strongly influences posterior inference when $k>k^*$.
Under  $p(\bm{\eta}\mid\Mk)=\mbox{Dir}(\bm{\eta}; q_{1},...,q_{k})$
with max$_{j}$$q_{j}<d/2$ where $d=\mbox{dim}(\Theta)$
the posterior of $\bm{\eta}$  collapses to 0
for redundant components, but if $\min_{j} q_{j}>d/2$ then it collapses on a solution where at least two
components $i \neq j$ have identical parameters $\btheta_i=\btheta_j$ and non-zero weights $\eta_i>0$, $\eta_j>0$.
That is, the posterior shrinkage induced by $q_j<d/2$ helps discard spurious components.
\cite{gelman4} set $q_{1}=...=q_{k}=1/k$,
but \cite{zoe} argued that this leads to insufficient shrinkage and proposed smaller $q_j$.
\cite{Dunson} argued that faster shrinkage may be obtained via overfitted repulsive priors,
i.e. assigning vanishing density to $\btheta_i=\btheta_j$ for $i \neq j$.
\cite{Fox} and \cite{xu:2015} gave related determinantal point process frameworks,
and \cite{xie:2017} proposed extensions to non-parametric Gaussian mixtures.
A recent approach by \cite{Silviamm} resembling repulsive mixtures
is to encourage nearby components merging into groups at a first hierarchical level
and to then enforce between-group separation at the second level.
Interestingly, repulsive mixtures are a shrinkage counterpart to our framework,
but, as we shall see, NLPs penalize not only $\btheta_i=\btheta_j$ but also small weights.

In spite of their usefulness, overfitted mixtures (whether repulsive or not) also bear limitations.
On the practical side one can study the number of components but cannot address more general model selection questions,
say choosing equal versus different component-specific covariances.
Also inference may be sensitive to the chosen $q_j$, $k$,
or the threshold to discard unoccupied components (Section \ref{ssec:repulsive}).
In terms of interpretation, cluster occupancy probabilities given by overfitted mixtures
are different from model probabilities $p(\Mk \mid \by)$.
In Section \ref{sec:computation} we show that Bayes factors, and hence $p(\Mk \mid \by)$,
are given by ratios of posterior to prior empty cluster probabilities.
This result motivates a novel empty cluster probability (ECP) estimator to obtain $p(\Mk \mid \by)$
from standard MCMC output that is computationally convenient and applicable to very general mixtures,
both under local and non-local priors.
We remark that estimating $p(\Mk \mid \by)$ requires one to consider multiple $k$,
relative to overfitted mixtures where one sets a single large $k$,
however this is an easily parallelized problem.
Building upon \cite{david1,david2}, we formally define NLPs in the context of mixtures.

\begin{define}
Let $\Mk$ be the $k$-component mixture in (\ref{likelihood1}).
A continuous prior density $p(\bvartheta_k\mid\Mk)$ is a NLP iff
$$\mathop {\lim }\limits_{\bvartheta_k \to {\bf t}} p(\bvartheta_k\mid\Mk)= 0$$
for any ${\bf t} \in \Theta_k$ such that
$p(\by\mid{\bf t},\Mk)=p(\by\mid\bvartheta_{k'},\mathcal{M}_{k'})$
for some $\bvartheta_{k'} \in \Theta_{k'}$, $k'<k$.
\label{def:nlp}
\end{define}

A local prior (LP) is any $p(\bvartheta_k \mid \Mk)$ not satisfying Definition \ref{def:nlp}.
Intuitively for nested $\mathcal{M}_{k'} \subset \Mk$ a NLP
$p(\bvartheta_k \mid \Mk)$ penalizes any $\bvartheta_k$ that would be consistent with $\mathcal{M}_{k'}$,
in our setting any $k$-mixture with redundant components.
For instance an NLP under $\mathcal{M}_2$ must assign $p(\bvartheta_2 \mid \mathcal{M}_2)=0$
whenever $p(\by \mid \bvartheta_2,\mathcal{M}_2)$ reduces to a one-component mixture,
{\it e.g.} $\btheta_1=\btheta_2$ or $\eta_1 \in \{0,1\}$.
That is one must penalize situations where two components have the same parameters
(as in a repulsive mixture) and also when there are zero-weight components.
This intuition is made precise in Section \ref{sec:nlps}
for the wide class of generically identifiable mixtures.

Beyond their philosophical appeal in separating probabilistically the models
under consideration, \cite{david1} showed that
for asymptotically Normal models NLPs penalize spurious parameters
at a faster rate than (\ref{eq:asymp_marglhood_regular}).
\cite{david2} found that NLPs are necessary and sufficient to achieve posterior consistency
$P(\mathcal{M}_{k^*} \mid \by) \stackrel{P}{\longrightarrow} 1$ in certain high-dimensional linear regression
with $o(n)$ predictors, whereas \cite{shin:2015} showed a similar result with $o(e^n)$ predictors.
These authors also observed model selection gains relative to popular penalized likelihood methods.

Here we investigate
theoretical, computational and practical issues
to enable the use of NLPs in mixtures.
In Section \ref{sec:nlps} we formulate a general NLP class,
show how it leads to stronger parsimony than LPs,
and propose a particular choice leading to tractable expressions.
Importantly we consider a natural elicitation for prior parameters, a key issue
that defines what separation between components is deemed practically relevant.
Section \ref{sec:computation} outlines computational schemes for
model selection and parameter estimation,
including a novel ECP estimator of interest both for local and non-local priors.
In Section \ref{sec:results} we illustrate the performance of the BIC, AIC,
sBIC, overfitted mixtures, repulsive overfitted mixtures,
LPs and NLPs in synthetic and real examples.
Conclusions are presented in Section \ref{sec:conclusions}.
All proofs and further results are in the Supplementary material.
Our methodology is implemented in R packages mombf and NLPmix available at CRAN and
\url{www.warwick.ac.uk/go/msteel/steel_homepage/software}.

\section{Prior formulation and parsimony properties}
\label{sec:nlps}

A NLP under $\Mk$ assigns vanishing density to any $\bvartheta_k$
such that (\ref{likelihood1}) is equivalent to a mixture with $k'<k$ components.
A necessary condition is to avoid vanishing ($\eta_j=0$) and overlapping components ($\btheta_i=\btheta_j$)
but for this to also be a sufficient condition we need {\it generic identifiability}.
Definition \ref{genericident} is adapted from \cite{leroux}.

\begin{define}
\label{genericident}
Let
$p(\by \mid \bvartheta_k, \Mk)= \sum_{j=1}^{k} \eta_{j}p(\by
  \mid \btheta_j)$ and
$p(\by \mid \tbvartheta_{\tilde{k}}, \mathcal{M}_{\tilde{k}})=
\sum_{j=1}^{\tilde{k}} \tilde{\eta}_{j}p(\by \mid \tbtheta_j)$
be two mixtures as in (\ref{likelihood1}).
Assume that $\eta_j>0, \tilde{\eta}_j>0$ for all $j$
and that $\btheta_j \neq \btheta_{j'}$, $\tbtheta_j \neq \tbtheta_{j'}$
for all $j \neq j'$. The class $p(\by \mid \btheta)$
defines a generically identifiable mixture if
$p(\by \mid \bvartheta_k,\Mk)=p(\by \mid \tbvartheta_{\tilde{k}},\mathcal{M}_{\tilde{k}})$
for almost every $\by$ implies that $k=\tilde{k}$ and
$\bvartheta_k=\tilde{\bvartheta}_{\Psi(\tilde{k})}$ for some permutation
$\Psi(\tilde{k})$ of the component labels in $\mathcal{M}_{\tilde{k}}$.
\end{define}

That is, assuming that all components have non-zero weights and distinct parameters
the mixture is uniquely identified by its parameters up to label permutations.
\cite{teicher3} showed that mixtures of univariate Normal, Exponential and Gamma distributions
are generically identifiable. \cite{Yakowitz}
extended the result to several multivariate
distributions, including the Normal case.
See also \cite{grun2008finite} for a study of generic identifiability for mixtures of GLMs and \cite{allman:2009} for multivariate Bernoulli mixtures, finite and infinite product Binomial mixtures, hidden Markov Models and random graph mixture models. In particular Binomial mixtures are generically identifiable if and only if the number of Binomial trials $L\geq 2k -1$ (\cite{grun2008finite}), and product Binomial mixtures with $p\geq3$ are generically identifiable when $L$ is above a small threshold (\cite{allman:2009}, Theorem 4),
e.g. when the number of trials $L_{if}=L$ for all $(i,f)$ then it suffices that $3 L^{p/3} > 2(k+1)$.
Throughout we assume $p(\by \mid \bvartheta_k,\Mk)$ to be generically identifiable.
Then $p(\bvartheta_k \mid \Mk)$ defines a NLP
if and only if $\lim\; p(\bvartheta_k \mid \Mk) = 0$ as either
(i) $\eta_{j} \rightarrow 0$ for any $j=1,...,k$
or (ii) $\btheta_{i} \rightarrow \btheta_{j}$ for any $i \neq j$.
Let $d_\vartheta(\bvartheta_k)$ be a continuous penalty function
converging to 0 under (i) or (ii),
then a general NLP class is defined by
\begin{equation}\label{penalties}
p(\bvartheta_{k}\mid \Mk)= d_{\vartheta}(\bvartheta_k)
p^L(\bvartheta_{k}\mid \Mk),
\end{equation}
where $p^L(\bvartheta_{k}\mid \Mk)$ is an arbitrary LP with the restriction that
$p(\bvartheta_{k}\mid \Mk)$ is proper.
We consider
$p^L(\bvartheta_{k}\mid \Mk)=p^L(\btheta \mid\Mk) p^L(\bm{\eta} \mid \Mk)$
and $d_{\vartheta}(\bvartheta_k)= d_{\theta}(\btheta) d_{\eta}(\bm{\eta})$, where
\begin{equation}\label{penalties2}
d_{\theta}(\btheta)=
\frac{1}{C_{k}}\left(\prod_{1\leq i < j \leq k}d(\btheta_{i},\btheta_{j})\right),
\end{equation}
is a repulsive force between components akin to \cite{Dunson},
$C_{k}=\int p^{L}(\btheta\mid\Mk) \prod_{1\leq i < j \leq k}d(\btheta_{i},\btheta_{j})
d\btheta$
a prior normalization constant
and $d_{\eta}(\bm{\eta}) \propto \prod_{i=1}^{k} \eta_j^r$ with $r>0$.
Evaluating $C_k$ may require numerical approximations (e.g. Monte Carlo)
but below we give closed expressions for specific $d_{\btheta}(\btheta)$ and $p^L(\btheta \mid \Mk)$.
Regarding the weights, we set the symmetric Dirichlet
$p(\bm{\eta} \mid \Mk)= \mbox{Dir}(\bm{\eta}; q) \propto d_{\eta}(\bm{\eta}) \mbox{Dir}(\bm{\eta}; q-r)$,
where importantly one must set $q>1$ to satisfy (i) above and $r \in [q-1,q)$.
Summarising, we set
\begin{align}
p(\bvartheta_k \mid \Mk)= d_{\theta}(\btheta) p^L(\btheta \mid \Mk) \mbox{Dir}(\bm{\eta}; q),
\label{eq:nlp_dirichlet}
\end{align}
where $q>1$ and $d_{\theta}(\btheta)$ is as in \eqref{penalties2}.

The specific form of $d(\btheta_i,\btheta_j)$ depends on the model under consideration.
For instance consider $\btheta_i=(\bmu_i,\Sigma_i)$ for a location parameter $\bmu_i$
and scale matrix $\Sigma_i$. Then one may adapt earlier proposals for variable selection
and define MOM penalties \citep{david1} $d(\btheta_i,\btheta_j)=(\bmu_i-\bmu_j)^{'} A^{-1}(\bmu_i-\bmu_j)/g$
where $A$ is a symmetric positive-definite matrix and $g$ is a prior dispersion parameter, or alternatively eMOM penalties \citep{david4}
$d(\btheta_{i},\btheta_j)=\exp\{-g/(\bmu_{i}-\bmu_j)^{'} A^{-1}(\bmu_{i}-\bmu_j)\}$, also adopted by \cite{Dunson} for repulsive mixtures.
Note that $C_{k}$ is guaranteed to be finite for eMOM penalties as $d(\btheta_i,\btheta_j) \leq 1$.
The main difference between MOM and eMOM is that the
latter induce a stronger model separation that give faster sparsity rates.
However,  empirical results in \cite{david1,david2} and \cite{david3} suggest that by setting $g$
adequately both MOM and eMOM are often equally satisfactory.
We now offer theoretical results for both penalties,
but in our implementations we focus on the MOM for the practical
reasons that $C_k$ has closed form and leads to simple prior elicitation.
Both MOM and eMOM remain applicable when $\btheta_i$ is a vector of probabilities,
as we illustrate for Binomial and product Binomial mixtures.
More generally $d(\btheta_i,\btheta_j)$ can be based on any distance or divergence between
probability measures, see Section \ref{ssec:mom}.
We defer discussion of prior elicitation to Section \ref{ssec:priorelicitation}.

\subsection{Parsimony enforcement}
\label{ssec:sparsity}

We show that NLPs induce extra parsimony via the penalty term $d_\vartheta(\bvartheta_k)$,
which  specifically affects overfitted mixtures.
We first lay out technical conditions for the result to hold.
Recall that $k^*$ is the true number of components and
$\bvartheta_{k^*}^*$
the true parameter value.
Let $p_k^*(\by)$ be the density
minimising Kullback-Leibler (KL) divergence between the data-generating
$p(\by \mid \bvartheta_{k^*}^*,\mathcal{M}_{k^*})$
and the class $\{p(\by \mid \bvartheta_k,\Mk), \bvartheta_k \in \Theta_k\}$.
When $k \leq k^*$ for generically identifiable mixtures $p_k^*(\by)$
is defined by a unique parameter $\bvartheta_k^* \in \Theta_k$ (up to label permutations).
When $k>k^*$ there are multiple minimizers giving
$p_k^*(\by)=p(\by \mid \bvartheta_{k^*}^*,\mathcal{M}_{k^*})$.
$p^L(\bvartheta_k \mid\Mk)$ denotes a LP
and $p(\bvartheta_k \mid \Mk)$ a NLP as in \eqref{penalties}.
$P^L(\cdot \mid \by,\Mk)$ and
$E^L(\cdot \mid \by,\Mk)$ are the posterior probability and
expectation under $p^L(\bvartheta_k \mid \by, \Mk)$.

\vspace{2mm}
\noindent
{\bf NLP parsimony conditions}

\begin{enumerate}[label=\bfseries B\arabic*,leftmargin=*]
\item {\it $L_1$ consistency.}
For all fixed $\epsilon>0$ as $n \rightarrow \infty$
$$
P^L \left( \int \left| p(\bz \mid \bvartheta_k,\Mk) - p_k^*(\bz) \right| d\bz > \epsilon
  \mid \by, \Mk \right) \rightarrow 0
$$
in probability with respect to $p(\by \mid
\bvartheta_{k^*}^*,\mathcal{M}_{k^*})$.

\item {\it Continuity.} $p(\by \mid \bvartheta_k,\Mk)$ is a continuous function in
  $\bvartheta_k$.

\item {\it Penalty boundedness.}
There is a constant $c_k$ such that $d_{\vartheta}(\bvartheta_k)
\leq c_k$ for all $\bvartheta_k$.
Alternatively, if $p(\bvartheta_k \mid \Mk)$ involves the MOM-IW prior \eqref{priorN2NL} and $k>k^*$ then there exist finite $\epsilon,U>0$ such that
$$\mathop {\lim }\limits_{n \to \infty} P\left(E^L \left[ \exp \left\{ \frac{1}{2g} \sum_{j=1}^{k} \bmu_j'A^{-1}\bmu_j \frac{\epsilon}{1+\epsilon} \right\} \mid \by,\Mk \right] < U\right)=1.$$
\end{enumerate}

Condition B1 amounts to posterior $L_1$ consistency of $p(\by \mid \bvartheta_k,\Mk)$ to the data-generating truth when
$k \geq k^*$ and to the KL-optimal density when $k < k^*$.
Note that B1 is assumed under the underlying local $p^L$ and hence follows from standard theory. Specifically,
B1 is a milder version of Condition A1 in \cite{judith} where
rather than fixed $\epsilon$ one has $\epsilon= \sqrt{\log n}/\sqrt{n}$.
See the discussion therein and \cite{van} for results on finite Normal mixtures,
\cite{rousseau:2007} for Beta mixtures and
\cite{ghosal:2007} for infinite Normal mixtures.
For strictly positive $p^L(\bvartheta_k \mid \Mk)>0$
Condition B1 is intimately connected to MLE consistency \citep{ghosal:2002},
proven for fairly general mixtures by \cite{redner} for $k \leq k^*$ and by \cite{leroux} for $k>k^*$.
The $L_1$ consistency results above focus on the case where the data-generating truth lies in the assumed family,
but see \cite{ramamoorthi2015posterior} (Theorem 2)
for posterior concentration results under model misspecification for independent and identically distributed data.
Condition B2 holds when the kernel
$p(\by \mid \btheta)$ is continuous in $\btheta$,
as in the vast majority of common models.
B3 is trivially satisfied when NLPs are defined using bounded penalties ({\it e.g.} eMOM
or MOM-Beta priors in Section \ref{ssec:mom}).
For the MOM-IW (Section \ref{ssec:mom}) we require the technical condition that the posterior exponential moment in B3
is bounded in probability when $k>k^*$.
To gain intuition, B3 requires that under the posterior distribution $p^L(\bmu\mid\Mk,\by)$ none of the elements in $\bmu$
diverges to infinity, and in particular is satisfied if $\bmu$ is restricted to a compact support.

Theorem 1 below states that $d_\vartheta(\bvartheta_k)$ imposes a complexity penalty
concentrating on 0 when $k>k^*$ and on a constant when $k \leq k^*$.
Part (i) applies to any model, Part (ii) only requires B1-B3
and Part (iii) holds under the mild conditions A1-A4 in \cite{judith} (Supplementary Section \ref{supplsec:cond_judith}),
hence the result applies to an ample class of mixtures.
The proof of Part(iii) only requires posterior contraction of the sum of redundant weights at a $n^{-1/2}$ rate,
and can be trivially adjusted when this rate is slower.
\cite{judith} showed that the $n^{-1/2}$ rate is achieved under Conditions A1-A3 and a strong identifiability condition A4.
Interestingly, \cite{ho:2016} showed that strong identifiability can be expressed in terms of partial differential
equations involving the kernel $p(\by \mid \btheta)$ and its first and second derivatives.
In particular location-scale Gaussian and Gamma mixtures are not strongly identifiable
for certain problematic $\bvartheta_k$.
When the data-generating $\bvartheta_k^*$ is one of those problematic values then the MLE of the component parameters $\hat{\btheta}$
is slower than $n^{-1/2}$,
however remarkably the MLE of the mixing weights $\hat{\bm{\eta}}$ does still contract at the $n^{-1/2}$ rate required by Part(iii).

\begin{defin2}

Let $p(\by \mid \bvartheta_k,\Mk)$ be a generically identifiable mixture,
$p(\by\mid\Mk)$ and $p^L(\by \mid \Mk)$ the integrated likelihoods under
$p(\bvartheta_k \mid \Mk)$ and $p^L(\bvartheta_k \mid\Mk)$.
Then
\begin{enumerate}[label=(\roman*),leftmargin=*]
\item $p(\by\mid\Mk)=p^{L}(\by\mid\Mk)
E^{L}\left(d_{\vartheta}(\bvartheta_k)\mid\by\right),
$
where $$E^{L}\left(d_{\vartheta}(\bvartheta_k \mid \by)\right)=
\int d_{\vartheta}(\bvartheta_k)
p^{L}(\bvartheta_{k}\mid\by,\Mk)
d\bvartheta_{k}.$$

\item If B1-B2 are satisfied then as $n \rightarrow \infty$
$$P^L\left( |d_{\vartheta}(\bvartheta_k) - d_k^*| > \epsilon \mid
  \by,\Mk \right) \rightarrow 0$$
where $d_k^*=0$ for $k>k^*$ and $d_k^*=d_{\vartheta}(\bvartheta_k^*)$
for $k \leq k^*$.

If B3 also holds then
$E^{L}\left(d_{\vartheta}(\bvartheta_k)\mid\by\right) \xrightarrow{P} d_k^*$.

\item Let $k>k^*$ and $p(\bvartheta_k \mid \Mk) \propto d_{\theta}(\btheta) p^L(\btheta \mid \Mk) \mbox{Dir}(\bm{\eta};q)$,
where $q>1$.
If B3 and A1-A4 in \cite{judith} hold for $p^L(\btheta \mid \Mk)$ then
for all $\epsilon>0$ and all $\delta \in (0,\mbox{dim}(\Theta)/2)$ there exists a finite $\tilde{c}_k>0$ such that
$$
P^L \left( d_{\vartheta}(\bvartheta_k) > \tilde{c}_k n^{- \frac{k-k^*}{2}(q-\delta) + \epsilon} \mid \by,\Mk \right) \rightarrow 0
$$
in probability as $n \rightarrow \infty$.
\end{enumerate}

\label{thm:sparsity}
\end{defin2}

Part (i) extends Theorem 1 in \cite{david3} to mixtures and
shows that $p(\by\mid\Mk)$ differs from $p^L(\by\mid\Mk)$ by
a term $E^{L}\left(d_{\vartheta}(\bvartheta_k)\mid\by\right)$ that intuitively should converge to 0 for overfitted models.
Part (i) also eases computation as $E^L(d_{\vartheta}(\bvartheta_k)\mid\by)$ can be estimated from standard MCMC output from $p^L(\bvartheta_k\mid\by,\Mk)$,
as we exploit in Section \ref{sec:computation}.
Part (ii) confirms that the posterior of $d_{\vartheta}(\bvartheta_k)$
under $p^L(\bvartheta_k\mid\by,\Mk)$
concentrates around 0 for overfitted models and a finite constant otherwise,
and that its expectation also converges.
Part (iii) states that for overfitted models this concentration rate
is essentially $n^{-(k-k^*)q/2}$, leading to an accelerated sparsity-inducing Bayes factor
$B_{k,k^*}(\by)= E^L(O_p(n^{-(k-k^*)q/2})) B_{k,k^*}^L(\by)$.
Recall that the LP-based $B_{k,k^*}^L(\by)= O_p(n^{-(\lambda - p_{k^*}/2)})$ for some $\lambda \in [p_{k^*}/2,p_k/2]$
under the conditions in \cite{watanabe:2013}.
For instance, one might set $q$ such that $(k-k^*)q/2=\lambda - p_{k^*}/2$ so that
$B_{k,k^*}(\by)$ converges to 0 at twice the rate for $B_{k,k^*}^L(\by)$.
As $\lambda$ is unknown in general one could conservatively take its upper bound
$\lambda=p_{k}/2$, then $q=(p_{k} - p_{k^*})/(k-k^*)$
is the number of parameters per component.
See Section \ref{ssec:priorelicitation} for further discussion on prior elicitation.

\subsection{Choice of penalty function}
\label{ssec:mom}

Although our theory holds for fairly general $d(\btheta_i,\btheta_j)$ in \eqref{penalties2},
we now propose choices that simplify interpretation and obtaining $C_k$.
Consider first the case where $\btheta_i=(\bmu_i,\Sigma_i)$,
$\bmu_i$ is a location parameter and $\Sigma_i$ a positive-definite matrix, as in Normal or T mixtures.
Then in (\ref{eq:nlp_dirichlet}) we may set the MOM-Inverse Wishart (MOM-IW) prior
$p(\btheta \mid \Mk)=$
\begin{align}\label{priorN2NL}
 d_{\theta}(\btheta) p^L(\btheta \mid \Mk)=\frac{1}{C_k} \prod_{1\leq i < j \leq k} \frac{(\bmu_{i}-\bmu_{j})^{'}A_{\Sigma}^{-1}(\bmu_{i}-\bmu_{j})}{g}
\prod_{j=1}^{k}N\left(\bmu_{j} \mid \boldsymbol{0},gA_{\Sigma} \right)
\text{IW}(\Sigma_j \mid \nu,S),
\end{align}
where $A_{\Sigma}^{-1}$ is a symmetric positive-definite matrix and $(g,\nu,S)$ are fixed prior hyperparameters.
A trivial choice is $A_{\Sigma}^{-1}=I$ but it has the inconvenience
of not being invariant to changes in scale of $\by$.
Instead we use
$A_{\Sigma}^{-1}=\frac{1}{k}\sum_{j=1}^{k}\Sigma_{j}^{-1}$, which is symmetric, positive-definite
and is related to the $L_2$ distance between Normal distributions.
In the particular case where $\Sigma_1=\ldots=\Sigma_k=\Sigma$, a parsimonious model sometimes considered to borrow
information across components, clearly $A_{\Sigma}=\Sigma$.
In our model-fitting algorithms and examples we consider both the equal and unequal covariance cases.
We remark that for unequal covariances the NLP in \eqref{priorN2NL}
penalizes $\bmu_i=\bmu_j$ even when $\Sigma_i \neq \Sigma_j$.
We do not view this as problematic, given that in most applications the interest is to identify
components with well-separated locations.
However, if one is interested in detecting components that differ only in $\Sigma_i \neq \Sigma_j$
then $d(\btheta_i,\btheta_j)$ should be adjusted,
e.g. $d(\btheta_i,\btheta_j)$ could be any measure of distance or divergence between probability distributions.
As illustration, consider the squared Hellinger distance between Normal distributions
\begin{align}\label{penaltyr2}
d_{\theta}(\btheta)= \frac{1}{C_k}
\prod_{1\leq i < j \leq k} 1 - \frac{ \det (\Sigma_i)^{1/4} \det (\Sigma_j) ^{1/4}} { \det \left( (\Sigma_i + \Sigma_j)/2\right)^{1/2} }
              \exp\left\{-\frac{1}{8}\frac{(\bmu_{i}-\bmu_{j})^{'}2(\Sigma_i + \Sigma_j)^{-1}(\bmu_{i}-\bmu_{j})}{g},
              \right\}.
\end{align}
For this choice $d_{\theta}(\btheta)=0$ if and only if $\bmu_{i}=\bmu_{j}$ and $\Sigma_i=\Sigma_j$.

We now consider binary data, specifically for product Binomial mixtures
(Binomial mixtures are the particular case where $p=1$). We define the MOM-Beta prior
\begin{align}\label{mombetaprior}
p(\btheta \mid \Mk)= \frac{1}{C_k} \prod_{1\leq i < j \leq k} (\btheta_i-\btheta_j)'(\btheta_i-\btheta_j)
\prod_{j=1}^{k}\prod_{f=1}^{p}\text{Beta}(\theta_{jf}; ag,(1-a)g),
\end{align}
where $\theta_{jf}>0$ is the success probability for outcome $f$ in component $j$ and $a>0$, $g>0$ are known prior parameters.
In our parameterization $a>0$ is the prior mean and $g>0$ the prior sample size for the underlying Beta prior.
In \eqref{mombetaprior} $g$ determines the prior separation in the binomial success probabilities across components
and the prior informativeness.
As discussed in Section \ref{ssec:priorelicitation} large $g$ leads to informative priors with little separation across components,
and there is a range of $g$ values that can be interpreted as being minimally informative in a fairly robust manner across $k$.
See also \cite{consonni:2013} for strategies to set MOM prior parameters when comparing Binomial probabilities
and \cite{collazo:2016} for their use in Chain Event Graphs.

An issue in \eqref{priorN2NL} and \eqref{mombetaprior} is the computation of the normalising constant $C_k$,
a non-trivial expectation of a product of quadratic forms.
Lemma \ref{lemma:generic_momprior_constant} (supplementary material) gives a recursive formula for $C_k$ for any prior with the generic form
\begin{align}\label{priorgeneral}
p(\bzeta \mid \Mk)= \frac{1}{C_k} \prod_{1\leq i < j \leq k} (\bzeta_i-\bzeta_j)'(\bzeta_i-\bzeta_j) \prod_{j=1}^{k} \prod_{f=1}^{p} p^L(\zeta_{jf})
\end{align}
where $\bzeta=(\bzeta_1,\ldots,\bzeta_k) \in \mathbb{R}^{pk}$.
Note that \eqref{priorN2NL} is the particular case where $\bzeta_i= (gA_{\Sigma})^{-1/2} \bmu_i$ and that \eqref{mombetaprior} corresponds to $\bzeta_i= \btheta_i$.
An interesting alternative to \eqref{mombetaprior} suggested by a referee is to consider a MOM-Normal prior on the
Binomial logit-probabilities, which can be achieved by setting $\zeta_{jf}=\log(\theta_{jf}/(1-\theta_{jf}))$.
We focus on the MOM-Beta for its simplicity and easy prior elicitation (Section \ref{ssec:priorelicitation}),
but we note that a logit parameterization would be particularly natural in settings where one wishes to regress $\theta_{jf}$ on covariates.
When $p^L$ is a Normal prior Lemma \ref{lemma:generic_momprior_constant} can be simplified,
see Corollary \ref{corollary1}.
Further simplifications are possible when $p=1$ or $k=2$,
these are given for Normal and product Binomial mixtures in Corollaries \ref{corollary2} and \ref{corollary3} respectively.

\begin{defin1}{MOM-IW, general $(p,k)$.}
\label{corollary1}
The normalization constant in (\ref{priorN2NL}) is
\begin{align}\label{formB1}
C_k=\frac{1}{s!}\sum_{\upsilon_{(1,2)}=0}^{1}...
\sum_{\upsilon_{(k-1,k)}=0}^{1}(-1)^{\sum\limits_{i, j}^{s}\upsilon_{(i,j)}}
\mathcal{Q}_{s}(B_{\upsilon}),
\end{align}
where $v_{(i,j)}\in \{0,1\}$, $s=\binom {k} {2}$,
$\mathcal{Q}_{s}(B_{\upsilon})=s!2^{s}d_{s}(B_{\upsilon})$,
$d_{s}(B_{\upsilon})=\frac{1}{2s}\sum_{i=1}^{s}tr(B_{\upsilon}^{i})d_{s-i}(B_{\upsilon})$,
$d_{0}(B_{\upsilon})=1$
 and $B_{\upsilon}$ is a $pk\times pk$ matrix with element $(l,m)$ given by
\begin{equation*}
\left\{
      \begin{array}{ll}
      b_{ll}=\dfrac{1}{2}(k-1) - \sum_{i<j}\upsilon_{(i,j)}, \;\;\; l=1+p(i-1),\ldots,pi \\
      b_{lm}=b_{ml}=-\dfrac{1}{2} + \sum_{i<j}\upsilon_{(i,j)}, \;\;\; (l,m)=(1+p(i-1),1+p(j-1)),\ldots,(pi,pj)\\
      \end{array}
\right.
\end{equation*}
where $i \neq j$, $i=1,\ldots,k$, $j=1,\ldots,k$ and $b_{lm}=0$ otherwise.
\end{defin1}

\begin{defin1}{MOM-IW, univariate or two-component mixtures.}
\label{corollary2}
Let $C_k$ be as in (\ref{priorN2NL})
\begin{enumerate}[label=(\roman*)]
\item If $p=1$, then $C_k=\prod_{j=1}^{k}\Gamma(j+1)$.

\item If $k=2$, then $C_k=2p$.
\end{enumerate}
\end{defin1}

\begin{defin1}{MOM-Beta, univariate or two-component mixtures.}
\label{corollary3} Let $C_k$ be as in (\ref{mombetaprior})
\begin{enumerate}[label=(\roman*)]
\item If $p=1$, then
  $${C_k}=\left(\dfrac{\Gamma(g)}{\Gamma(ag)\Gamma((1-a)g)}\right)^{k}\prod_{j=1}^{k}\dfrac{\Gamma(ag+k-j)\Gamma((1-a)g+k-j)\Gamma(j+1)}{\Gamma(g+2k-j-1)}.$$
\item If $k=2$, then ${C_k}=2p a (1-a) / (g+1)$.
\end{enumerate}
\end{defin1}

Despite having closed-form $C_k$ its evaluation for general $(p,k$) can be cumbersome,
e.g. $S_k$ in Lemma \ref{lemma:generic_momprior_constant} is the set of partitions of $k(k-1)/2$ and has size exponential in $k$
\citep{andrews1998theory}.
The sum in \eqref{formB1} is simpler but contains $k(k-1)/2$ terms, still prohibitive for large $k$.
A practical option for large $k$ is to evaluate $C_k$ via Monte Carlo as the prior mean of $d_k(\btheta)$ under $p^L$
and tabulate it upfront, prior to data analysis.
This is particularly convenient in Corollary \ref{corollary1} where $C_k$ does not depend on the prior parameter $g$.
To facilitate implementation Tables \ref{tab:MCnormal}-\ref{tab:MCbeta} provide $C_k$
for \eqref{priorN2NL} and \eqref{mombetaprior} (respectively) and various $(p,k)$.
$C_k$ is also implemented in the R package mombf, function \texttt{bfnormmix}.

\subsection{Prior elicitation}
\label{ssec:priorelicitation}

A critical aspect in a NLP is its induced separation between components,
driven by $g$ and $q$ in (\ref{eq:nlp_dirichlet}).
We propose defaults that can be used in the absence of a priori knowledge,
whenever the latter is available we naturally recommend to include it in the prior.
To facilitate use these defaults are included in the R package mombf.

\begin{figure}[ht]
\begin{center}
\begin{tabular}{ccc}
\includegraphics[width=0.55\textwidth]{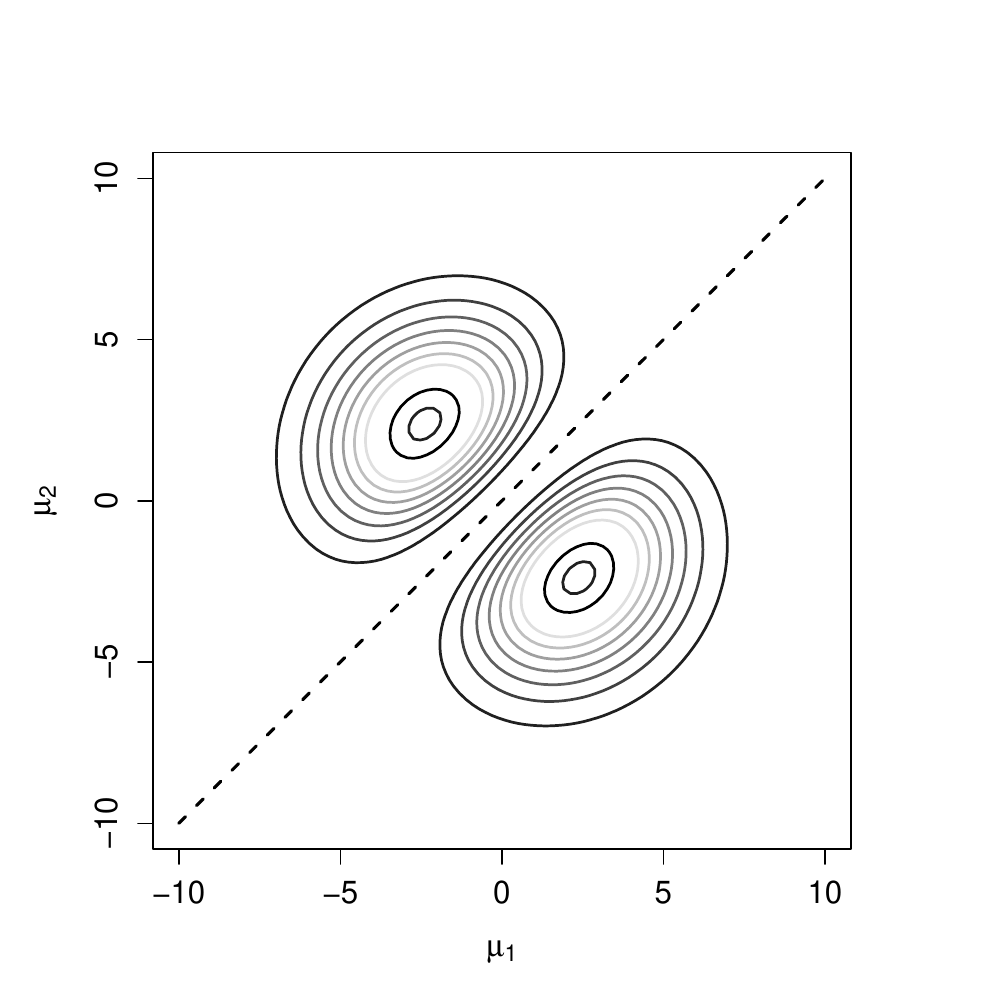} & \hspace{-1.5cm}
\includegraphics[width=0.55\textwidth]{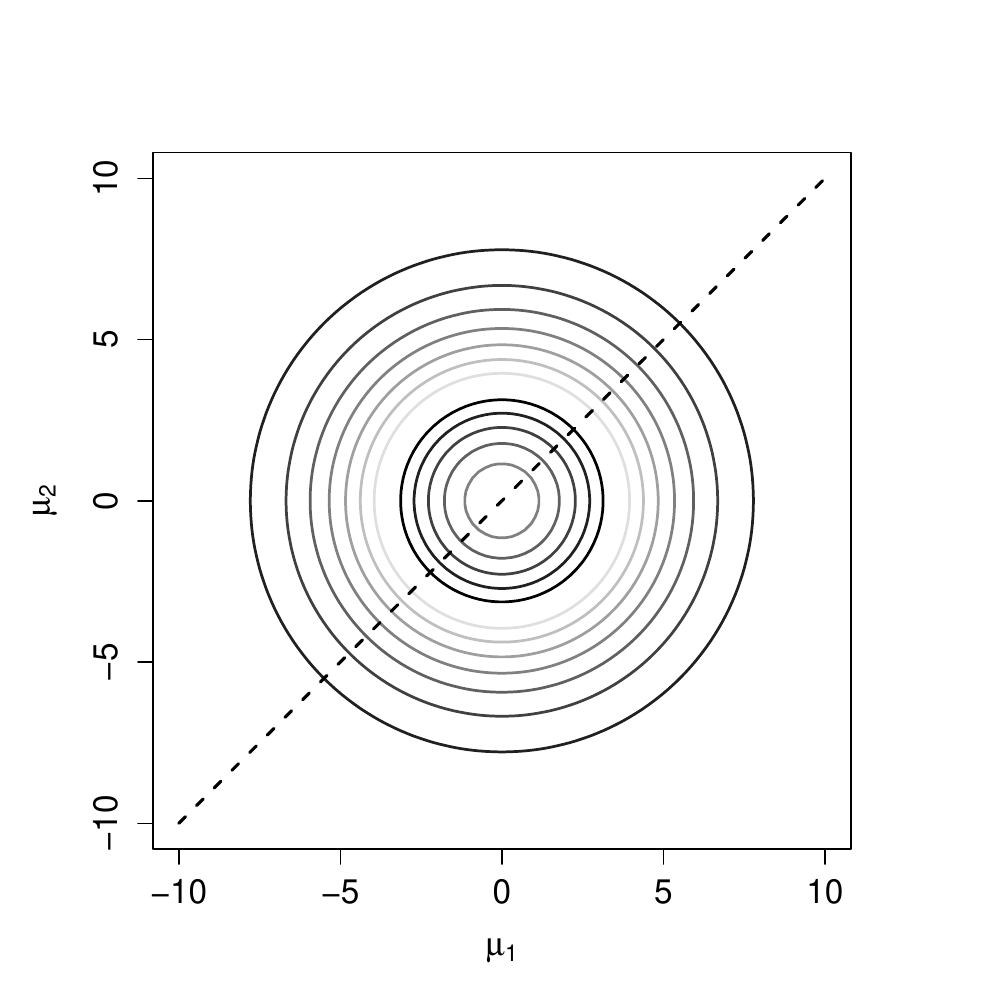} \\
\end{tabular}
\end{center}
\caption{Default MOM-IW $p(\mu_{1},\mu_{2}\mid\sigma^{2}=1,g=5.68,\mathcal{M}_{2})$ (left)
and Normal-IW $p^{L}(\mu_{1},\mu_{2}\mid\sigma^{2}=1,g^{L}=11.56,\mathcal{M}_{2})$ (right)}
\label{fig:contours}
\end{figure}

We start by discussing $g$, first for Normal 
and T mixtures and subsequently for Binomial and product Binomial mixtures.
The main idea for Normal and T mixtures is that we wish to find clearly-separated components,
so we can interpret the data-generating process in terms of distinct sub-populations.
We thus set $g$ such that there is small prior probability
that any two components are poorly-separated, i.e.~give rise to a unimodal density.
In Normal mixtures the number of modes depends on non-trivial parameter combinations \citep{Ray},
but when $\eta_1=\eta_2=0.5$ and $\Sigma_1=\Sigma_2$
the mixture is bimodal if and only if $\kappa=(\bmu_1-\bmu_2)^{'}\Sigma^{-1}(\bmu_1-\bmu_2)>4$.
Thus we set $g$ such that
$P(\kappa<4\mid \mathcal{M}_2)=0.05$.
This is trivial, the prior on $\kappa$ implied by \eqref{priorN2NL} is
$p(\kappa\mid \mathcal{M}_2)=\text{Gamma}(\kappa; p/2+1,1/(4g))$.
For instance in a univariate Normal mixture $g=5.68$,
Figure \ref{fig:contours} (left) portrays the associated prior.
For comparison the right panel shows a Normal prior with $g^L=11.56$,
which also assigns $P^L(\kappa < 4 \mid \mathcal{M}_2)=0.05$.
To assess sensitivity we considered $g$ such that $P(\kappa<4\mid \mathcal{M}_2)=0.1$,
finding that $P(\kappa<4 \mid \mathcal{M}_2)=0.05$ is slightly preferable for
balancing parsimony vs. sensitivity (Supplementary Section \ref{supplsec:results}).

For T mixtures \cite{dovsla2009conditions} showed that a univariate mixture with two components
and equal degrees of freedom $\upsilon$
is bimodal if $\kappa>4\upsilon/(\upsilon+2)$.
For multivariate T mixtures,
again with $\eta_1=\eta_2=0.5$ and $\Sigma_1=\Sigma_2$,
it is easy to develop the arguments in \cite{Ray} (Theorem 1 and Remark 4)
to show that the mixture is bimodal if and only if $\kappa>4\upsilon/(\upsilon+p+1)$.
This matches the result from \cite{dovsla2009conditions} for $p=1$
and for Normal mixtures in \cite{Ray} as $\upsilon \rightarrow \infty$.
Summarising, we set $g$ such that $P(\kappa< 4\upsilon/(\upsilon+p+1)\mid \upsilon, \Mk)=0.05$,
where we recall that $p(\kappa \mid \upsilon,\mathcal{M}_2)=\text{Gamma}(\kappa; p/2+1,1/(4g))$.

Consider now the MOM-Beta prior (\ref{mombetaprior}). In contrast to continuous mixtures here one cannot use multi-modality to set the prior
parameters $(a,g)$.
Instead we set $(a,g)$ such that the amount of prior information (measured by the variance) is comparable to that in
\begin{align}
  p^L(\btheta \mid \Mk)= \prod_{j=1}^k \prod_{f=1}^{p} \mbox{Beta}(\theta_{jf}; g_L a_L,g_L (1-a_L)),
\label{eq:betaprior}
\end{align}
where $a_L=0.5$ and $g_L=2$ are typically viewed as minimally informative.
Specifically we recommend $a=0.5$ and $g$ as listed in Table \ref{tab:defaultg_mombeta} for $p \in [1,20]$, and $g=2$ for $p>20$.
We briefly outline the reasoning, further details are in Supplementary Section \ref{sec:priorvar_mombeta}.
Simple algebra shows that the variance under $p^L$ is
$$\mbox{Var}_{p^L}(\theta_{j f} - \theta_{j' f} \mid \Mk)= 2 \left[\frac{a_L(g_L a_L +1)}{g_L+1} - a_L^2 \right],$$
and for $(a_L,g_L)=(0.5,2)$ this variance is 1/6.
We seek $g$ such that the variance under the MOM-Beta prior
$\mbox{Var}_{p}(\theta_{j f} - \theta_{j' f} \mid \Mk)=1/6$.
Although such $g$ depends on $(k,p)$
the dependence on $k$ is mild (in fact for large $k$ the variance grows less sensitive to $g$, Figure \ref{supfig:priorsd_mombeta})
and one can focus on the $k=2$ case.
See Supplementary Section \ref{sec:priorvar_mombeta} for the variance under general $(a,g)$ and $k=2$.
Interestingly as $p$ grows one may simply set $(a,g)=(0.5,2)$, since then
\begin{align}
  \mbox{Var}_p(\theta_{jf} - \theta_{j'f} \mid \mathcal{M}_2)=
  \frac{1}{p} \left( \frac{2}{5} + \frac{p-1}{6} \right),
\nonumber
\end{align}
which converges to $1/6$ as $p \rightarrow \infty$, thus for large $p$ one may simply set $g=2$.
Figure \ref{supfig:contours2} displays the default MOM-Beta and Beta(1,1) priors for $k=2$.
See also Section \ref{supplsec:sensMOMBeta} for an illustration of the sensitivity of results to various $g$ in an application.

Regarding $q$, as discussed earlier $q>1$ is required for (\ref{eq:nlp_dirichlet}) to define a NLP.
One option is to set $q=3$ so that $p(\bm{\eta} \mid \Mk) \propto \prod_{j=1}^{k} \eta_j^2$ induces
a quadratic penalty comparable to the MOM prior on $\bmu$ given in \eqref{priorN2NL}.
Alternatively from the discussion after Proposition \ref{thm:sparsity}
setting $q=(p_{k} - p_{k^*})/(k-k^*)$, the number of parameters per component,
seeks to (at least) double the Bayes factor sparsity rate of the underlying LP.
For instance, for Normal mixtures with common covariances this leads to $q=p+1$,
and under unequal covariances to $q=p + 0.5p(p+1) +1$.
These are the values we used in our examples with $p=1$ or $p=2$ (Section \ref{sec:results}),
but we remark that for larger $p$ such $q$ may lead to an overly informative prior on $\bm{\eta}$.
In our experience $q \in [2,4]$ (Supplementary Section \ref{supplsec:results})
gives fairly robust results and satisfactory sparsity, thus larger values do not seem warranted.

The prior distribution on the remaining parameters,
which may be thought of as nuisance parameters, will typically reduce to a standard form for which defaults
are available.
For example, for location-scale mixtures we set
$p(\Sigma_1,\ldots,\Sigma_k \mid \Mk)= \prod_{j=1}^{k} \mbox{IW}(\Sigma_j;\nu,S)$.
We follow the recommendation in \cite{Hathaway2}
that eigenvalues of $\Sigma_i \Sigma_j^{-1}$ for any $i \neq j$ should be bounded away from 0 to prevent
the posterior from becoming unbounded,
which is achieved if $\nu \geq p+4$ \citep[Ch.~6]{book_silvia}.
We assume that the data are standardized to have mean 0
and variance 1 and set a default $S=(p+4)^{-1}I$ and $\nu=p+4$, so that $E(\Sigma_{j}^{-1})=I$. For T mixtures, we also consider a prior the degrees of freedom $\upsilon$. We refer to \cite{mark_david} for a review of popular options.

\section{Computational algorithms}
\label{sec:computation}

Computation for mixtures is challenging, and potentially more so when embarking upon non-standard formulations such as ours.
Fortunately, Theorem 1(i) allows estimating the integrated likelihood $p(\by \mid \Mk)$
for arbitrary mixtures through direct extensions of existing algorithms.
Intuitively, one can use any algorithm to estimate a local prior integrated likelihood $\tilde{p}(\by \mid \Mk)$
and the mean of $d_\vartheta(\bvartheta_k)$ under the local posterior.
In Section \ref{ssec:comp_marglhood} we outline the main idea.
Section \ref{ssec:comp_marglhood_local} gives two algorithms to estimate $\tilde{p}(\by \mid \Mk)$ from MCMC output.
The first one was proposed by \cite{Robert1} and, while we found it to be reasonably accurate,
it is limited to conjugate models and requires an MCMC post-processing step that can have non-negligible cost.
The second algorithm is novel (to our knowledge), applicable to non-conjugate models and only requires cluster
probabilities readily available as an MCMC by-product.
This algorithm is based on a novel result connecting Bayes factors with the ratio of posterior to prior empty cluster probabilities, i.e. a type of Savage-Dickey ratio,
hence we named it the empty cluster probability (ECP) estimator.
We found that in some situations the ECP estimator can increase precision by an order of magnitude relative to that in \cite{Robert1} (Supplementary Section \ref{supplsec:comparison2}, Figures \ref{supfig:BFs}-\ref{supfig:BFs2}) using the same number of MCMC iterations. Further, as explained below the ECP estimator can have a substantially smaller per-iteration cost.
We remark that the result has interest beyond purely computational purposes,
e.g. to set thresholds on empty cluster probabilities in overfitted mixtures.
Also note that computations are easily done in parallel,
e.g. to consider multiple $k$ or process MCMC output in batches.
See Section \ref{ssec:syntheticexamples} for further discussion and empirical results on the run time required by our algorithms.
Although our main interest is to infer $k$, in Section \ref{ssec:comp_postmode}
we discuss posterior mode parameter estimates
via an Expectation-Maximimation (EM) algorithm (\cite{dempster2}).
Relative to local priors the EM algorithm only requires an extra gradient evaluation,
which typically has negligible cost.

\subsection{Approximation of $p(\by \mid \Mk)$}
\label{ssec:comp_marglhood}

Theorem 1(i) suggests the estimator
\begin{align}\label{estimator4}
\hat{p}(\by\mid \Mk)=\tilde{p}(\by\mid \Mk)\dfrac{1}{T}\sum_{t=1}^{T}
\omega(\bvartheta_{k}^{(t)}),
\end{align}
where $\omega(\bvartheta_{k})=p(\bvartheta_{k}\mid \Mk)/ \tilde{p}(\bvartheta_{k}\mid \Mk)$
and $\tilde{p}(\bvartheta_k \mid \Mk)$ is an arbitrary LP conveniently chosen so that MCMC algorithms to sample
$\bvartheta_k^{(t)} \sim \tilde{p}(\bvartheta_k \mid \by, \Mk) \propto p(\by \mid \bvartheta_k,\Mk) \tilde{p}(\bvartheta_k \mid \Mk)$
are readily available.
See Supplementary Section \ref{supplsec:gibbs} for standard Gibbs algorithms for Normal and product Binomial mixtures.
For the MOM-IW in \eqref{priorN2NL} we used
\begin{align}
\tilde{p}(\bvartheta_{k}\mid \Mk)= \text{Dir}(\boldsymbol{\eta}; q)
\prod_{j=1}^{k}N\left(\bmu_{j} \mid \boldsymbol{0},g \Sigma_j
\right)\text{IW}(\Sigma_j \mid \nu,S),
\nonumber
\end{align}
with $q>1$, which gives
\begin{align}
\omega(\bvartheta_k)=\dfrac{1}{C_k} \prod_{1\leq
i < j \leq
k}\frac{(\bmu_{i}-\bmu_{j})^{'}A_\Sigma^{-1}(\bmu_{i}-\bmu_{j})}{g}
\prod_{j=1}^{k} \frac{N(\bmu_j \mid {\bf 0}, gA_\Sigma)}{N(\bmu_j \mid {\bf 0}, g\Sigma_j)}.
\nonumber
\end{align}
For the MOM-Beta in \eqref{mombetaprior} we used
$\tilde{p}(\bvartheta_k \mid \Mk)= \text{Dir}(\boldsymbol{\eta}; q) \prod_{j=1}^{k}\prod_{f=1}^{p}\text{Beta}(\theta_{jf}; ag,(1-a)g)$,
hence
$$
\omega(\bvartheta_k)= \frac{1}{C_k} \prod_{1\leq i < j \leq k} (\btheta_i-\btheta_j)'(\btheta_i-\btheta_j).
$$

Our strategy is admittedly simple but has convenient advantages.
After obtaining $\tilde{p}(\by \mid \Mk)$ one need only compute a posterior average.
Furthermore, only posterior sampling under $\tilde{p}(\bvartheta_k \mid \by,\Mk)$ is required.
As a caveat the posterior variance of $\omega(\bvartheta_k)$ has an effect on $\hat{p}(\by \mid \Mk)$,
specifically when the local and non-local posteriors differ substantially this variance can potentially be large.
However from Theorem \ref{thm:sparsity} these posteriors differ mainly in overfitted mixtures ($k>k^*$),
and only the numerator but not the denominator in $\omega(\bvartheta_k)$ may vanish
(provided $\tilde{p}$ is positive over its domain, as is the case),
hence in practice we found \eqref{estimator4} to be quite stable
(Supplementary Section \ref{supplsec:comparison}).
We remark that $\omega(\bvartheta_k)$ is not a reweighting
to convert samples from $\tilde{p}(\bvartheta_k \mid \by,\Mk)$
into samples from $p(\bvartheta_k \mid \by,\Mk)$,
but a direct approximation to the posterior mean of $d(\bvartheta_k)$ under $\tilde{p}(\bvartheta \mid \by,\Mk)$.
However, if interested in posterior samples from $p(\bvartheta_k \mid \by,\Mk)$ one could clearly use such a reweighting.
Alternatively one can devise a sampler directly for the non-local
$p(\bvartheta_k \mid \by,\Mk)$, {\it e.g.} using slice sampling \citep{Dunson},
latent truncations \citep{david3} or collapsed Gibbs \citep{xie:2017},
but we do not pursue this as our main interest is model selection.

\subsection{Approximation of $\tilde{p}(\by \mid \Mk)$}
\label{ssec:comp_marglhood_local}

There are a number of proposals to estimate $\tilde{p}(\by\mid \mathcal{M}_{k})$ in the literature,
e.g. trans-dimensional MCMC \citep{Richardson},
Bridge sampling \citep{silvia3}, dual importance sampling \citep{lee2016importance}
and collapsed Gibbs sampling \citep{xie:2017}.
Each of these has its own set of advantages and limitations,
but ultimately obtaining $\tilde{p}(\by\mid \mathcal{M}_{k})$ in a truly scalable fashion remains an open problem.

We focus on two algorithms that are simple to implement and we found to attain a good cost versus precision tradeoff.
We denote by $z_i \in \{1,\ldots,k\}$ the latent cluster indicators, i.e. $z_i=j$ if observation $i$ is assigned to component $j$,
$\bz=(z_1,\ldots,z_n)$ and $n_j=\sum_{i=1}^{n} \mbox{I}(z_i=j)$ is the number of individuals in cluster $j$.
The first algorithm is a refinement proposed by \cite{Robert1} of an algorithm by \cite{chib}.
The strategy uses the identity
\begin{equation}\label{bayes222}
\tilde{p}(\by\mid \Mk)=
\frac{p(\by\mid \hat{\bvartheta}_{k},\Mk)\tilde{p}(\hat{\bvartheta}_{k}\mid \Mk)}
{\tilde{p}(\hat{\bvartheta}_{k}\mid \by,\Mk)}=
\frac{p(\by\mid \hat{\bvartheta}_{k},\Mk)\tilde{p}(\hat{\bvartheta}_{k}\mid \Mk)}
{\sum_{\psi\;\in\;\mathfrak{N}(k)}
\tilde{p}(\psi(\hat{\bvartheta}_{k})\mid \by,\Mk)/(k!)},
\end{equation}
where $\hat{\bvartheta}_{k}$ is the posterior mode
and $\mathfrak{N}(k)$ the set of $k!$ possible permutations of $\{1,...,k\}$.
The right-hand side holds for exchangeable $\tilde{p}(\bvartheta_{k}\mid \Mk)$,
as then the posterior is invariant to label-switching.
The numerator in (\ref{bayes222}) simply requires
evaluating the likelihood and prior at $\hat{\bvartheta}_{k}$.
\cite{Robert1} 
propose estimating the denominator by
\begin{equation}\label{perm3}
\dfrac{1}{Tk!}\sum_{\psi\;\in\;\mathfrak{N}(k)}\sum_{t=1}^{T}
\tilde{p}(\psi(\hat{\bvartheta}_{k})\mid \by,\bz^{(t)},\Mk),
\end{equation}
where $\bz^{(t)}=(z_1^{(t)},\ldots,z_n^{(t)})$ are samples from
$\tilde{p}(\bz, \bvartheta_k \mid \by,\Mk)$.
The estimator (\ref{bayes222})-(\ref{perm3}) can be applied as long as
the posterior density $\tilde{p}(\psi(\hat{\bvartheta}_{k})\mid \by,\bz^{(t)},\Mk)$ has closed-form,
e.g.~in conjugate models.
Specifically for Normal mixtures
\begin{align*}
\tilde{p}(\psi(\hat{\bvartheta}_{k})\mid \by,\boldsymbol{z}^{(t)},\Mk)&=
\prod_{j=1}^{k}N\left(\psi(\hat{\bmu}_{j});\frac{gn_{j}^{(t)}\bar{\by}_{j}^{(t)}}{1+gn_{j}^{(t)}},\dfrac{g}{1+gn_{j}^{(t)}} \Sigma_j^{(t)}\right)
\text{IW}\left(\psi(\hat{\Sigma}_{j});\nu+n_{j}^{(t)},S_j^{(t)}\right)\\
&\times  \text{Dir}(\psi(\hat{\boldsymbol{\eta}});q+n_{1}^{(t)},...,q+n_{k}^{(t)}),
\end{align*}
and for product Binomial mixtures
\begin{align*}
\tilde{p}(\psi(\hat{\bvartheta}_{k})\mid \by,\boldsymbol{z}^{(t)},\Mk)=&
\prod_{j=1}^{k}\prod_{f=1}^{p}
\text{Beta}\left(\psi(\hat{\theta}_{jf});ag + \sum_{z_i^{(t)}=j}^{}y_{if}, (1-a)g + \sum_{z_i^{(t)}=j}^{}(L_{if}-y_{if})\right)\\
&\times  \text{Dir}(\psi(\hat{\boldsymbol{\eta}});q+n_{1}^{(t)},...,q+n_{k}^{(t)}).
\end{align*}

We now outline our ECP algorithm, which relies on Proposition \ref{prop:bf_oneemptyclus} below
expressing Bayes factors as a ratio of posterior to prior empty cluster probabilities.
This representation can be viewed as a Savage-Dickey probability ratio, a natural extension of the familiar density ratio.
The result applies to any mixture and prior satisfying the minimal conditions C1-C4 below.
In the remainder of this section $p(\bvartheta_k \mid \Mk)$ denotes an arbitrary prior for which one wants to obtain
posterior model probabilities,
e.g. in our examples this is the local prior $\tilde{p}(\bvartheta_k \mid M_k)$
and then non-local posterior probabilities are obtained from \eqref{estimator4}.

\begin{enumerate}[label=\bfseries C\arabic*,leftmargin=*]
\item Conditional independence. $p(\by \mid \bz,\bvartheta_k,\Mk)= \prod_{j=1}^{k} \prod_{z_i=j}^{} p(\by_i \mid \btheta_j,\Mk)$

\item Invariance to label permutations. $p(\by \mid \bvartheta_k)= p(\by \mid \psi(\bvartheta_k))$
and $p(\bz \mid \Mk)=p(\varrho(\bz) \mid \Mk)$ for any permutation of component parameters $\psi$ and component indexes $\varrho$.

\item Coherence of prior on cluster allocations.
$p(\bz \mid n_k=0, \Mk)= p(\bz \mid \mathcal{M}_{k-1})$

\item Coherence of prior on parameters.
For any $\btheta_1,\ldots,\btheta_{k-1}$ and any $\bz$ such that $n_k=0$, it holds that
$$p(\btheta_1,\ldots,\btheta_{k-1} \mid \bz, \mathcal{M}_{k-1})=\int p(\btheta_1,\ldots,\btheta_k, \mid \bz,\Mk) d\btheta_k$$
\end{enumerate}

Conditions C1-C2 hold for the vast majority of mixtures, including
mixtures of regressions and most hidden Markov models. 
Conditions C3-C4 hold for most common priors.
For instance C3 holds when $p(\bm{\eta} \mid \Mk)$ and $p(\bm{\eta} \mid \mathcal{M}_{k-1})$ are both
symmetric Dir$(q)$ distributions  and C4 is satisfied by
priors that factor across components,
e.g. $p(\bvartheta_k \mid \Mk)= \mbox{Dir}(\bm{\eta};q) \prod_{j=1}^{k} p(\btheta_j \mid \Mk)$.

\begin{prop}
Suppose that C1-C4 hold. Then the Bayes factor for ${\mathcal{M}_{k-1}}$ versus $\Mk$ is
$$
B_{k-1,k}(\by)=
\frac{\sum_{j=1}^{k} P(n_j=0 \mid \by,\Mk) /k}{P(n_j=0 \mid \Mk)}.
$$
\label{prop:bf_oneemptyclus}
\end{prop}

Once $B_{k-1,k}(\by)$ for $k \in \{2,\ldots,K\}$ are available then $P(\Mk \mid \by)$
are obtained as usual.
Proposition \ref{prop:bf_oneemptyclus} is easy to implement, e.g.~if
$p(\bm{\eta} \mid M_j)=\mbox{Dir}(\bm{\eta};q)$ for all $j$ then
$$P(n_j=0 \mid \Mk)= \frac{\Gamma(kq) \Gamma(n+(k-1)q)}{\Gamma((k-1)q) \Gamma(n+kq)}.$$

Further, given draws $\bvartheta_k^{(t)} \sim p(\bvartheta_k \mid \by,\Mk)$
one can obtain Rao-Blackwellised estimates
\begin{align}
\hat{P}(n_j=0 \mid \by,\Mk)=
\frac{1}{T} \sum_{t=1}^{T} P(n_j=0 \mid \by,\bvartheta_k^{(t)},\Mk)=
\frac{1}{T} \sum_{t=1}^{T} \prod_{i=1}^{n} P(z_i \neq j \mid \by,\bvartheta_k^{(t)},\Mk).
\label{eq:ecp_raoblack_num}
\end{align}
That is, the Bayes factor under local priors $\tilde{B}_{k-1,k}(\by)$ is obtained
dividing \eqref{eq:ecp_raoblack_num} by $P(n_j=0 \mid \Mk)$.
To estimate Bayes factors under NLPs $\hat{B}_{k-1,k}(\by)$ we use the estimator in \ref{estimator4}
\begin{align*}
\hat{B}_{k-1,k}(\by)=\tilde{B}_{k-1,k}(\by) \dfrac{\sum_{t=1}^{T}
\omega(\bvartheta_{k-1}^{(t)})}{\sum_{t=1}^{T}
\omega(\bvartheta_{k}^{(t)})},
\end{align*}
Note that ECP only requires cluster probabilities, hence it remains valid for non-conjugate models.

Proposition \ref{prop:bf_oneemptyclus} is of independent interest to help discard unoccupied
clusters in overfitted mixtures.
It suggests that the threshold on posterior empty cluster probabilities should depend on the corresponding prior empty
cluster probabilities. The latter are a function of $n$, $k$ and $q$, hence using fixed thresholds may be suboptimal.
Note also that Proposition \ref{prop:bf_oneemptyclus} can be used to compare structurally different models.
For instance let $B_{k1}$ be the Bayes factor between a $k$-component unequal-covariance Normal mixture
vs. a one-component Normal, and $B^c_{k1}$ that for a $k$-component common-covariance Normal mixture
vs. a one-component Normal.
Then $B_{k1}/B^c_{k1}$ is the Bayes factor comparing $k$ components with unequal vs. equal covariances.
Similarly one could combine the Bayes factor between a one-component Normal vs. a one-component T (which is easy to compute)
with Proposition \ref{prop:bf_oneemptyclus} to obtain Bayes factors between any $k$-component Normal vs. T mixture.
That is, the ECP estimator is connected to empty cluster probabilities but really is a tool to obtain $P(\Mk \mid \by)$
and hence remains applicable in more general settings.

\subsection{Posterior modes}
\label{ssec:comp_postmode}

The EM algorithm provides a fast way to obtain posterior modes
$\hat{\bvartheta}_k=\arg\max_{\bvartheta_k} p(\bvartheta_k \mid \by,\Mk)$
or cluster assigments $\hat{z}_i= \arg\max_{j \in \{1,\ldots,K\}} p(z_i=j \mid \by,\hat{\bvartheta}_k,\Mk)$.
    This optimization problem is conceptually related to maximizing a penalized likelihood,
    e.g. setting fused LASSO penalties on the separation between means \citep{heinzl:2014},
    although we remark that the latter shrink components closer to each other rather than pushing them apart as is the case for non-local priors.

We briefly describe our algorithm, which is derived in Supplementary Sections \ref{supplsec:em_algorithm} and \ref{supplsec:em_algorithm2}.
At iteration $t$ the E-step computes
$$\bar{z}_{ij}^{(t)}=P(z_i=j\mid \by_{i},\bvartheta_{j}^{(t-1)})=
\eta_{j}^{(t-1)}p(\by_{i}\mid \btheta_{j}^{(t-1)})/\sum_{j=1}^{k}\eta_{j}^{(t-1)}
p(\by_{i}\mid \btheta_{j}^{(t-1)})
$$
and is trivial to implement.
The M-step requires updating $\bvartheta_k^{(t)}$ in a manner that increases
the expected complete log-posterior, which we denote by $\xi(\bvartheta_k)$,
but under our prior $p(\bvartheta_k \mid \Mk)=d_{\vartheta}(\bvartheta_k) \tilde{p}(\bvartheta_k \mid \Mk)$ this cannot be done in closed-form.
A key observation is that if $\tilde{p}(\bvartheta_k \mid \Mk)$ leads to closed-form updates,
the corresponding target $\tilde{\xi}(\bvartheta_k)$ only differs from $\xi(\bvartheta_k)$ by a term $d_{\vartheta}(\bvartheta_k)$,
thus one may approximate $\xi(\bvartheta_k)$ via a first order Taylor expansion of $d_{\vartheta}(\bvartheta_k)$.
These approximate updates need not lead to an increase in $\xi(\bvartheta_k)$
(although they typically do since $d_{\vartheta}(\bvartheta_k)$ has a mild influence for moderately large $n$), and
whenever this happens we use gradient algorithm updates.
Algorithms \ref{alg:em} and S\ref{alg:em1} detail the steps for Normal and product Binomial mixtures
(extensions to other models follow similar lines),
for simplicity outlining the approximate updates
(see Supplementary Section \ref{supplsec:em_algorithm} for the gradient algorithm).
In our implementation 
we initialize $\bvartheta_{k}^{(0)}$ to the MLE
and stop when the increase in $\xi(\bvartheta_k)$ is below a tolerance $\epsilon^*=0.0001$
or a maximum number of iterations $T=10,000$ is reached.
For ease of notation in Algorithm \ref{alg:em} we define
$d_{ij}=(\bmu_{i}-\bmu_{j})' A_{\Sigma}^{-1}(\bmu_{i}-\bmu_{j})$
evaluated at the current value of $\bmu_1,\ldots,\bmu_k,\Sigma_1,\ldots,\Sigma_k$.

\RestyleAlgo{boxruled}

\vspace{2mm}
\begin{algorithm}
\small
Set $t=1$. \While{$\zeta>\epsilon^{*}$ and $t<T$}{
\For{$t\geq 1$ and $j=1,...,k$}{
E-step. Let
$\bar{z}^{(t)}_{ij}=\dfrac{\eta_{j}^{(t-1)}N(\by_{i};\bmu_{j}^{(t-1)},\Sigma_{j}^{(t-1)})}{\sum_{j=1}^{k}\eta_{j}^{(t-1)}
N(\by_{i}; \bmu_{j}^{(t-1)},\Sigma_{j}^{(t-1)})}
$
and $n_{j}^{(t)}=\sum_{i=1}^{n}\bar{z}^{(t)}_{ij}$.

M-step. Let $\bar{\by}_{j}^{(t)}=\sum_{i=1}^{n}\bar{z}^{(t)}_{ij}\by_{i}/n_{j}^{(t)}$.
Update
\begin{align*}
&\hspace{-2.0cm}\bmu_{j}^{(t)}=\left(\left(\Sigma_{j}^{-1}\right)^{(t-1)}n_{j}^{(t)} + A_{\Sigma^{(t-1)}}^{-1} \left(\dfrac{1}{g} +
\sum_{\substack{i\neq j}} \frac{2}{d_{ij}}\right)\right)^{-1}
\\&\hspace{-1.0cm}\times\left(\Sigma^{-1(t-1)}n_{j}^{(t)}\bar{\by}_{j}^{(t)} +  A_{\Sigma^{(t-1)}}^{-1}
\left(
\sum_{\substack{ i\neq j}}
\frac{\bmu_{j}^{(t-1)}-(\bmu_{i}^{(t-1)}-\bmu_{j}^{(t-1)})}{d_{ij}}\right)\right),
\end{align*}

Update $(\nu-p+n_{j}^{(t)})\Sigma^{(t)}_{j}=$
\begin{align*}
\hspace{-2.0cm}
S^{-1} + \dfrac{\bmu_{j}^{(t)} (\bmu_{j}^{(t)})'}{kg}+\sum_{i=1}^{n}\bar{z}^{(t)}_{ij}
(\by_{i}-\bmu_{j}^{(t)})(\by_{i}-\bmu_{j}^{(t)})^{'}
-\dfrac{1}{k}
\sum_{\substack{i\neq j
}}\dfrac{2(\bmu_{j}^{(t)}-\bmu_{k}^{(t)})(\bmu_{j}^{(t)}-\bmu_{k}^{(t)})^{'}}{d_{ij}}.
\end{align*}
Update
$\eta^{(t)}_{j}=\dfrac{n_{j}^{(t)}+q-1}{n+k(q-1)}$.

}
Compute  $\zeta=|\xi(\bvartheta_{k}^{(t)})-\xi(\bvartheta_{k}^{(t-1)})|$ and set $t=t+1$.
}
\caption{EM under MOM-IW-Dir priors.}
\label{alg:em}
\end{algorithm}

\section{Empirical Results}
\label{sec:results}

We compared our MOM-IW-Dir and MOM-Beta-Dir priors
with default parameters (Section \ref{ssec:priorelicitation}) to their local counterparts, Normal-IW-Dir
and Beta(1,1)-Dir respectively.
As described in Section \ref{ssec:priorelicitation}
the Normal-IW-Dir prior parameter $g^{L}$ was set to match the 95\% percentile for the separation parameter $\kappa$.
Throughout we use uniform model priors $P(\mathcal{M}_j)=1/K, j=1\dots,K$.
Unless otherwise stated we estimated the integrated likelihoods using Algorithm S\ref{alg:marginal_lhood}  and
S\ref{alg:marginal_lhoodBin} based on 5,000 MCMC draws after a 2,500 burn-in.
We also considered the BIC, AIC, sBIC, overfitted mixtures and repulsive overfitted mixtures.
We only found an sBIC implementation for Normal mixture with $\Sigma_{i}\neq\Sigma_{j}$.
For Normal mixtures we used the function GaussianMixtures and for Binomial mixtures we used the function BinomialMixtures from the R package sBIC \citep{sBICp}.
In GaussianMixtures we used the default real canonical threshold value given by $\lambda\leq \frac{1}{2}(jd+j-1+(k-j)\varphi)$ with $\varphi$ chosen in relation to a prior on mixture weights and we denote this by sBIC.
In BinomialMixtures we tried two sBIC versions, named $\overline{\text{sBIC}}$ and $\overline{\text{sBIC}}_{05}$ corresponding to setting the real canonical threshold to $\lambda\leq \frac{1}{2}(k+j-1)$  and $\lambda\leq \frac{1}{4}(j+3k) - \frac{1}{2}$ respectively.

In Sections \ref{ssec:syntheticexamples}-\ref{ssec:repulsive} we use Normal mixtures.
Section \ref{ssec:syntheticexamples} presents a simulation study for univariate and bivariate Normal mixtures.
Section \ref{ssec:mispec} explores model misspecification by simulating data from T mixtures.
In Sections \ref{ssec:Cytometry}-\ref{ssec:iris} we analyse several datasets,
including a flow cytometry experiment and Fisher's Iris data for which there is a known ground truth.
Section \ref{ssec:repulsive} offers a comparison to applying overfitted mixtures to these datasets.
Section \ref{ssec:binomials} reproduces a Binomial mixture example used by \cite{drton2017bayesian}
to illustrate the sBIC, and Section \ref{ssec:LCA_a} analyses a US political blog dataset via product Binomial mixtures.
We used R package NLPmix for the EM algorithm and the estimate $\hat{p}(\by \mid \Mk)$ from  \cite{Robert1},
and for our ECP estimator we used \texttt{bfnormmix} from R package \texttt{mombf}.
As illustration, the code for a simulation in Section \ref{ssec:syntheticexamples}
is provided in Supplementary Section \ref{supplsec:rcode}
and that for Section \ref{ssec:mispec} in a supplementary file.
See also Supplementary Section \ref{supplsec:EMMCMC} for a simulation experiment
for product Binomial mixtures to illustrate the usage of diagnostics for multiple EM and MCMC runs.

\subsection{Simulation study with Normal mixtures}
\label{ssec:syntheticexamples}

We consider choosing amongst the three competing models
\begin{align*}
\mathcal{M}_{1}&: N(\by_{i};\bmu,\Sigma),\\
\mathcal{M}_{2}&: \eta_{1}N(\by_{i};\bmu_{1},\Sigma)+
(1-\eta_{1})N(\by_{i};\bmu_{2},\Sigma)\\
\mathcal{M}_{3}&: \eta_{1}N(\by_{i};\bmu_{1},\Sigma)+
\eta_{2}N(\by_{i};\bmu_{2},\Sigma)+
(1-\eta_{1}-\eta_{2})N(\by_{i};\bmu_{3},\Sigma),
\end{align*}
where independence is assumed across $i=1,\ldots,n$.
We simulated 100 datasets under each of the 8 data-generating truths with Normal components
depicted in Figure \ref{supfig:densities} for univariate (Cases 1-4) and bivariate outcomes (Cases 5-8).
Case 1 corresponds to $k^*=1$ components,
Cases 2-3 to $k^*=2$ moderately and strongly-separated components respectively,
and Case 4 to $k^*=3$ with two strongly overlapping components and a third component with smaller weight.
Cases 5-8 are analogous for the bivariate outcome.

\begin{figure}[ht]
\begin{center}
\begin{tabular}{cc}
Case 1 & Case 2 \\
\includegraphics[width=0.30\textwidth]{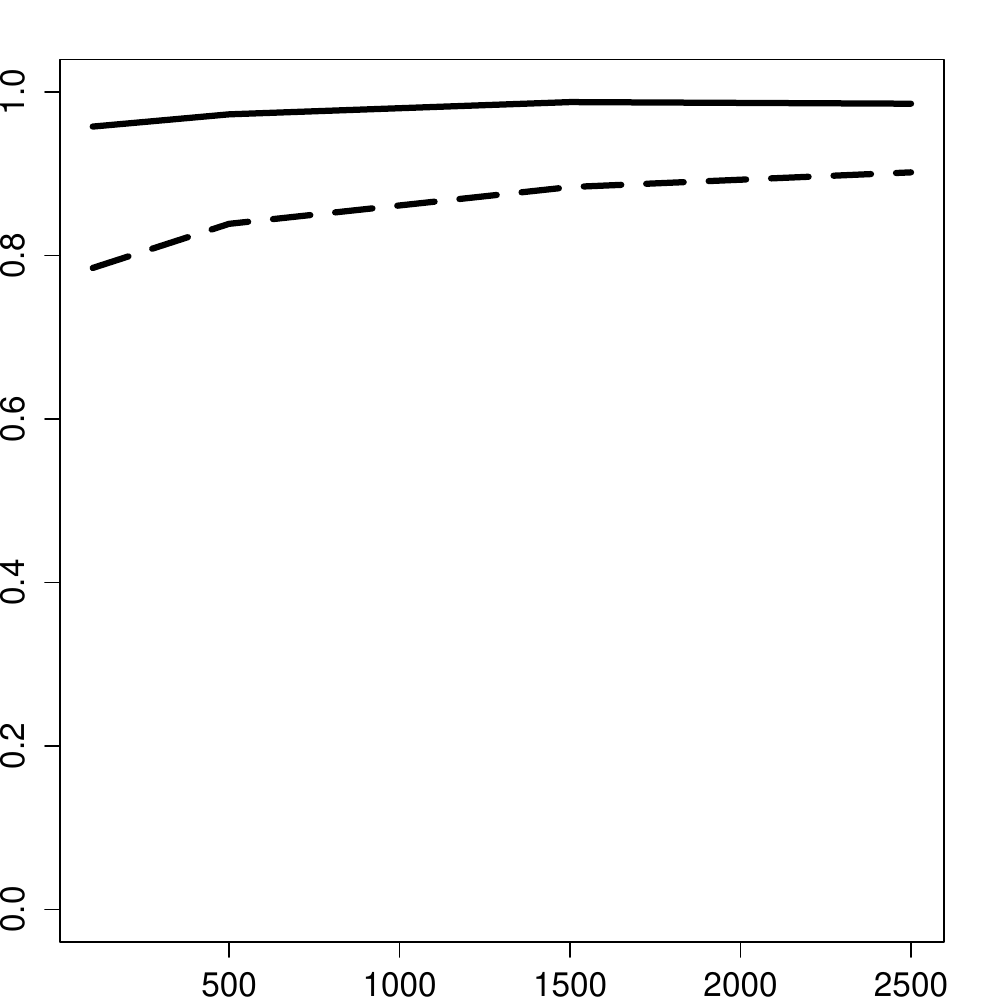} &
\includegraphics[width=0.30\textwidth]{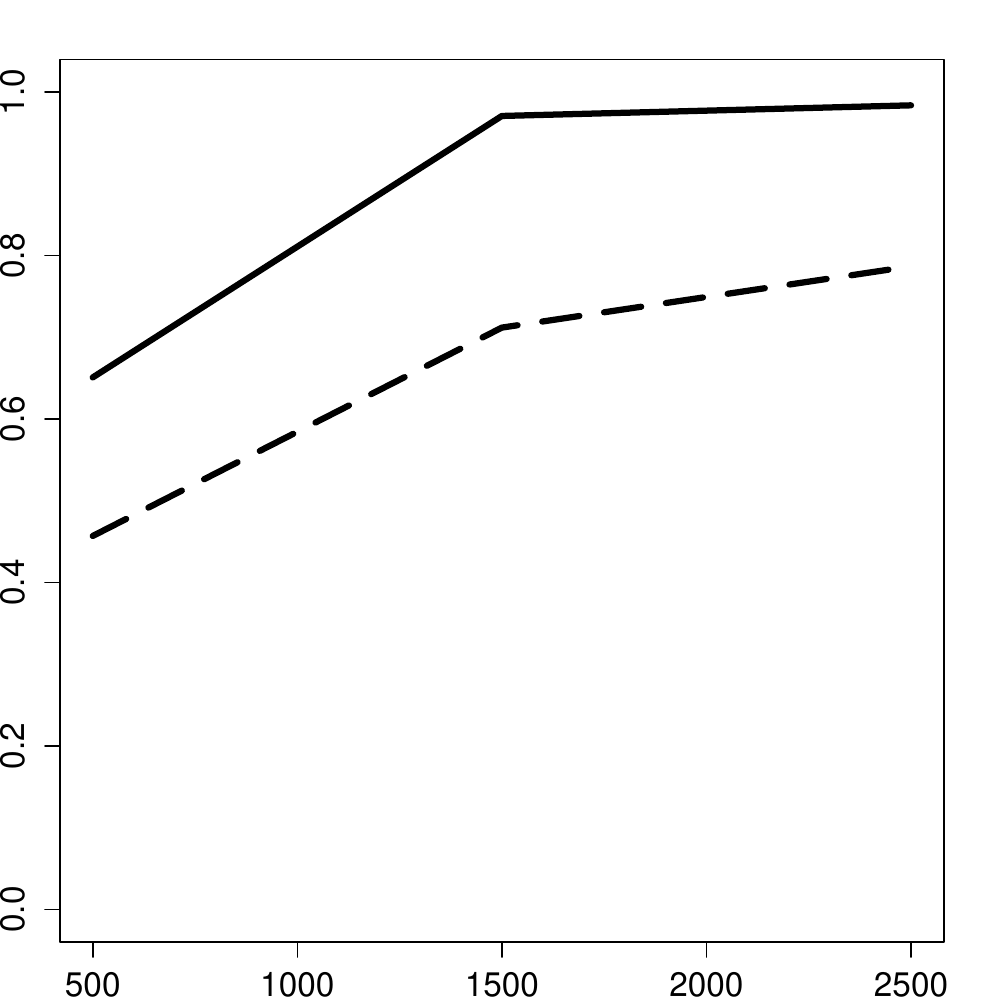} \\
Case 3 & Case 4 \\
\includegraphics[width=0.30\textwidth]{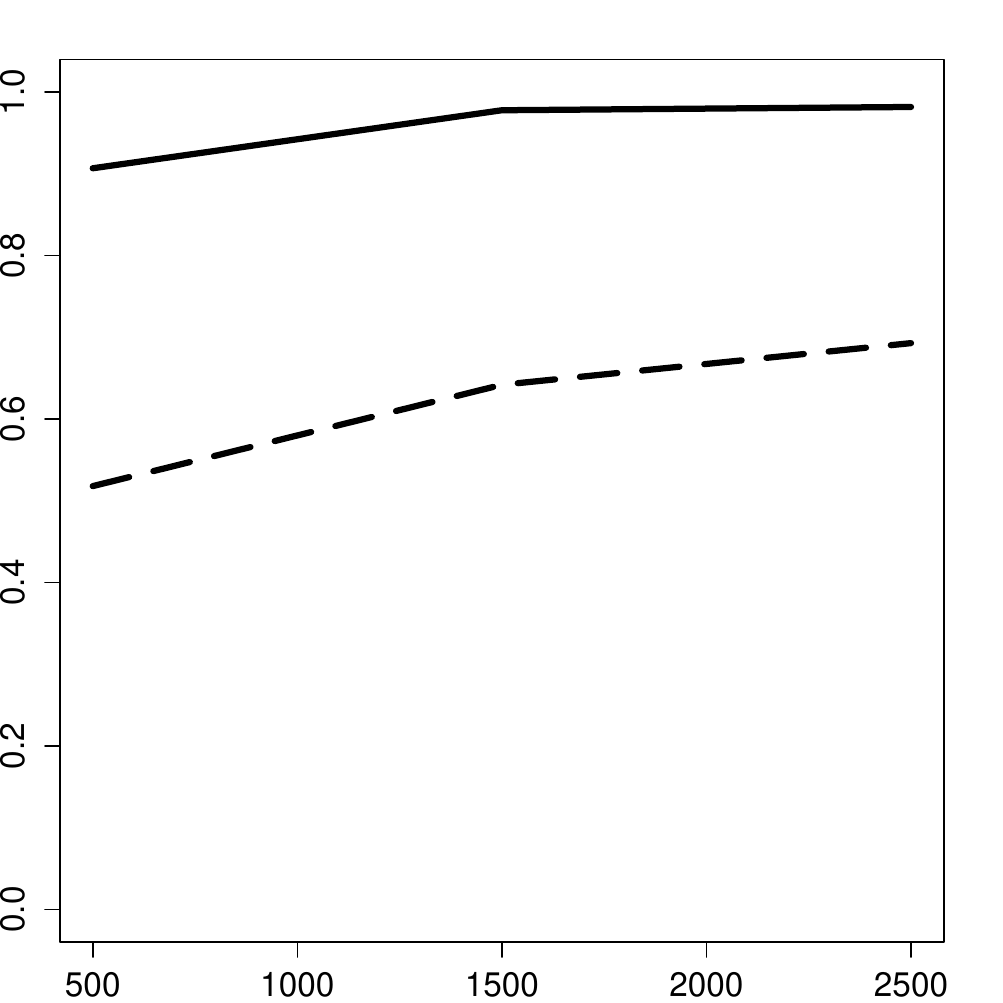} &
\includegraphics[width=0.30\textwidth]{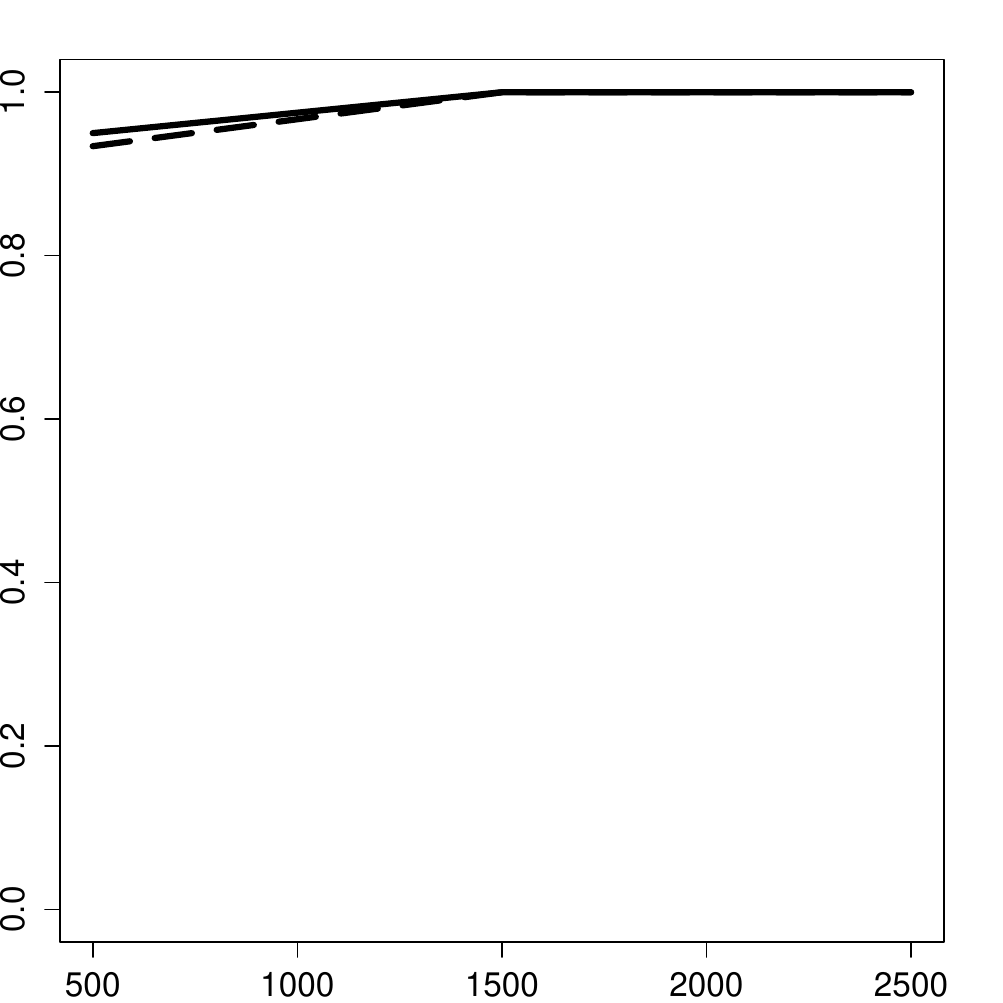} \\
Case 5 & Case 6 \\
\includegraphics[width=0.30\textwidth]{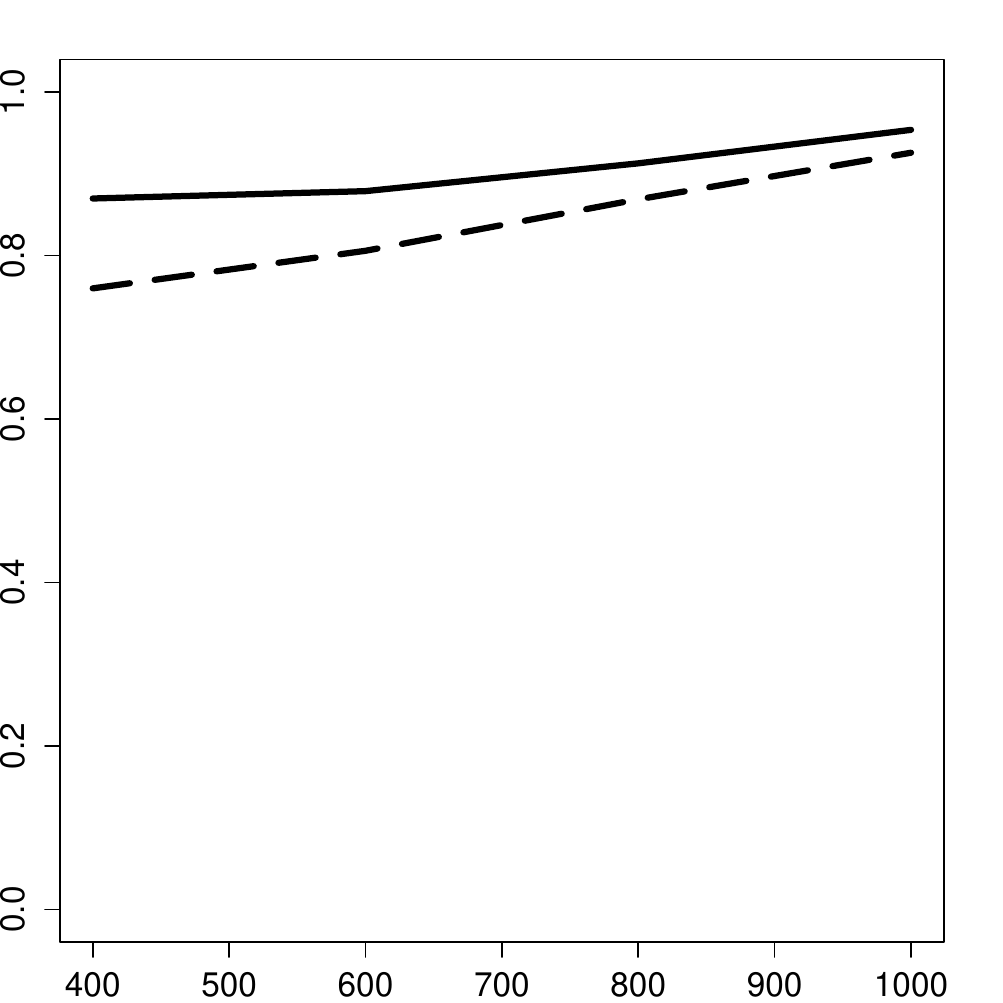} &
\includegraphics[width=0.30\textwidth]{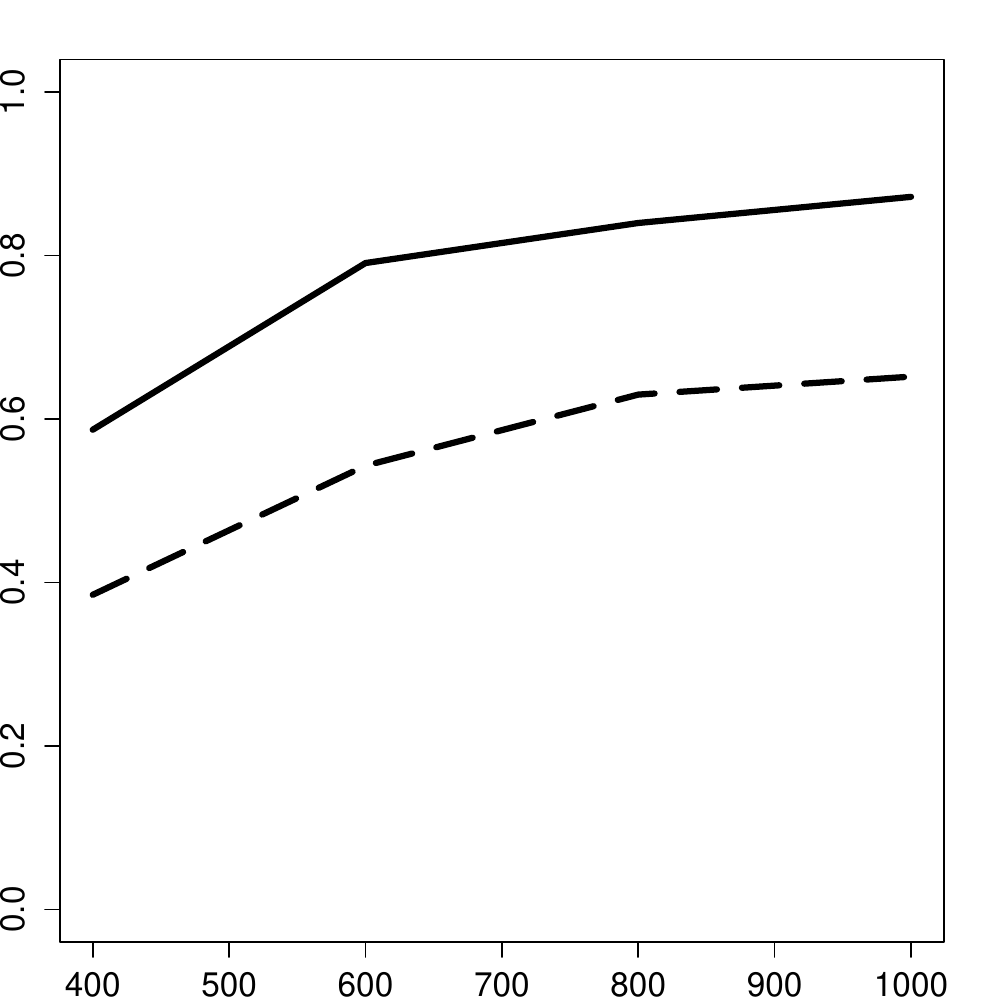} \\
Case 7 & Case 8 \\
\includegraphics[width=0.30\textwidth]{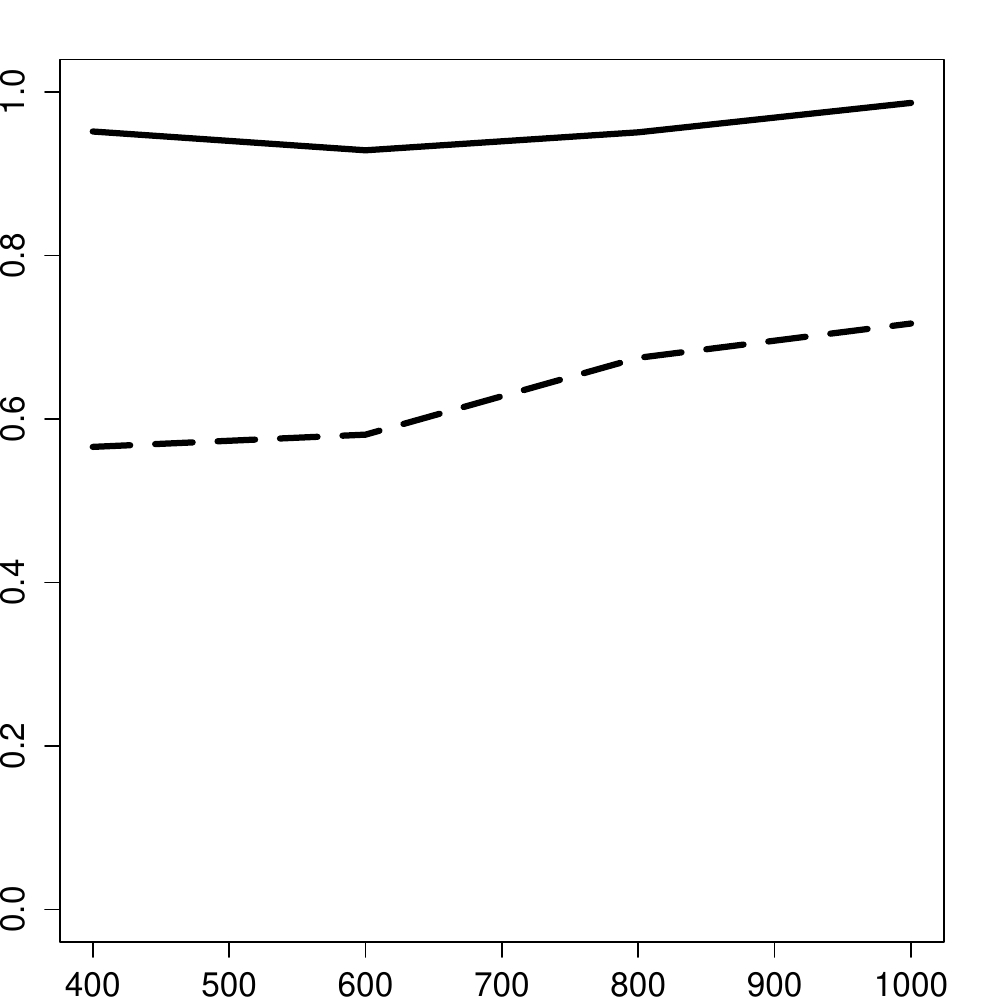} &
\includegraphics[width=0.30\textwidth]{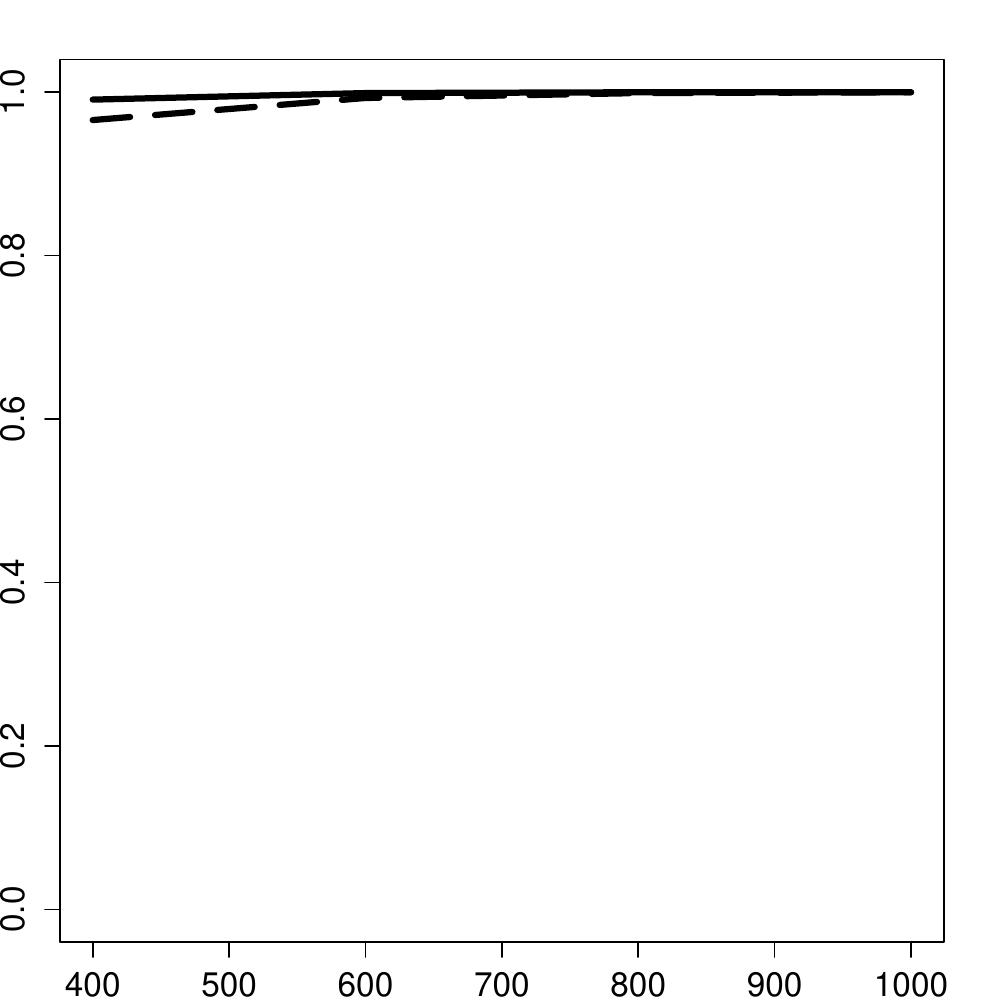} \\
\end{tabular}
\end{center}
\caption{Simulation study. $P(\mathcal{M}_{k^*} \mid \by)$ versus $n$
for the MOM-IW (solid line) and Normal-IW (dashed line).}
\label{supfig:synthetic_truth}
\end{figure}

\begin{figure}[ht]
\begin{center}
\begin{tabular}{cc}
Case 1 & Case 2\\
\includegraphics[width=0.30\textwidth]{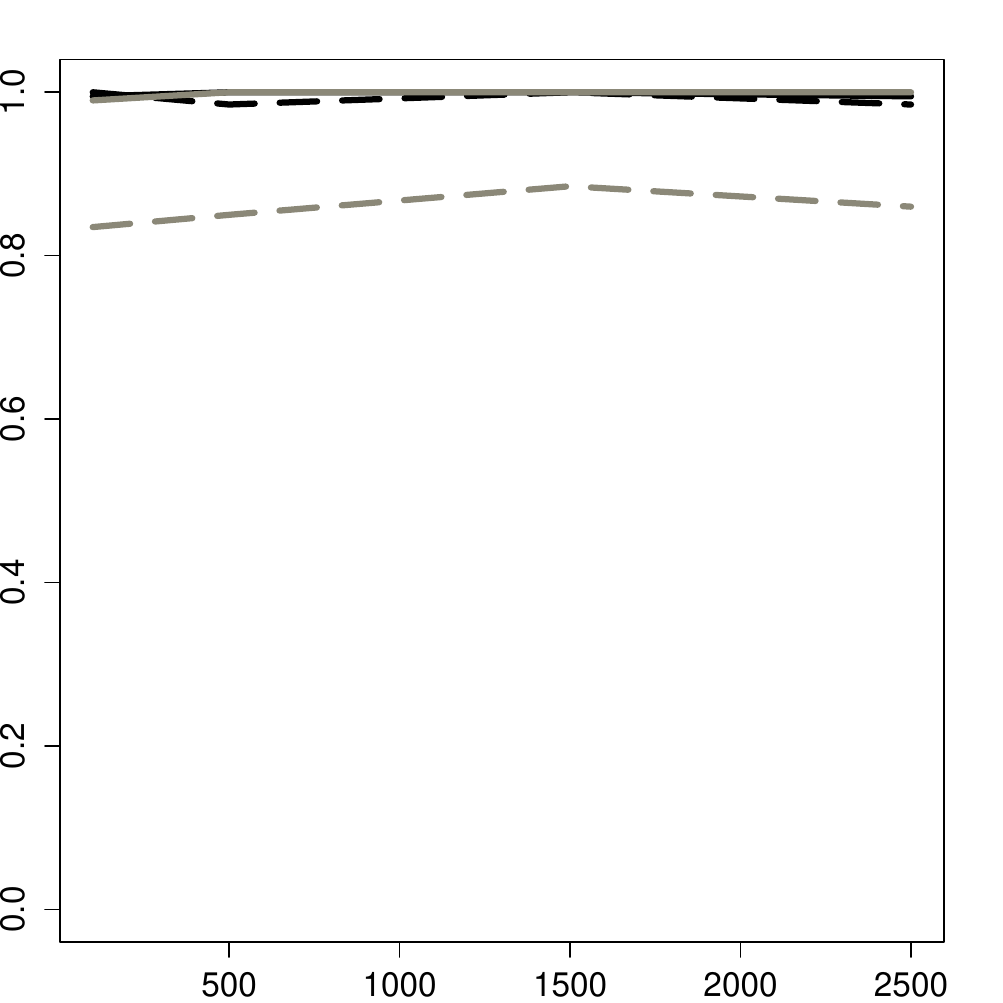} &
\includegraphics[width=0.30\textwidth]{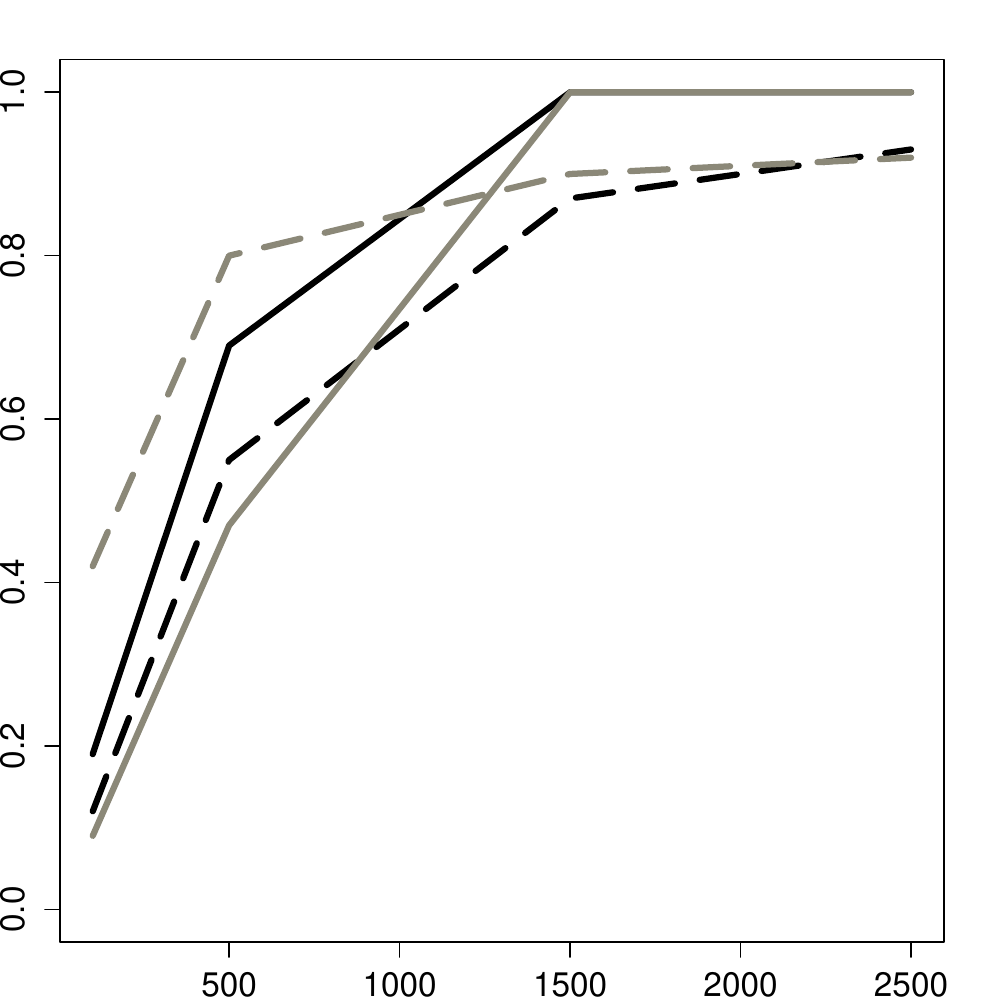} \\
Case 3 & Case 4\\
\includegraphics[width=0.30\textwidth]{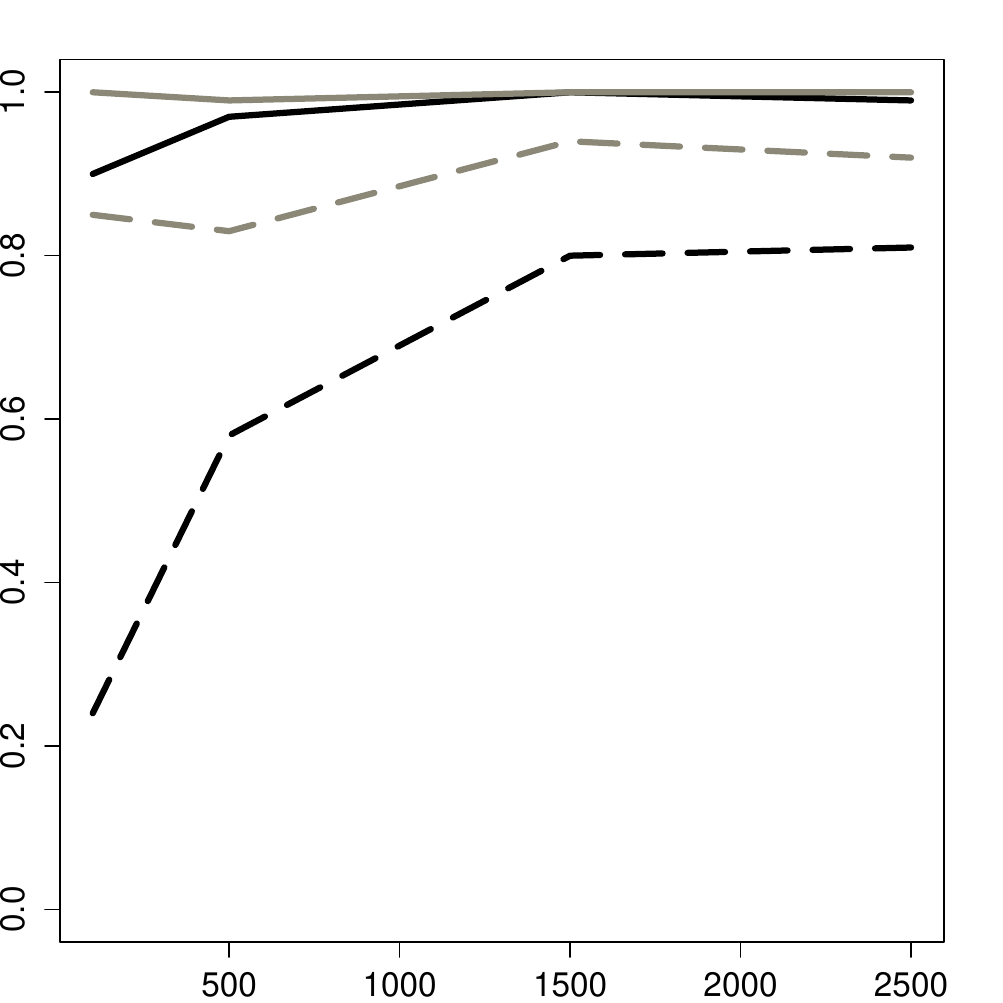} &
\includegraphics[width=0.30\textwidth]{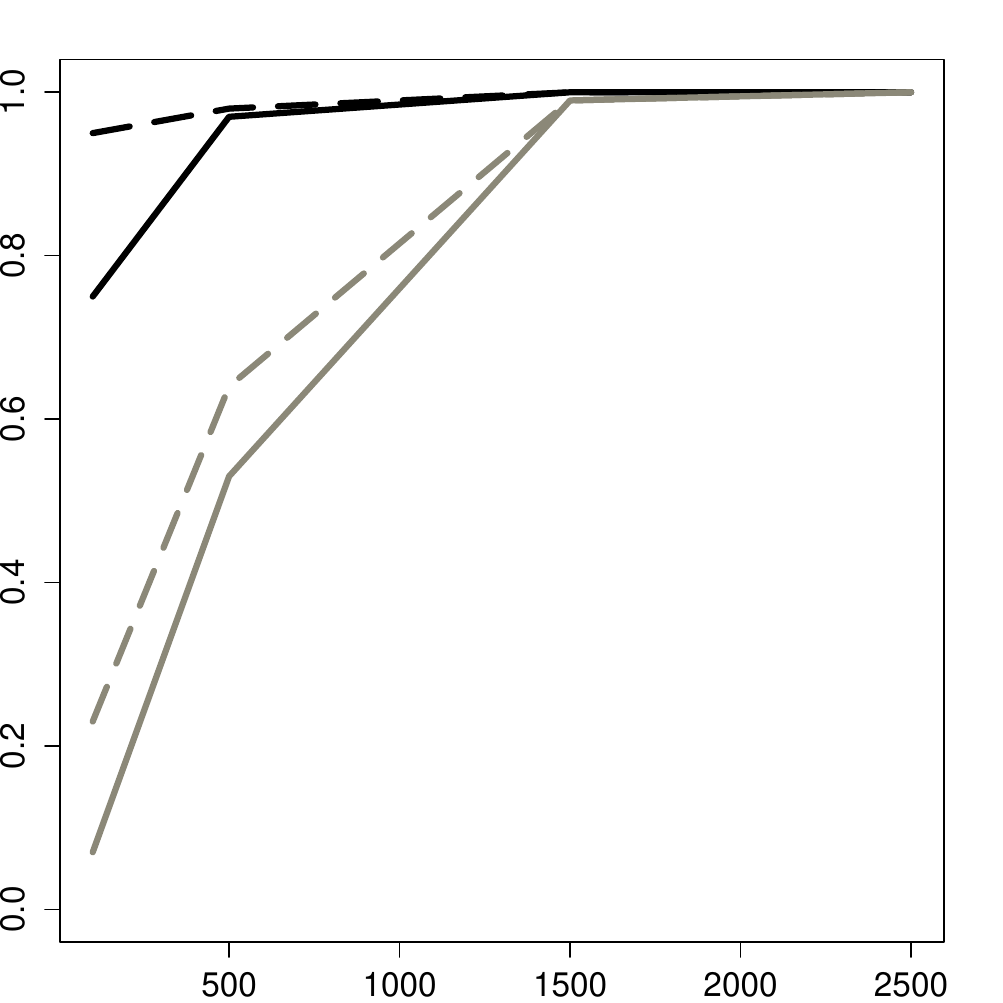}\\
Case 5 & Case 6\\
\includegraphics[width=0.30\textwidth]{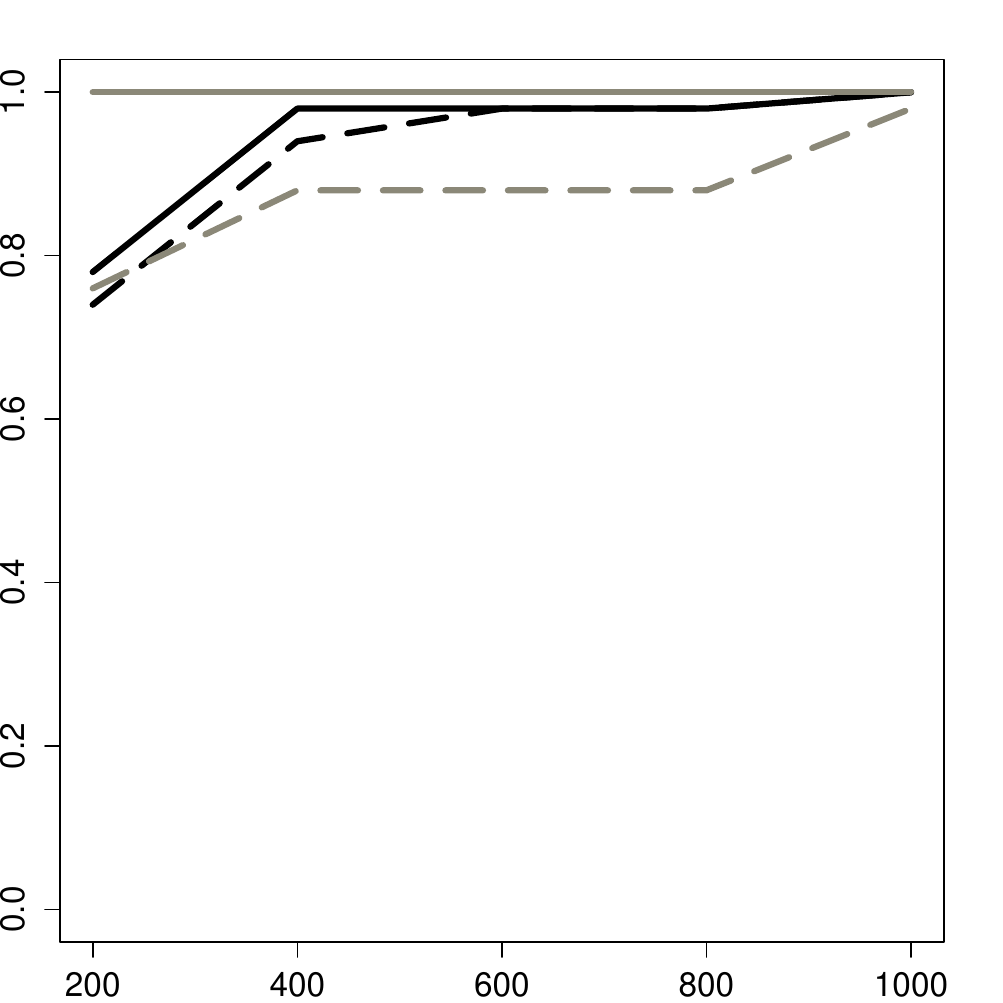} &
\includegraphics[width=0.30\textwidth]{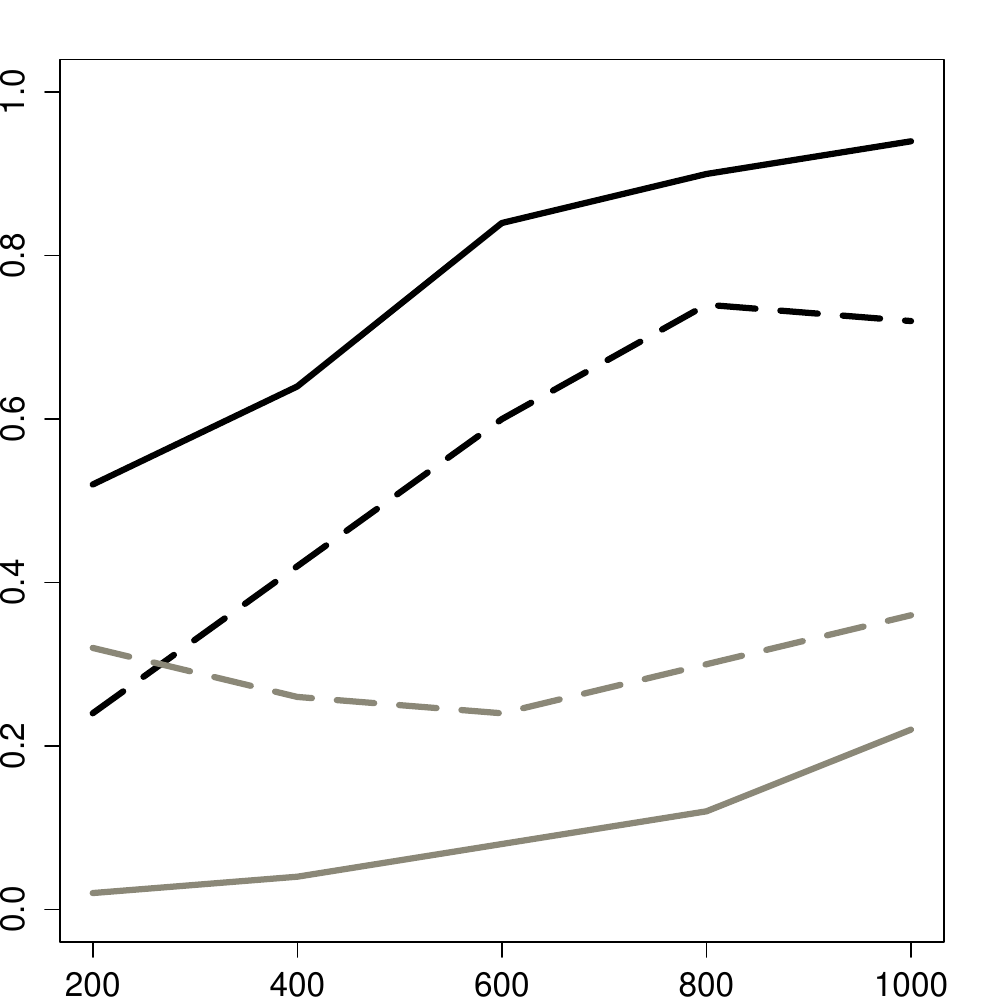} \\
Case 7 & Case 8\\
\includegraphics[width=0.30\textwidth]{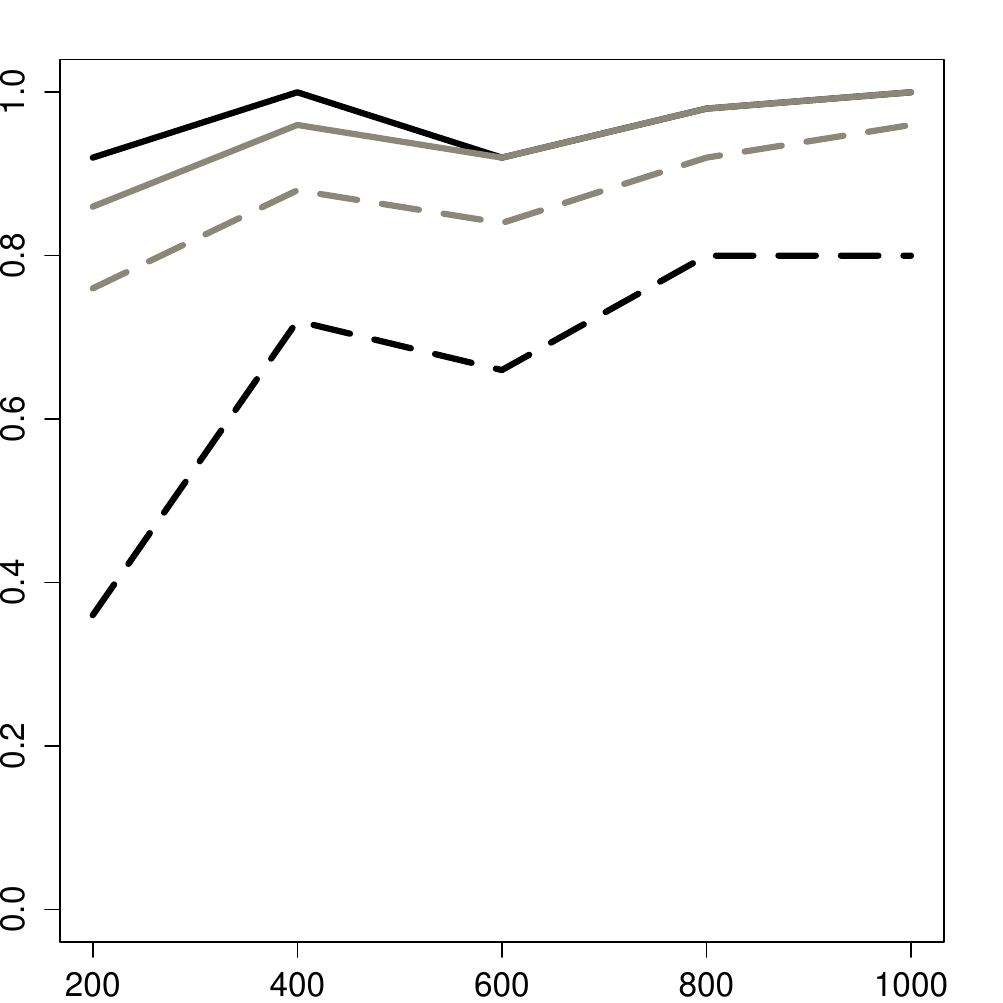} &
\includegraphics[width=0.30\textwidth]{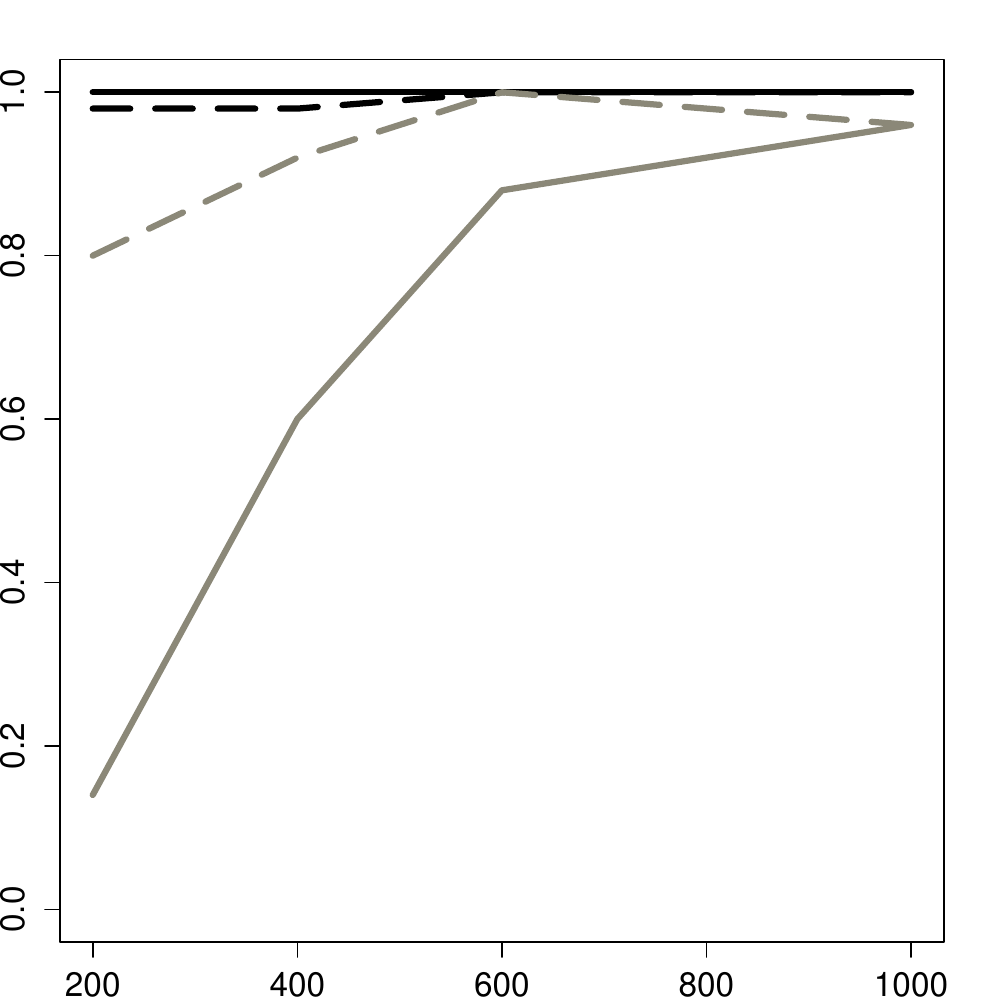}\\
\end{tabular}
\end{center}
\caption{Simulation study. Proportion of correct $\hat{k}=k$ vs. $n$
for MOM-IW (solid black), Normal-IW (dashed black), AIC (dashed gray) and BIC (solid gray)}
\label{supfig:unicorrect1}
\end{figure}

Figure \ref{supfig:synthetic_truth} shows the average posterior probability assigned to the data-generating model $P(\mathcal{M}_{k^*} \mid \by)$ as a function of $n$ under NLP and LP.
To compare frequentist and Bayesian methods Figure \ref{supfig:unicorrect1}
reports the (frequentist) proportion of correct model selections,
{\it i.e.} the proportion of simulated datasets in which $\hat{k}=k^*$, where $\hat{k}$ is the selected
number of components by any given method (for Bayesian methods $\hat{k}=\arg\max_k p(\Mk \mid \by)$).

Figures \ref{supfig:synthetic_allmodels}-\ref{supfig:synthetic_allmodels3}
show the corresponding posterior expected model size $E(k \mid \by)$ and average $\hat{k}$.
For $E(k \mid \by)$ we set $q=p+1$ as the default prior specification of $q$ and we perform a sensitivity prior analysis (Subsection \ref{ssec:priorelicitation}) with another $q$ suggested in \cite{book_silvia}.
Figure \ref{supfig:synthetic_truth2} plots $P(\mathcal{M}_{k^*} \mid \by)$ setting $g$ so that $P(\kappa < 4 \mid \Mk)=0.1$ instead of 0.05.
Overall a similar behavior is observed in the univariate and bivariate cases.
The BIC adequately favoured sparse solutions (Cases 1,3,5,7)
but showed an important lack of sensitivity to detect some truly present components (Cases 2,4,6,8).
AIC was suboptimal in almost all scenarios.
As seen in Figure \ref{supfig:synthetic_truth},
the Normal-IW led to substantially less posterior concentration of $P(\mathcal{M}_{k^*}\mid \by)$ than our MOM-IW
in all cases except the non-sparse Cases 4 and 8, where results were practically indistinguisable.
As predicted by theory, the LP put too much posterior mass on overfitted models.
Interestingly, Cases 2 and 6 illustrate that additionally to enforcing parsimony NLPs
can sometimes also increase sensitivity to detect moderately-separated components.
This is due to assigning a prior $p(\bvartheta_k \mid \Mk)$  with that degree of separation between the component parameters.
Figures \ref{supfig:synthetic_allmodels} and \ref{supfig:synthetic_truth2}
show similar results, but $P(\kappa < 4 \mid \Mk)=0.05$
led to slightly better parsimony than $P(\kappa < 4 \mid \Mk)=0.10$.

\begin{table}
\begin{center}
\begin{tabular}{|lr|rrr|rrr|r|} \hline
       &   & \multicolumn{3}{c|}{MOM-IW-Dir} & \multicolumn{3}{c|}{Normal-IW-Dir} & \\
       & n & k=1 & k=2 & k=3 & k=1 & k=2 & k=3 & CPU time \\  \hline
Case 1 & 200  & 0.860 & 0.061 & 0.079 & 0.701 & 0.190 & 0.109 & 1.8  sec.\\
       & 1000 & 0.989 & 0.010 & 0.001 & 0.893 & 0.089 & 0.018 & 8.3  sec.\\ \hline
Case 3 & 200  & 0.000 & 0.727 & 0.273 & 0.000 & 0.592 & 0.408 & 1.8  sec.\\
       & 1000 & 0.000 & 0.933 & 0.067 & 0.000 & 0.776 & 0.224 & 8.4  sec.\\  \hline
Case 5 & 200  & 0.937 & 0.060 & 0.003 & 0.871 & 0.110 & 0.019 & 2.7  sec.\\
       & 1000 & 0.994 & 0.006 & 0.000 & 0.925 & 0.070 & 0.006 & 12.9 sec. \\ \hline
Case 7 & 200  & 0.611 & 0.343 & 0.046 & 0.675 & 0.277 & 0.048 & 2.7  sec.\\
       & 1000 & 0.000 & 0.955 & 0.045 & 0.000 & 0.879 & 0.121 & 13.3 sec. \\ \hline
\end{tabular}
\end{center}
\caption{Simulation study. Mean $P(\Mk \mid \by)$ for $k \in \{1,2,3\}$ and Cases 1, 3, 5 and 7 under MOM-IW-Dir and Normal-IW-Dir priors. Median CPU time (seconds) to compute $P(\Mk \mid \by)$ for both priors and all $k$}
\label{tab:sim_unequal}
\end{table}

We extended the study to the case where the model (wrongly) assumes unequal covariances.
Here we used the ECP estimator in Proposition \ref{prop:bf_oneemptyclus} (9,000 iterations after a 1,000 burnin), to study its precision and scalability.
We generated 50 simulations under Cases 1, 3, 5 and 7 for $n \in \{200,1000\}$, and
for each dataset we obtained $\hat{P}(\Mk \mid \by)$ for $k \in \{1,2,3\}$,
under MOM-IW-Dir and Normal-IW-Dir priors.
Table \ref{tab:sim_unequal} reports the average $P(\Mk \mid \by)$ and computing time on a laptop
running OS X 10.11.6 with 1.6 GHz processor and 8Gb 1600MHz DDR3.
This is the total time of obtaining $\hat{P}(\Mk \mid \by)$ for all $k$ and both priors,
using function \texttt{bfnormmix} in R package \texttt{mombf}, and no parallel processing.
It ranged from 1.8 seconds for Case 1 where $n=200$ and $p=1$ to 13.3 seconds for Case 7, where $n=1000$ and $p=2$.
Analogously to the earlier results we observed a higher posterior concentration around $k^*$ for the MOM-IW-Dir
than for the Normal-IW-Dir prior.

We conducted further experiments to assess the computational scalability of the ECP estimator
as a function of $n$, $p$ and the upper bound $K$ on the number of clusters.
We simulated data for $n \in \{200,1000\}$, $p\in \{1,2,5\}$ and $K=\{2,5,10\}$
under a single-component multivariate Normal with zero mean and identity covariance matrix.
Figure \ref{supfig:cputime} shows the median run times.
The time increased linearly with $n$ and slightly supra-linearly in $k$ and $p$.
It is easy to show that the per-iteration computational complexity of the Gibbs sampler is linear in $n$,
whereas to sample the covariance matrix (via a Cholesky decomposition) it is cubic in $p$.
Regarding $k$ the Gibbs per-iteration complexity under a Normal-IWishart prior grows linearly with $k$
(hence so does obtaining ECP-based posterior model probabilities),
however the post-processing step in \eqref{estimator4}
to evaluate the MOM-IW penalty contains $k(k-1)/2$ terms.
Despite such supra-linear complexity our results show that for moderately large $(k,p)$ computations remain feasible.
We reported total times for $k=1,\ldots,K$;
one could use parallel computing or stop at the smallest $k$ such that $P(\Mk\mid \by)<P(\mathcal{M}_{k-1} \mid \by)$
which typically happens well before reaching $K$ \citep{chambaz:2008}.

\subsection{Inference under a misspecified model}
\label{ssec:mispec}

In practice the data-generating density may present non-negligible departures from the assumed class.
An important case we investigate here is the presence of heavy tails,
which under an assumed Normal mixture likelihood may affect
both the chosen $k$ and the parameter estimates.
We generated $n=600$ observations from $k^*=3$ bivariate T components with $4$ degrees of freedom,
means $\bmu_{1}=(-1,1)'$, $\bmu_{2}=-\bmu_{1}$, $\bmu_{3}=(6,6)'$,
a common scale matrix with elements
$\sigma_{11}=\sigma_{22}=2$ and $\sigma_{12}=\sigma_{21}=-1$ and
$\eta_{1}=\eta_{2}=\eta_{3}=1/3$. We considered up to $K=6$ components with either homogeneous $\Sigma_1=\ldots=\Sigma_k$ or heterogeneous covariances,
giving a total of 11 models.

\begin{figure}[ht]
\begin{center}
\begin{tabular}{cc}
BIC/sBIC  & AIC\\
\includegraphics[width=0.5\textwidth]{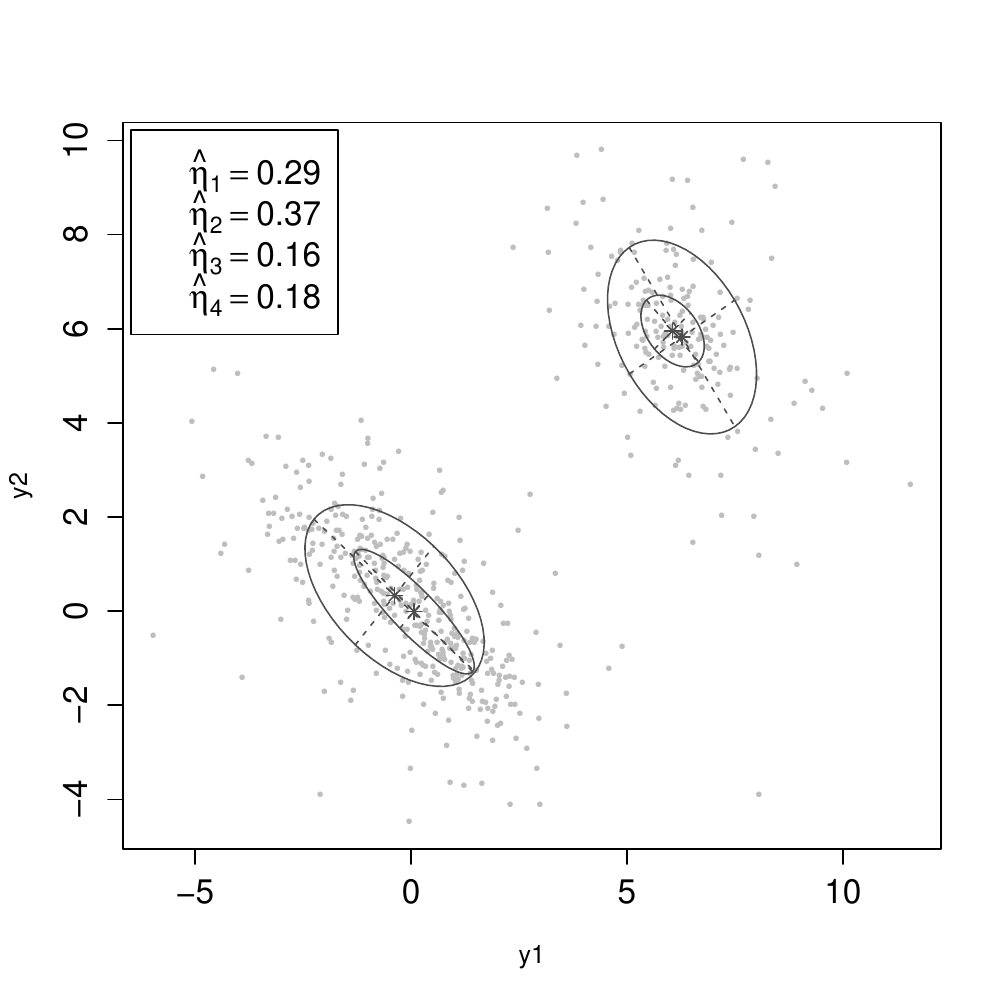} &
\includegraphics[width=0.5\textwidth]{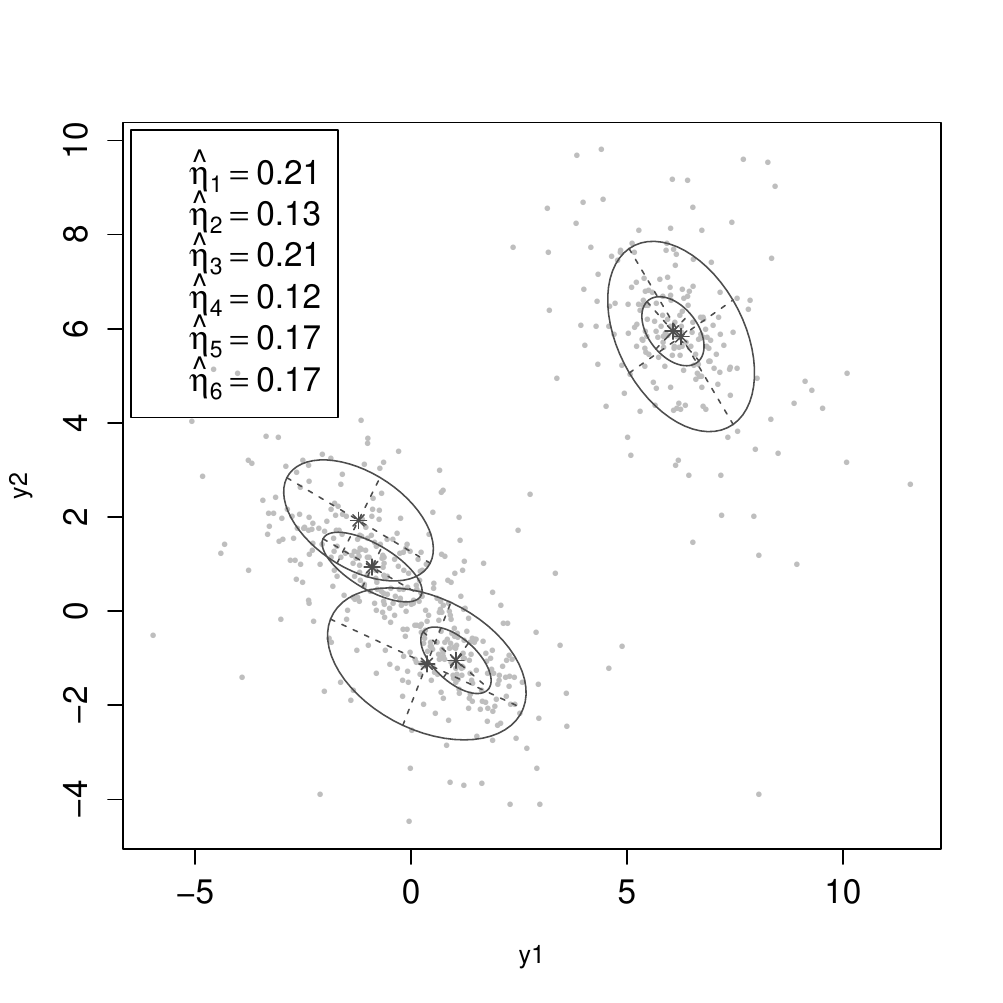} \\
Normal-IW & MOM-IW\\
\includegraphics[width=0.5\textwidth]{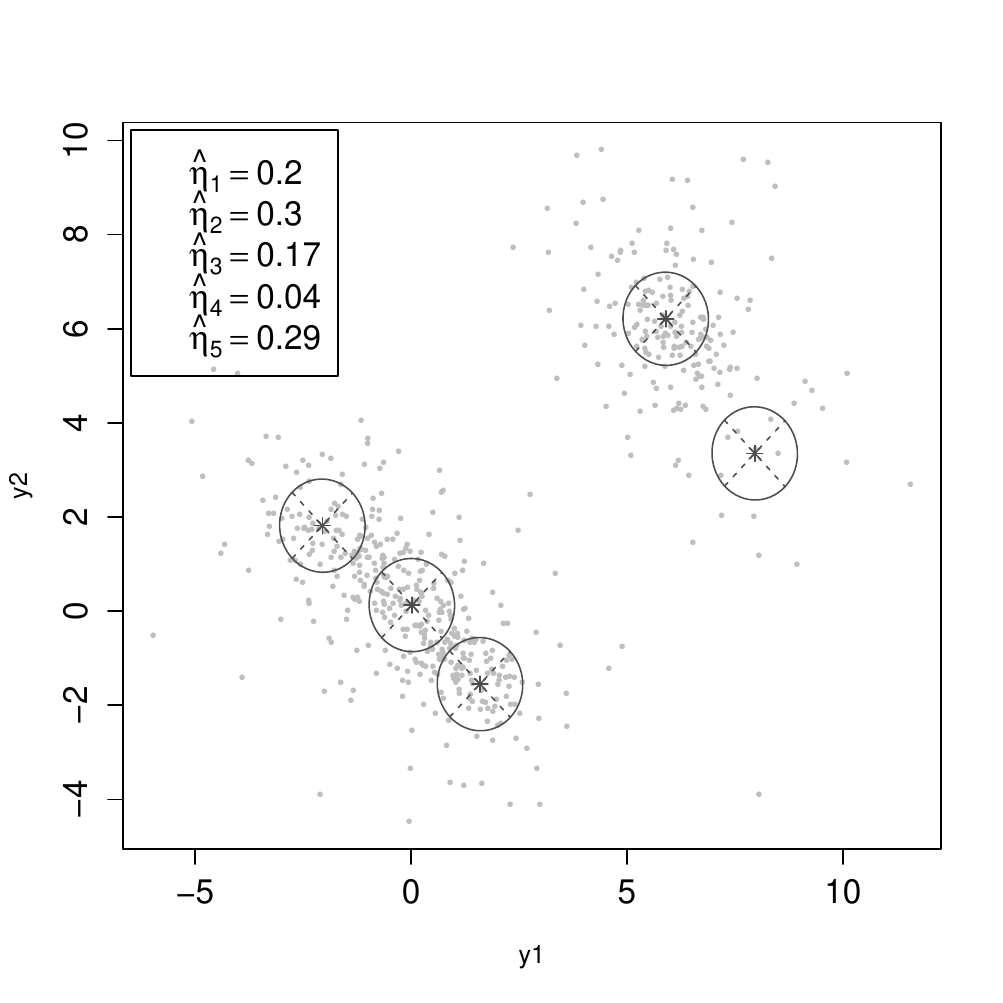} &
\includegraphics[width=0.5\textwidth]{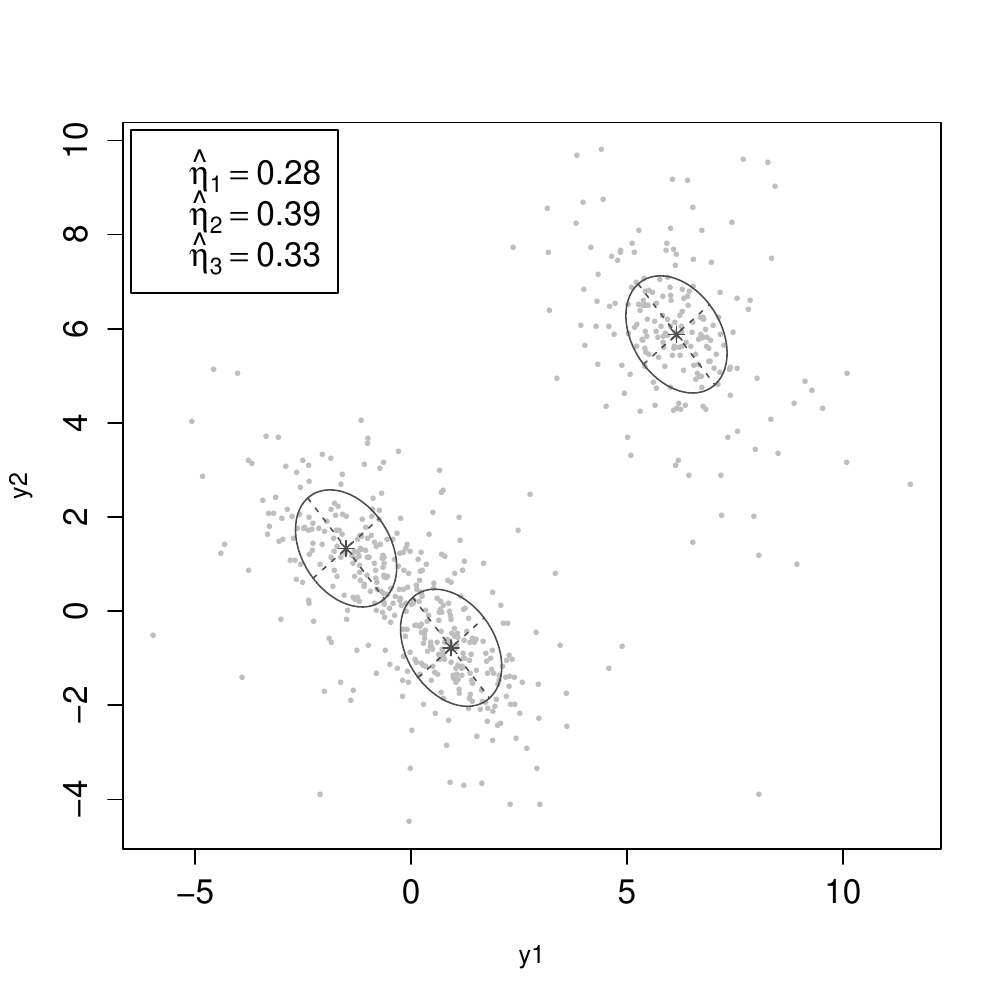}\\
\end{tabular}
\end{center}
\caption{Misspecified model. Estimated contours for BIC and sBIC (top left), AIC (top right),
Normal-IW (bottom left) and MOM-IW (bottom right). Points indicate the simulated data.}
\label{fig:mispec_model}
\end{figure}

Table \ref{tab:resultsmiss} summarises the results.
BIC and sBIC strongly favoured $\hat{k}=4$ components with unequal covariances,
AIC chose $\hat{k}=6$ components with unequal covariances,
and the Normal-IW prior placed most posterior probability on $k \in \{5,6\}$ with common covariances.
In contrast, our MOM-IW assigned posterior probability 1 (up to rounding) to $k=3$ with equal covariances.
To provide further insight Figure \ref{fig:mispec_model} shows the component contours for $\hat{k}$ under each method,
estimating $\hat{\bvartheta}_{\hat{k}}$ via maximum likelihood (AIC, BIC, sBIC) or posterior modes (Normal-IW, MOM-IW).
The means of the three MOM-IW components matched those of the true T components.
The BIC and sBIC approximated the two mildly-separated components with two normals centered roughly at (0,0),
whereas the AIC split the components even further.
The two extra components in the Normal-IW solution essentially account for heavy tails.
This example illustrates how by penalizing poorly-separated or low-weight components
NLPs may induce a form of robustness to model misspecification,
although we remark that this is a finite-sample effect and would eventually vanish as $n \rightarrow \infty$.


\subsection{Cytometry data}
\label{ssec:Cytometry}

We analysed the Graf-versus-Host flow cytometry data in \cite{brinkman2007high},
an experiment used for cell counting, {\it e.g.} to diagnose diseases.
The data contain $p=4$ variables called CD3, CD4, CD8b and CD8.
The study goal was to find cell subpopulations with positive CD3, CD4 and CD8b (CD3+/CD4+/CD8b+), i.e. high values in the first three variables.
Interestingly, the authors created a control sample designed not to contain any CD4+/CD8b+ cells.
Following the analysis in \cite{baudry2012combining},
we selected the $n=1,126$ cells in the control sample for which CD3 $>280$.

\begin{figure}[ht]
\begin{center}
\begin{tabular}{cc}
BIC & AIC/sBIC \\
\includegraphics[width=0.5\textwidth]{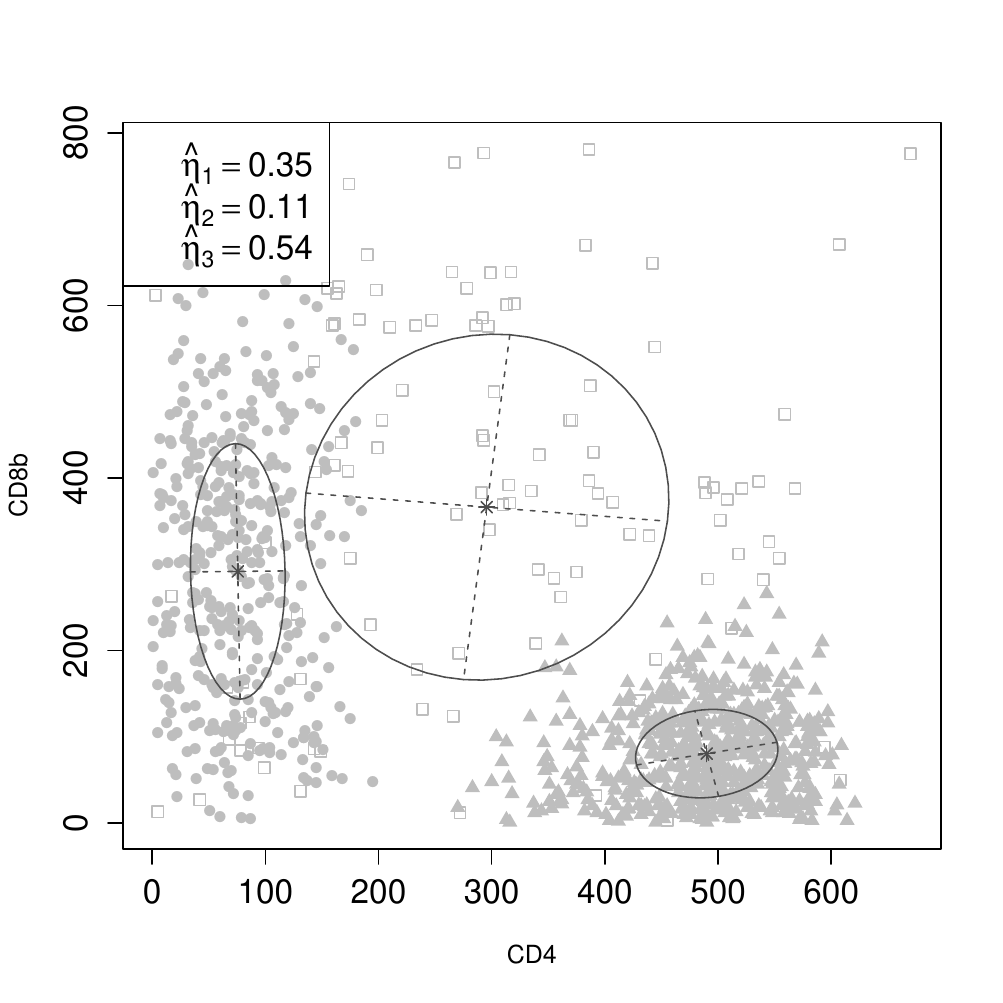} &
\includegraphics[width=0.5\textwidth]{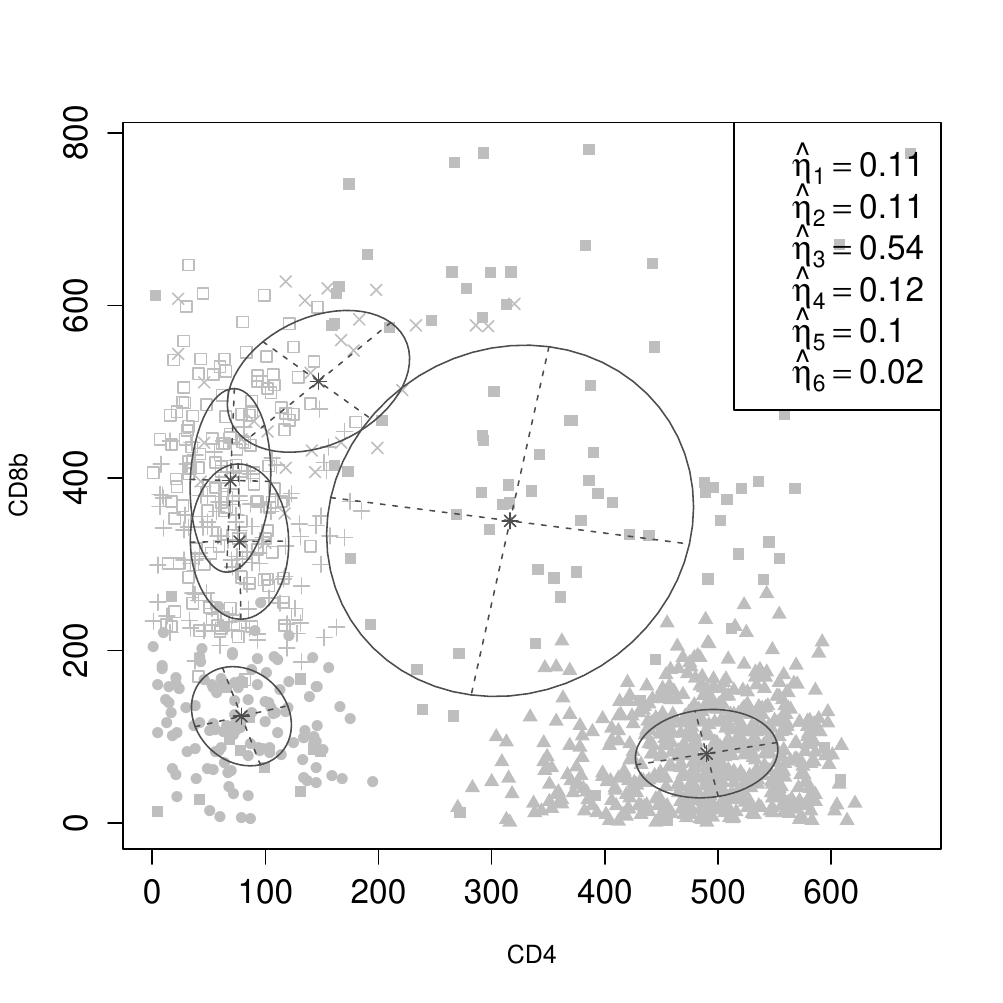} \\
Normal-IW & MOM-IW\\
\includegraphics[width=0.5\textwidth]{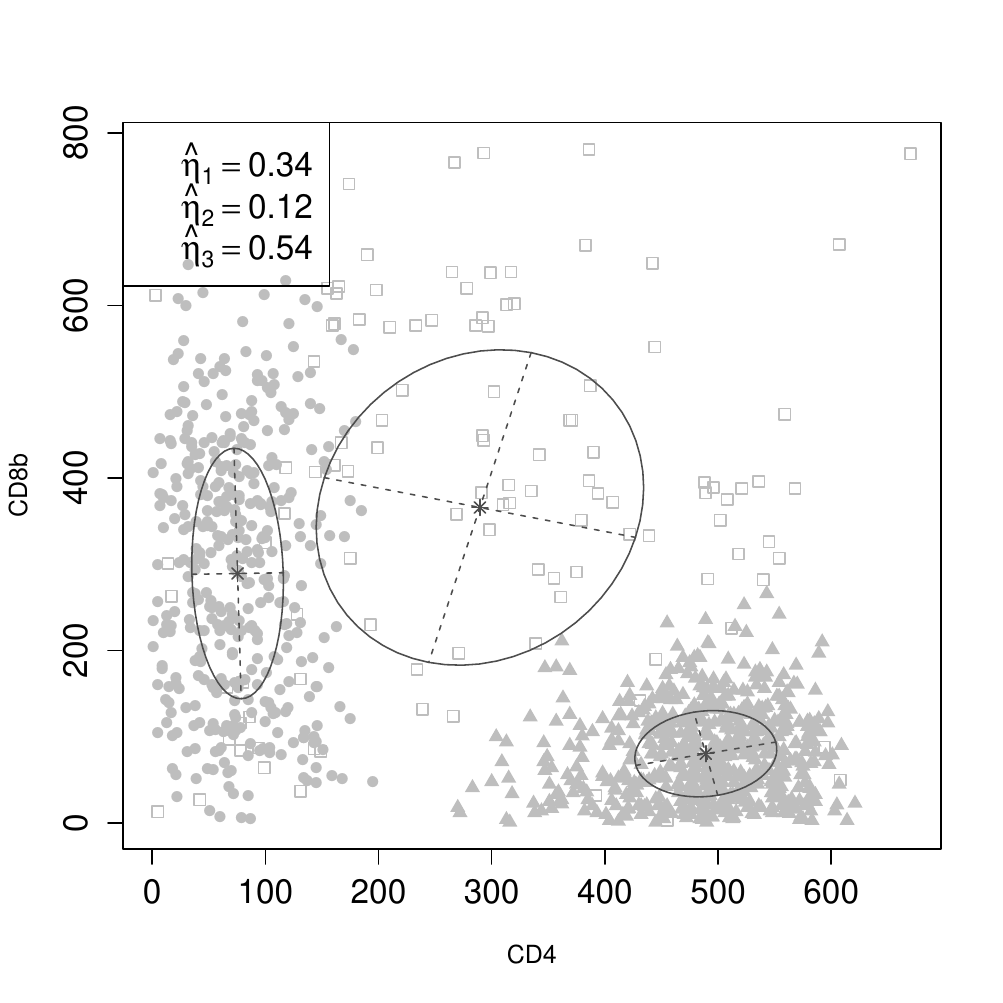} &
\includegraphics[width=0.5\textwidth]{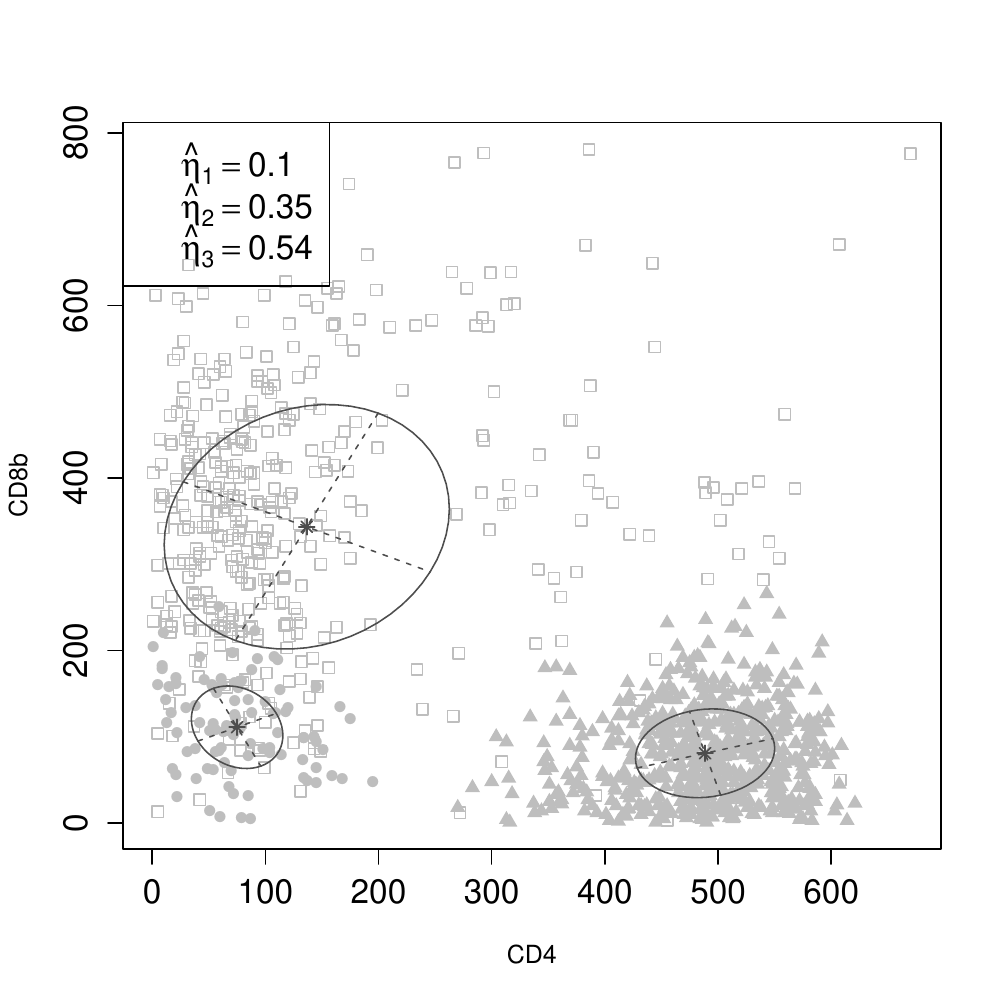}\\
\end{tabular}
\end{center}
\caption{Projection of the variables CD4 and CD8b for the Cytometry data-set, classification of observations
and contours using EM algorithm for BIC and AIC/sBIC (top), and under
Normal-IW and MOM-IW (bottom).}
\label{fig:cytometry}
\end{figure}

Figure \ref{fig:cytometry} plots (CD4,CD8b) values and the solution
chosen by BIC, AIC, sBIC, Normal-IW and MOM-IW.
The first four methods identified a CD4+/CD8b+ subpopulation that, as discussed,
is not there by design, whereas it was not present in the MOM-IW solution.
Intuitively, the spurious CD4+/CD8b+ cluster contains a few outlying observations, and
our MOM-IW penalises such a low-weight component.
These results illustrate the benefits of jointly penalising small weights and overlapping components.
See Table \ref{tab:Cytometry} for further details,
e.g. the Normal-IW and MOM-IW chose $k=3$ with 0.928 and 0.995 posterior probability, respectively.

\subsection{Old Faithful}
\label{ssec:oldfaithful}

We briefly describe this classical example to illustrate potential issues with poorly-separated components.
The results are in Table \ref{tab:Faithful} and Figure \ref{supfig:faithful}.
The Old Faithful is a cone-type geyser in the Yellowstone National Park.
We seek clusters in a dataset with $n=272$ eruptions recording their duration and the time to the next eruption (dataset faithful in R).
We considered up to $K=6$ Normal components either with equal or unequal covariance matrices.
Our MOM-IW selected $k=3$ equal-covariance components with 0.967 posterior probability.
The Normal-IW chose $k=4$ with 0.473 posterior probability,
this resulted from splitting an MOM-IW component in the lower-left corner into two.
The sBIC and BIC chose $\hat{k}=3$ components with roughly the similar location as the MOM-IW,
though their shapes were slightly different, whereas AIC returned $\hat{k}=4$.

\subsection{Fisher's Iris data}
\label{ssec:iris}

We present another classical dataset by \cite{fisher}
for the practical reason that there is a ground truth for the underlying number of subpopulations.
The data contain four variables ($p=4$) measuring the dimensions of $n=150$ iris flowers.
The plants are known to belong to $k^*=3$ species,
setosa, versicolor and virginica, each with 50 observations.
We compare the ability of the various methods to recover these three species in an unsupervised fashion.
We considered up to $K=6$ Normal components with either equal or unequal covariances.

Table \ref{tab:Iris} provides posterior model probabilities.
The BIC and sBIC supported $\hat{k}=2$ and $\hat{k}=4$ components with unequal covariances, respectively.
Upon inspection the BIC solution merged the versicolor and virginica species into a single component,
akin to its lack of sensitivity observed in Section \ref{ssec:syntheticexamples},
whereas the sBIC split the versicolor species into two components.
The AIC supported $\hat{k}=6$ with unequal covariances.
Both the Normal-IW and our MOM-IW  chose $\hat{k}=3$,
but the evidence under the former was weaker ($P^{L}(\mathcal{M}_{3}\mid \by)=0.81$ and $P(\mathcal{M}_{3}\mid \by)=1$ respectively).
Figure \ref{supfig:iris} shows the MOM-IW solution contours for the first two principal components
(accounting for 96.0\% of the variance), which closely resemble the three species.

\subsection{Comparison to overfitted mixtures}
\label{ssec:repulsive}

Table \ref{tab:many13} and Table \ref{tab:many23}
summarise the results from analysing the datasets from Sections \ref{ssec:mispec}-\ref{ssec:iris}
with overfitted mixtures and repulsive overfitted mixtures (respectively).
Here repulsion was induced by a pMOM penalty where $g$ is set to its default in Section \ref{ssec:priorelicitation}.
We set $k=6$ and report the posterior distribution of the number of empty components (with no assigned observations) from the MCMC output.
Note that $k=6$ implies overfitted mixtures
as our analyses in Sections \ref{ssec:mispec}-\ref{ssec:iris} suggested less than 6 components.
To assess sensitivity we tested prior parameter values $q=1$ (no shrinkage),
$q=0.01$ (satisfying \cite{judith} and \cite{gelman4}) and $3\times 10^{-8}$ (proposed by \cite{zoe}).
We observed little differences between overfitted and repulsive overfitted mixtures.
As expected, in general smaller $q$ led to less occupied components in the posterior,
except in the cytometry data where the posterior focused on 6 components for all $q$.
Note that $q=3\times 10^{-8}$ recovered the true $k^*=3$ in the misspecified example from Section \ref{ssec:mispec},
whereas this was not the case in the Iris and Cytometry data that truly contain 3 subpopulations.
The results for the faithful data matched those of the MOM-IW.

\begin{table}[ht]
\caption{Posterior distribution on the number of non-empty components $m=\sum_{j=1}^k \mbox{I}(n_j>0)$
in overfitted mixtures with common  $\Sigma_{j}=\Sigma$. The Misspecified,
Faithful, Iris and Cytometry data.}
\centering
\begin{tabular}{rrrrrrrrr}\\  \hline
& \multicolumn{6}{c}{$\hat{P}(m\mid \by,\mathcal{M}_{6})$} \\
& $m=1$ & $m=2$ & $m=3$ & $m=4$ & $m=5$ & $m=6$  \\ \hline \hline
&  &  & $q=1$ &  &  &   \\ \hline \hline
Misspecified & 0.00 & 0.00 & 0.00 & 0.00 & 0.07 & 0.93 \\
Faithful & 0.00 & 0.00 & 0.00 & 0.01 & 0.15 & 0.85 \\
Fisher's Iris & 0.00 & 0.99 & 0.01 & 0.00 & 0.00 & 0.00 \\
Cytometry & 0.00 & 0.00 & 0.00 & 0.00 & 0.00 & 1.00 \\ \hline \hline
&  &  & $q=0.01$ &  &  &   \\ \hline \hline
Misspecified & 0.00 & 0.00 & 0.03 & 0.35 & 0.56 & 0.06 \\
Faithful & 0.00 & 0.00 & 0.63 & 0.31 & 0.04 & 0.02 \\
Fisher's Iris & 0.00 & 1.00 & 0.00 & 0.00 & 0.00 & 0.00 \\
Cytometry & 0.00 & 0.00 & 0.00 & 0.00 & 0.00 & 1.00 \\
&  &  & $q=3.10^{-8}$ &  &  &   \\ \hline \hline
Misspecified & 0.00 & 0.00 & 0.95 & 0.00 & 0.00 & 0.05 \\
Faithful & 0.00 & 0.00 & 0.96 & 0.00 & 0.01 & 0.03 \\
Fisher's Iris & 0.00 & 1.00 & 0.00 & 0.00 & 0.00 & 0.00 \\
Cytometry & 0.00 & 0.00 & 0.00 & 0.00 & 0.00 & 1.00 \\ \hline
\end{tabular}
\label{tab:many13}
\end{table}

\subsection{Simulation with Binomial mixtures}
\label{ssec:binomials}

To assess the performance of MOM-Beta (default $a=1/2$, $g=7.11$) and Beta(1,1) priors as well as the BIC and sBIC,
we reproduced the Binomial mixture example used by \cite{drton2017bayesian} to illustrate the sBIC.
We generated 200 data sets of sample sizes $n=50$, 200 and 500 from a $k^{*}=4$ component Binomial mixture
with $L_{if}=30$ trials for all $i=1,\ldots,n$, equal component weights $\eta_{j}=1/4$ and
component-specific success probabilities $\theta_{j}=j/5$ for $j=1,\ldots,4$.
We computed the two sBIC versions $\overline{\text{sBIC}}$ and $\overline{\text{sBIC}}_{05}$.

\begin{figure}[ht]
\begin{center}
\begin{tabular}{cccc}
n=50 \\
\includegraphics[width=0.65\textwidth]{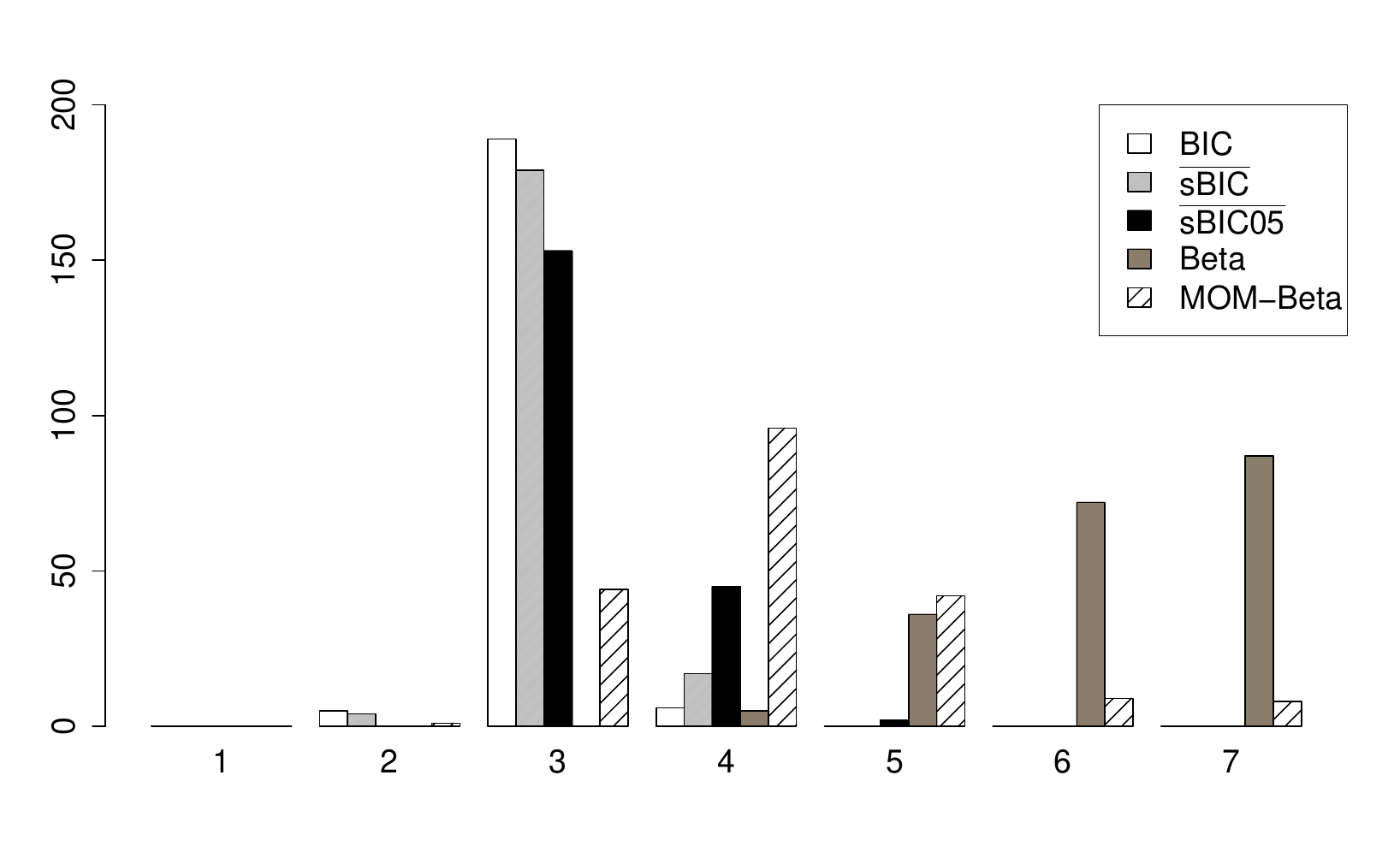} \\
n=200\\
\includegraphics[width=0.65\textwidth]{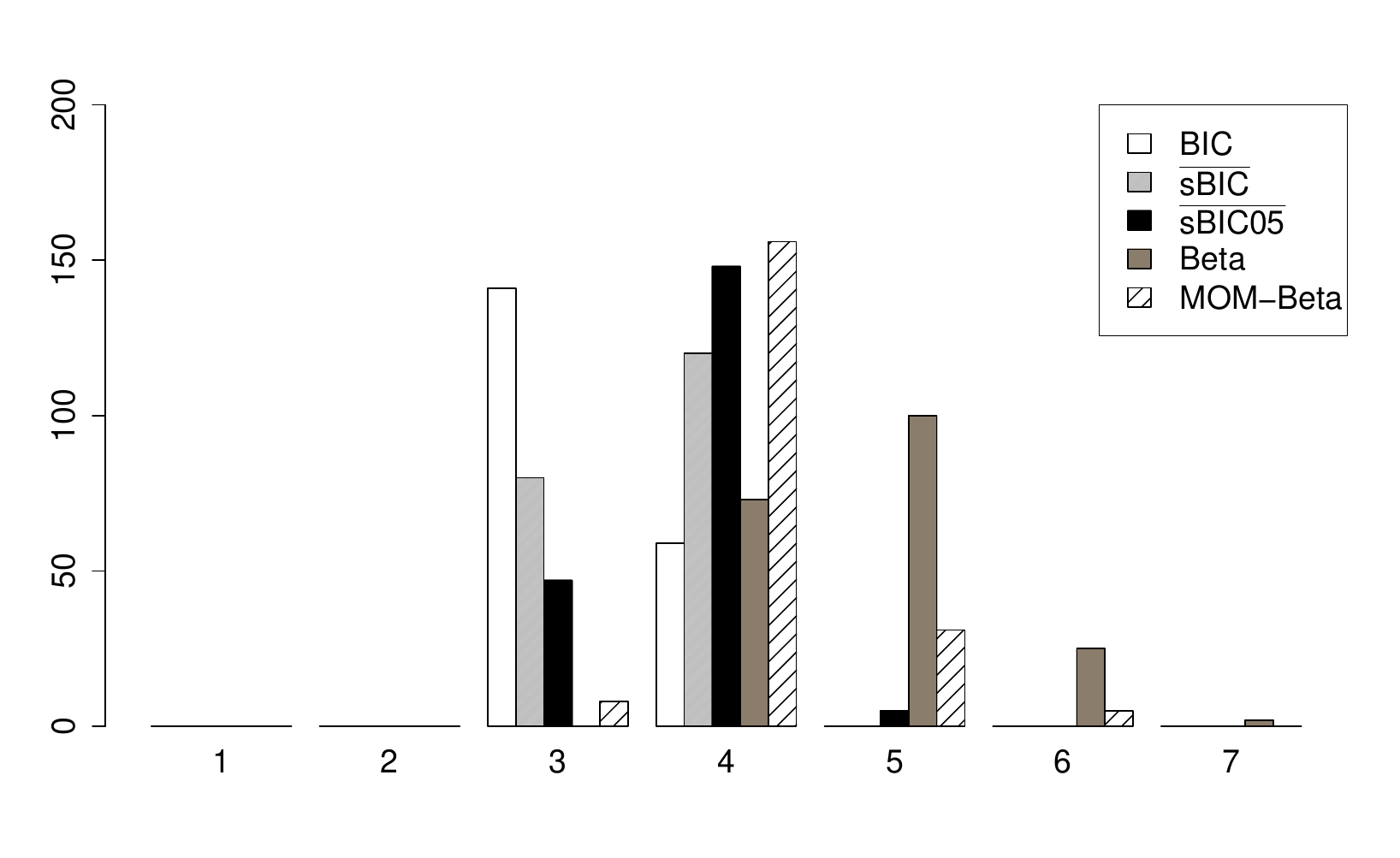} \\
n=500\\
\includegraphics[width=0.65\textwidth]{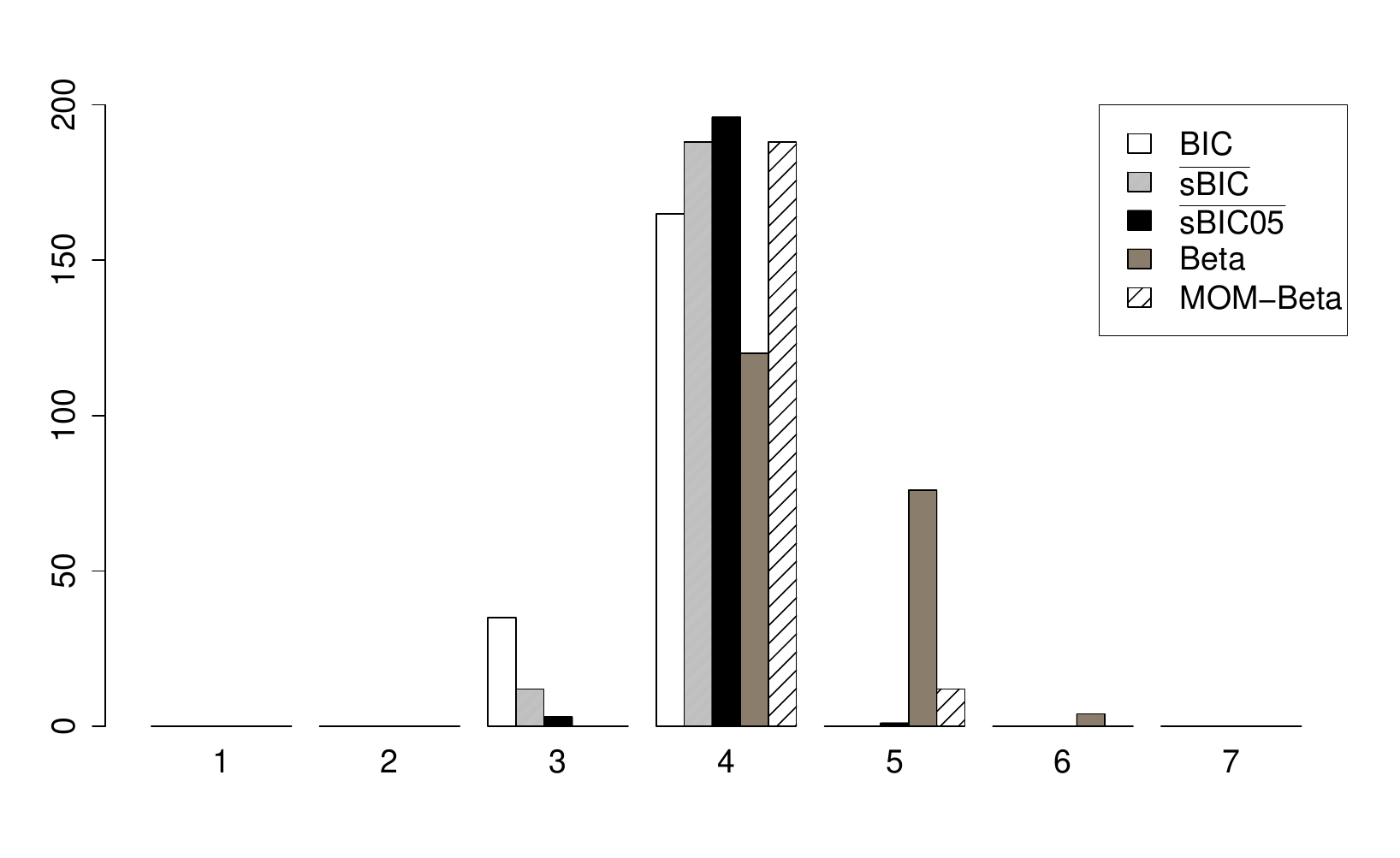} \\
\end{tabular}
\end{center}
\caption{Binomial mixture. Frequencies of $\hat{k}$ for BIC, $\overline{\text{sBIC}}$, $\overline{\text{sBIC}}_{05}$, Beta and MOM-Beta.
Results from 200 data sets with $n=50$, 200 and 500, $L = 30$ and $k^{*}=4$}
\label{fig:sBICBinomial}
\end{figure}

Figure \ref{fig:sBICBinomial} shows the results.
The two sBIC versions ameliorated the BIC's overpenalization as reported in \cite{drton2017bayesian},
whereas the Beta prior often returned too many components.
The proportion of correct model selections was generally highest for the MOM-Beta, particularly for smaller $n$
(roughly 50\% of the simulations when $n=50$, relative to 25\% for $\overline{\text{sBIC}}_{05}$).
To assess sensitivity Figure \ref{supfig:sensitivityBinomial1}
shows the results for alternative prior parameter settings $g=16.09$ and  $g=29.99$ discussed in Supplementary Section \ref{supplsec:sensMOMBeta}.
These larger $g$ values result in more informative priors that adversely affect inference,
reinforcing our recommendation for $g=7.11$.

\subsection{Political blog data}
\label{ssec:LCA_a}

We illustrate product Binomial mixtures using a dataset on $n=773$ USA political blogs from 2008 \citep{lda}.
Each blog provides word counts (how many times a given word was used).
To facilitate interpretation we combined similar words (e.g. america, american and americans, see Table \ref{tab:blogs2})
and selected the $p=234$ words with overall frequency above 100.
We fitted a product Binomial mixture $y_{if} \mid  z_i=j,\theta_{jf} \sim \text{Bin}(\theta_{jf}, L_i)$,
where $L_{i}$ is the total number of words in blog $i=1,...,773$.
We considered MOM-Beta and Beta priors, the BIC and AIC.
The MOM-Beta parameters were set to the default $(a,g)=(0.5,2)$
whereas as a local prior we chose the Beta$(1,1)$ (Section \ref{ssec:priorelicitation}).

\begin{figure}[ht]
\begin{center}
  \begin{tabular}{cc}
Component 1 & Component 2\\
\includegraphics[width=0.5\textwidth]{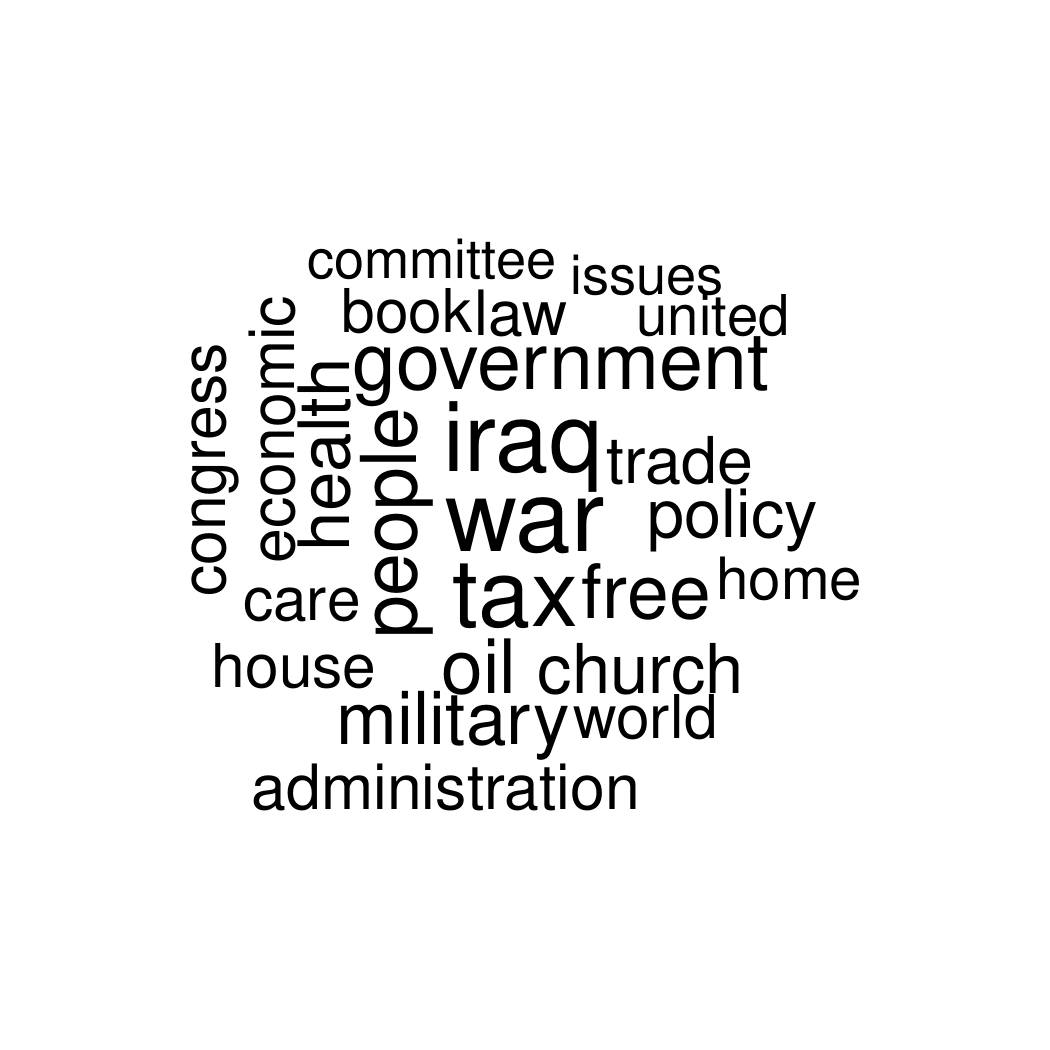} &
\includegraphics[width=0.5\textwidth]{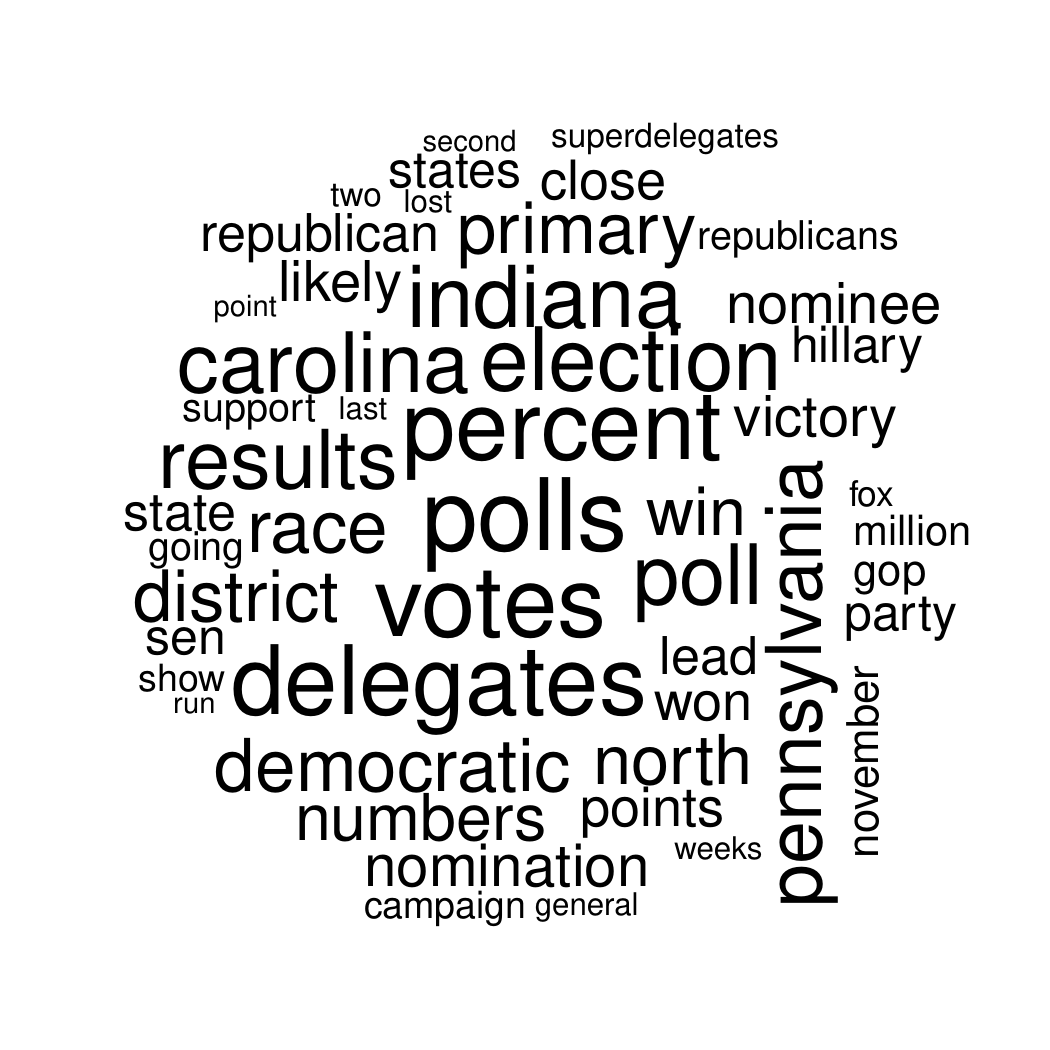}
\end{tabular}
\end{center}
\caption{Political blog data. Each word was assigned to its most probable component under a MOM-Beta prior. Word sizes based on $\chi^2$ residuals from cross-tabulating word frequency versus assigned component}
\label{fig:blogsw}
\end{figure}

The MOM-Beta and Beta selected $\hat{k}=2$ and $\hat{k}=4$ both with posterior probability one (up to rounding),
whereas BIC and AIC chose $\hat{k}=3$ and $\hat{k}=6$ respectively (Table \ref{tab:blogs}).
To assess the inferred components,
these data contain an independent labeling that classifies blogs either as liberal or conservative.
Figure \ref{supfig:postcNLP} displays the estimated cluster probabilities (Algorithm S\ref{alg:em1}) for liberal and conservative blogs.
Interestingly under the MOM-Beta prior conservative blogs fell mostly in Component 1.
Figure \ref{fig:blogsw} shows the most characteristic words for each MOM-Beta component
($\chi^2$ residual$>2$ when cross-tabulating word count versus assigned component).
For instance, ``war, iraq, tax, government" are representative of Component 1
whereas ``polls, votes, percent, delegates" are representative of Component 2.
In contrast,
under the Beta prior Components 2 and 4 showed a similar distribution for liberal and conservative blogs
and 
the clusters returned by the AIC did not show appreciable differences  between conservative or liberal blogs.

\section{Conclusions}
\label{sec:conclusions}

The use of NLPs for selecting mixture components
leads to solutions that balance parsimony and sensitivity,
and also facilitates interpretation in terms of well-separated subpopulations.
From a theoretical standpoint the formulation asymptotically enforces parsimony under
the wide class of generically identifiable mixtures, which we confirmed in finite $n$ examples.
As another practical issue defining prior dispersion is often regarded as an inconvenience,
here we showed how it can be advantageously calibrated to detect well-separated components resulting in multimodality.
We also showed that the computations required to implement NLPs are comparable to those for standard local priors and,
although not exploited here, they can easily be parallelized for multiple $k$.
In particular the ECP estimator provides a convenient strategy to estimate posterior model probabilities
for non-local and local priors, by utilizing readily available MCMC output.

Our examples showed that BIC may pathologically miss components, in some instances even with large $n$.
The AIC and local priors tended to add spurious components in simulations and in datasets with known subgroup structure.
The sBIC showed a mixed behavior that was similar to the BIC in some instances
and to local priors or the AIC in others.
Overfitted and repulsive overfitted mixtures
proved useful in several examples, but prior parameters and the choice of $k$ need to be carefully calibrated.
Our framework can also be sensitive to prior specification,
but as we illustrated there are natural defaults based on multi-modality and minimal
informativeness that result in competitive behaviour.
Despite the resemblance between NLPs and repulsive overfitted mixtures we emphasise that the former require
not only a repulsive force but also penalising low-weight components,
and that this was found to improve inference in our examples.
A related intriguing observation was that, by penalizing poorly-separated and low-weight components,
NLPs showed robustness to model misspecification in an example.
It would be interesting to study the combined effect of NLPs and robust likelihoods.

\section*{Acknowledgments}

David Rossell was partially funded by the NIH grant R01 CA158113-01,
EPSRC First Grant EP/N011317/1 and RyC-2015-18544, Plan Estatal PGC2018-101643-B-I00,
and Ayudas Fundaci\'on BBVA a equipos de investigaci\'on cient\'ifica 2017.

\bibliographystyle{plainnat}
\bibliography{bibliop}

\clearpage

\section*{Supplementary material}


\newtheorem{lemma2}[theorem]{Lemma}

\setcounter{table}{0}
\setcounter{figure}{0}
\setcounter{equation}{0}
\setcounter{section}{0}

\renewcommand{\thetable}{S\arabic{table}}
\renewcommand\thefigure{S\arabic{figure}}
\renewcommand{\theequation}{S\arabic{equation}}%
\renewcommand{\thesection}{S\arabic{section}}%

\renewcommand{\theHtable}{Supplement.\thetable}
\renewcommand{\theHfigure}{Supplement.\thefigure}
\renewcommand{\theHequation}{Supplement.\theequation}
\renewcommand{\theHsection}{Supplement.\thesection}

\RestyleAlgo{boxruled}
\LinesNumbered
\SetAlgorithmName{Algorithm S\hspace{-1.5mm}}{}

\section{Conditions A1-A4 in \cite{judith}}
\label{supplsec:cond_judith}

For convenience we reproduce verbatim Conditions  A1-A4 in \cite{judith}, adjusted to the notation we used in this
paper. Their Condition A5 is trivially satisfied by our $\bm{\eta} \sim \mbox{Dir}(q)$ prior,
hence is not reproduced here.
Recall that we defined $p_{k^*}^*(\by)=p(\by \mid \bvartheta_{k^*}^*,\mathcal{M}_{k^*})$ to be the data-generating truth.

We denote $\Theta_k^*=\{\bvartheta_k\in \Theta_k; p(\by \mid \bvartheta_{k},\mathcal{M}_{k}) = p_{k^*}^*(\by)\}$ and
let $\log(p(\by \mid \bvartheta_k,\Mk))$ be the log-likelihood calculated at $\bvartheta_k$.
Denote $F_{0}(g)=\int p(\by \mid \bvartheta_{k^*}^*,\mathcal{M}_{k^*})g(\by)d\by$ where $g(\cdot)$
is a probability density function, denote by $\text{Leb}(A)$ the Lebesgue measure of a set $A$ and let
$\nabla p(\by\mid \btheta)$ be the vector of derivatives of $p(\by\mid \btheta)$
with respect to $\btheta$, and $\nabla^{2} p(\by\mid \btheta)$ be the second derivatives
with respect to $\btheta$. Define for $\epsilon\geq 0$

\begin{align*}
\bar{p}(\by\mid \btheta)&= \sup_{|\btheta^{l}-\btheta|\leq \epsilon} p(\by\mid \btheta^{l}), &
\underline{p}(\by\mid \btheta)&= \inf_{|\btheta^{l}-\btheta|\leq \epsilon} p(\by\mid \btheta^{l})
\end{align*}

We now introduce some notation that is useful to characterize $\Theta_k^*$, following \cite{judith}.
Let $\bw=(w_{i})_{i=0}^{k^*}$ with $0=w_{0}<w_{1}<...<w_{k^{*}}\leq k$
be a partition of $\{1,...,k\}$. For all $\bvartheta_{k} \in \Theta_{k}$ such that $p(\by \mid \bvartheta_{k},\mathcal{M}_{k}) = p_k^*(\by)$ there exists
$\bw$ as defined above such that, up to a permutation
of the labels,

\begin{align*}
\forall i=1,...,k^*, \;\;\; \btheta_{w_{i-1}+1}=...=\btheta_{w_{i}}=\btheta_{i}^*,
\;\;\; \eta(i)=\sum_{j=w_{i-1}+1}^{w_{i}}\eta_{j}=\eta_{i}^*,  \;\;\; \eta_{w_{k^*+1}}=...=\eta_{k}=0.
\end{align*}

In other words, $I_{i}=\{w_{i-1}+1,...,w_{i}\}$ represents the cluster of components in $\{1,...,k\}$ having the same parameter
as $\btheta_{i}^*$. Then define the following parameterisation  of $\bvartheta_{k} \in \Theta_{k}$
(up to permutation)

\begin{align*}
\boldsymbol{\iota_{w}}&=\left((\btheta_{j})_{j=1}^{w_{k^{*}}},(r_{i})_{i=1}^{k^{*}-1},(\eta_{j})_{j=w_{k^{*}+1}}^{k}\right) \in \mathbb{R}^{pw_{k^{*}}+k^{*}+k-w_{k^{*}}-1},
  & r_{i}&=\eta(i)-\eta_{i}^{*}, & i&=1,...,k^{*},
\end{align*}

and

\begin{align*}
\boldsymbol{\varpi_{w}}&=\left((f_{j})_{j=1}^{w_{k^{*}}} ,\btheta_{w_{k^{*}}+1},...,\btheta_{{k}}\right), &  f_{j}&=\dfrac{\eta_{j}}{\eta(i)}, \;\;\;
\text{when} \;\;\; j \in I_{i}=\{w_{i-1}+1,...,w_{i}\},
\end{align*}

note that for $p(\by \mid \bvartheta_{k^*}^*,\mathcal{M}_{k^*})$

\begin{align*}
\boldsymbol{\iota_{w}}^{*}=(\btheta_{1}^*,...,\btheta_{1}^*,\btheta_{2}^*,...,\btheta_{2}^*,...,\btheta_{k^*}^*,...,\btheta_{k^*}^*
,0...0...0)
\end{align*}

where $\btheta_{i}^{*}$  is repeated $w_{i}-w_{i-1}$ times in the above vector for any $\boldsymbol{\varpi_{w}}$. Then we parameterize $(\boldsymbol{\iota_{w}},\boldsymbol{\varpi_{w}})$,
so that $p(\by \mid
\bvartheta_{k},\mathcal{M}_{k})=p(\by \mid
(\boldsymbol{\iota_{w}},\boldsymbol{\varpi_{w}}),\mathcal{M}_{k})$ and we denote

$\nabla p(\by \mid
(\boldsymbol{\iota_{w}^*},\boldsymbol{\varpi_{w}}),\mathcal{M}_{k})$ and
$\nabla^{2} p(\by \mid
(\boldsymbol{\iota_{w}^*},\boldsymbol{\varpi_{w}}),\mathcal{M}_{k})$ the first and second derivatives of
$p(\by \mid
(\boldsymbol{\iota_{w}},\boldsymbol{\varpi_{w}}),\mathcal{M}_{k})$ with respect to $\boldsymbol{\iota_{w}}$
and computed at $\bvartheta_{k^*}^*=(\boldsymbol{\iota_{w}^*},\boldsymbol{\varpi_{w}})$. We also denote by
$P^L (\cdot \mid \by, \Mk )$ the posterior distribution using a LP.  \\

\textbf{Conditions}\\

\begin{enumerate}[label=\bfseries A\arabic*,leftmargin=*]
\item {\it $L_1$ consistency.}
For all $\epsilon= (\log n)^{e}/\sqrt{n}$ with $e\geq 0$ as $n \rightarrow \infty$
$$
P^L \left( \int \left| p(\bz \mid \bvartheta_k,\Mk) - p_k^*(\bz) \right| d\bz > \epsilon
  \mid \by, \Mk \right) \rightarrow 0
$$
in probability with respect to $p(\by \mid
\bvartheta_{k^*}^*,\mathcal{M}_{k^*})$.

\item {\it Regularity.} The component density $p(\by\mid \btheta)$
indexed by a parameter $\btheta\in\Theta$ is three times differentiable and regular in the sense that
for all $\btheta\in\Theta$ the Fisher information matrix associated with $p(\by\mid \btheta)$ is positive definite at
$\btheta$. Denote $\nabla^{3} p(\by\mid \btheta)$ the array whose components are

\begin{align*}
\frac{\partial^{3} p(\by\mid \btheta)}{\partial_{\btheta_{i1}}\partial_{\btheta_{i2}}\partial_{\btheta_{i3}}}
\end{align*}

For all $i \leq k^*$, there exists $\epsilon>0$ such that

\begin{align*}
F_{0}&\left(\dfrac{\bar{p}(\by\mid \btheta_{i}^{*})^{3}}{\underline{p}(\by\mid \btheta_{i}^{*})^{3}}\right)<\infty, &
F_{0}&\left(\dfrac{  \sup_{|\btheta-\btheta^{*}|\leq \epsilon} |\nabla p(\by|\btheta)|^{3}}{\underline{p}(\by\mid \btheta_{i}^{*})^{3}}\right)<\infty, &
F_{0}&\left(\frac{|p(\by|\btheta_{i}^{*})|^{4}}{(p(\by \mid
\bvartheta_{k^*}^*,\mathcal{M}_{k^*}))^{4}}\right)<\infty,
\end{align*}

\begin{align*}
F_{0}&\left(\dfrac{  \sup_{|\btheta-\btheta^{*}|\leq \epsilon} |\nabla^{2} p(\by\mid \btheta)|^{2}}{\underline{p}(\by\mid \btheta_{i}^{*})^{2}}\right)<\infty, &
F_{0}&\left(\dfrac{  \sup_{|\btheta-\btheta^{*}|\leq \epsilon} |\nabla^{3} p(\by\mid \btheta)|^{2}}{\underline{p}(\by\mid \btheta_{i}^{*})}\right)<\infty.
\end{align*}

Assume also that for all $i=1,...,k^*$,  $\btheta_{i}^* \in \text{int}(\Theta^{k})$ the interior of $\Theta^{k}$.

\item {\it Integrability.} There exists $\Theta^{k^*}\subset\Theta^{k}$ satisfying
 $\text{Leb}(\Theta^{k^*})>0$ and for all
$i\leq k^*$

\begin{align*}
d(\btheta_{i}^*,\Theta^{k^*})=  \inf_{\btheta \in \Theta^{k^*}} |\btheta-\btheta_{i}^*|>0
\end{align*}

and such that for all $\btheta \in \Theta^{k^*}$,

\begin{align*}
F_{0}&\left(\dfrac{p(\by\mid \btheta)^{4}}{(p(\by \mid
\bvartheta_{k^*}^*,\mathcal{M}_{k^*}))^{4}}\right)<\infty, &
F_{0}&\left(\dfrac{p(\by\mid \btheta)^{3}}{\underline{p}(\by\mid \btheta_{i}^{*})^{3}}\right)<\infty, &
& \forall i\leq k^*.
\end{align*}

\item {\it Stronger identifiability.}

For all  $\bw$ partitions of $\{1,...,k\}$ as defined above, let $\bvartheta_{k} \in \Theta_{k}$ and write
$\bvartheta_{k}$ as $(\boldsymbol{\iota_{w}},\boldsymbol{\varpi_{w}})$; then

\begin{align*}
&(\boldsymbol{\iota_{w}}-\boldsymbol{\iota_{w}}^{*})^{'}\nabla p(\by \mid
(\boldsymbol{\iota_{w}^{*}},\boldsymbol{\varpi_{w}}),\mathcal{M}_{k}) + \dfrac{1}{2}(\boldsymbol{\iota_{w}}-\boldsymbol{\iota_{w}}^{*})^{'}
\nabla^{2}p(\by \mid
(\boldsymbol{\iota_{w}^{*}},\boldsymbol{\varpi_{w}}),\mathcal{M}_{k})(\boldsymbol{\iota_{w}}-\boldsymbol{\iota_{w}}^{*})=0 \Leftrightarrow \\
&\forall i\leq k^{*}, r_{i}=0 \;\; \text{and} \;\; \forall j\in I_{i} \;\; f_{j}(\btheta_{j}-\btheta_{j}^{*})=0, \;\; \forall i\geq w_{k^{*}}+1, \;\; p_{i}=0.
\end{align*}

\end{enumerate}

Assuming also that if $\btheta \notin \{\btheta_{1},...,\btheta_{k}\}$ then for all functions $h_{\btheta}$
which are linear combinations of derivatives of $p(\by\mid \btheta)$ of order less than or equal to 2 with respect to
$\btheta$, and all functions $h_{1}$ which are also linear combinations of derivatives of the
$p(\by\mid \btheta_{j})$'s $j=1,2,..,k$ and its derivatives of order less or equal to 2, then $\alpha h_{\btheta} + \beta h_{1}=0 $
if and only if $\alpha h_{\btheta} = \beta h_{1}=0 $.\\

Extension to non compact spaces: If $\subset\Theta^{k}$ is not compact then we also assume
that for all sequences $\btheta_{n}$ converging to a point in $\partial \Theta^{k}$ the frontier
of $\Theta^{k}$, considered as a subset of $\Re \cup \left\{-\infty, \infty\right\}^{p}$, $p(\by\mid \btheta_{n})$
converges pointwise either to a degenerate function or to a proper density $p(\cdot)$ such that $p(\cdot)$
is linearly independent of any null combinations of $p^{*}(\by\mid \btheta_{i})$, $\nabla p^{*}(\by\mid \btheta_{i})$ and $\nabla^{2} p^{*}(\by\mid \btheta_{i})$, $i=1,...,k^*$.

\section{Prior normalization constant for MOM priors}
\label{sec:priornorm_mompriors}

\begin{lemma2}
Let $p(\bzeta \mid \Mk)$ be as in (\ref{priorgeneral}). Then
\begin{align}
C_k=
\sum_{s \in S_k}^{} \left( \prod_{l=1}^{pk} \kappa_s \right)  \sum_{v(1,2)=0}^{1} \ldots \sum_{v(k-1,k)=0}^{1} (-1)^{\sum_{i<j}^{} v(i,j)}
 \left( \prod_{l=1}^{pk}\prod_{m=1}^{pk} \frac{b_{lm}^{s_{l,m}}(v)}{s_{l,m}!} \right)
\nonumber
\end{align}
where $\kappa_s= E^L( \theta_{11}^{\sum_{m=1}^{pk} s_{lm} + s_{ml}})$,
$S_k= \left\{ (s_{1,1},s_{1,2},\ldots,s_{pk,pk}): \sum_{l=1}^{pk} \sum_{m=1}^{pk} s_{l,m} = k(k-1)/2 \right\}$
with non-negative integers $0 \leq s_{l,m} \leq k(k-1)/2$,
and $b_{lm}(v)$ is the $(l,m)$ element of the $pk \times pk$ matrix $B_v$ given by
$$
\begin{cases}
b_{ll}= \frac{1}{2}(k-1) - \sum_{i<j}^{} v(i,j) \mbox{, }l=1+p(i-1),\ldots,pi \\
b_{lm}= b_{ml}= -\frac{1}{2} + \sum_{i<j}^{} v(i,j) \mbox{, } (1+p(i-1),1+p(j-1)),\ldots,(pi,pj)
\end{cases}.
$$
\label{lemma:generic_momprior_constant}
\end{lemma2}

We remark that Lemma \ref{lemma:generic_momprior_constant} holds for any $p^L(\bzeta)$ composed by independent and identically-distributed $p^L(\zeta_{jf})$
and that $\kappa_s$ requires raw moments up to order $k(k-1)/2$, which can be pre-computed.
Specifically, for the MOM-Beta prior in \eqref{mombetaprior}
$$
\kappa_s=
\left(\frac{\Gamma(g)}{\Gamma(ag)}\right)^{pk}\frac{\Gamma \left(ag + \sum_{m=1}^{pk} s_{lm} + s_{ml} \right)}{\Gamma \left( g + \sum_{m=1}^{pk} s_{lm} + s_{ml}  \right)}.
$$
For certain common settings with MOM-IW and MOM-Beta priors,  Lemma \ref{lemma:generic_momprior_constant} can be simplified,
see Corollaries \ref{corollary1}-\ref{corollary3}.

\begin{table}
\begin{center}
  \begin{tabular}{|c|c|cc|} \hline
        & & \multicolumn{2}{c|}{$\mbox{Var}(\theta_{jf}-\theta_{j'f} \mid \mathcal{M}_2)$} \\
    $p$ & Default $g$ & Default $g$ & $g=2$ \\ \hline
  1 & 7.11 & 1/6 & 0.400 \\
  2 & 4.39 & 1/6 & 0.283 \\
  3 & 3.54 & 1/6 & 0.244 \\
  4 & 3.13 & 1/6 & 0.225 \\
  5 & 2.89 & 1/6 & 0.213 \\
  6 & 2.74 & 1/6 & 0.206 \\
  7 & 2.63 & 1/6 & 0.200 \\
  8 & 2.55 & 1/6 & 0.196 \\
  9 & 2.48 & 1/6 & 0.193 \\
  10 & 2.43 & 1/6 & 0.190 \\
  11 & 2.39 & 1/6 & 0.188 \\
  12 & 2.36 & 1/6 & 0.186 \\
  13 & 2.33 & 1/6 & 0.185 \\
  14 & 2.31 & 1/6 & 0.183 \\
  15 & 2.29 & 1/6 & 0.182 \\
  16 & 2.27 & 1/6 & 0.181 \\
  17 & 2.25 & 1/6 & 0.180 \\
  18 & 2.24 & 1/6 & 0.180 \\
  19 & 2.23 & 1/6 & 0.179 \\
  20 & 2.21 & 1/6 & 0.178 \\
\hline
\end{tabular}
\end{center}
\caption{Default $g$ in MOM-Beta$(0.5g,0.5g)$ prior giving $\mbox{Var}(\theta_{jf}-\theta_{j'f}\mid \mathcal{M}_2)=1/6$ as a function of $p$, and variance for $g=2$. For $p>20$ we recommend the default $g=2$}
\label{tab:defaultg_mombeta}
\end{table}

\section{Prior variance under MOM-Beta priors}
\label{sec:priorvar_mombeta}

    Let $p(\btheta \mid \Mk)$ be the MOM-Beta prior in \eqref{mombetaprior}
    and $p^L(\btheta \mid \Mk)$ be the Beta prior in \eqref{eq:betaprior} with parameters $(a_L,g_L)=(0.5,2)$.
    Our suggested defaults are setting $a=0.5$ and $g$ \eqref{mombetaprior} such that
$$
\mbox{Var}_p(\theta_{jf} - \theta_{j'f} \mid \Mk)= \mbox{Var}_{p^L}(\theta_{jf} - \theta_{jf'} \mid \Mk)= \frac{1}{6}.
$$
It is in principle possible to find such $g$ by noting that, due to $(\theta_{jf},\theta_{jf'})$ being exchangeable a priori we have
$E_p(\theta_{jf} - \theta_{j'f} \mid \Mk)=0$. Hence $\mbox{Var}_p(\theta_{jf} - \theta_{j'f} \mid \Mk)=$
\begin{align}
\int  \frac{(\theta_{jf} - \theta_{j'f})^2}{C_k} \prod_{1\leq i < j \leq k} \sum_f (\theta_{if}-\theta_{jf})^2
\prod_{j=1}^{k}\prod_{f=1}^{p}\text{Beta}(\theta_{jf}; ag,(1-a)g) d \btheta
\nonumber
\end{align}
and one may expand the product within the integral as a sum involving products of polynomials.
As illustration for $p=1$ and $a=0.5$ simple algebra shows
\begin{align}
  \mbox{Var}_p(\theta_{j1} - \theta_{j'1} \mid \mathcal{M}_2)= 1.5 \frac{2+g}{(1+g)(3+g)}
  \nonumber \\
  \mbox{Var}_p(\theta_{j1} - \theta_{j'1} \mid \mathcal{M}_3)= 2 \frac{4+g}{(3+g)(5+g)}
\nonumber
\end{align}
and so on for larger $k$.
For instance if $k=2$ then the desired defaults are $(a,g)=(0.5,7.11)$, and for $k=3$ they would be $(a,g)=(0.5,8.08)$.
This strategy grows tedious for larger $k$ as the expressions become less manageable and even evaluating $C_k$ becomes non-trivial for general $p$.

\begin{figure}[ht]
\begin{center}
\includegraphics[width=0.5\textwidth]{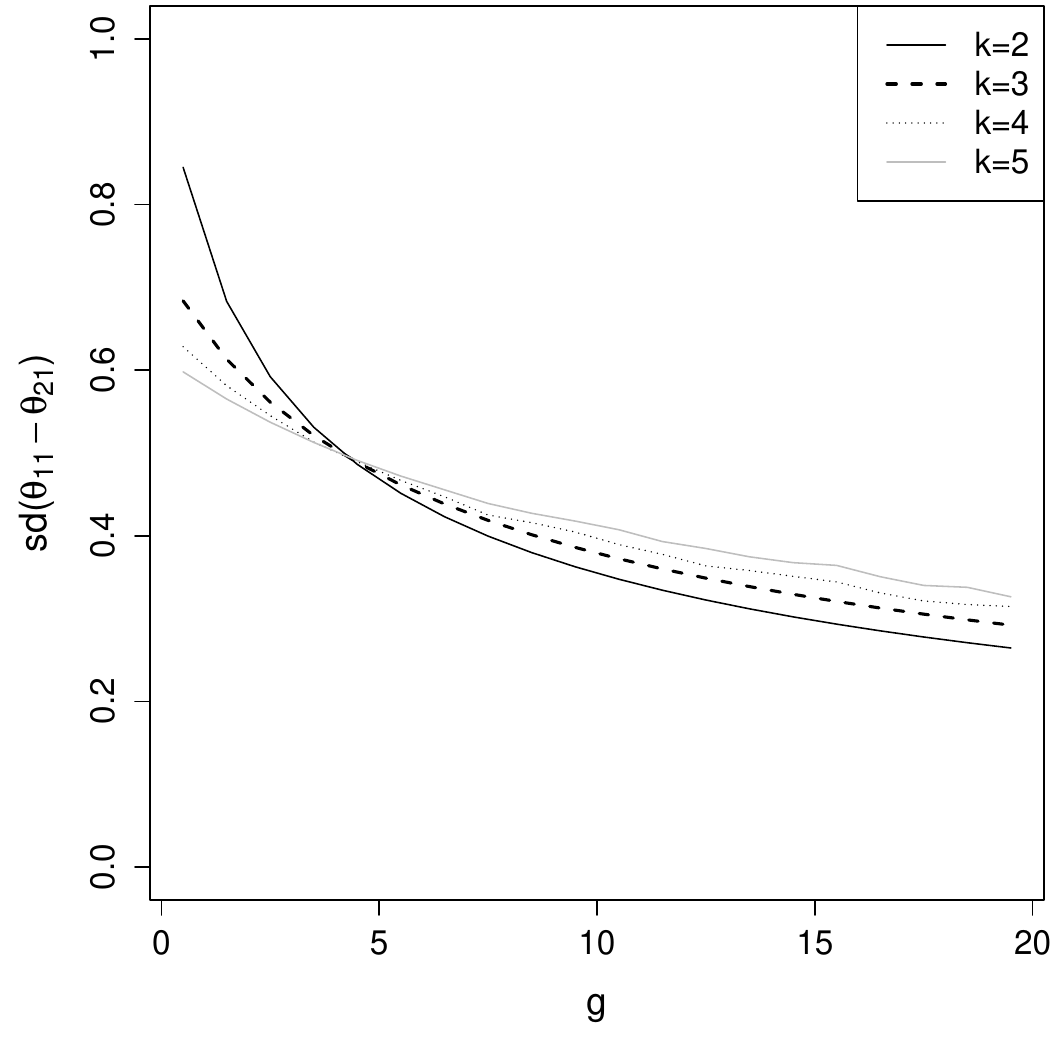}
\end{center}
\caption{Prior standard deviation $\mbox{SD}(\theta_{11}-\theta_{21} \mid \Mk)$ under a MOM-Beta$(0.5g,0.5(1-g))$}
\label{supfig:priorsd_mombeta}
\end{figure}

We adopt the simpler alternative strategy of focusing on the $k=2$ case,
this is analogous to the approach to set the MOM-Normal prior dispersion in continuous mixtures
and in practice we observed that the target prior variance becomes more robust to $g$ as $k$ grows
(Figure \ref{supfig:priorsd_mombeta}).
That is, our strategy is
\begin{align}
 \frac{1}{6}=  \mbox{Var}_p(\theta_{jf} - \theta_{j'f} \mid \mathcal{M}_2)=
2 \left[ E_p(\theta_{11}^2 \mid \mathcal{M}_2) - E_p(\theta_{11} \theta_{21} \mid \mathcal{M}_2) \right],
\nonumber
\end{align}
where the right-hand side follows from exchangeability.
We first state the result and subsequently outline its derivation.
\begin{align}
  E_p(\theta_{11}^2 \mid \mathcal{M}_2)&=
                                         \frac{ag+1}{2p(1-a)}
      \left[ \frac{(ag+2)(ag+3)}{(g+2)(g+3)} + \frac{a(ag+1) + 2(p-1)a(1-a)}{g+1} - \frac{2a(ag+2)}{g+2} \right]
  \nonumber \\
  E_p(\theta_{11} \theta_{21} \mid \mathcal{M}_2)&=
                                                   \frac{a}{p (1-a)}
                                                   \left[\frac{(ag+1)(ag+2)}{g+2} - \frac{(ag+1)^2}{(g+1)} + (p-1)a(1-a) \right]
\nonumber
\end{align}
For the particular case $(a,g)=(0.5,2)$ one obtains
\begin{align}
  \mbox{Var}_p(\theta_{jf} - \theta_{j'f} \mid \mathcal{M}_2)=
  \frac{1}{p} \left( \frac{2}{5} + \frac{p-1}{6} \right),
\nonumber
\end{align}
which clearly converges to $1/6$ as $p \rightarrow \infty$.
Table \ref{tab:defaultg_mombeta} lists the default $g$ giving $\mbox{Var}_p(\theta_{jf} - \theta_{j'f} \mid \mathcal{M}_2)=1/6$ for various $p$.
For $p=20$ setting $g=2$ already gives $\mbox{Var}_p(\theta_{jf} - \theta_{j'f} \mid \mathcal{M}_2)=0.178$
and thus for $p > 20$ we recommend $(a,g)=(0.5,2)$.

    The remainder of this section outlines the derivation of $E_p(\theta_{11}^2 \mid \mathcal{M}_2)$
    and $E_p(\theta_{11} \theta_{21} \mid \mathcal{M}_2)$.
\begin{align}
  E_p(\theta_{11}^2 \mid \mathcal{M}_2)=
  \int \frac{\theta_{11}^2 (\theta_{11}-\theta_{21})^2}{C_2} \prod_{j=1}^2 \mbox{Beta}(\theta_{j1};ag,(1-a)g) d\theta_{j1}
  \nonumber \\
  + \int \frac{\theta_{11}^2}{C_2} \sum_{f=2}^{p} (\theta_{1f} - \theta_{2f})^2 \prod_{j=1}^2 \prod_{f=1}^{p} \mbox{Beta}(\theta_{jf};ag,(1-a)g) d\btheta
  \nonumber \\
  = \frac{a (ag+1)}{C_2 (g+1)} \left[ \frac{(ag+2)(ag+3)}{(g+2)(g+3)} + \frac{a(ag+1)}{g+1} - \frac{2a(ag+2)}{(g+2)} \right]
  - \frac{a(ag+1) \tilde{C}_2}{C_2(g+1)},
\nonumber
\end{align}
where the right-hand side follows from the moments of a Beta distribution
and $\tilde{C}_2$ is the prior normalization constant for $p-1$ variables.
Using that $C_2=2p a(1-a)/(g+1)$ and $\tilde{C}_2=2(p-1) a(1-a)/(g+1)$ gives the desired expression for $E_p(\theta_{11}^2 \mid \mathcal{M}_2)$.
Similarly,
\begin{align}
  E_p(\theta_{11} \theta_{21} \mid \mathcal{M}_2)=
  \int \frac{\theta_{11}\theta_{21}(\theta_{11}^2+\theta_{21}^2-2 \theta_{11}\theta_{21})}{C_2}
  \prod_{j=1}^2 \mbox{Beta}(\theta_{j1};ag,(1-a)g) d\theta_{j1}
  \nonumber \\
  +  \int \frac{\theta_{11} \theta_{21}}{C_2} \sum_{f=2}^p (\theta_{1f} - \theta_{2f})^2 \prod_{j=1}^2 \prod_{f=1}^p \mbox{Beta}(\theta_{jf}; ag, (1-a)g) d\theta_{jf}
  \nonumber \\
  \frac{2 a^2 (ag+1)}{C_2 (g+1)} \left[ \frac{ag+2}{g+2} - \frac{ag+1}{g+1} \right] - \frac{a^2 \tilde{C}_2}{C_2}.
  \nonumber
\end{align}
The result is obtained by plugging in $C_2=2p a(1-a)/(g+1)$, $\tilde{C}_2=2(p-1) a(1-a)/(g+1)$ and rearranging terms.

\section{Proofs}
\label{supplsec:proofs}

\subsection{Auxiliary lemmas to prove Theorem 1}

We state and prove two auxiliary lemmas that will be used in the proof of Theorem 1.

\begin{lemma2}
Let  $p(\bvartheta_k \mid \Mk)=d_{\theta}(\btheta)p^L(\btheta \mid  \Mk) p(\bm{\eta} \mid  \Mk)$
be the MOM prior in \eqref{priorN2NL}.
Then $p(\bvartheta_k \mid \Mk)=
\tilde{d}_{\theta}(\btheta) \tilde{p}^L(\btheta \mid  \Mk)
p(\bm{\eta} \mid  \Mk)$,
where $\tilde{d}_{\theta}(\btheta) \leq c_k$ for some finite $c_k$,
$$\tilde{p}^L(\bvartheta_k \mid  \Mk)=
\prod_{j=1}^{k}N\left(\bmu_{j}; {\bf 0},(1+\epsilon)g A_{\Sigma}\right),$$
and $\epsilon \in (0,1)$ is an arbitrary constant.
\label{lem:bounded_mom}
\end{lemma2}

\noindent
\textbf{Proof.} The MOM prior
has an unbounded penalty
\begin{equation*}
d_{\theta}(\btheta)=\frac{1}{C_k}\prod_{1\leq i < j \leq k}\left((\bmu_{i}-\bmu_{j})^{'} A_{\Sigma}^{-1}(\bmu_{i}-\bmu_{j})/g\right)^{t},
\end{equation*}
however we may rewrite $d_{\theta}(\btheta) p^L(\btheta \mid  \Mk)$
\begin{align}
\notag
=&d_{\theta}(\btheta)
\prod_{j=1}^{k}N\left(\bmu_{j}; {\bf 0},g A_{\Sigma}
\right)\dfrac{N\left(\bmu_{j}; {\bf 0},(1+\epsilon)g A_{\Sigma}
\right)}{N\left(\bmu_{j}; {\bf 0},(1+\epsilon)g A_{\Sigma}
\right)},\\
=&
\tilde{d}_{\theta}(\btheta)
\prod_{j=1}^{k}N\left(\bmu_{j}; {\bf 0},(1+\epsilon)g A_{\Sigma}\right),
\end{align}
where $\epsilon \in (0,1)$ is an arbitrary constant and
$\tilde{d}_{\theta}(\btheta)=$
\begin{align*}
&d_{\theta}(\btheta)
\prod_{j=1}^{k} \frac{N\left(\bmu_{j}; {\bf 0},g
  A_{\Sigma}\right)}{N\left(\bmu_{j}; {\bf 0},(1+\epsilon)g
  A_{\Sigma}\right)}=
d_{\theta}(\btheta)
\prod_{j=1}^{k}(1+\epsilon)^{1/2}\exp\left\{-\dfrac{1}{2}\dfrac{\epsilon
  \bmu_{j}^{'}A_{\Sigma}^{-1} \bmu_{j}}{(1+\epsilon)
g}\right\}.
\end{align*}
The fact that $\tilde{d}_{\theta}(\btheta)$ is bounded follows from the fact that
the product term is a Normal kernel and hence bounded,
whereas $d_{\theta}(\btheta)$ can only become unbounded when
$\bmu_jA_{\Sigma}^{-1}\bmu_j \rightarrow \infty$ for some $j$,
but this polynomial increase is countered by the exponential decrease in
$\exp\left\{-\dfrac{1}{2}\dfrac{\epsilon
  \bmu_{j}^{'}A_{\Sigma}^{-1} \bmu_{j}}{(1+\epsilon)
g}\right\}$.
\qed

\begin{lemma2}
Let $d_{\vartheta}(\bvartheta_k) \in [0,c_k]$ be a bounded continuous function in
$\bvartheta_k$, where $c_k$ is a finite constant. Let
\begin{equation*}
g_{k}(\by)=E^{L}(d_{\vartheta}(\bvartheta_k) \mid \by, \Mk)=
\int d_{\vartheta}(\bvartheta_k) p^{L}(\bvartheta_k\mid \by,\Mk)d\bvartheta_k.
\end{equation*}
If for any $\epsilon>0$ we have that
$P^{L}(d_{\vartheta}(\bvartheta)>\epsilon \mid  \by,\Mk)
\stackrel{P}{\longrightarrow} 0$
then $g_k(\by) \stackrel{P}{\longrightarrow} 0$.
Alternatively, if there exists some $d_k^*>0$ such that for any $\epsilon>0$
$P^{L}(|d_{\vartheta}(\bvartheta_k) - d_k^*| >\epsilon \mid  \by,\Mk)
\stackrel{P}{\longrightarrow} 1$,
then $g_k(\by) \stackrel{P}{\longrightarrow} d_k^*$.
\label{lem:meanconv_boundedpen}
\end{lemma2}

\noindent
\textbf{Proof.} Consider the case
$P^{L}(d_{\vartheta}(\bvartheta)>\epsilon \mid  \by,\Mk)
\stackrel{P}{\longrightarrow} 0$, then $g_{k}(\by)=$

\begin{align}
&\int_{d_{\vartheta}(\bvartheta_k)<\epsilon}
            d_{\vartheta}(\bvartheta_k)
            p^{L}(\bvartheta_k\mid \by,\Mk)d\bvartheta_k
+ \int_{d_{\vartheta}(\bvartheta_k)>\epsilon}
            d_{\vartheta}(\bvartheta_k) p^{L}(\bvartheta_k\mid \by,\Mk)d\bvartheta_k
\notag \\
&\leq \epsilon P^{L}(d_{\vartheta}(\bvartheta_k)<\epsilon\mid \by,\Mk)+
c_k P^{L}(d_{\vartheta}(\bvartheta_k)>\epsilon\mid \by,\Mk)
\notag \\
&\leq \epsilon + c_k  P^{L}(d_{\vartheta}(\bvartheta_k)>\epsilon\mid \by,\Mk)
\stackrel{P}{\longrightarrow} \epsilon,
\notag
\end{align}
where $\epsilon>0$ is arbitrarily small.
Hence $g_k(\by) \stackrel{P}{\longrightarrow} 0$.

Next consider the case $P^{L}(|d_{\vartheta}(\bvartheta_k) - d_k^*| >\epsilon \mid  \by,\Mk)
\stackrel{P}{\longrightarrow} 1$.
Then
\begin{align}
g_k(\by) &>
\int_{d_{\vartheta}(\bvartheta_k)>d_k^* - \epsilon} d_{\vartheta}(\bvartheta_k) p^{L}(\bvartheta_k\mid \by)d\bvartheta_k
\nonumber \\
&\geq (d_k^*-\epsilon)
P^L\left(d_{\vartheta}(\bvartheta_k)>d_k^* - \epsilon \mid  \by,\Mk\right)
\stackrel{P}{\longrightarrow} d_k^*-\epsilon,
\nonumber
\end{align}
and analogously $g_k(\by)=$
\begin{align}
&\int_{d_{\vartheta}(\bvartheta_k)<d_k^* + \epsilon}
            d_{\vartheta}(\bvartheta_k)
            p^{L}(\bvartheta_k\mid \by,\Mk)d\bvartheta_k
+ \int_{d_{\vartheta}(\bvartheta_k)>d_k^* + \epsilon}
            d_{\vartheta}(\bvartheta_k) p^{L}(\bvartheta_k\mid \by,\Mk)d\bvartheta_k
\notag \\
&\leq (d_k^*+\epsilon) +
c_k P^{L}(d_{\vartheta}(\bvartheta_k)>d_k^*+\epsilon\mid \by,\Mk)
\stackrel{P}{\longrightarrow} d_k^* + \epsilon,
\notag
\end{align}
for any $\epsilon>0$ and hence $g_k(\by) \stackrel{P}{\longrightarrow} d_k^*$.
\qed

\subsection{Proof of Theorem \ref{thm:sparsity}}

\noindent
\underline{Part (i).} The result is straightforward.
Briefly,
$p(\by\mid \Mk)=$
\begin{align}
&\int
d_{\vartheta}(\bvartheta_k)
p(\by\mid \bvartheta_{k},\Mk) p^L(\bvartheta_k \mid \Mk)
d\bvartheta_k
\notag \\
&=\int
d_{\vartheta}(\bvartheta_k)
\frac{p(\by\mid \bvartheta_{k},\Mk) p^L(\bvartheta_k \mid \Mk)}{p^{L}(\by\mid \Mk)}p^{L}(\by\mid \Mk)
d\bvartheta_k \notag \\
&=p^{L}(\by\mid \Mk)
  E^L(d_{\vartheta}(\bvartheta_k) \mid \by),
\nonumber
\end{align}
as desired.

\vspace{3mm}
\noindent
\underline{Part (ii). Posterior concentration.}
We need to prove that
$$P^L\left( |d_{\vartheta}(\bvartheta_k) - d_k^*| > \epsilon \mid
  \by,\Mk \right) \rightarrow 0$$
where $d_k^*=0$ for $k>k^*$ and $d_k^*=d_{\vartheta}(\bvartheta_k^*)$
for $k \leq k^*$.
Intuitively, the result follows from the fact that by the $L_1$ posterior concentration
assumption B1 the posterior concentrates on the KL-optimal model
$p_k^*(\by)$, but for generically identifiable mixtures this
corresponds to parameter values satisfying $d(\bvartheta_k)=0$ if $k>k^*$ and $d(\bvartheta_k)>0$
if $k \leq k^*$.

More formally, let $A_k$ be the set of $\bvartheta_k \in \Theta_k$ defining
$p_k^*(\by)$,
{\it i.e.} minimizing KL divergence between the data-generating
$p(\by \mid \bvartheta_{k^*}^*, \mathcal{M}_{k^*})$
and $p(\by \mid \bvartheta_k, \Mk)$.
Consider first the overfitted model case $k>k^*$,
then generic identifiability gives that
$$A_k= \left\{\bvartheta_k \in \Theta_k: \eta_j=0 \mbox{ for some }
  j=1,\ldots,k \mbox{ or } \btheta_i=\btheta_j \mbox{ for some } i
  \neq j \right\}.$$
This implies that for all $\bvartheta_k \in A_k$ we have that
$d_{\vartheta}(\bvartheta_k)=0$
and also that the $L_1$ distance
$$
l(\bvartheta_k)= \int_{}^{}\!\, \left| p_k^*(\by) - p(\by \mid
  \bvartheta_k,\Mk) \right| d\by= 0.
$$
Thus $d_{\vartheta}(\bvartheta_k)>0 \Rightarrow \bvartheta_k \not\in
A_k \Rightarrow l(\bvartheta_k)>0$.
Given that by assumption $p(\by \mid \bvartheta_k,\Mk)$ and
$d_{\vartheta}(\bvartheta_k)$ are continuous in $\bvartheta_k$,
for all $\epsilon'>0$ there is an $\epsilon>0$ such that
$d_{\vartheta}(\bvartheta_k)>\epsilon'$ implies
$l(\bvartheta_k)> \epsilon$ and hence that the probability of the
former event must be smaller.
That is,
\begin{align}
P^L\left(d_{\vartheta}(\bvartheta_k)>\epsilon' \mid \by,\Mk \right)
\leq P^L(l(\bvartheta_k)>\epsilon \mid \by,\Mk)
\nonumber
\end{align}
and the right hand side converges to 0 in probability for an arbitrary $\epsilon$
by Condition B1, proving the result for the case $k>k^*$.

The proof for the $k \leq k^*$ case proceeds analogously. Briefly,
when $k \leq k^*$ generic identifiability gives that
$A_k=\{ \bvartheta_k^* \}$ is a singleton with positive weights
$\eta_j^*>0$ for all $j=1,\ldots,k$ and $\btheta_i^* \neq \btheta_j^*$
for $i \neq j$.
Thus $d_k^*=d_{\vartheta}(\bvartheta_k^*)>0$.
By continuity of $p(\by \mid \bvartheta_k,\Mk)$ and
$d_{\vartheta}(\bvartheta_k)$ with respect to $\bvartheta_k$ this
implies that for all $\epsilon'>0$ there exists an $\epsilon>0$ such
that
$|d_{\vartheta}(\bvartheta_k)-d_k^*|>\epsilon' \Rightarrow l(\bvartheta_k)>\epsilon$,
and thus that
\begin{align}
P^L \left( |d_{\vartheta}(\bvartheta_k)-d_k^*|>\epsilon' \mid  \by,\Mk \right) \leq
P^L \left( l(\bvartheta_k) > \epsilon \mid \by,\Mk \right),
\nonumber
\end{align}
where the right hand side converges to 1 in probability by Condition
B1, proving the result.

\vspace{3mm}
\noindent
\underline{Part (ii). Convergence of $E^L(d_{\vartheta}(\bvartheta_k) \mid  \by)$}

Consider first the case where $d_{\vartheta}(\bvartheta_k) \in [0,c_k]$ is bounded
below some finite constant $c_k$.
Then Part (ii) above and Lemma \ref{lem:meanconv_boundedpen} below give that
\begin{align}
E^L\left(d_{\vartheta}(\bvartheta) \mid \by,\Mk \right) \stackrel{P}{\longrightarrow} 0
&\mbox{, for } k>k^* \nonumber \\
E^L\left(d_{\vartheta}(\bvartheta) \mid \by,\Mk \right) \stackrel{P}{\longrightarrow} d_k^*>0
&\mbox{, for } k\leq k^*
\label{eq:boundedpen_meanconv}
\end{align}
as we wished to prove.
Next consider the MOM prior case $d_{\vartheta}(\bvartheta)=$
$$
d_{\eta}(\bm{\eta}) \frac{1}{C_k} \prod_{1\leq i<j \leq k}^{} \left( (\bmu_i-\bmu_j)' A_{\Sigma}^{-1} (\bmu_i-\bmu_j) \right),
$$
where $d_{\eta}(\bm{\eta})$ is bounded by assumption.
From Lemma \ref{lem:bounded_mom}
\begin{align}
E^L\left(d_{\vartheta}(\bvartheta) \mid \by,\Mk \right)=
\int_{}^{}\!\, \tilde{d}_{\theta}(\btheta) d_{\eta}(\bm{\eta})
\frac{p(\by \mid  \bvartheta_k,\Mk) \tilde{p}(\bvartheta_k\mid \Mk)}{p^L(\by\mid \Mk)}
\frac{\tilde{p}^L(\by\mid \Mk)}{\tilde{p}^L(\by\mid \Mk)} d\bvartheta_k
\nonumber \\
=\frac{\tilde{p}^L(\by\mid \Mk)}{p^L(\by\mid \Mk)}
\int_{}^{}\!\, \tilde{d}_{\theta}(\btheta) d_{\eta}(\bm{\eta})
\tilde{p}^L(\bvartheta_k \mid  \by,\Mk) d\bvartheta_k,
\label{eq:mom_meanconv}
\end{align}
where $\tilde{d}_{\theta}(\btheta) d_{\eta}(\bm{\eta})$ is bounded and hence by Part (ii)
and Lemma \ref{lem:meanconv_boundedpen} the integral in \eqref{eq:mom_meanconv}
converges to 0 in probability when $k>k^*$ and to a non-zero finite constant when $k \leq k^*$.
Therefore it suffices to show that $\tilde{p}^L(\by\mid \Mk)/p^L(\by\mid \Mk)$ is bounded in probability,
as this would then immediately imply the desired result \eqref{eq:boundedpen_meanconv}.
From Lemma \ref{lem:bounded_mom} $\tilde{p}^L(\by\mid \Mk)=$
\begin{align}
&\int_{}^{}\!\, p(\by \mid  \bvartheta_k,\Mk) \tilde{p}^L(\bvartheta_k\mid \Mk) d\bvartheta_k= \nonumber \\
&\int_{}^{}\!\, p(\by \mid  \bvartheta_k,\Mk) p^L(\bvartheta_k\mid \Mk) \frac{\tilde{p}^L(\bvartheta_k\mid \Mk)}{p^L(\bvartheta_k\mid \Mk)} d\bvartheta_k= \nonumber \\
&\int_{}^{}\!\,  p(\by \mid  \bvartheta_k,\Mk) p^L(\bvartheta_k\mid \Mk) \prod_{j=1}^{k} \frac{N(\bmu_j;{\bf 0},(1+\epsilon)g \Sigma_j)}{N(\bmu_j;{\bf 0},g \Sigma_j)} d\bvartheta_k
\nonumber \\
&\int_{}^{}\!\,  p(\by \mid  \bvartheta_k,\Mk) p^L(\bvartheta_k\mid \Mk) \frac{1}{(1+\epsilon)^{kp/2}} \exp \left\{ \frac{1}{2g} \sum_{j=1}^{k} \bmu_j'A_{\Sigma}^{-1}\bmu_j \frac{\epsilon}{1+\epsilon} \right\} d\bvartheta_k
\nonumber \\
&
=\frac{p^L(\by \mid \Mk)}{(1+\epsilon)^{kp/2}}
E^L \left( \exp \left\{ \frac{1}{2g} \sum_{j=1}^{k} \bmu_j'A_{\Sigma}^{-1}\bmu_j \frac{\epsilon}{1+\epsilon} \right\} \mid \by,\Mk \right)
\nonumber \\
&\geq
\frac{p^L(\by \mid  \Mk)}{(1+\epsilon)^{kp/2}},
\label{eq:mom_ratio}
\end{align}
thus $\tilde{p}^L(\by\mid \Mk)/p^L(\by\mid \Mk) \geq \frac{1}{(1+\epsilon)^{kp/2}}$.
From \eqref{eq:mom_meanconv} this implies that when $k\leq k^*$
we obtain
$E^L\left(d_{\vartheta}(\bvartheta) \mid \by,\Mk \right) \stackrel{P}{\longrightarrow} d_k^*>0$.
Further, by Condition B3 the $E^L()$ term in \eqref{eq:mom_ratio} is bounded above in probability
when $k>k^*$, implying that
$E^L\left(d_{\vartheta}(\bvartheta) \mid \by,\Mk \right) \stackrel{P}{\longrightarrow} 0$.
\qed

\vspace{3mm}
\noindent
\underline{Part (iii).}

By assumption $p(\bm{\eta}\mid \Mk)=\mbox{Dir}(\bm{\eta};q) \propto d_{\eta}(\bm{\eta}) \mbox{Dir}(\bm{\eta};q-r)$,
where $d_{\eta}(\bm{\eta})= \prod_{j=1}^{k} \eta_j^r$
and $q>1$, $q-r<1$.
Consider the particular choice $q-r<\mbox{dim}(\Theta)/2$ and without loss of generality let $k^*+1,\ldots,k$ be
the labels for the spurious components.
Theorem 1 in \cite{judith} showed that under the assumed A1-A4
and a further condition A5 trivially satisfied by $p^L(\bm{\eta}\mid \Mk)=\mbox{Dir}(\bm{\eta};q-r)$ the corresponding posterior distribution
of the spurious weights concentrates around 0, specifically
\begin{align}
P^L \left( \sum_{j=k^*+1}^{k} \eta_j > n^{-\frac{1}{2} + \tilde{\epsilon}} \mid  \by,\Mk \right) \rightarrow 0
\label{eq:boundweights_rousseau}
\end{align}
in probability for all $\tilde{\epsilon}>0$ as $n \rightarrow \infty$.
Now, the fact that the geometric mean is smaller than the arithmetic mean gives that
$$
(k-k^*) \left( \prod_{j=k^*+1}^{k} \eta_j \right)^{\frac{1}{k-k^*}} \leq \sum_{j=k^*+1}^{k} \eta_j,
$$
and thus
\begin{align}
P^L\left(\sum_{j=k^*+1}^{k} \eta_j > n^{-\frac{1}{2} + \tilde{\epsilon}} \mid  \by,\Mk \right) \geq
\nonumber \\
P^L \left( (k-k^*) \left( \prod_{j=k^*+1}^{k} \eta_j \right)^{\frac{1}{k-k^*}} > n^{-\frac{1}{2} + \tilde{\epsilon}} \mid  \by,\Mk \right)=
\nonumber \\
P^L \left( \prod_{j=k^*+1}^{k} \eta_j^r > \frac{1}{(k-k^*)^r} n^{-\frac{r(k-k^*)}{2} + \epsilon} \mid  \by,\Mk \right),
\label{eq:boundweights_nlp}
\end{align}
where $\epsilon=r(k-k^*)\tilde{\epsilon}$ is a constant.
Thus \eqref{eq:boundweights_rousseau} implies that \eqref{eq:boundweights_nlp} also converges to 0 in probability.
Finally, given that  by assumption $d_{\vartheta}(\bvartheta)= d_{\theta}(\btheta) d_{\eta}(\bm{\eta}) \leq c_k \prod_{j=k^*+1}^{k} \eta_j^r$
we obtain
\begin{align}
P^L \left( d_{\vartheta}(\bvartheta) > n^{-\frac{r(k-k^*)}{2} + \epsilon} \mid  \by,\Mk \right) \leq
P^L \left( \prod_{j=k^*+1}^{k} \eta_j^r > \frac{1}{c_k} n^{-\frac{r(k-k^*)}{2} + \epsilon} \mid  \by,\Mk \right),
\label{eq:boundpenalty_nlp}
\end{align}
where the right hand side converges in probability to 0
given that \eqref{eq:boundweights_nlp} converges to 0 in probability
and $c_k,k,k^*,r$ are finite constants.
As mentioned earlier this result holds for any $r>0$ satisfying
$q-r < \mbox{dim}(\Theta)/2$, in particular we may set
$q-r= \delta < \mbox{dim}(\Theta)/2$
(where $\delta>0$ can be arbitrarily small) so that plugging $r=q-\delta$ into the left hand side
of \eqref{eq:boundpenalty_nlp} gives the desired result.
\qed

\subsection{Proof of Lemma \ref{lemma:generic_momprior_constant}}

Let $D_{ij}$ be a $pk \times pk$ matrix where the $i^{th}$ and $j^{th}$ diagonal blocks are equal to the $p \times p$ identity matrix,
and the $(i,j)$ off-diagonal block is minus the identity matrix,
so that $(\bzeta_i - \bzeta_j)' (\bzeta_i - \bzeta_j)=\bzeta' D_{ij} \bzeta$.
Then a direct application of Lemma 1 in \cite{kan} gives that
\begin{align}
&d_k(\bzeta)= \prod_{i<j}^{} (\bzeta_i - \bzeta_j)' (\bzeta_i - \bzeta_j)=
\prod_{i<j}^{} \btheta' D_{ij} \bzeta=
\nonumber \\
= &\frac{1}{[k(k-1)/2]!} \sum_{v(1,2)=0}^{1} \sum_{v(k-1,k)=0}^{1} (-1)^{\sum_{i<j}^{} v(i,j)}
\left[ \bzeta' \left( \sum_{i<j}^{} \left( \frac{1}{2}-v(i,j) \right) D_{ij} \right) \bzeta \right]^{\frac{k(k-1)}{2}}
\nonumber \\
= &\frac{1}{[k(k-1)/2]!} \sum_{v(1,2)=0}^{1} \sum_{v(k-1,k)=0}^{1} (-1)^{\sum_{i<j}^{} v(i,j)}
\left[ \bzeta' B_v \bzeta \right]^{\frac{k(k-1)}{2}}
\label{eq:expand_mompenalty}
\end{align}
where $B_v= \left( \sum_{i<j}^{} \left( \frac{1}{2}-v(i,j) \right) D_{ij} \right)$ is a matrix
with element $(l,m)$ given by
$$
\begin{cases}
b_{ll}= \frac{1}{2}(k-1) - \sum_{i<j}^{} v(i,j) \mbox{, }l=1+p(i-1),\ldots,pi \\
b_{lm}= b_{ml}= -\frac{1}{2} + \sum_{i<j}^{} v(i,j) \mbox{, } (1+p(i-1),1+p(j-1)),\ldots,(pi,pj)
\end{cases}.
$$
Let $\zeta_l$ be the $l^{th}$ element in $\bzeta$, then following Expression (6.1) in \citep{mohsenipour}
\begin{align}
\left[ \bzeta' B_v \bzeta \right]^{\frac{k(k-1)}{2}}=
\sum_{s \in S_k}^{} [k(k-1)/2]! \left( \prod_{l=1}^{pk}\prod_{m=1}^{pk} \frac{b_{lm}^{s_{lm}}}{s_{lm}!} \right) \prod_{l=1}^{pk} \bzeta_{l}^{\sum_{m=1}^{pk} s_{lm} + s_{ml}}
\label{eq:expand_quadform}
\end{align}
where
$s=(s_{1,1},s_{1,2},\ldots,s_{pk,pk})$ is a $(pk)^2$ integer vector,
$S_k$ denotes the set of partitions of $k(k-1)/2$
such that $\sum_{l=1}^{pk} \sum_{m=1}^{pk} s_{l,m} = k(k-1)/2$
with $0 \leq s_{l,m} \leq k(k-1)/2$.
Plugging \eqref{eq:expand_quadform} into \eqref{eq:expand_mompenalty} gives that the prior normalization constant is
\begin{align}
E^L(d_k(\bzeta))=
\sum_{v(1,2)=0}^{1} \sum_{v(k-1,k)=0}^{1} (-1)^{\sum_{i<j}^{} v(i,j)}
\sum_{s \in S_k}^{} \left( \prod_{l=1}^{pk}\prod_{m=1}^{pk} \frac{b_{lm}^{s_{lm}}}{s_{lm}!} \right) \prod_{l=1}^{pk}
\kappa_s
\label{eq:generic_momprior_constant}
\end{align}
where $\kappa_s= E^L( \bzeta_{jf}^{\sum_{m=1}^{pk} s_{lm} + s_{ml}})$.\qed

\subsection{Proof of Corollary \ref{corollary1}}

In order to compute the normalization, $C_k$ we need to find the expectation:

\begin{align*}
C_k=E\left(\prod_{1\leq
i < j \leq
k}\left(\frac{(\bmu_{i}-\bmu_{j})^{'}A_{\Sigma}^{-1}(\bmu_{i}-\bmu_{j})}{g}\right)\right).
\end{align*}
with respect to $(\bmu_{1},...,\bmu_{k}\sim N(\boldsymbol{0},A_{\Sigma})$. Moreover consider the Cholesky decomposition
$A_{\Sigma}=\boldsymbol{L}\boldsymbol{L}^{'}$ where
$A_{\Sigma}^{-1}=(\boldsymbol{L}^{'})^{-1}\boldsymbol{L}^{-1}$,
by setting
$\sqrt{g}\boldsymbol{L}\bmu_{j}^{*}=\bmu_{j}$ the jacobian of the transformation is the determinant
of the block diagonal matrix:
$$
|J(\bmu_{1}^{*},...,\bmu_{k}^{*})_{}|=\left|\left(
\begin{array}{ccc}
   \sqrt{g}\boldsymbol{L} & \cdots & 0 \\
   \vdots & \ddots & \vdots \\
   0 & \cdots & \sqrt{g}\boldsymbol{L}
\end{array}
\right)\right|=g^{k/2}(\text{det}(\boldsymbol{L}))^{k},
$$
where
$(\text{det}(\boldsymbol{L}))^{k}=(\text{det}(A_{\Sigma}))^{k/2}$.
The normalization constant $C_k$ can be found by using the
following expectation
\begin{align}
C_k=E\left(\prod_{1\leq
i < j \leq
k}((\bmu_{i}^{*}-\bmu_{j}^{*})^{'}(\bmu_{i}^{*}-\bmu_{j}^{*}))\right),
\end{align}
where $\bmu_{k}^{*} \sim
N_{p}\left(\bmu_{k}^{*}; \boldsymbol{0},\boldsymbol{I_{p}}\right)$.

To obtain the result we apply the adapted Proposition 4 in \cite{kan}
to the $p \times k$ vector $\bmu^{*}=(\bmu^{*}_{1},...,\bmu^{*}_{k})$,
where $k$ is the number of components and $\bmu^{*}_j \in \mathbb{R}^p$
for $j=1,\ldots,k$, which for convenience we reproduce below as Proposition \ref{propo1}.

\begin{defin}
\label{propo1}
Suppose $\bmu^{*}=(\mu_{1}^{*},...,\mu_{k}^{*})^{'}\sim
N_{k}(\boldsymbol{0},\boldsymbol{I}_{k})$, for symmetric matrices
$A_{(1,2)},...,A_{(k-1,k)}$, we have
\begin{align}\label{formB1propo1}
E\left(\prod_{1\leq i < j \leq k }(\bmu^{*'}A_{(i,j)}\bmu^{*})\right)=\frac{1}{s!}\sum_{\upsilon_{(1,2)}=0}^{1}...
\sum_{\upsilon_{(k-1,k)}=0}^{1}(-1)^{\sum\limits_{i, j}^{\binom
{k} {2}}\upsilon_{(i,j)}}
\mathcal{Q}_{s}(B_{\upsilon}),
\end{align}
where $s=\binom {k} {2}$,
$B_{\upsilon}=(\frac{1}{2}-\upsilon_{(1,2)})A_{(1,2)}+,...,+(\frac{1}{2}-\upsilon_{(k-1,k)})A_{(k-1,k)}$
and $\mathcal{Q}_{s}(B_{\upsilon})$ is given by the
recursive equation:
$\mathcal{Q}_{s}(B_{\upsilon})=s!2^{s}d_{s}(B_{\upsilon})$ where
$d_{s}(B_{\upsilon})=\frac{1}{2s}\sum_{i=1}^{s}tr(B_{\upsilon}^{i})d_{s-i}(B_{\upsilon})$ and
$d_{0}(B_{\upsilon})=1$ and $A_{(i,j)}$ is a $pk\times pk$ matrix $(l,m)$
element

\begin{equation*}
\left\{
      \begin{array}{ll}
      a_{ll}=1, \;\;\; l=1+p(i-1)...p_{i} \;\;\; \text{and} \;\;\; l=1+p(j-1)...p_{j}.\\
      a_{lm}=a_{ml}=-1, \;\;\; (l,m)=(1+p(i-1),1+p(j-1))...(pi,pj).\\
      a_{lm}=0 \;\;\; \text{otherwise}.
      \end{array}
\right.
\end{equation*}

\end{defin}

We define now the
$A_{(1,2)},...,A_{(k-1,k)}$ matrices
with dimensions $pk\times pk$. These matrices can be found using $p\times p$ identity matrices in the diagonal blocks
corresponding to the $i$ and $j$ components minus the
identity matrix in the ``cross-blocks'' corresponding to $(i,j)$. Finally using the $A_{(i,j)}$ matrices,
$B_{\upsilon}$ can be expressed as a $pk\times pk$ matrix with element $(l,m)$ as follows
\begin{equation*}
\left\{
      \begin{array}{ll}
      b_{ll}=\dfrac{1}{2}(k-1) - \sum_{i<j}\upsilon_{(i,j)}, \;\;\; l=1+p(i-1)...p_{i} \;\;\; \text{and} \;\;\; l=1+p(j-1)...p_{j}.\\
      b_{lm}=b_{ml}=-\dfrac{1}{2} + \sum_{i<j}\upsilon_{(i,j)}, \;\;\; (l,m)=(1+p(i-1),1+p(j-1))...(pi,pj).\\
      \end{array}
\right.
\end{equation*}
\qed
\subsection{Proof of Corollary \ref{corollary2}}

Using Corollary 2.2 in \cite{lu1993},
if $z>-1/n$, then
\begin{equation}
(2\pi)^{-n/2}\int_{-\infty}^{\infty}...\int_{-\infty}^{\infty}
\prod_{1\leq i < j \leq
n}(x_{i}-x_{j})^{2z}\prod_{j=1}^{n}\exp\{-x_{j}^{2}/2\}dx_{j}=
\prod_{j=1}^{n}\dfrac{\Gamma(jz+1)}{\Gamma(z+1)},
\end{equation}
and using $x_{i}=(\mu_{i}-m)/(\sqrt{a_{\sigma^{2}}g})$ with
$i=1,...,k$, we have that the normalization constant for a Normal mixture $(p=1)$ is
\begin{equation}
C_{k}=E_{\mu_{1},...,\mu_{k}}\mid a_{\sigma^{2}}\left(\prod_{1\leq i < j
\leq k}\left(\frac{\mu_{i}-\mu_{j}}{
\sqrt{a_{\sigma^{2}}g}}\right)^{2t}\right)=
\prod_{j=1}^{k}\dfrac{\Gamma(jt+1)}{\Gamma(t+1)},
\end{equation}
and for $k=2$ is straightforward to show that $C_k=E(\bmu_i-\bmu_j)'(\mu_i-\bmu_j)=2tr(I_{p})$.

\qed
\subsection{Proof of Corollary \ref{corollary3}}

For $p=1$ $C_k$ is computed
using (3.10) in \cite{lu1993} and for $k=2$
is straightforward to show that $C_k=E(\btheta_i-\btheta_j)'(\btheta_i-\btheta_j)=2\sum_{f=1}^{p}V(\btheta_{jf})$.
\qed

\subsection{Proof of Proposition \ref{prop:bf_oneemptyclus}}

We start by noting that
\begin{align}
p(\by \mid \Mk) =
\sum_{\bz: n_k=0}^{} p(\by \mid \bz,\Mk) p(\bz \mid \Mk) + \sum_{\bz: n_k>1}^{} p(\by \mid \bz,\Mk) p(\bz \mid \Mk)
\label{eq:marglhood_fromz1}
\end{align}

From C1, for any $\bz$ such that $n_k=0$ we have that $p(\by \mid \bz,\Mk)=$
\begin{align}
&\int p(\by \mid \bvartheta_k, \bz, \Mk) p(\bvartheta_k \mid \bz, \Mk) d\bvartheta_k= 
  \int \left( \prod_{j=1}^{k-1} \prod_{z_i=j}^{} p(\by_i \mid \btheta_j) \right) p(\bvartheta_k \mid \bz, \Mk) d\bvartheta_k= \nonumber \\
& \int \left( \prod_{j=1}^{k-1} \prod_{z_i=j}^{} p(\by_i \mid \btheta_j) \right) p(\btheta_1,\ldots,\btheta_{k-1} \mid \bz, \Mk) d\btheta_1\ldots d\btheta_{k-1}= \nonumber \\
& \int \left( \prod_{j=1}^{k-1} \prod_{z_i=j}^{} p(\by_i \mid \btheta_j) \right) p(\btheta_1,\ldots,\btheta_{k-1} \mid \bz, \mathcal{M}_{k-1}) d\btheta_1 \ldots d\btheta_{k-1}=
p(\by \mid \bz,\mathcal{M}_{k-1})
\label{eq:constant_marglhood_emptyclus}
\end{align}
where the third line in \eqref{eq:constant_marglhood_emptyclus}  follows from C4.
Further, from Condition C3, for any $\bz$ such that $n_k=0$ we have
\begin{align}
p(\bz \mid \mathcal{M}_{k-1})= p(\bz \mid n_k=0, \Mk)= \frac{p(\bz \mid \Mk)}{P(n_k=0 \mid \Mk)}
\Rightarrow
p(\bz \mid \Mk)= p(\bz \mid \mathcal{M}_{k-1}) P(n_k=0 \mid \Mk).
\label{eq:coherency_clusterprob}
\end{align}
Plugging \eqref{eq:constant_marglhood_emptyclus} and \eqref{eq:coherency_clusterprob} into \eqref{eq:marglhood_fromz1}
gives that $p(\by \mid \Mk) =$
\begin{align}
P(n_k=0 \mid \Mk) \sum_{\bz: n_k=0}^{} p(\by \mid \bz,\mathcal{M}_{k-1}) p(\bz \mid \mathcal{M}_{k-1}) + \sum_{\bz: n_k>1}^{} p(\by \mid \bz,\Mk) p(\bz \mid \Mk)= \nonumber \\
P(n_k=0 \mid \Mk) p(\by \mid \mathcal{M}_{k-1}) + \sum_{\bz: n_k>1}^{} p(\by \mid \bz,\Mk) p(\bz \mid \Mk)
\label{eq:marglhood_fromz2}
\end{align}
That is, $p(\by \mid \Mk)$ is a linear combination of $p(\by \mid \mathcal{M}_{k-1})$
and a sum of $p(\by,\bz \mid \Mk)$ over cluster configurations such that the last cluster $k$ is occupied.
This recursive relation is an extension of Theorem 3.1 in \cite{nobile:2004},
who proved a similar result under more restrictive conditions than our C1-C4.
Dividing both sides of \eqref{eq:marglhood_fromz2} by $p(\by \mid \Mk)$ and rearranging terms gives
\begin{align}
B_{k-1,k}(\by)= \frac{1}{P(n_k=0 \mid \Mk)} \left( 1 - \sum_{\bz: n_k>1} \frac{p(\by,\bz \mid \Mk)}{p(\by \mid \Mk)} \right)=
\frac{P(n_k=0 \mid \by,\Mk)}{P(n_k=0 \mid \Mk)}.
\nonumber
\end{align}
Finally, from Condition C2 both the likelihood and prior are invariant to label permutations and thus
$P(n_j=0 \mid \by,\Mk) = P(n_k=0 \mid \by,\Mk)$ for any $j \neq k$, hence
$$
B_{k-1,k}(\by)= \frac{1}{k P(n_k=0 \mid \Mk)} \sum_{j=1}^{k} P(n_j=0 \mid \by,\Mk),
$$
as we wished to prove.

For completeness we derive $P(n_k=0 \mid \Mk)$ when $\bm{\eta} \sim \mbox{Dir}(q)$.
From \eqref{eq:coherency_clusterprob}, $P(n_k=0 \mid \Mk)=$
\begin{align}
\frac{p(\bz \mid \Mk)}{p(\bz \mid \mathcal{M}_{k-1})}= \frac{\Gamma(kq) \prod_{j=1}^{k} \Gamma( n_j+q )}{\Gamma(q)^k \Gamma(n+kq)}
\frac{\Gamma(q)^{k-1} \Gamma(n+(k-1)q)}{\Gamma((k-1)q) \prod_{j=1}^{k-1} \Gamma( n_j+q )}
=\frac{\Gamma(kq) \Gamma(n+(k-1)q)}{\Gamma(n+kq) \Gamma((k-1)q)} \nonumber
\end{align}

\section{Monte Carlo estimation of the normalization constant}
\label{supplsec:montecarlock}

\begin{table}[ht]
\caption{Estimation of $\log(C_k)$ and associated standard error (se) via Monte Carlo for the MOM-IW prior
where $k=2,...,10$ and $p=1,...,10$. Values for $p=1$ and $k=2$ are based on the exact formulas in  Corollary \ref{corollary2} }
\small
\centering
\begin{tabular}{rrrrrrrrrrrrrrrrrrrrr}
  \hline
    &  &  & & $p$   &  &  &    \\ \hline
  \hline
 & 1 && 2 && 3 && 4 && 5  \\
  \hline
$k$ & $\log(C_k)$ & se & $\log(C_k)$ & se & $\log(C_k)$ & se & $\log(C_k)$ & se & $\log(C_k)$ & se  \\
 \hline
2 & 0.693  & 0 & 1.386 & 0    & 1.792 & 0    & 2.079 & 0    & 2.303 & 0    \\
3 & 2.485  & 0 & 4.57 & $<$0.01 & 5.70 & $<$0.01 & 6.51 & $<$0.01 & 7.14 & $<$0.01 \\
4 & 5.663  & 0 & 9.83 & $<$0.01 & 11.98 & $<$0.01 & 13.51 & $<$0.01 & 14.70 & $<$0.01 \\
5 & 10.451 & 0 & 17.36 & $<$0.01 & 20.83 & $<$0.01 & 23.25 & $<$0.01 & 25.16 & $<$0.01 \\
6 & 17.030 & 0 & 27.27 & 0.04 & 32.26 & 0.02 & 35.99 & 0.03 & 38.58 & $<$0.01 \\
7 & 25.555 & 0 & 38.81 & 0.07 & 46.33 & 0.04 & 51.11 & 0.02 & 55.01 & 0.02 \\
8 & 36.160 & 0 & 53.01 & 0.10 & 62.05 & 0.05 & 69.70 & 0.07 & 74.51 & 0.04 \\
9 & 48.961 & 0 & 66.46 & 0.08 & 80.73 & 0.11 & 89.83 & 0.08 & 96.35 & 0.05 \\
10 & 64.066 & 0 & 82.71 & 0.10 & 100.43 & 0.08 & 111.81 & 0.09 & 120.87 & 0.10 \\
    &  & & & $p$   &  &  &    \\ \hline
  \hline
 & 6 && 7 && 8 && 9 && 10  \\
  \hline
$k$ & $\log(C_k)$ & se & $\log(C_k)$ & se & $\log(C_k)$ & se & $\log(C_k)$ & se & $\log(C_k)$ & se  \\
 \hline
2 & 2.485 & 0    & 2.639 & 0   & 2.773 & 0   & 2.890 & 0    & 2.996 & 0 \\
3 & 7.66 & $<$0.01 & 8.09 & $<$0.01 & 8.48 & $<$0.01 & 8.82 & $<$0.01 & 9.12 & $<$0.01 \\
4 & 15.68 & $<$0.01 & 16.51 & $<$0.01 & 17.25 & $<$0.01 & 17.90 & $<$0.01 & 18.49 & $<$0.01 \\
5 & 26.72 & $<$0.01 & 28.04 & $<$0.01 & 29.23 & $<$0.01 & 30.26 & $<$0.01 & 31.22 & $<$0.01 \\
6 & 40.81 & $<$0.01 & 42.78 & 0.01 & 44.47 & $<$0.01 & 45.99 & $<$0.01 & 47.35 & $<$0.01 \\
7 & 58.21 & 0.04 & 60.78 & 0.02 & 63.05 & 0.01 & 65.15 & 0.01 & 67.08 & 0.01 \\
8 & 78.44 & 0.04 & 82.13 & 0.04 & 84.96 & 0.02 & 88.01 & 0.04 & 90.19 & 0.02 \\
9 & 101.82 & 0.05 & 106.15 & 0.05 & 110.12 & 0.05 & 113.81 & 0.04 & 116.87 & 0.03 \\
10 & 127.88 & 0.07 & 133.19 & 0.05 & 138.22 & 0.05 & 143.08 & 0.06 & 146.70 & 0.04 \\
   \hline
   \hline
\end{tabular}
\label{tab:MCnormal}
\end{table}

\begin{table}[ht]
  \caption{Estimation of $\log(C_k)$ for $k=2,...,10$, $p=1,...,10$
    via Monte Carlo and its standard error (se) for the MOM-Beta prior with $a=0.5$ and default $g$ in Table \ref{tab:defaultg_mombeta}. Values for $p=1$ and $k=2$ are based on the exact formulas in  Corollary \ref{corollary3} }
\small
\centering
\begin{tabular}{rrrrrrrrrrrrrrrrrrrrr}
  \hline
    &  &  & & $p$   &  &  &    \\ \hline
  \hline
 & 1 && 2 && 3 && 4 && 5  \\
  \hline
$k$ & $\log(C_k)$ & se & $\log(C_k)$ & se & $\log(C_k)$ & se & $\log(C_k)$ & se & $\log(C_k)$ & se  \\
 \hline
     2 & -2.786    & 0 & -1.685 & 0 & -1.107 & 0 & -0.725 & 0 & -0.442 & 0 \\
  3 & -8.305  & 0 & -4.88 & $<$0.01 & -3.18 & $<$0.01 & -2.05 & $<$0.01 & -1.23 & $<$0.01 \\
  4 & -16.539 & 0 & -9.47 & $<$0.01 & -6.09 & $<$0.01 & -3.89 & $<$0.01 & -2.26 & $<$0.01 \\
  5 & -27.481 & 0 & -15.40 & 0.02 & -9.75 & 0.02 & -6.15 & 0.01 & -3.48 & $<$0.01 \\
  6 & -41.130 & 0 & -22.55 & 0.06 & -14.15 & 0.04 & -8.75 & 0.02 & -4.82 & 0.02 \\
  7 & -57.488 & 0 & -31.45 & 0.11 & -19.43 & 0.06 & -11.74 & 0.05 & -6.29 & 0.03 \\
  8 & -76.556 & 0 & -41.92 & 0.12 & -25.40 & 0.13 & -15.39 & 0.07 & -7.90 & 0.05 \\
  9 & -98.337 & 0 & -54.12 & 0.20 & -32.34 & 0.22 & -19.11 & 0.16 & -9.35 & 0.16 \\
  10 & -122.834 & 0 & -67.71 & 0.30 & -40.06 & 0.29 & -23.74 & 0.19 & -11.81 & 0.17 \\
    &  & & & $p$   &  &  &    \\ \hline
  \hline
 & 6 && 7 && 8 && 9 && 10  \\
  \hline
$k$ & $\log(C_k)$ & se & $\log(C_k)$ & se & $\log(C_k)$ & se & $\log(C_k)$ & se & $\log(C_k)$ & se  \\
 \hline
     2 & -0.220 & 0 & -0.036 & 0 & 0.119 & 0 & 0.257 & 0 & 0.377 & 0 \\
  3 & -0.57 & $<$0.01 & -0.03 & $<$0.01 & 0.43 & $<$0.01 & 0.83 & $<$0.01 & 1.19 & $<$0.01 \\
  4 & -0.98 & $<$0.01 & 0.07 & $<$0.01 & 0.98 & $<$0.01 & 1.78 & $<$0.01 & 2.48 & $<$0.01 \\
  4 & -1.39 & $<$0.01 & 0.36 & $<$0.01 & 1.84 & $<$0.01 & 3.15 & $<$0.01 & 4.30 & $<$0.01 \\
  6 & -1.74 & $<$0.01 & 0.84 & $<$0.01 & 3.05 & $<$0.01 & 4.99 & $<$0.01 & 6.69 & $<$0.01 \\
  7 & -1.97 & 0.03 & 1.64 & 0.03 & 4.67 & 0.02 & 7.34 & $<$0.01 & 9.70 & $<$0.01 \\
  8 & -2.20 & 0.06 & 2.66 & 0.06 & 6.59 & 0.04 & 10.24 & 0.03 & 13.34 & 0.03 \\
  9 & -2.21 & 0.10 & 3.78 & 0.07 & 9.00 & 0.07 & 13.60 & 0.06 & 17.52 & 0.05 \\
  10 & -2.45 & 0.15 & 5.22 & 0.11 & 11.61 & 0.09 & 17.41 & 0.10 & 22.31 & 0.07 \\
   \hline
   \hline
\end{tabular}
\label{tab:MCbeta}
\end{table}

\section{Sensitivity of choosing $g$ for MOM-Beta priors}
\label{supplsec:sensMOMBeta}

\begin{figure}[ht]
\begin{center}
\begin{tabular}{ccc}
\includegraphics[width=0.5\textwidth]{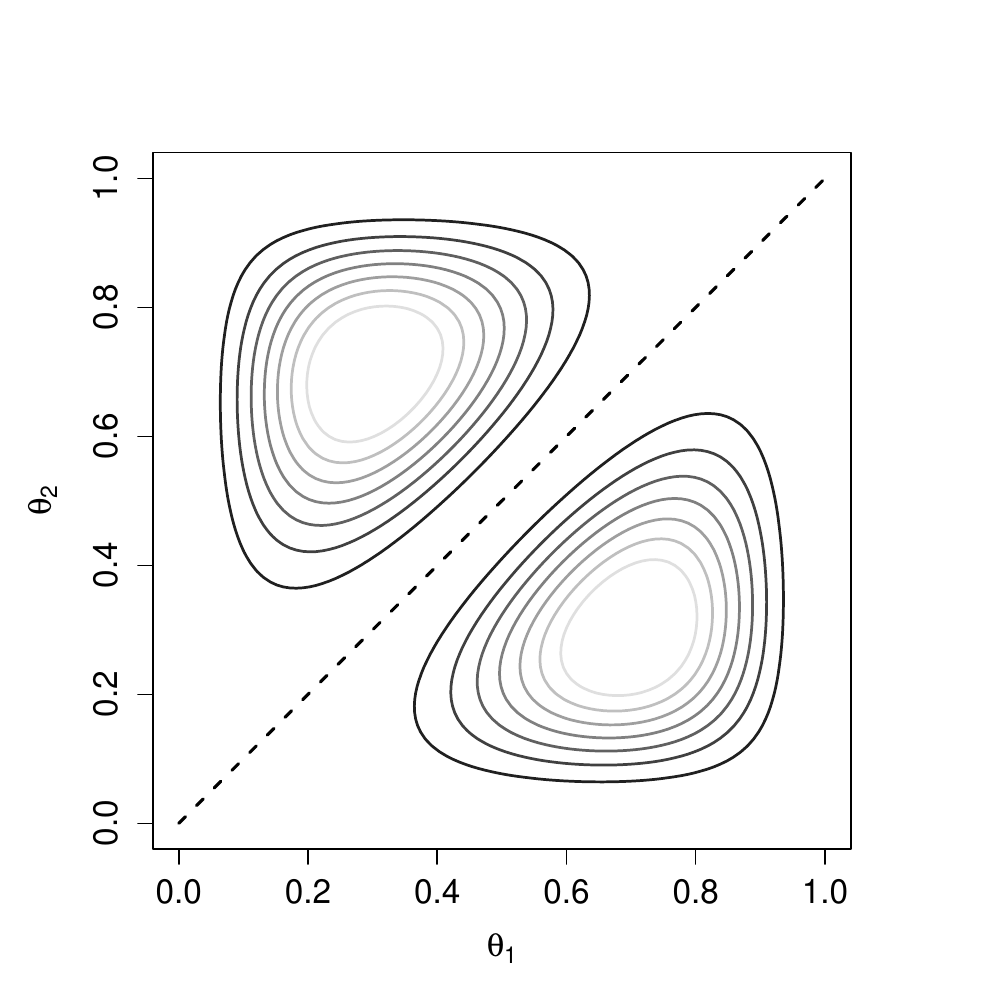} & \hspace{-1.5cm}
\includegraphics[width=0.5\textwidth]{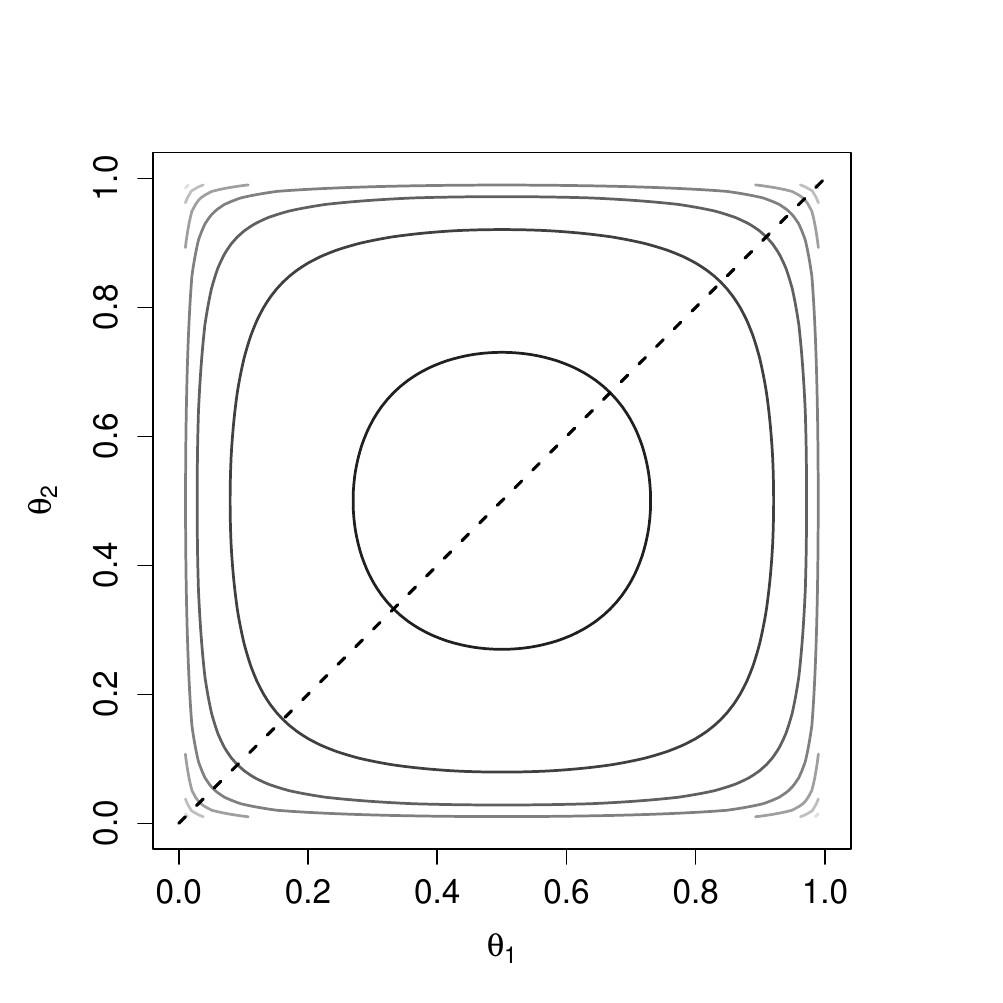} \\
\end{tabular}
\end{center}
\caption{Default MOM-Beta ($a=0.5$, $g=7.11$) (left) and Beta(1,1) (right)}
\label{supfig:contours2}
\end{figure}

\begin{figure}[ht]
\begin{center}
\begin{tabular}{cccc}
n=50 \\
\includegraphics[width=0.7\textwidth]{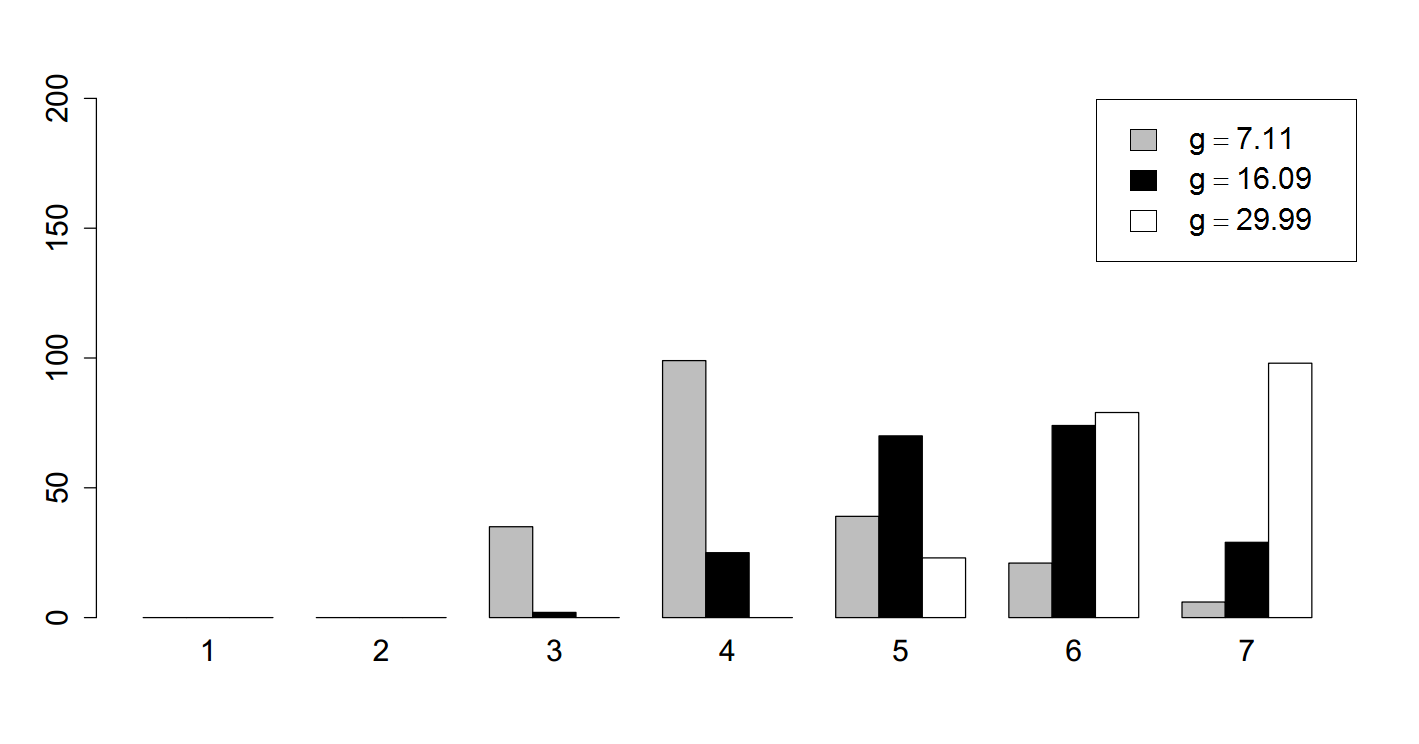} \\
n=200\\
\includegraphics[width=0.7\textwidth]{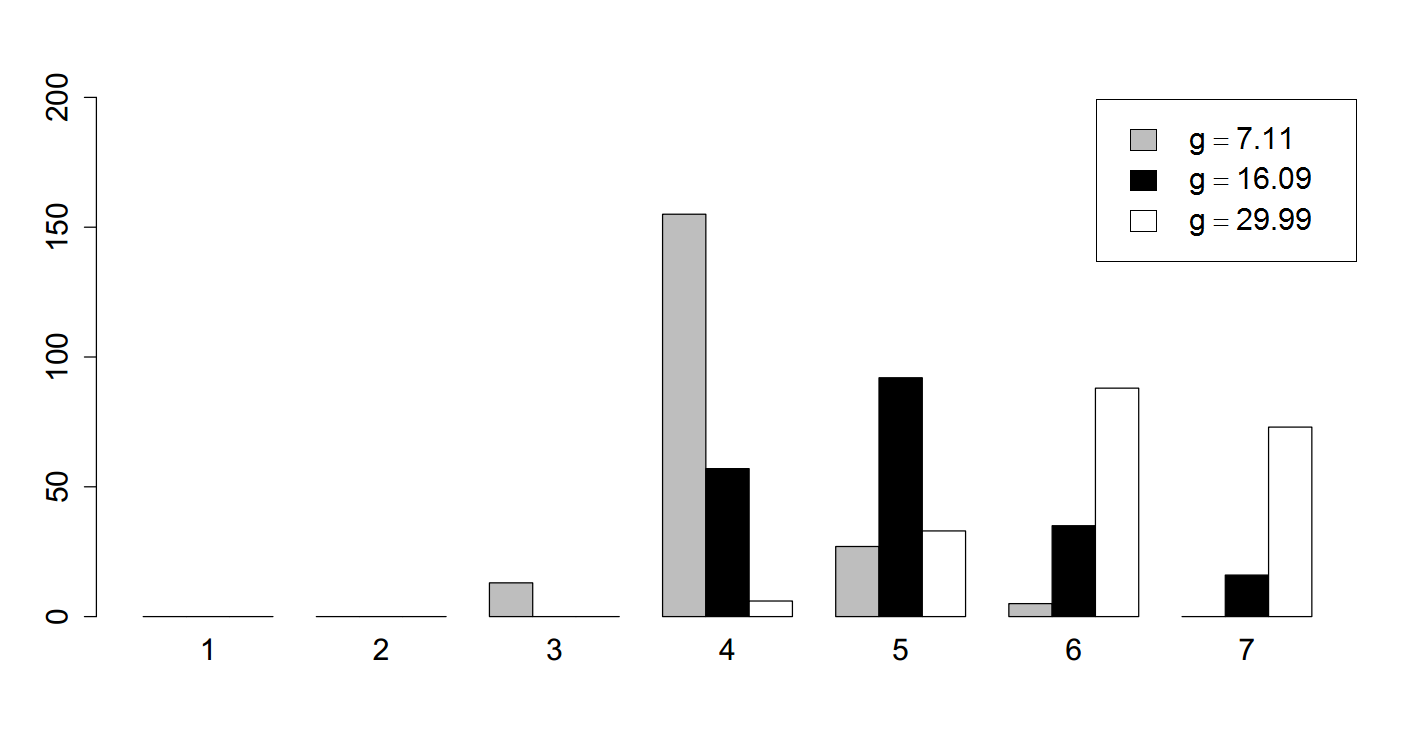} \\
n=500\\
\includegraphics[width=0.7\textwidth]{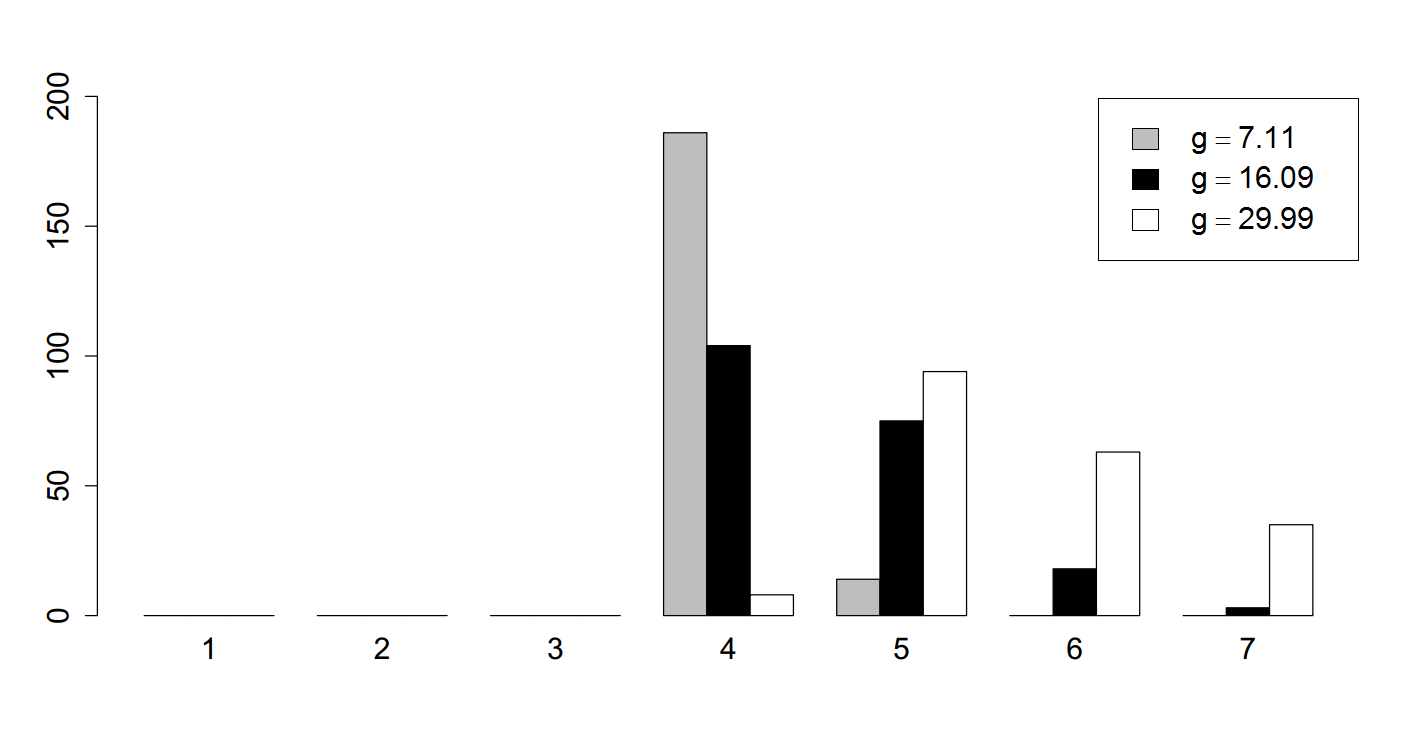} \\
\end{tabular}
\end{center}
\caption{Binomial mixture. Frequencies of $\hat{k}$ for MOM-Beta for $g=7.11$, $g=16.09$ and $g=29.99$ with $q=2$.
Results from 200 data sets with $n=50$, $n=200$ and $n=500$, $L = 30$ and $k^{*}=4$}
\label{supfig:sensitivityBinomial1}
\end{figure}

    Figure \ref{supfig:sensitivityBinomial1} reproduces the Binomial mixture simulations from Section \ref{ssec:binomials}.
    Additionally to the default $g=7.11$ for the MOM-Beta prior, we now considered larger (more informative) $g=16.09,29.99$.
    Under these larger $g$ the performance remains competitive but does suffer, suggesting that the default $g=7.11$ is preferable.
    For a comparison to the BIC, sBIC and Beta priors see Figure \ref{fig:sBICBinomial}.

\section{Comparison of ECP with other alternatives}
\label{supplsec:comparison2}

\begin{figure}[ht]

\begin{center}
\begin{tabular}{cccc}
\hspace{-1cm}$\hat{P}(\mathcal{M}_{1} \mid \by )$ & \hspace{-3cm} $\hat{P}(\mathcal{M}_{2} \mid \by )$ \vspace{-1cm}\\
\hspace{-1cm}\includegraphics[width=0.6\textwidth]{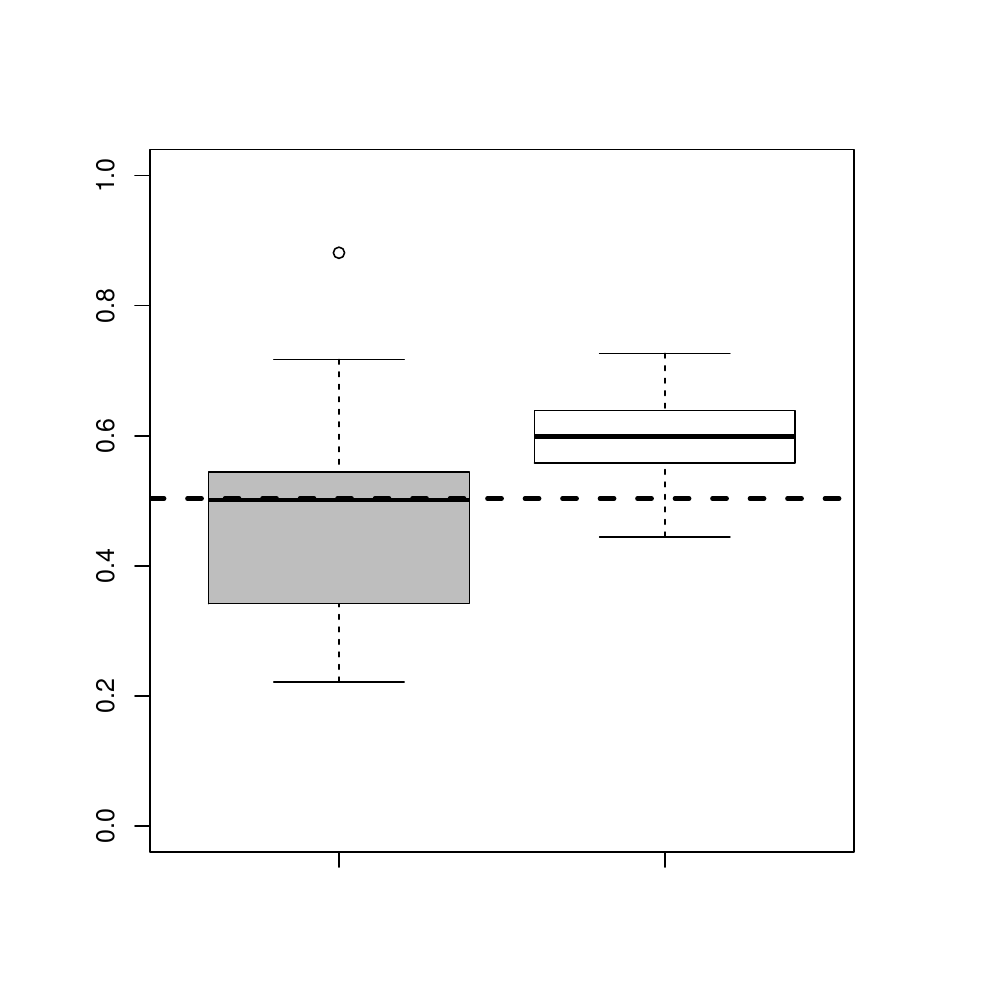} & \hspace{-2cm}
\includegraphics[width=0.6\textwidth]{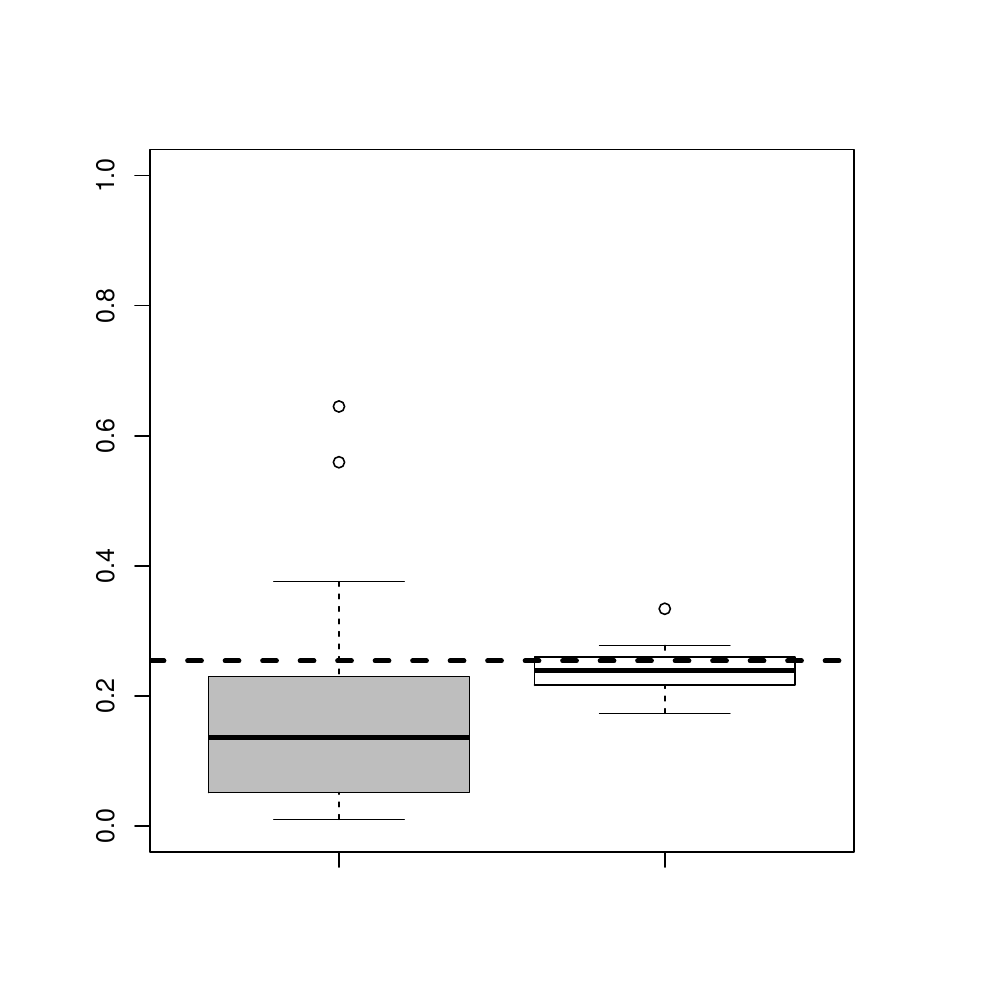} \\
\hspace{-1cm}$\hat{P}(\mathcal{M}_{3} \mid \by )$ & \hspace{-3cm} $\hat{P}(\mathcal{M}_{4} \mid \by )$ \vspace{-1cm}\\
\hspace{-1cm}\includegraphics[width=0.6\textwidth]{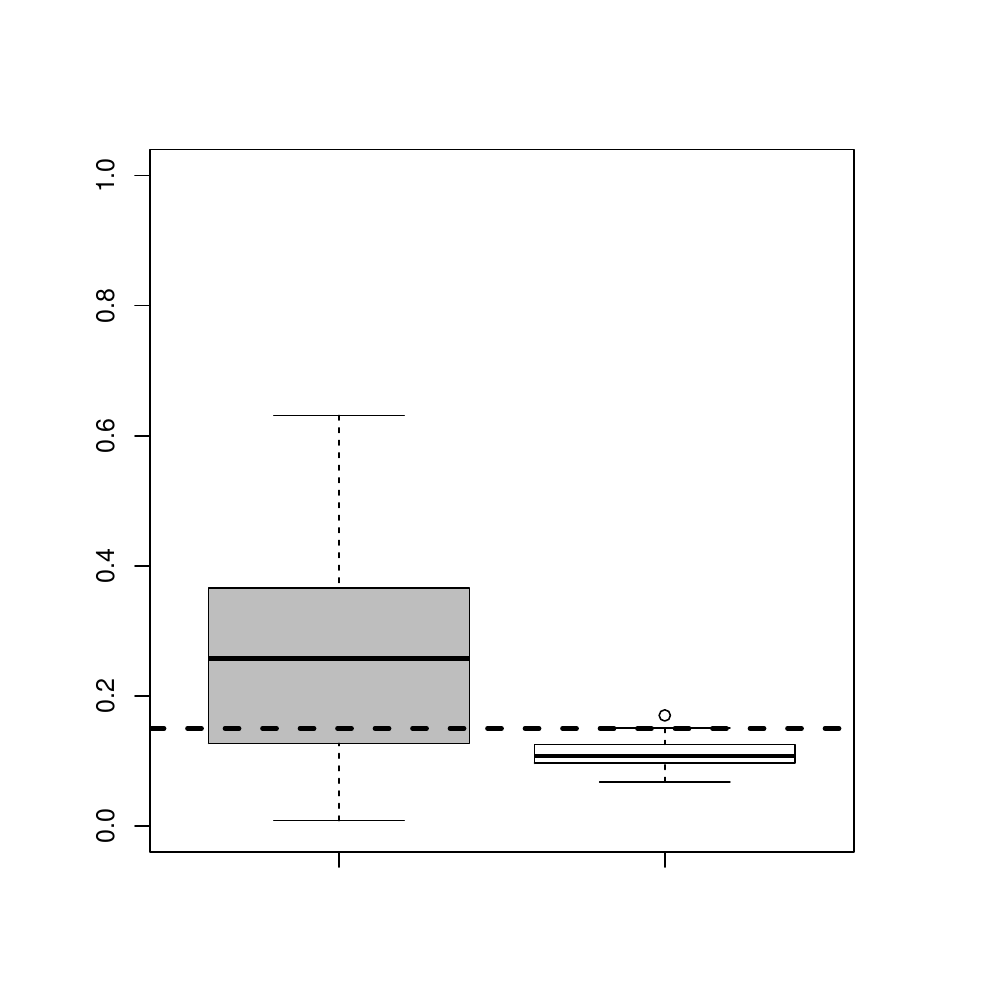}& \hspace{-2cm}
\includegraphics[width=0.6\textwidth]{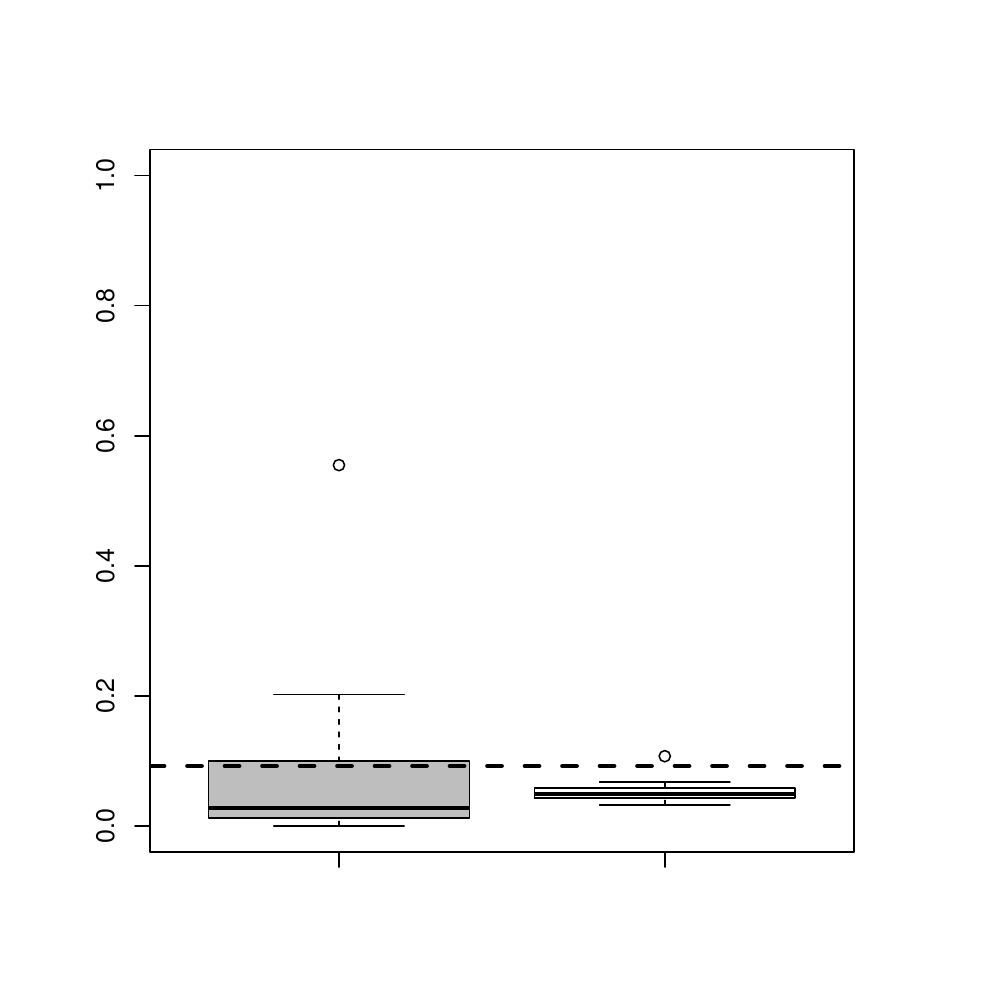} \vspace{-1cm} \\
\end{tabular}
\end{center}
\caption{Boxplots
display 50 independent estimates based on separate MCMC runs ($T=10,000$
iterations after a $T/10$ burn-in each). Precision
of $\hat{P}(\mathcal{M}_{k} \mid \by )$ under Normal-IW-Dir using the \cite{Robert1} estimator (gray)
and ECP estimator  (white) for $n=200$ observations in simulation Case 1. Dashed line indicate $\hat{P}(\mathcal{M}_{k} \mid \by )$ under
Normal-IW-Dir
obtained by simulating $1,000,000$ values from the prior and averaging the likelihood.}
\label{supfig:BFs}
\end{figure}

\begin{figure}[ht]

\begin{center}
\begin{tabular}{cccc}
\hspace{-1cm}$\hat{P}(\mathcal{M}_{1} \mid \by )$ & \hspace{-3cm} $\hat{P}(\mathcal{M}_{2} \mid \by )$ \vspace{-1cm}\\
\hspace{-1cm}\includegraphics[width=0.6\textwidth]{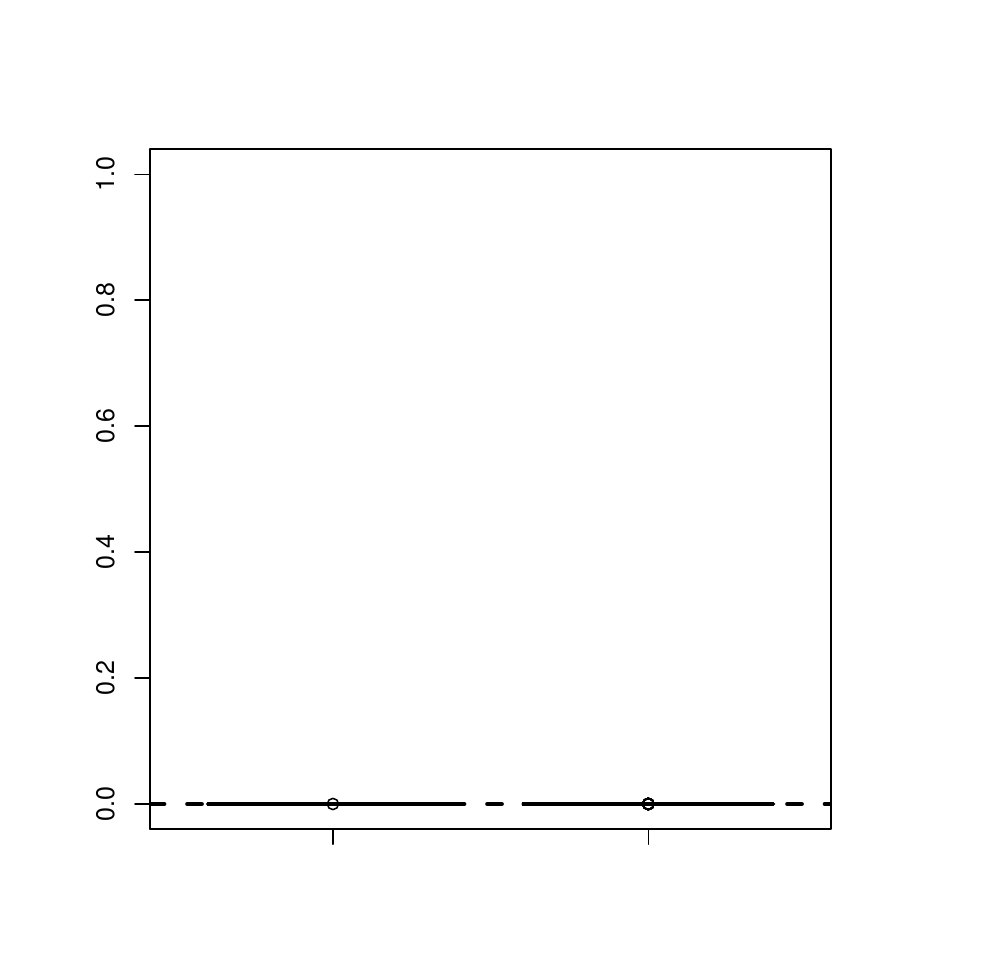} & \hspace{-2cm}
\includegraphics[width=0.6\textwidth]{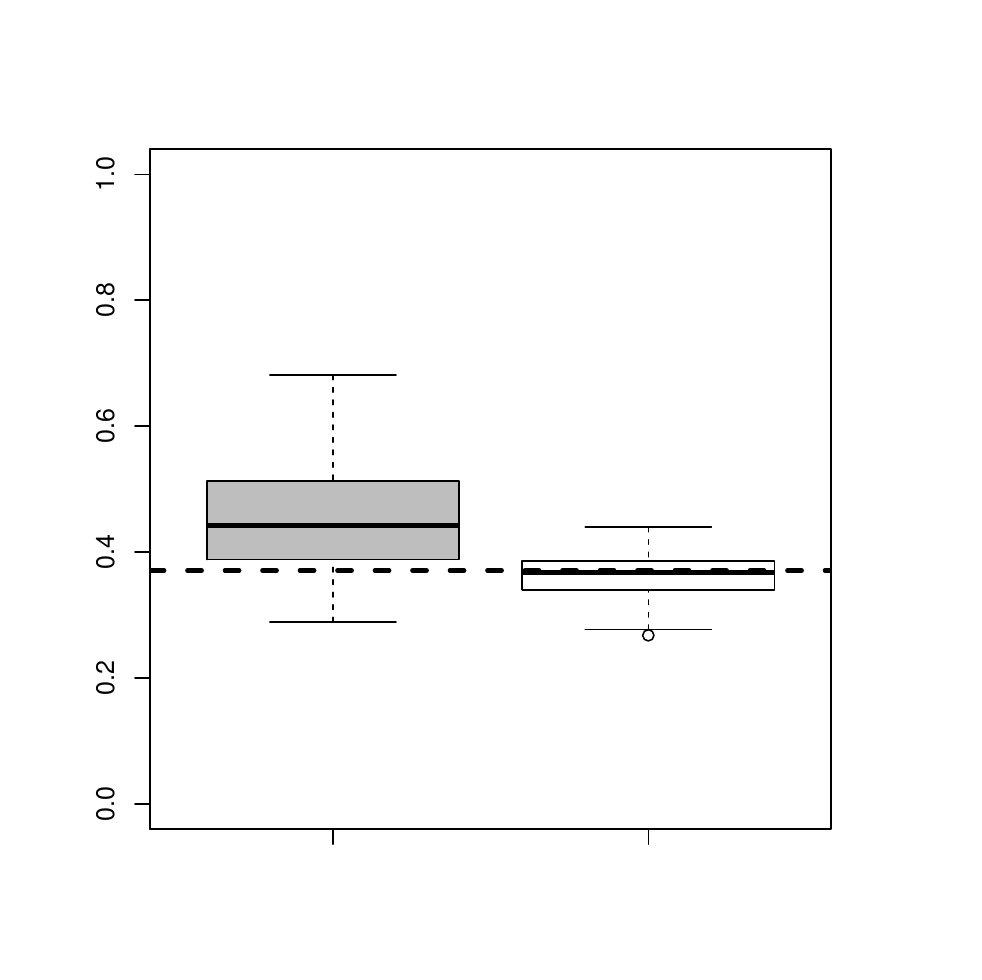} \\
\hspace{-1cm}$\hat{P}(\mathcal{M}_{3} \mid \by )$ & \hspace{-3cm} $\hat{P}(\mathcal{M}_{4} \mid \by )$ \vspace{-1cm}\\
\hspace{-1cm}\includegraphics[width=0.6\textwidth]{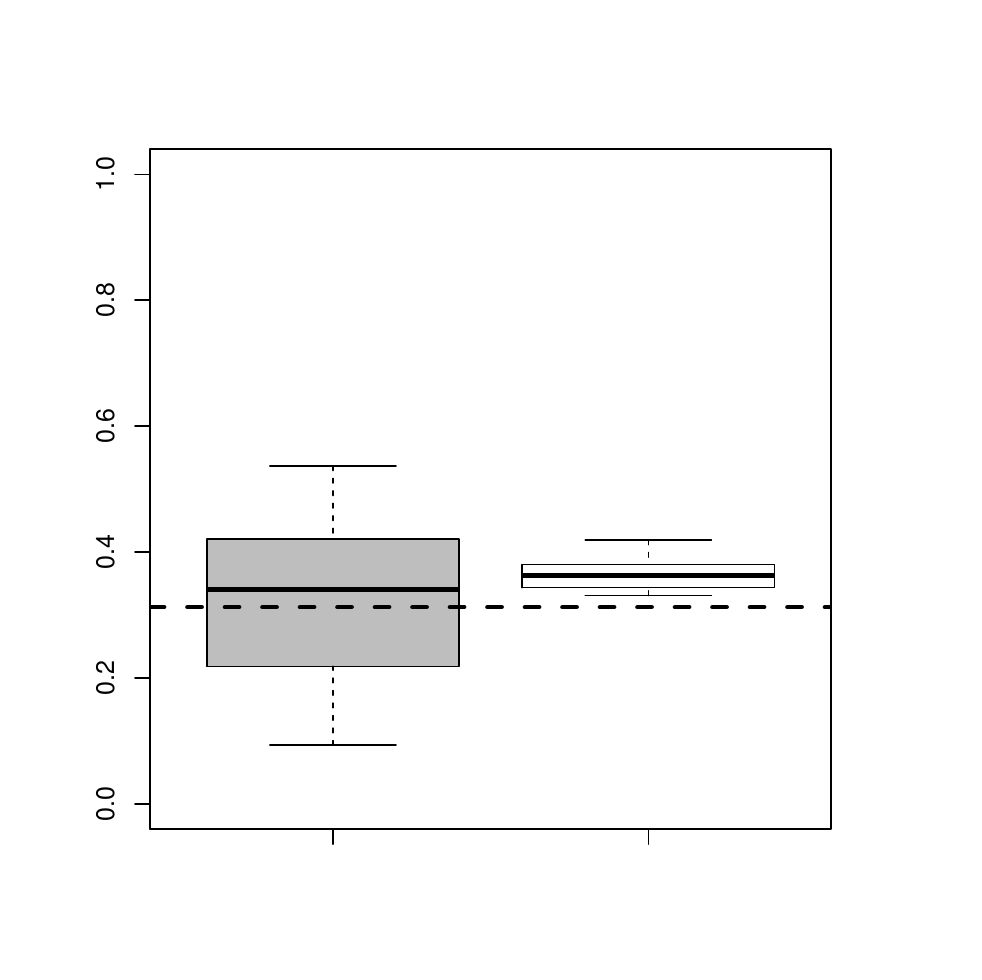}& \hspace{-2cm}
\includegraphics[width=0.6\textwidth]{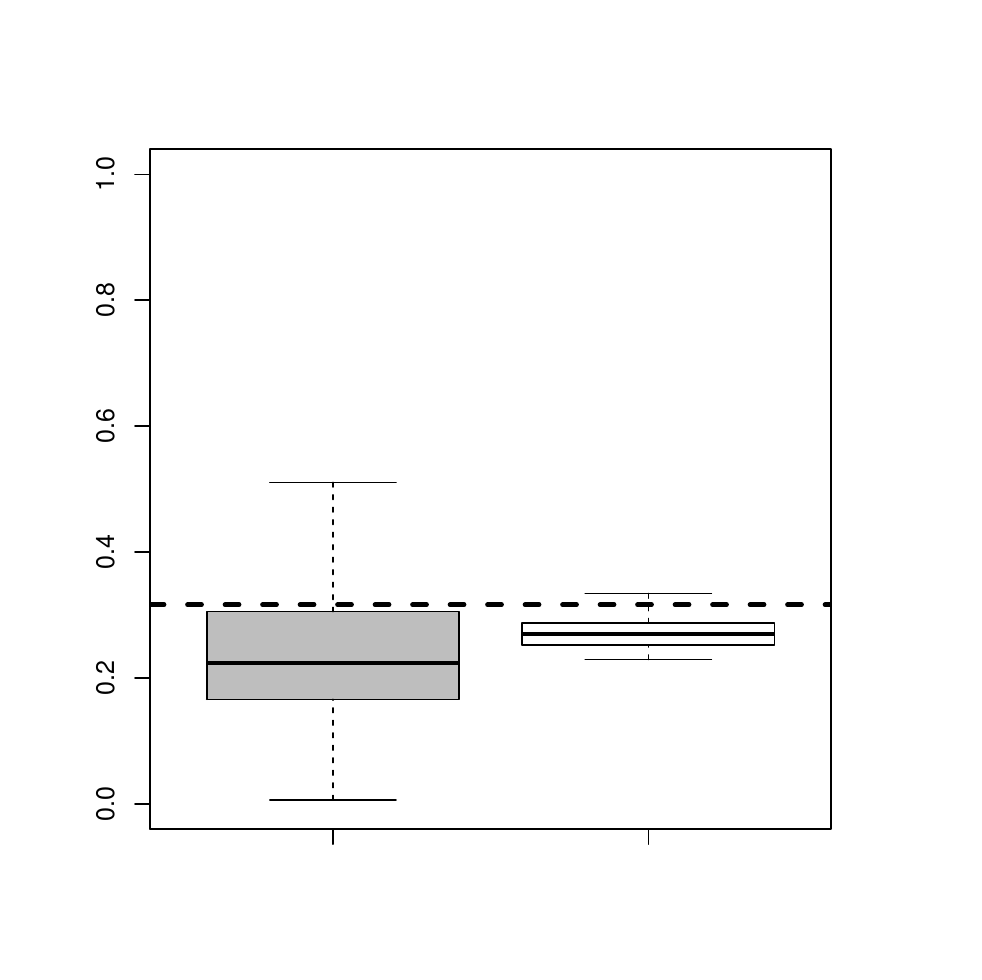} \vspace{-1cm} \\
\end{tabular}
\end{center}
\caption{Boxplots
display 50 independent estimates based on separate MCMC runs ($T=10,000$
iterations after a $T/10$ burn-in each). Precision
of $\hat{P}(\mathcal{M}_{k} \mid \by )$ under Normal-IW-Dir using the \cite{Robert1} estimator (gray)
and ECP estimator  (white) for $n=200$ observations in simulation Case 3. Dashed line indicate $\hat{P}(\mathcal{M}_{k} \mid \by )$ under
Normal-IW-Dir
obtained by simulating $1,000,000$ values from the prior and averaging the likelihood.}
\label{supfig:BFs2}
\end{figure}

We simulated a single data set of $n = 200$ observations from Cases 1 and 3 in Section \ref{ssec:syntheticexamples}
and computed 50 times $\hat{P}(\mathcal{M}_{k} \mid \by )$ under Normal-IW-Dir priors using the ECP estimator
and the \cite{Robert1} estimator given by
(\ref{perm3}). Figures \ref{supfig:BFs}-\ref{supfig:BFs2} show that the medians of
the ECP estimator and the \cite{Robert1} estimator  with $k=\{1,...,4\}$ are similar, but that the ECP estimator produces higher precision estimates.
To compute $\hat{P}(\mathcal{M}_{k} \mid \by )$ using the
ECP estimator we
implement the bfnormmix function given in the R package \texttt{mombf} \citep{mombf}.

\section{Precision of the Monte Carlo-estimated integrated likelihood}
\label{supplsec:comparison}

\begin{figure}[!ht]
\begin{center}
\includegraphics[width=0.95\textwidth]{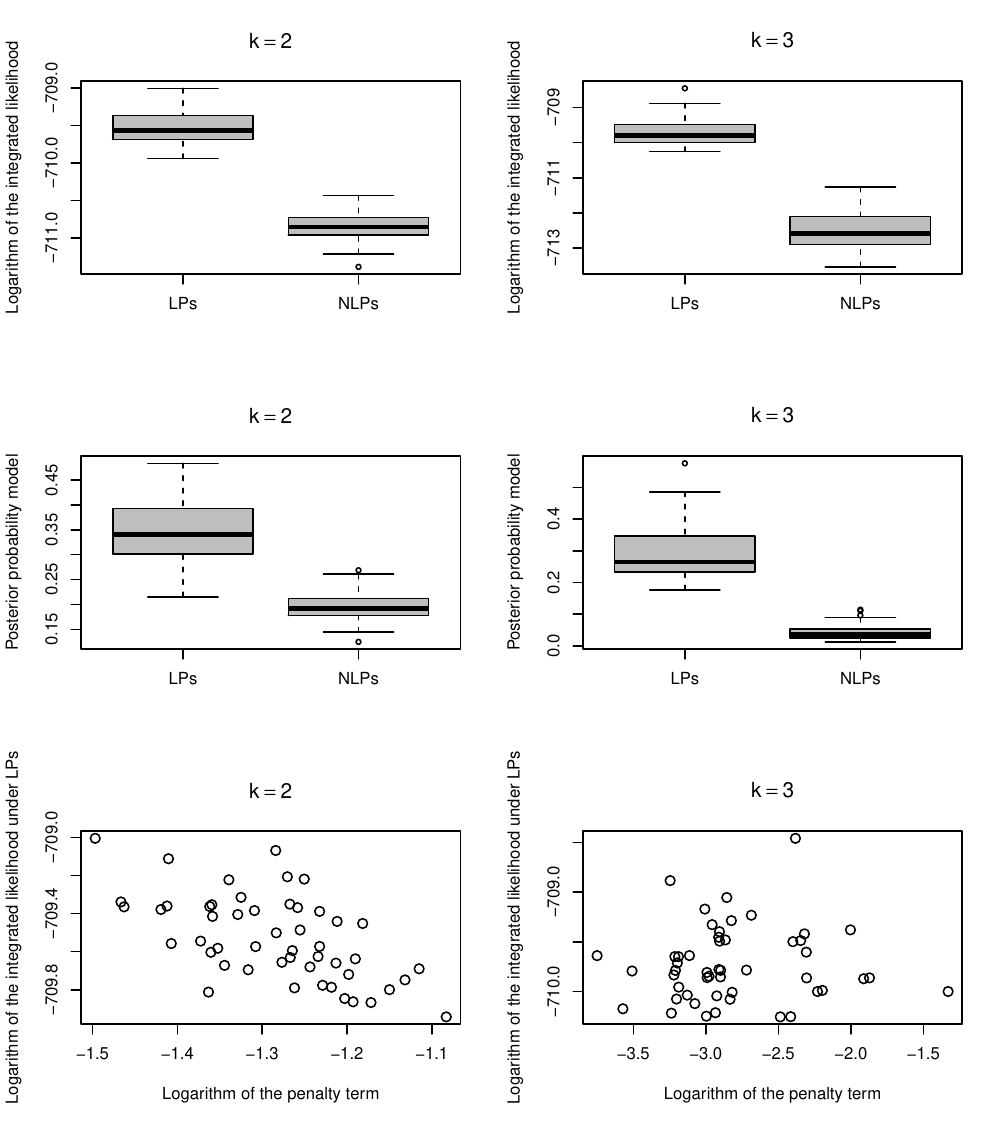}
\caption{Precision of $\hat{p}(\by \mid \Mk)$ in 100 univariate simulations, $k^*=1$.
Top: $\log\hat{p}(\by\mid \Mk)$. Middle: $\hat{p}(\Mk\mid \by)$.
Bottom: $\log\tilde{p}(\by\mid \Mk)$ vs. $\log \hat{E}(d_\vartheta(\bvartheta_k) \mid \by)$}
\label{supfig:comp_univ}
\end{center}
\end{figure}

\begin{figure}[!ht]
\begin{center}
\includegraphics[width=0.95\textwidth]{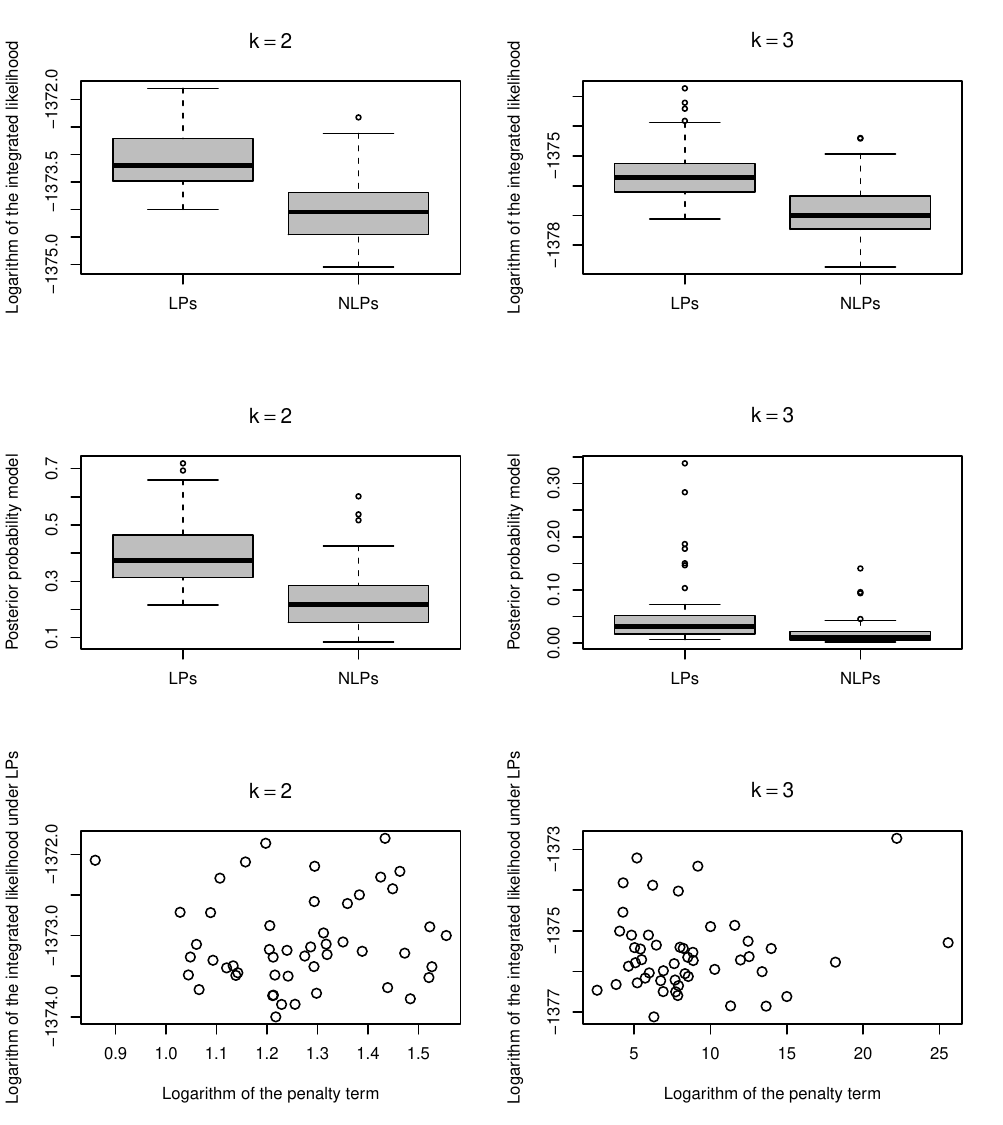}
\caption{Precision of $\hat{p}(\by \mid \Mk)$ in 100 bivariate simulations, $k^*=1$.
Top: $\log\hat{p}(\by\mid \Mk)$. Middle: $\hat{p}(\Mk\mid \by)$.
Bottom: $\log\tilde{p}(\by\mid \Mk)$ vs. $\log \hat{E}(d_\vartheta(\bvartheta_k) \mid \by)$}
\label{supfig:comp_biv}
\end{center}
\end{figure}

We compared empirically the precision of $\hat{p}(\by \mid \Mk)$ vs. the local prior-based $\tilde{p}(\by \mid \Mk)$ (Section \ref{ssec:comp_marglhood})
for univariate and bivariate Normal mixture and $k=2,3$ components
(if $k=1$ then $p(\by \mid \Mk)=\tilde{p}(\by \mid \Mk)$ has closed form).
To inspect whether the precision of $\hat{p}(\by \mid \Mk)$ suffers under overfitted mixtures
we simulated a single data set of $n=500$ observations from a $k^*=1$ component mixture
and computed 100 times both $\hat{p}(\by \mid \Mk)$ and $\tilde{p}(\by \mid \Mk)$.
Figures \ref{supfig:comp_univ} and \ref{supfig:comp_biv} show the results for a univariate and bivariate outcome respectively.
The precision of $\hat{p}(\by \mid \Mk)$ was comparable to that of $\tilde{p}(\by \mid \Mk)$,
in fact in some situations the former was more precise
(this is due to
$\mbox{Var}(\log {\hat{p}})= \mbox{Var}(\log \tilde{p}) + \mbox{Var}(\log \hat{\omega}) + 2\mbox{cov}(\log \tilde{p},\log \hat{\omega})$
where the latter covariance may be negative).
More importantly, posterior model probabilities $\hat{p}(\Mk \mid \by)$ (middle panels)
were more precise than $\tilde{p}(\Mk \mid \by)$, as in our experience tends to be the case
due to $p(\Mk \mid \by)$ having a higher concentration around 0 or 1 (Theorem 1).
The lower panels show that as $k$ grows larger than $k^*$ the precision in $\hat{w}$ tends to degrade,
however as mentioned this is compensated by the fact that $p(\Mk \mid \by)$ is small for large $k$ (middle panels),
thus it does not appear to be a practical concern.

\begin{figure}
\begin{center}
\begin{tabular}{cc}
$n=200$ & $n=1000$ \\
\includegraphics[width=0.5\textwidth]{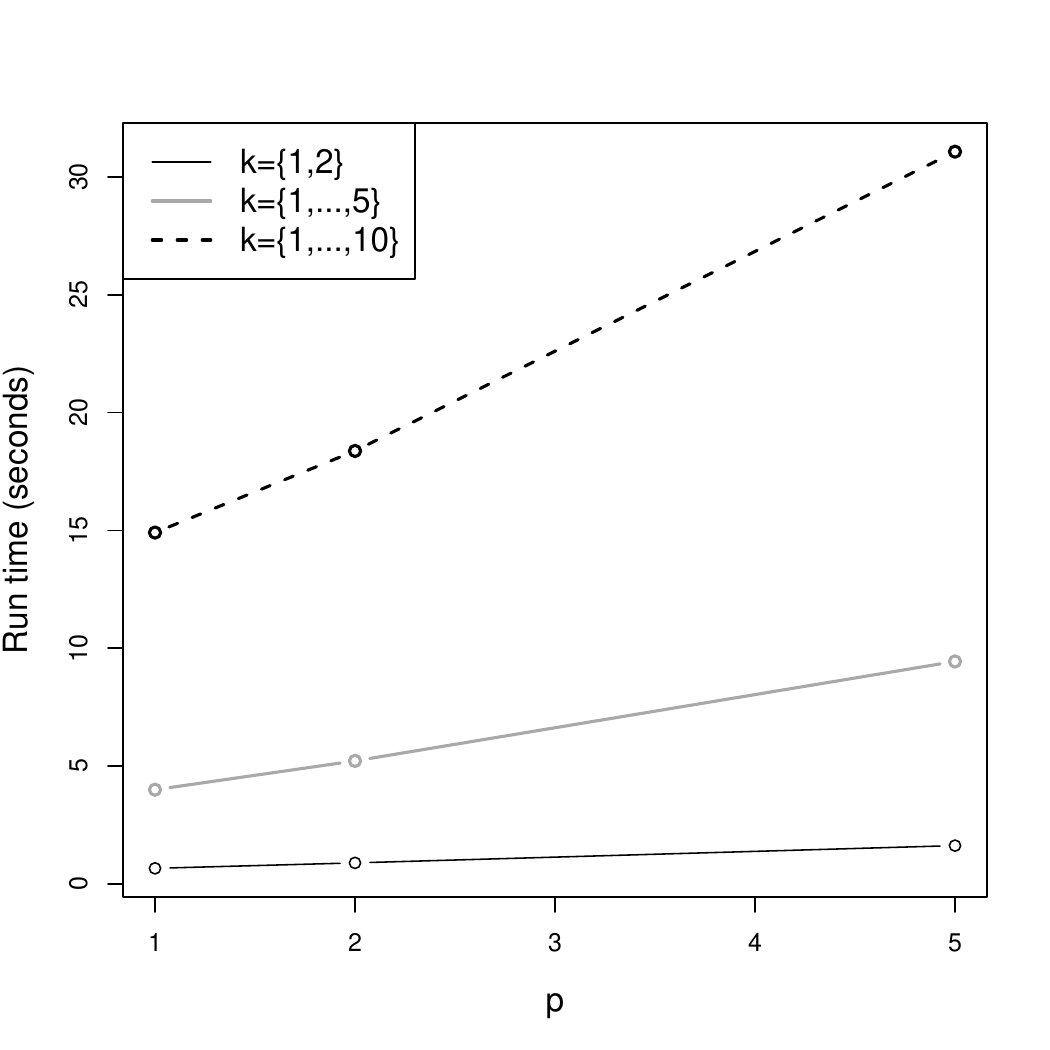} &
\includegraphics[width=0.5\textwidth]{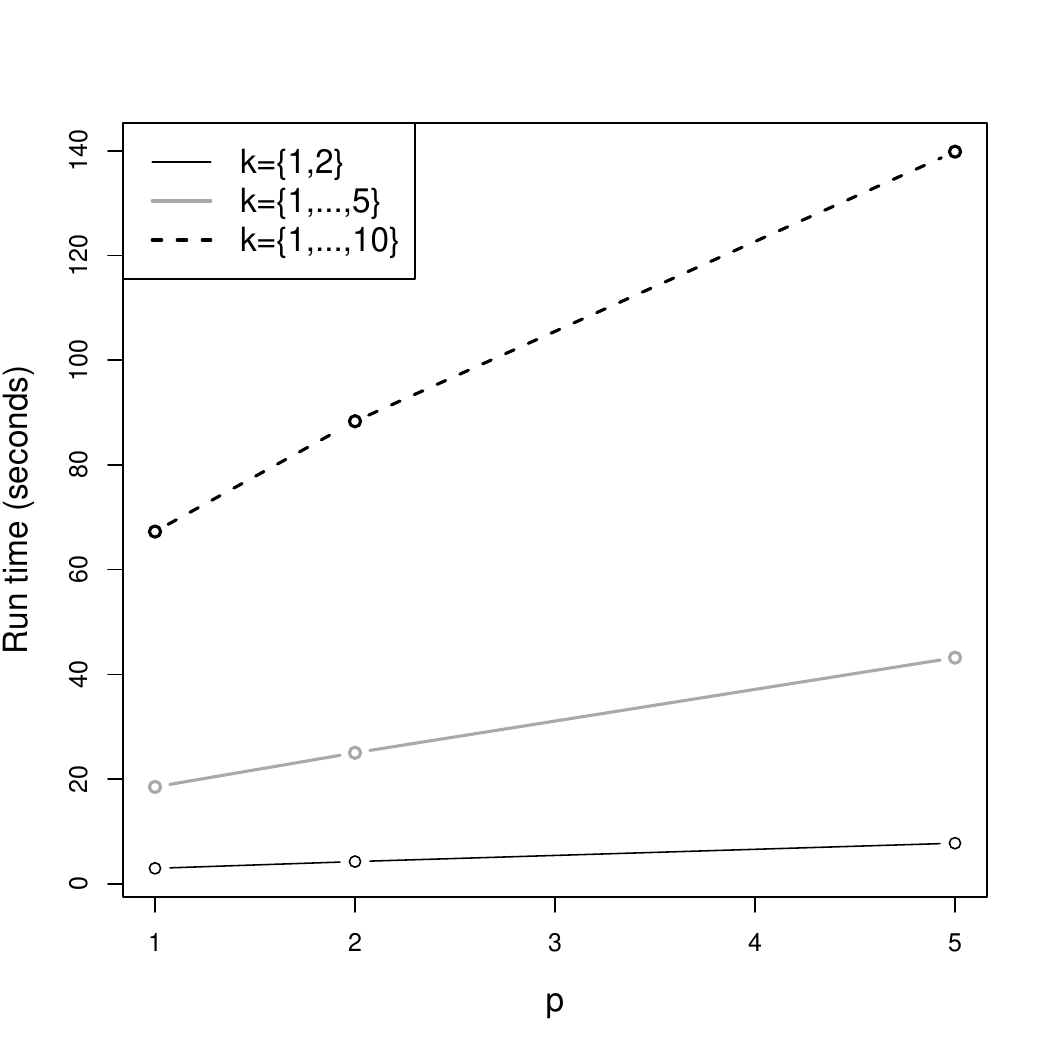}
\end{tabular}
\end{center}
\caption{Median run time (seconds) to compute $P(\Mk \mid \by)$ for all $k=1,\ldots,K$ via the ECP estimator
(function \texttt{bfnormmmix} in R package \texttt{mombf}). Data was generated from a one-component standard multivariate Normal}
\label{supfig:cputime}
\end{figure}

\section{Gibbs sampling algorithms}
\label{supplsec:gibbs}

Algorithm S\ref{alg:marginal_lhood} outlines a Gibbs sampling algorithm for Normal mixtures under the Normal-IW-Dir prior
$$\mbox{Dir}(\bm{\eta};q) \prod_j N(\bmu_j; 0, g \Sigma_j) \mbox{IW}(\Sigma_j;\nu,S).$$

Analogously, Algorithm S\ref{alg:marginal_lhoodBin} outlines a Gibbs sampling algorithm
for product Binomial mixtures under the Beta-Dirichlet prior
$$\mbox{Dir}(\bm{\eta};q) \prod_{jf} \mbox{Beta}\left(\theta_{jf}; ag, (1-a)g \right).$$

\setcounter{algocf}{0}
\begin{algorithm}[H]
\small
Initialize $\bvartheta_{k}^{(0)}=(\btheta_{1}^{(0)},...,\btheta_{k}^{(0)}, \boldsymbol{\eta}^{(0)})$
with $\btheta_{j}^{(0)}=(\bmu_{j}^{(0)},\Sigma_{j}^{(0)})$.
\For{$t=1,...,T$}{
Draw $z_i^{(t)}=j$ with probability:
\begin{align*}
\dfrac{\eta^{(t-1)}_{k}N(\by_{i};\bmu_{j}^{(t-1)},\Sigma_{j}^{(t-1)})}
{\sum_{j=1}^{k}\eta^{(t-1)}_{j}N(\by_{i};\bmu_{j}^{(t-1)},\Sigma_{j}^{(t-1)})}.
\end{align*}

Let $n_{j}^{(t)}=\sum_{i=1}^{n} \mbox{I}(z_i^{(t)}=j)$ and
$\bar{\by}_{j}^{(t)}=\dfrac{1}{n_{j}}\sum_{z_i^{(t)}=j}^{}\by_{i}$
if $n_{j}^{(t)}>0$, else $\bar{\by}_{j}^{(t)}=0$.
Draw
\begin{align*}
\boldsymbol{\eta}^{(t)}&\sim
\text{Dir}(q+n_{1}^{(t)},...,q+n_{k}^{(t)}).
\end{align*}

Let $S_j=S^{-1}+\sum_{z_i=j}^{}(\by_{i}-\bar{\by}_{j}^{(t-1)})(\by_{i}-\bar{\by}_{j}^{(t-1)})^{'}+
\sum_{j=1}^{k}\dfrac{n_{j}/g}{n_{j}+1/g}\bar{\by}_{j}^{(t)}\bar{\by}_{j}^{'(t)}$.
Draw

\begin{equation*}
\Sigma^{(t)}_{j}\sim
\text{IW}\left(\nu+n_j,S_j\right),
\end{equation*}

Draw
\begin{equation*}
\bmu_{j}^{(t)}\sim
N\left(\dfrac{gn_{j}^{(t)}\bar{\by}_{j}^{(t)}}{1+gn_{j}^{(t)}},\dfrac{g}{1+gn_{j}^{(t)}} \Sigma_j^{(t)}\right),
\end{equation*}
}
\caption{Gibbs sampling for Normal mixtures under a Normal-IW-Dir prior.}
\label{alg:marginal_lhood}
\end{algorithm}

\RestyleAlgo{boxruled}
\vspace{2mm}
\begin{algorithm}
\small
Initialize $\bvartheta_{k}^{(0)}=(\btheta_{1}^{(0)},...,\btheta_{k}^{(0)}, \boldsymbol{\eta}^{(0)})$ where
$\btheta_{j}^{(0)}=(\theta_{j1}^{(0)},\ldots,\theta_{jp}^{(0)})$.
\For{$t=1,...,T$}{
Draw $z_i^{(t)}=j$ with probability:
\begin{align*}
\dfrac{\eta^{(t-1)}_{j}\prod_{f=1}^{p}\text{Bin}(y_{if};L_{if},\theta_{jf}^{(t-1)})}
{\sum_{j=1}^{k}\eta^{(t-1)}_{j}\prod_{f=1}^{p}\text{Bin}(y_{if};L_{if},\theta_{jf}^{(t-1)})}.
\end{align*}
Draw
\begin{align*}
\boldsymbol{\eta}^{(t)}&\sim
\text{Dir}(q+n_{1}^{(t)},...,q+n_{k}^{(t)}).
\end{align*}
where $n_{j}^{(t)}=\sum_{i=1}^{n} \mbox{I}(z_i^{(t)}=j)$.
Draw
\begin{equation*}
\theta_{jf}^{(t)}\sim
\text{Beta}\left(ag + \sum_{z_i^{(t)}=j}^{}y_{if}, (1-a)g + \sum_{z_i^{(t)}=j}^{}(L_{if}-y_{if}) \right),
\end{equation*}
}
\caption{Gibbs sampling for product Binomial mixtures under the Beta-Dir prior.}
\label{alg:marginal_lhoodBin}
\end{algorithm}

\section{EM algorithm for multivariate Normal mixtures under MOM-Wishart-Dirichlet priors}
\label{supplsec:em_algorithm}

The complete-data posterior can be written as follows
\begin{align}
p(\bvartheta_{k}\mid \by,\bz,\mathcal{M}_{k})=
\prod_{j=1}^{k}\prod_{i=1}^{n}(\eta_{j}N(\by; \bmu_{j},\Sigma_{j}))^{z_{ij}}
N\left(\bmu_{j}; \boldsymbol{0},gA_{\Sigma}
\right)\text{Wishart}(\Sigma_j^{-1}; \nu,S)\mbox{Dir}(\bm{\eta};q).
\end{align}
The E-step at iteration $t$ requires the expectation of
$\log p(\bvartheta_{k}\mid \by,\bz,\Mk)$
with respect to $p(\bz \mid \by,\bvartheta_k^{(t-1)},\Mk)$, where
$\bvartheta_k^{(t-1)}=(\boldsymbol{\eta}^{(t-1)},\boldsymbol{\mu}_1^{(t-1)},...,\boldsymbol{\mu}_k^{(t-1)},
\Sigma_1^{(t-1)},...,\Sigma_k^{(t-1)})$
are the parameter values at iteration $t-1$.
Let
\begin{align}
\bar{z}^{(t)}_{ij}=p(z_{i}=j\mid \by_{i},\bvartheta_k^{(t-1)}) =\dfrac{\eta_{j}^{(t-1)}N(\by_{i};\bmu_{j}^{(t-1)},\Sigma_{j}^{(t-1)})}{\sum_{j=1}^{k}\eta_{j}^{(t-1)}
N(\by_{i}; \bmu_{j}^{(t-1)},\Sigma_{j}^{(t-1)})},
\end{align}
then the M-step seeks $\bvartheta_k^{(t)}$ maximising
\begin{align}\label{obj1}
\small
\log(p(\bvartheta_{k}\mid \by,\bar{z}_{ij},\mathcal{M}_{k}))=&
\sum_{j=1}^{k}n_{j}\log(\eta_{j})+
\sum_{j=1}^{k}\sum_{i=1}^{n}\bar{z}_{ij}\log(N(\boldsymbol{y}_{i}; \boldsymbol{\mu_{j}},\Sigma_{j}))
+ \sum_{j=1}^{k}\log(N\left(\bmu_{j}; \boldsymbol{0},gA_{\Sigma}
\right))\\\notag
&+\sum_{1\leq i < j \leq
 k}\log((\boldsymbol{\mu}_{i}-\boldsymbol{\mu}_{j})^{'}A_{\Sigma}^{-1}(\boldsymbol{\mu}_{i}-\boldsymbol{\mu}_{j}))+
\sum_{j=1}^{k}\log(\text{Wishart}(\Sigma_j^{-1}; \nu,S))\\\notag
&+
\log(\mbox{Dir}(\bm{\eta};q))
\end{align}
where $n_{j}^{(t)}=\sum_{i=1}^{n}\bar{z}^{(t)}_{ij}$. We successively update
$\boldsymbol{\eta}^{(t)}$,
$\boldsymbol{\mu}_1^{(t)}$,...,$\boldsymbol{\mu}_k^{(t)}$ and
$\Sigma_{1}^{(t)},...,\Sigma_{k}^{(t)}$ in a fashion that guarantees
that (\ref{obj1}) increases at each step.
The update $\eta^{(t)}_{j}$ is
\begin{align}
\eta^{(t)}_{j}=\dfrac{n_{j}^{(t)}+q-1}{n+k(q-1)},
\end{align}
which maximizes (\ref{obj1}) with respect to $\boldsymbol{\eta}$ conditional on the current
$\boldsymbol{\mu}_{1}^{(t-1)}$,...,$\boldsymbol{\mu}_k^{(t-1)}$ and
$\Sigma_{1}^{(t-1)},...,\Sigma_{k}^{(t-1)}$.
To update $\boldsymbol{\mu}_{j}^{(t)}$ we seek to maximize
\begin{align*}
\xi(\boldsymbol{\mu}_{j}^{(t)})
=\sum_{i\neq j}\log(\boldsymbol{C}_{ij}^{(t)'}A_{\Sigma^{(t-1)}}^{-1}\boldsymbol{C}_{ij}^{(t)})
-\dfrac{1}{2
g}\boldsymbol{\mu}_{j}^{'(t)}A_{\Sigma^{(t-1)}}^{-1}\boldsymbol{\mu}_{j}^{(t)}-\dfrac{1}{2}\sum_{i=1}^{n}\bar{z}^{(t)}_{ij}(\boldsymbol{y}_{i}-\boldsymbol{\mu}_{j}^{(t)})^{'}
A_{\Sigma^{(t-1)}}^{-1}
(\boldsymbol{y}_{i}-\boldsymbol{\mu}_{j}^{(t)}),
\end{align*}
where $\boldsymbol{C}_{ij}=(\boldsymbol{\mu}_{i}-\boldsymbol{\mu}_{j})$.
The first derivative of $\xi(\boldsymbol{\mu}_{j}^{(t)})$ is
\begin{align*}
\nabla\xi(\boldsymbol{\mu}_{j}^{(t)})&=-2\sum_{i \neq j}\dfrac{A_{\Sigma^{(t-1)}}^{-1}\boldsymbol{C}_{ij}^{(t)}}{\boldsymbol{C}_{ij}^{(t)'}A_{\Sigma^{(t-1)}}^{-1}\boldsymbol{C}_{ij}^{(t)}}-\dfrac{1}{
g}(A_{\Sigma^{(t-1)}}^{-1}\boldsymbol{\mu}_{j}^{(t)})-\sum_{i=1}^{n}\bar{z}^{(t)}_{ij}(A_{\Sigma^{(t-1)}}^{-1}(\boldsymbol{y}_{i}-\boldsymbol{\mu}_{j}^{(t)})).
\end{align*}
Because an analytic solution of
$\nabla\xi(\boldsymbol{\mu}_{j}^{(t)})= \boldsymbol{0}$ in terms of
$\boldsymbol{\mu}_{j}^{(t)}$  is not feasible we resort to a first order Taylor's approximation for
$-2\sum_{i \neq j}(A_{\Sigma^{(t-1)}}^{-1}\boldsymbol{C}_{ij}^{(t)})/(\boldsymbol{C}_{ij}^{(t)'}A_{\Sigma^{(t-1)}}^{-1}\boldsymbol{C}_{ij}^{(t)})$
around $\boldsymbol{\mu}_{j}^{(t-1)}$.
Finding the maximum of this Taylor approximation gives the candidate update
\begin{align}
\bmu_{j}^{*}=&\left(\Sigma_{j}^{-1(t-1)}n_{j}^{(t)} + A_{\Sigma^{(t-1)}}^{-1} \left(\dfrac{1}{g} +
\sum_{\substack{j\neq k}} \frac{2}{d_{ij}^{(t-1)}}\right)\right)^{-1}
\\&\hspace{-1.0cm}\times\left(\Sigma^{-1(t-1)}n_{j}^{(t)}\bar{\by}_{j}^{(t)} +  A_{\Sigma^{(t-1)}}^{-1}
\left(
\sum_{\substack{ i\neq j}}
\frac{\bmu_{j}^{(t-1)}-(\bmu_{i}^{(t-1)}-\bmu_{j}^{(t-1)})}{d_{ij}^{(t-1)}}\right)\right), \notag
\end{align}
where $d_{ij}^{(t-1)}=(\bmu_{i}^{(t-1)}-\bmu_{j}^{(t-1)})' A_{\Sigma^{(t-1)}}^{-1}(\bmu_{i}^{(t-1)}-\bmu_{j}^{(t-1)})$.
If $\xi(\boldsymbol{\mu}_{j}^{*})>\xi(\boldsymbol{\mu}_{j}^{(t-1)})$ we set $\boldsymbol{\mu}_{j}^{(t)}=\boldsymbol{\mu}_{j}^{*}$,
else take the gradient step in Algorithm S\ref{alg:Gradient}.

Finally we describe updating $\Sigma_j$ for $j=1,\ldots,k$.
Redefine $\xi(\Sigma_j)$ to now be \eqref{obj1} viewed as a function of $\Sigma_j$.
Due to the terms
$\sum_{\substack{i\neq j}}\log(\boldsymbol{\mu}_{i}^{(t)}-\boldsymbol{\mu}_{j}^{(t)})^{'}A_{\Sigma^{(t)}}^{-1}(\boldsymbol{\mu}_{i}^{(t)}-\boldsymbol{\mu}_{j}^{(t)})$
and $-\frac{1}{2}\log(|A_{\Sigma^{(t)}}^{-1}|)$
an analytic solution of $\nabla\xi(\Sigma_{j})={\bf 0}$ is not available,
hence we use the Taylor expansion around $\Sigma_j^{(t-1)}$

\begin{align*}\small
\sum_{\substack{i\neq j
}}\log(\boldsymbol{\mu}_{i}^{(t)}-\boldsymbol{\mu}_{j}^{(t)})^{'}A_{\Sigma^{(t)}}^{-1}(\boldsymbol{\mu}_{i}^{(t)}-\boldsymbol{\mu}_{j}^{(t)})
-\frac{1}{2}\log(|A_{\Sigma^{(t)}}^{-1}|) \approx\\
\sum_{\substack{i\neq j
}}\frac{(\boldsymbol{\mu}_{i}^{(t)}-\boldsymbol{\mu}_{j}^{(t)})^{'}A_{\Sigma^{(t)}}^{-1}(\boldsymbol{\mu}_{i}^{(t)}-\boldsymbol{\mu}_{j}^{(t)})}
{(\boldsymbol{\mu}_{i}^{(t-1)}-\boldsymbol{\mu}_{j}^{(t-1)})^{'}A_{\Sigma^{(t-1)}}^{-1}(\boldsymbol{\mu}_{i}^{(t-1)}-\boldsymbol{\mu}_{j}^{(t-1)})}-
\frac{1}{2}\log(|\Sigma_{j}^{(t)}|).
\end{align*}

Note that when a common $\Sigma_1=\ldots=\Sigma_k$ is assumed then $A_{\Sigma^{(t)}}=\Sigma^{(t)}$ we only need a Taylor expansion of first term.
Summarising, the candidate update is
\begin{align*}
\hspace{-2.0cm}
(\nu-p+n_{j}^{(t)})\Sigma^*_j=
  S^{-1} + \dfrac{\bmu_{j}^{(t)} (\bmu_{j}^{(t)})'}{kg}+\sum_{i=1}^{n}\bar{z}^{(t)}_{ij}
(\by_{i}-\bmu_{j}^{(t)})(\by_{i}-\bmu_{j}^{(t)})^{'} \nonumber \\
-\dfrac{1}{k}
\sum_{\substack{i\neq j
}}\dfrac{2(\bmu_{j}^{(t)}-\bmu_{k}^{(t)})(\bmu_{j}^{(t)}-\bmu_{k}^{(t)})^{'}}{d_{ij}^{(t-1)}}.
\end{align*}

If $\xi(\Sigma_{j}^{*})>\xi(\Sigma_{j}^{(t-1)})$ we set $\Sigma_{j}^{(t)}=\Sigma_{j}^{*}$,
else take a gradient step (Algorithm S\ref{alg:Gradient})
with a small enough step size to ensure that $\Sigma_{j}^{(t)}$ remains positive-definite.

\begin{algorithm}[H]
\small

Initialization $\boldsymbol{\zeta}=\boldsymbol{\zeta}^{*}$, $\bar{k}=\sqrt{\dfrac{\|\boldsymbol{\zeta}^{*}-\boldsymbol{\zeta}^{(t-1)}\|}{\nabla\xi(\boldsymbol{\zeta}^{(t-1)})}}$
and $h=0$\;
\While{$(\xi(\boldsymbol{\zeta}^{(t-1)})> \xi(\boldsymbol{\zeta}^{*}))$}{
$\boldsymbol{\zeta}^{*}=\boldsymbol{\zeta}^{(t-1)} + \dfrac{\bar{k}}{2^{h}}\nabla\xi(\boldsymbol{\zeta}^{(t-1)})$\;
$h=h+1$
}
$\boldsymbol{\zeta}^{(t)}=\boldsymbol{\zeta}^{*}$
\caption{Gradient Ascend algorithm.}
\label{alg:Gradient}
\end{algorithm}

\section{EM algorithm for product Binomial mixture under MOM-Beta priors}
\label{supplsec:em_algorithm2}

The EM algorithm is derived analogously to that for Normal mixtures (Supplementary Section \ref{supplsec:em_algorithm}),
and is described in Algorithm S\ref{alg:em1}.

\vspace{2mm}
\begin{algorithm}[H]
\small
Set $t=1$. \While{$\zeta>\epsilon^{*}$ and $t<T$}{
\For{$t\geq 1$ and $j=1,...,k$}{
E-step. Let
$\bar{z}^{(t)}_{ij}=\dfrac{\eta^{(t-1)}_{j}\prod_{f=1}^{p}\text{Bin}(y_{if};L_{if},\theta_{jf}^{(t-1)})}
{\sum_{j=1}^{k}\eta^{(t-1)}_{j}\prod_{f=1}^{p}\text{Bin}(y_{if};L_{if},\theta_{jf}^{(t-1)})}.
$ \linebreak
M-step.
Update
\begin{equation*}
\btheta_{j}^{(t)}=\dfrac{ag + \sum_{i=1}^{n}\bar{z}^{(t)}_{ij}\by_{i} + \ell_{1}(\btheta_{j}^{(t)})}{(1-a)g + \sum_{i=1}^{n}\bar{z}^{(t)}_{ij}(L_{if}-\by_{i})+ 2\ell_{2}(\btheta_{j}^{(t)})},
\end{equation*}
$
\ell_{1}(\btheta_{j}^{(t)})=\dfrac{\btheta_{j}^{(t-1)}-(\btheta_{i}^{(t-1)}-\btheta_{j}^{(t-1)})}{(\btheta_{i}^{(t-1)}-\btheta_{j}^{(t-1)})'(\btheta_{i}^{(t-1)}-\btheta_{j}^{(t-1)})},
$
$
\ell_{2}(\btheta_{j}^{(t)})=\left[(\btheta_{i}^{(t-1)}-\btheta_{j}^{(t-1)})' (\btheta_{i}^{(t-1)}-\btheta_{j}^{(t-1)})\right]^{-1}.
$
Update
$\eta^{(t)}_{j}=\dfrac{n_{j}^{(t)}+q-1}{n+k(q-1)}$.
}
Compute  $\zeta=|\xi(\bvartheta_{k}^{(t)})-\xi(\bvartheta_{k}^{(t-1)})|$ and set $t=t+1$.
}
\caption{EM under MOM-Beta priors.}
\label{alg:em1}
\end{algorithm}

\section{Sensitivity to prior elicitation}
\label{supplsec:results}

\begin{figure}[ht]
\begin{center}
\begin{tabular}{cc}
\includegraphics[width=0.3\textwidth]{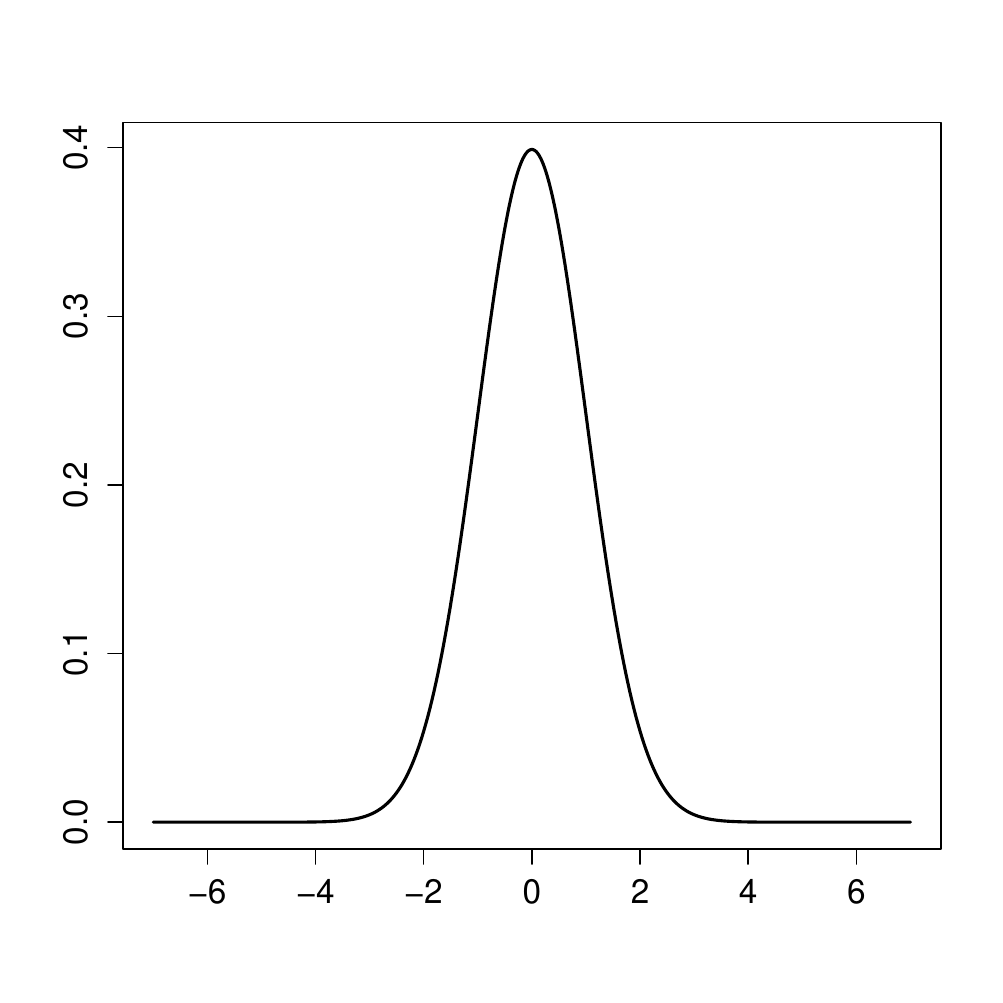} &
\includegraphics[width=0.3\textwidth]{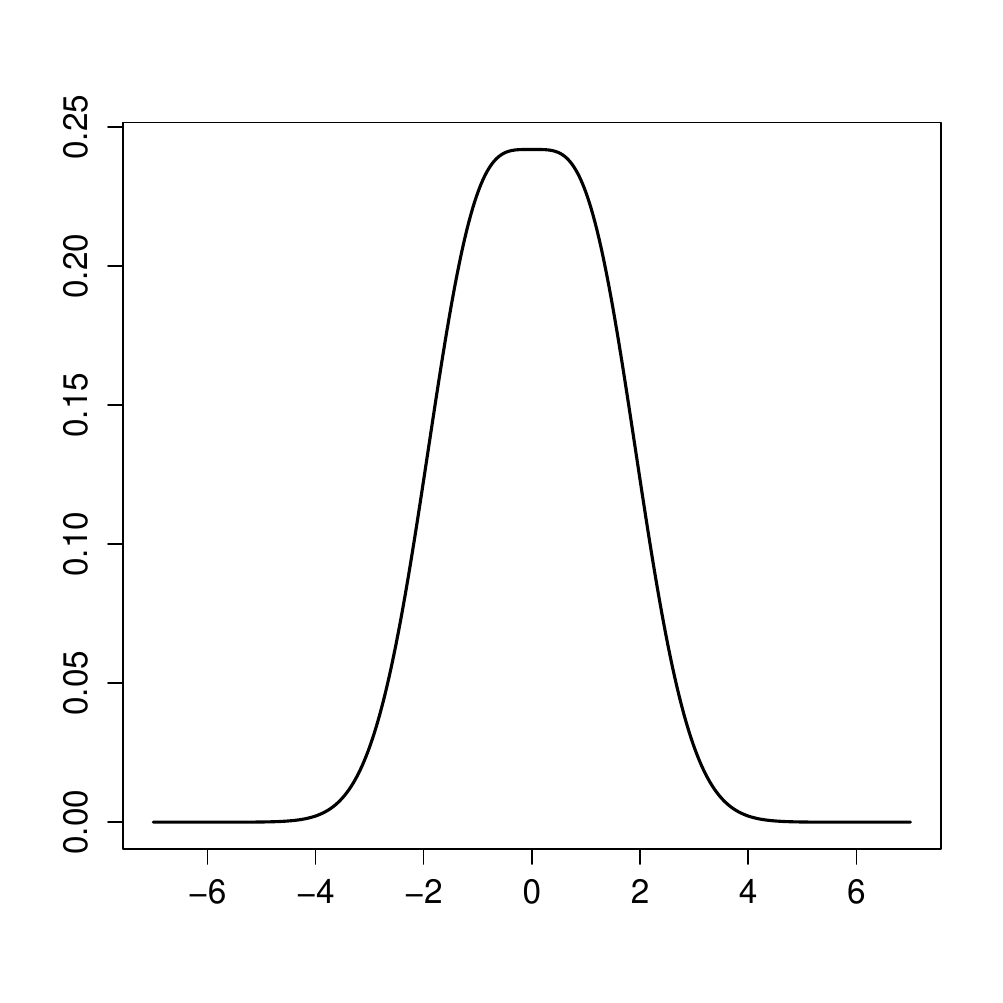}  \\
\includegraphics[width=0.3\textwidth]{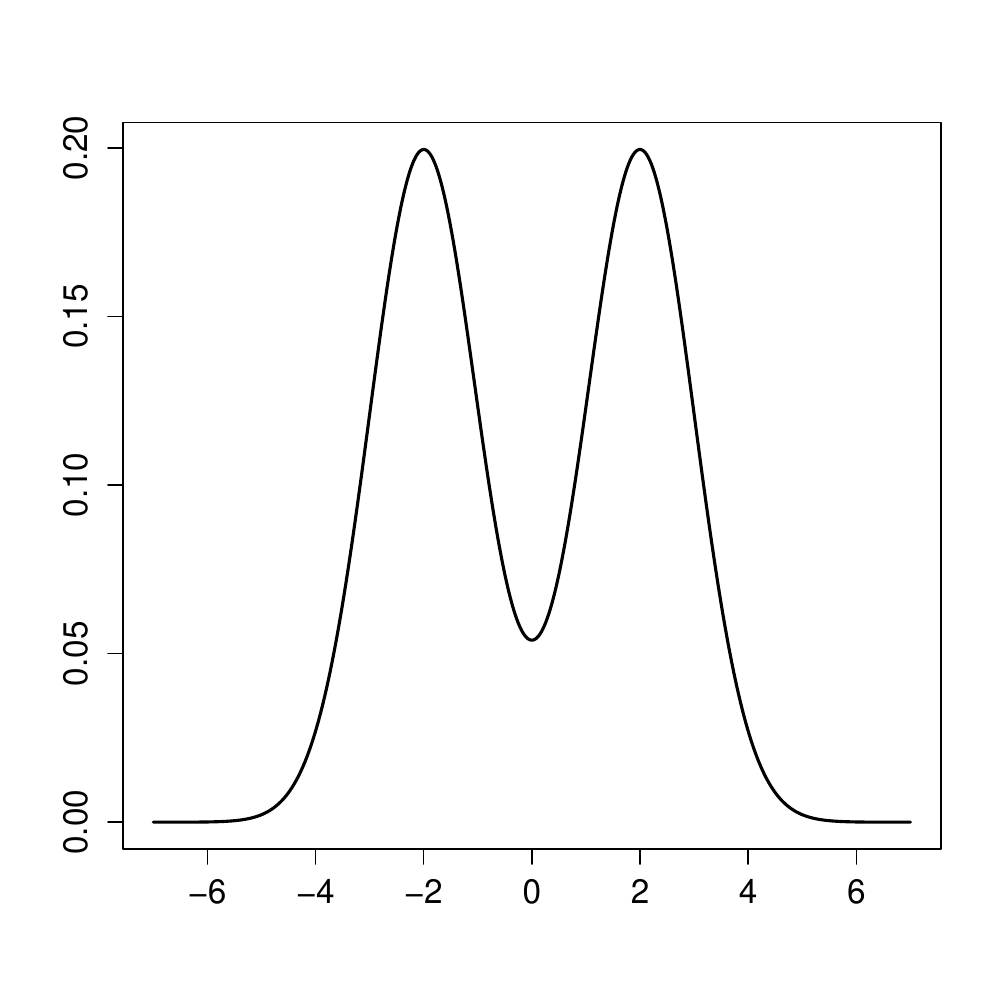} &
\includegraphics[width=0.3\textwidth]{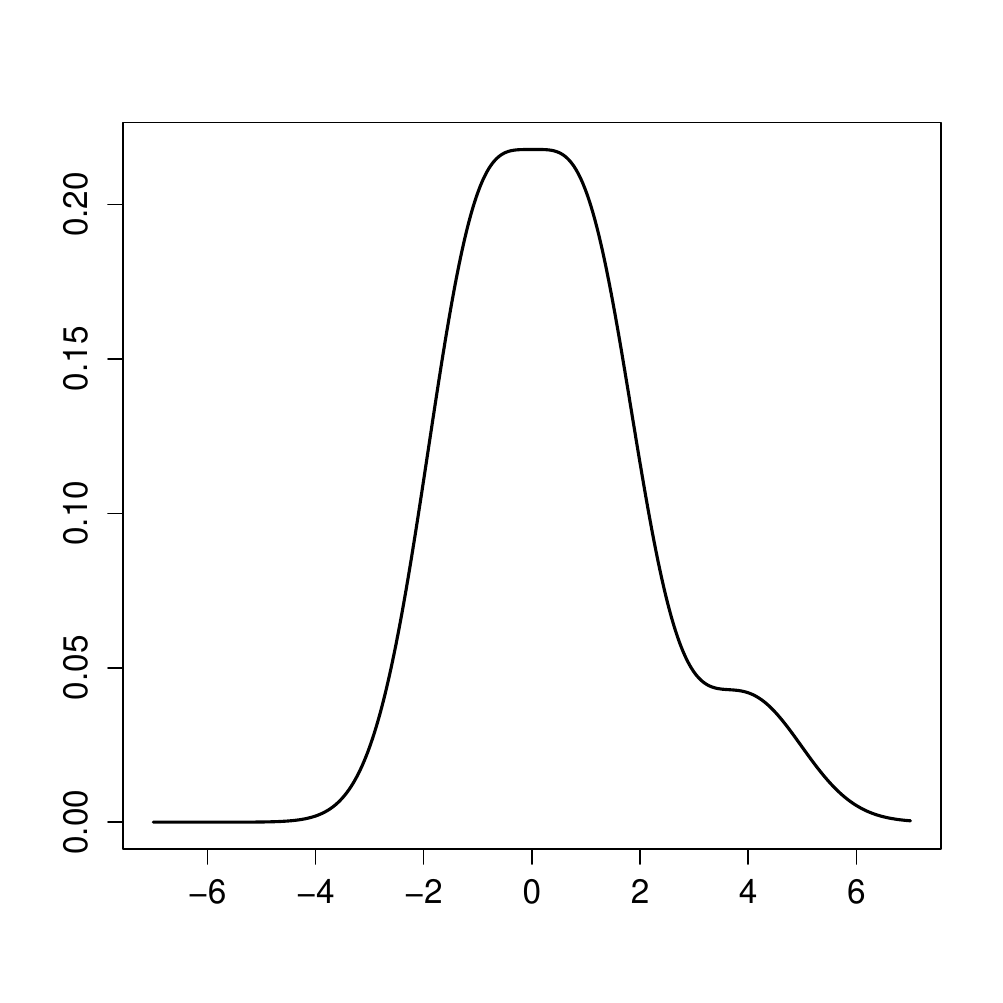}\\
\includegraphics[width=0.3\textwidth]{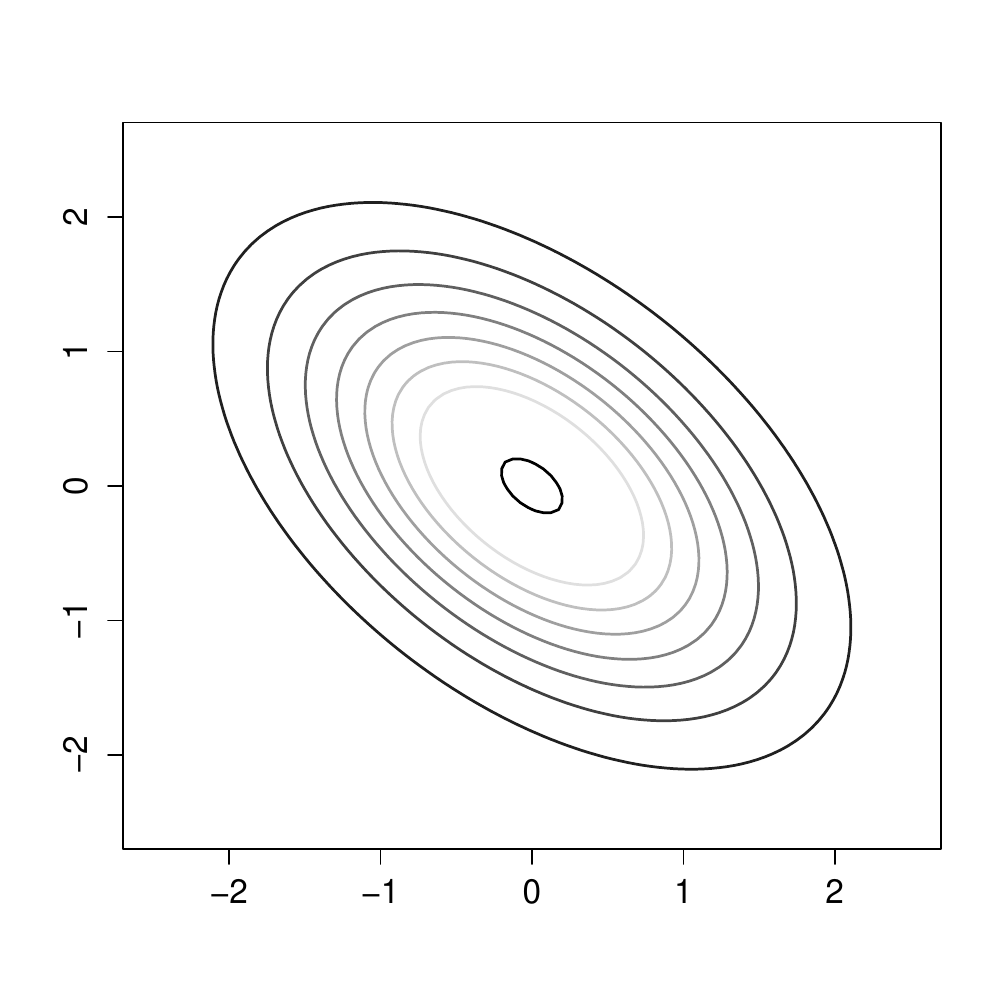} &
\includegraphics[width=0.3\textwidth]{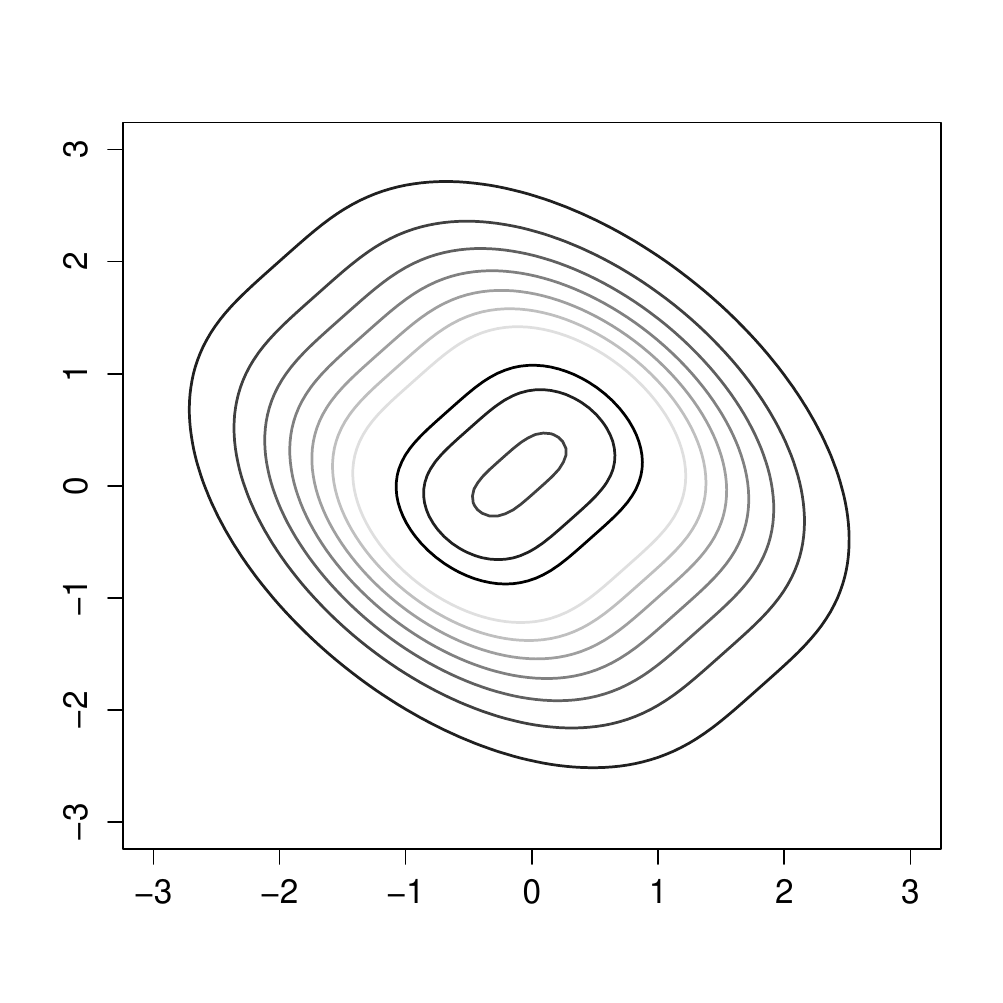} \\
\includegraphics[width=0.3\textwidth]{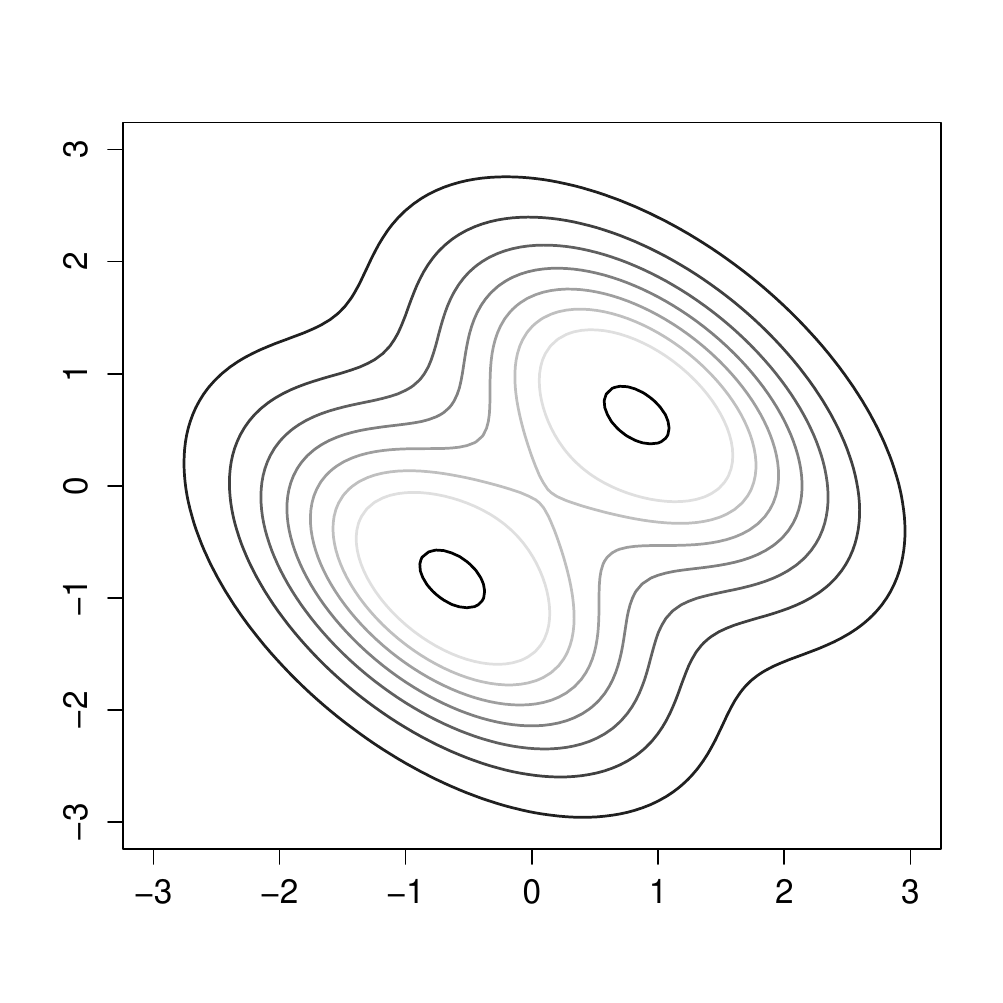} &
\includegraphics[width=0.3\textwidth]{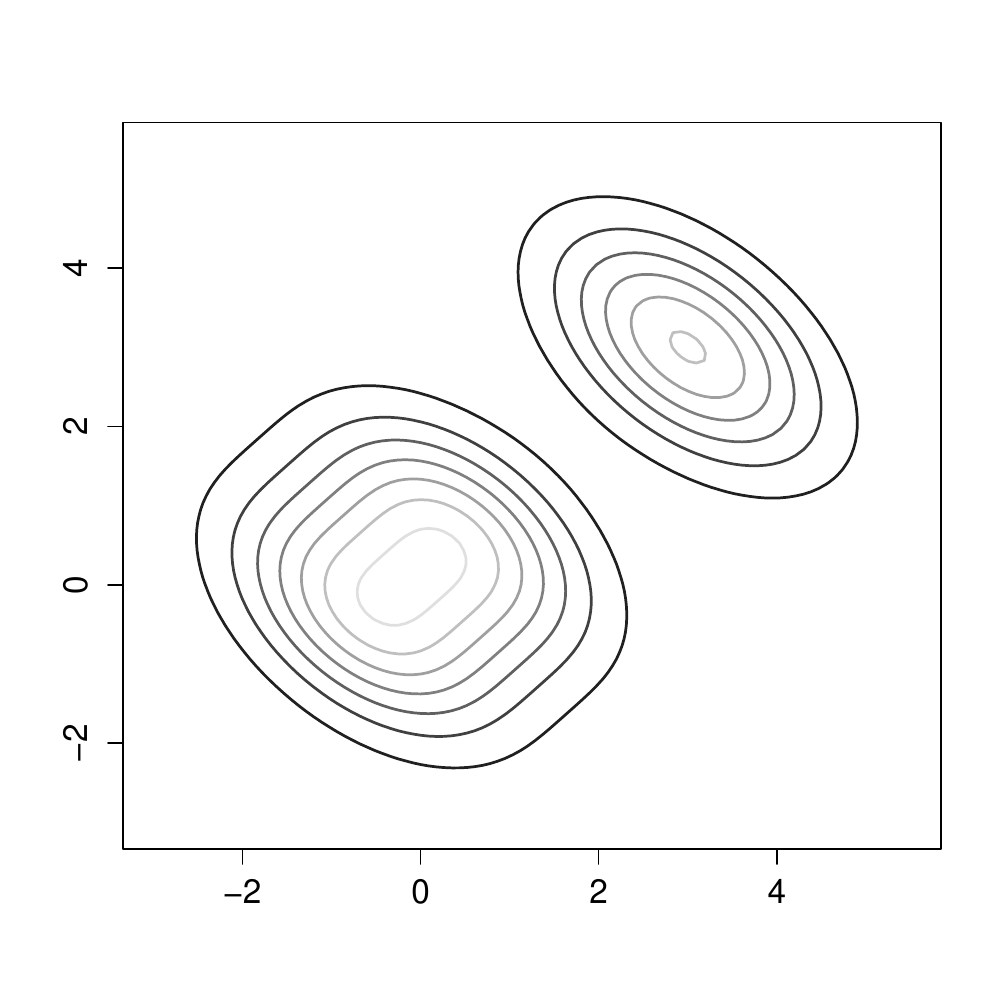}\\
\end{tabular}
\end{center}
\caption{\small Simulation study data-generating truth.
Case 1: $k^*$=1, $\bmu_1$=0;
Case 2: $k^*$=2, $\bmu_1$=-1, $\bmu_2$=1, $\bm{\eta}=(0.5,0.5)$;
Case 3: $k^*$=2, $\bmu_1=-2,\bmu_2=2$, $\bm{\eta}$=0.5;
Case 4: $k^*$=3, $\bmu_1=-1,\bmu_2=1,\bmu_3=4$, $\bm{\eta}=$(0.45,0.45,0.1);
Case 5: $k^*$=1, $\bmu=(0,0)'$;
Case 6: $k^*$=2, $\bmu_{1}=(-0.4,-0.6)'$, $\bmu_{2}=-\bmu_{1}$;
Case 7: $k^*$=2, $\bmu_{1}=(-0.65,-0.85)'$, $\bmu_{2}=-\bmu_{1}$;
Case 8: $k^*$=3, $\bmu_{1}=(-0.65,-0.85)'$, $\bmu_{2}=-\bmu_{1}$, $\bmu_{3}=(3,3)'$, $\bm{\eta}$=(0.35,0.35,0.3).
$\Sigma=1$ in Cases 1-4, $\sigma^{2}_{11}=\sigma^{2}_{22}=1$ and $\sigma^{2}_{12}=\sigma^{2}_{21}=-0.5$ in Cases 5-8.}
\label{supfig:densities}
\end{figure}

We provide additional results for the simulation study in Section \ref{ssec:syntheticexamples}.

\begin{figure}[ht]
\begin{center}
\begin{tabular}{cc}
Case 1 ($k^{*}=1$, q=2) & Case 2 ($k^{*}=2$, p=1, q=2) \\
\includegraphics[width=0.30\textwidth]{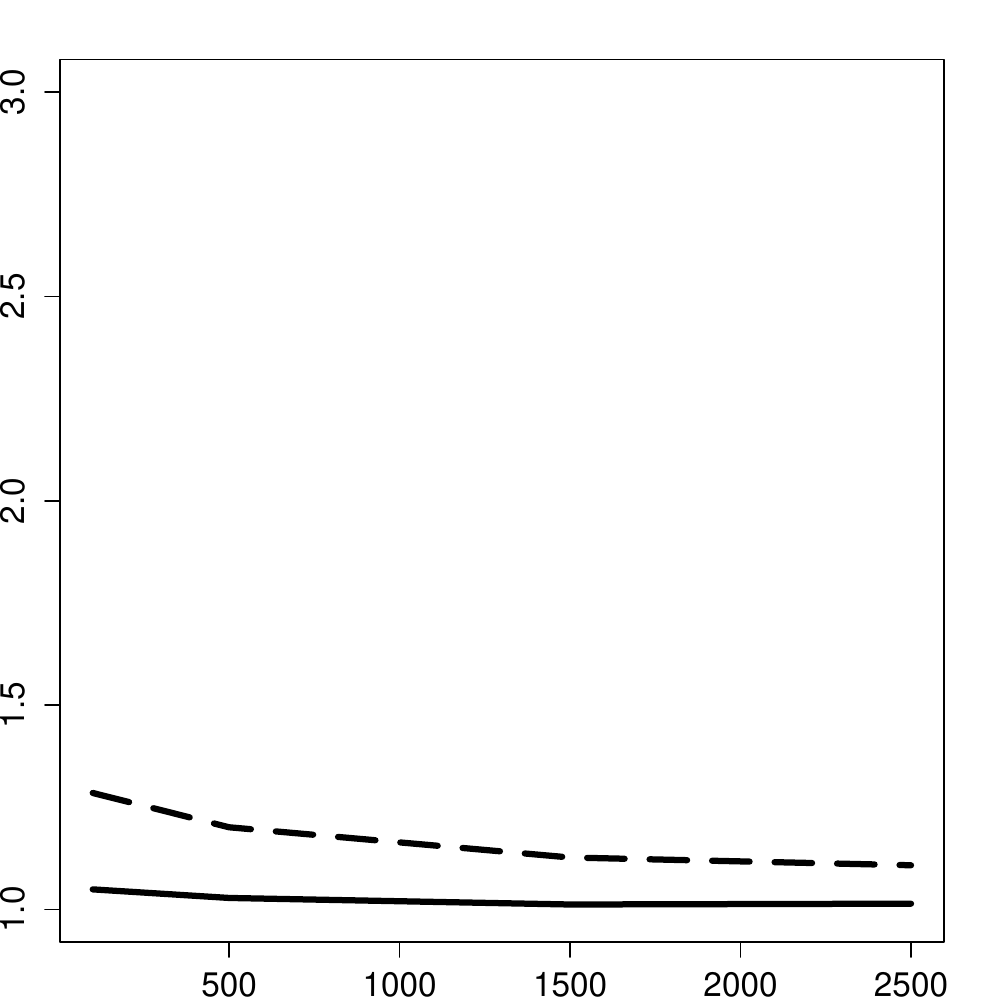} &
\includegraphics[width=0.30\textwidth]{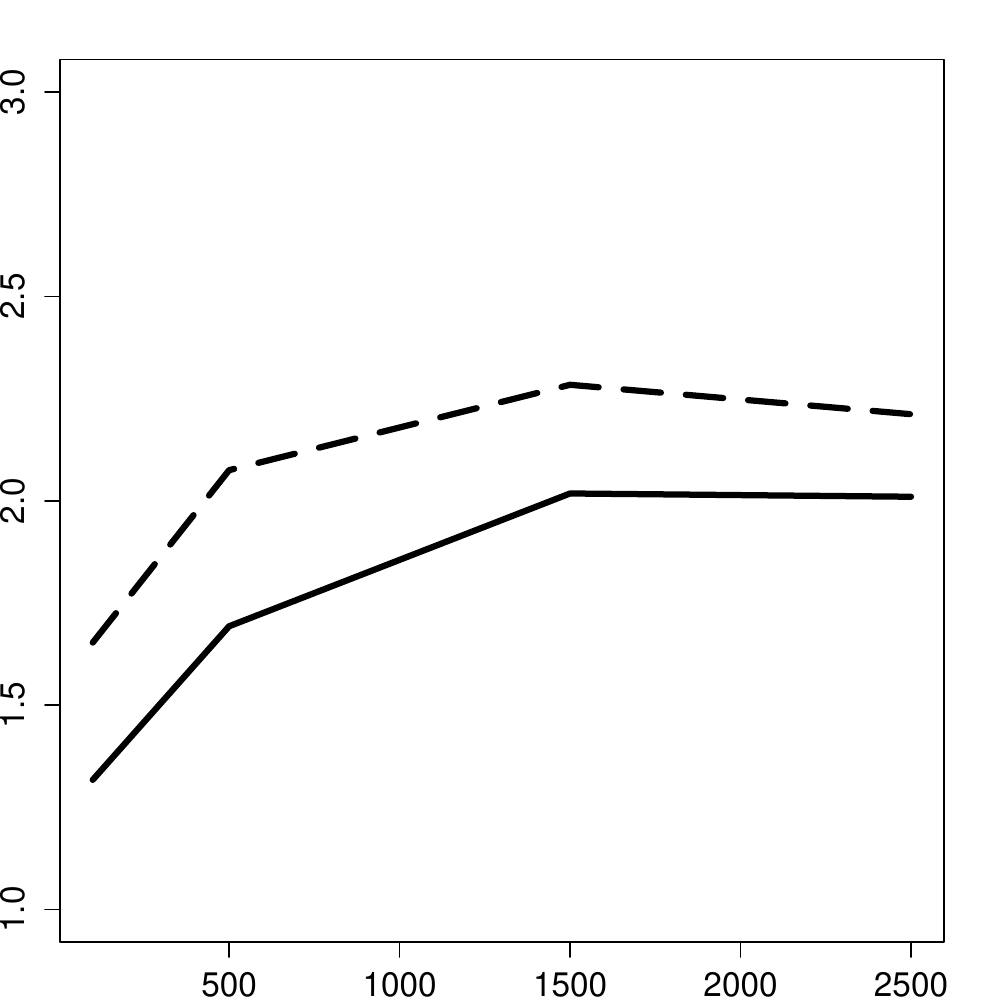} \\
Case 3 ($k^{*}=2$, p=1, q=2) & Case 4 ($k^{*}=3$, p=1, q=2) \\
\includegraphics[width=0.30\textwidth]{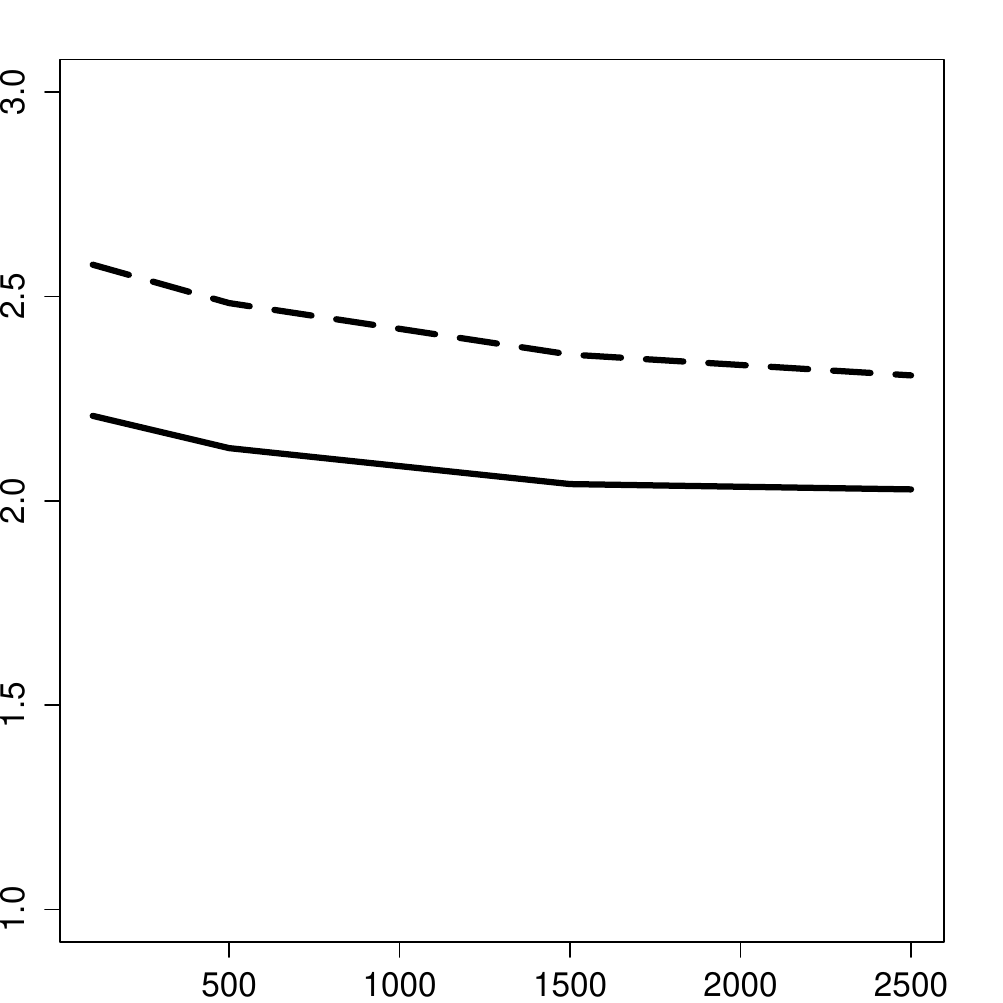} &
\includegraphics[width=0.30\textwidth]{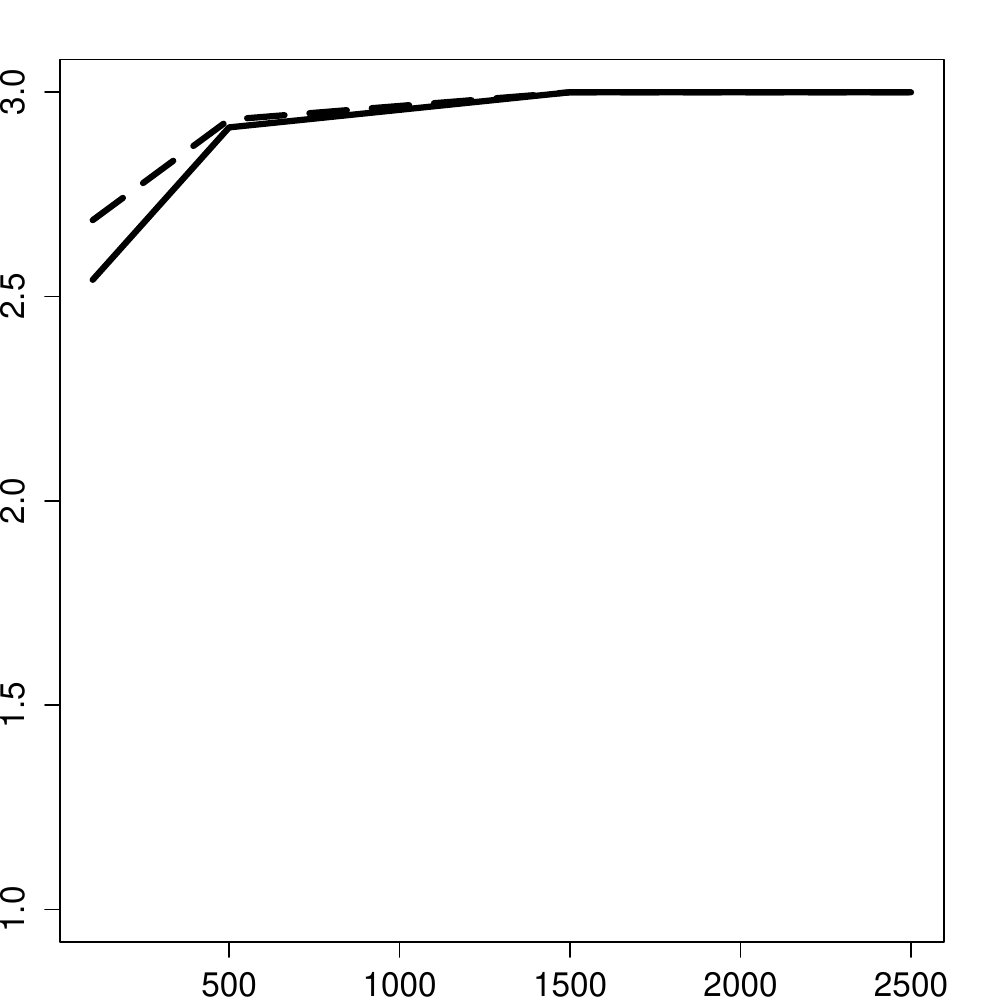} \\
Case 5 ($k^{*}=1$, p=2, q=3) & Case 6 ($k^{*}=2$, p=2, q=3) \\
\includegraphics[width=0.30\textwidth]{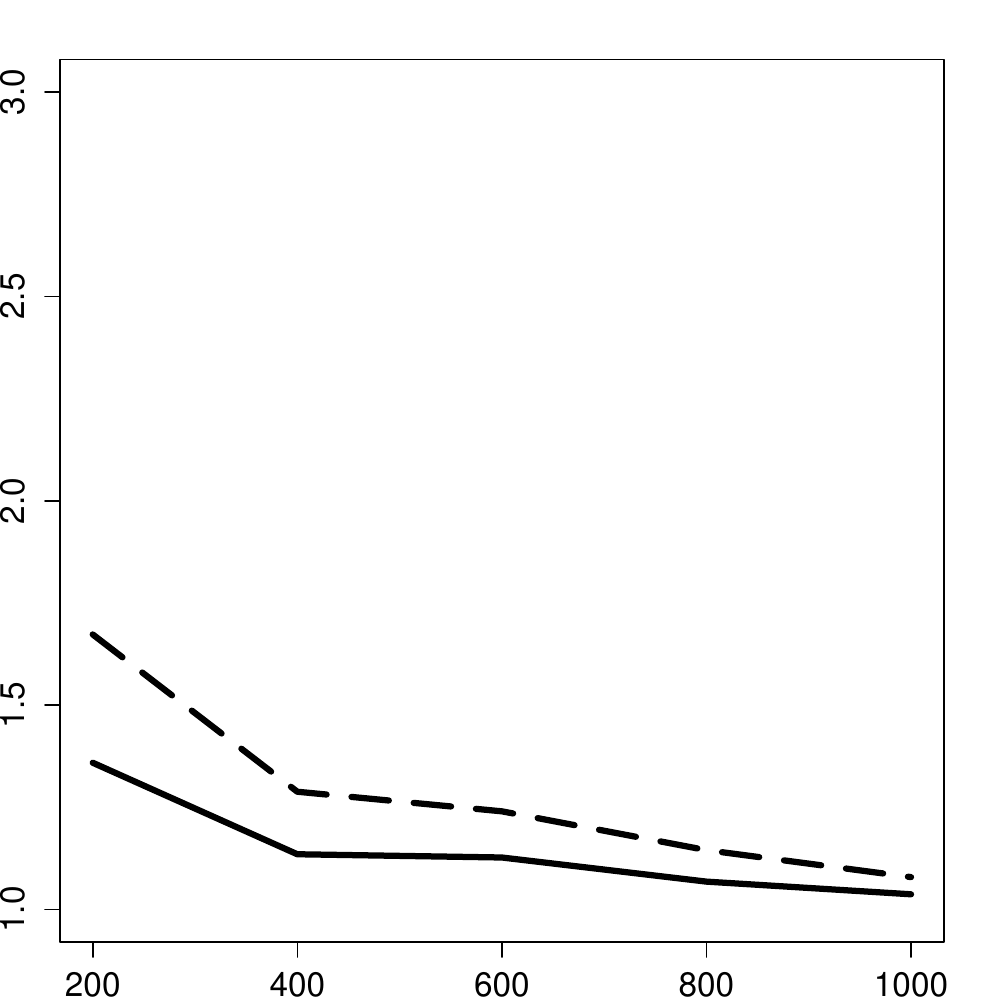} &
\includegraphics[width=0.30\textwidth]{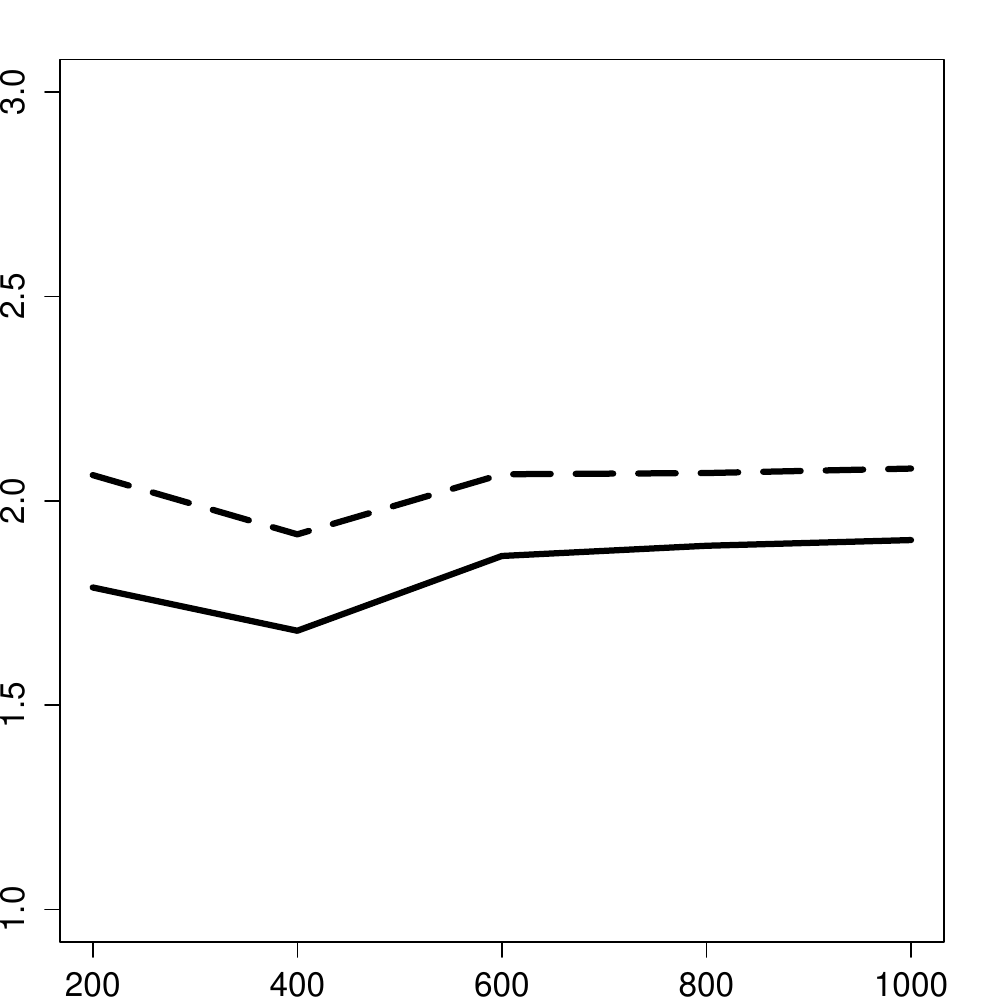} \\
Case 7 ($k^{*}=2$, p=2, q=3) & Case 8 ($k^{*}=3$, p=2, q=3) \\
\includegraphics[width=0.30\textwidth]{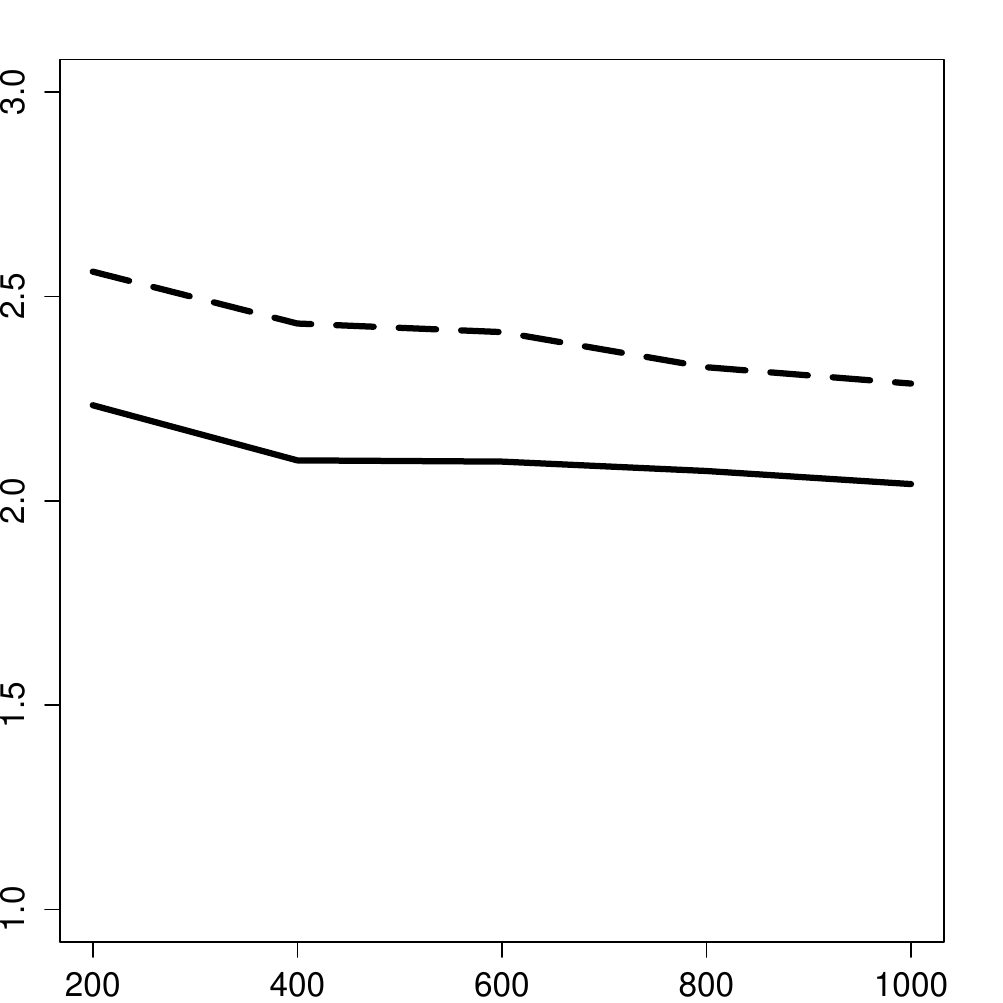} &
\includegraphics[width=0.30\textwidth]{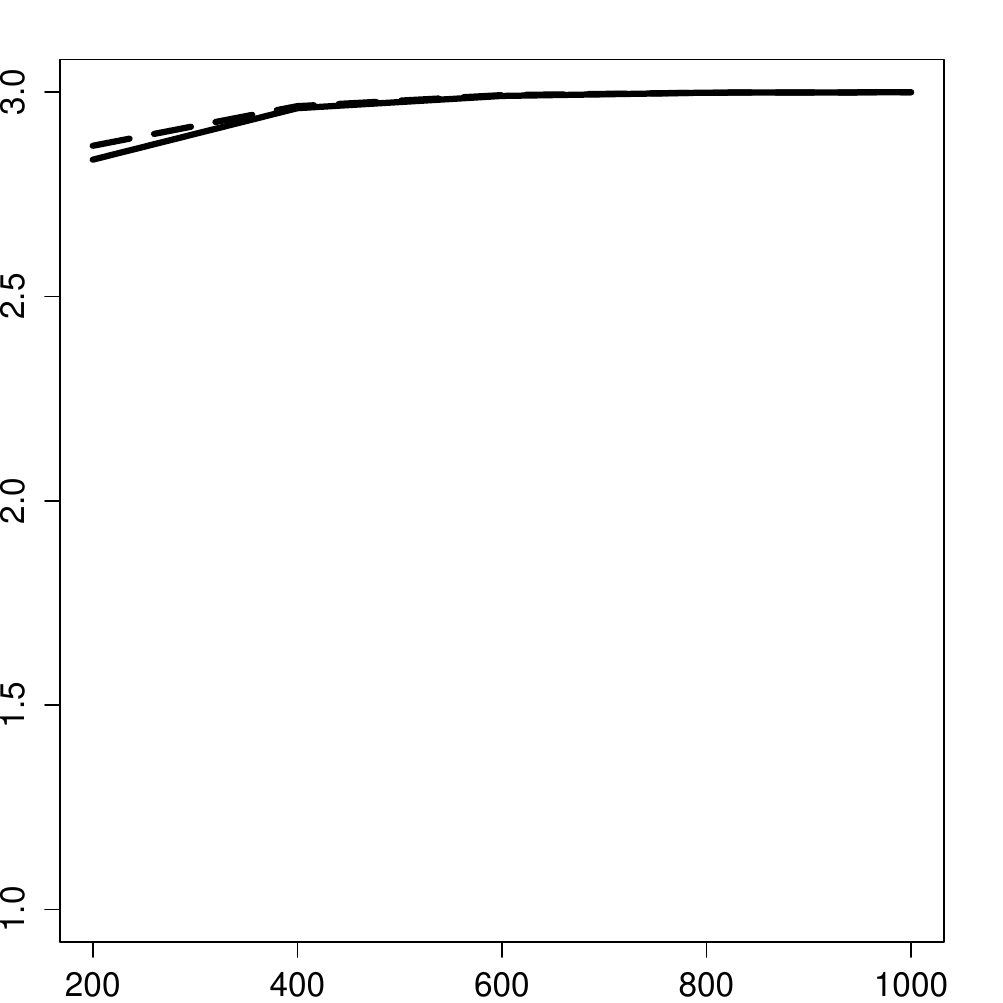} \\
\end{tabular}
\end{center}
\caption{Simulation study. Posterior expected model size $E(k \mid \by)$  versus $n$ for $q=p+1$ for the MOM-IW-Dir (solid line) and Normal-IW-Dir (dotted line).}
\label{supfig:synthetic_allmodels}
\end{figure}

\begin{figure}[ht]
\begin{center}
\begin{tabular}{cc}
Case 1 ($k^{*}=1$, q=4) & Case 2 ($k^{*}=2$, p=1, q=4) \\
\includegraphics[width=0.30\textwidth]{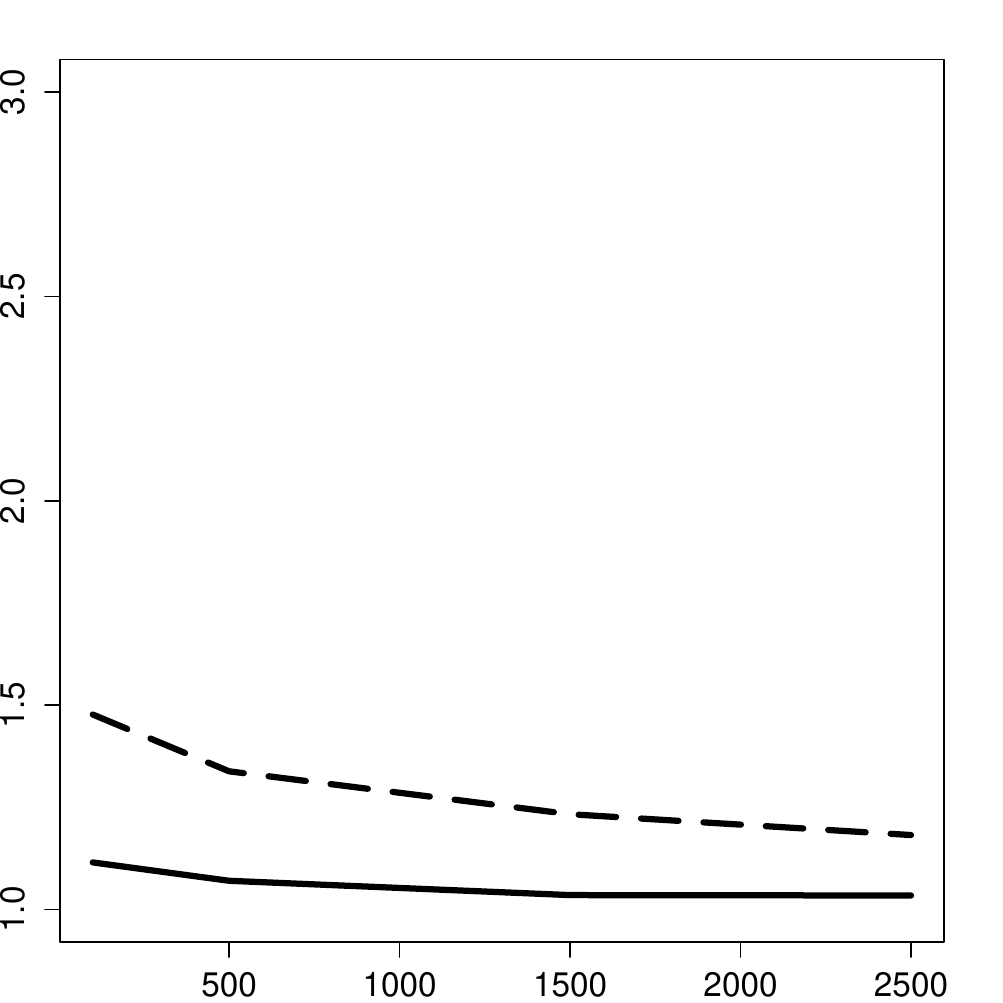} &
\includegraphics[width=0.30\textwidth]{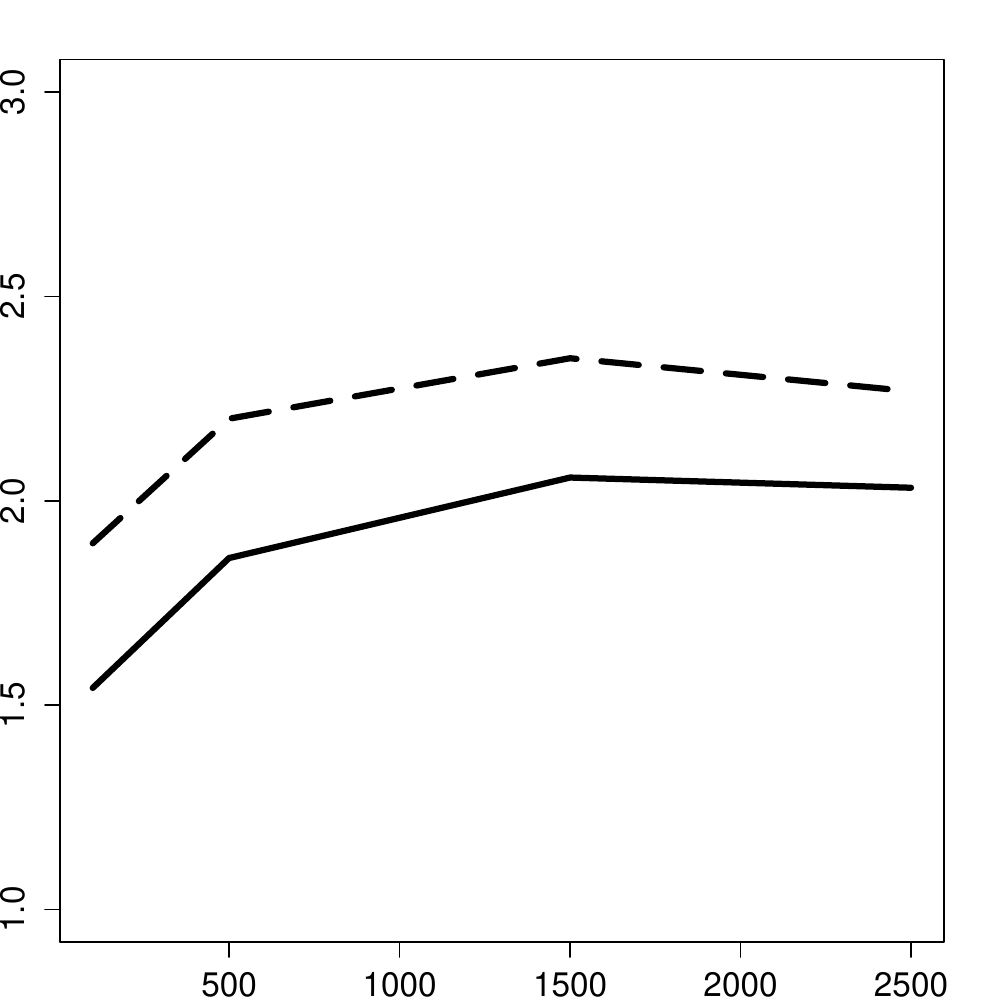} \\
Case 3 ($k^{*}=2$, p=1, q=4) & Case 4 ($k^{*}=3$, p=1, q=4) \\
\includegraphics[width=0.30\textwidth]{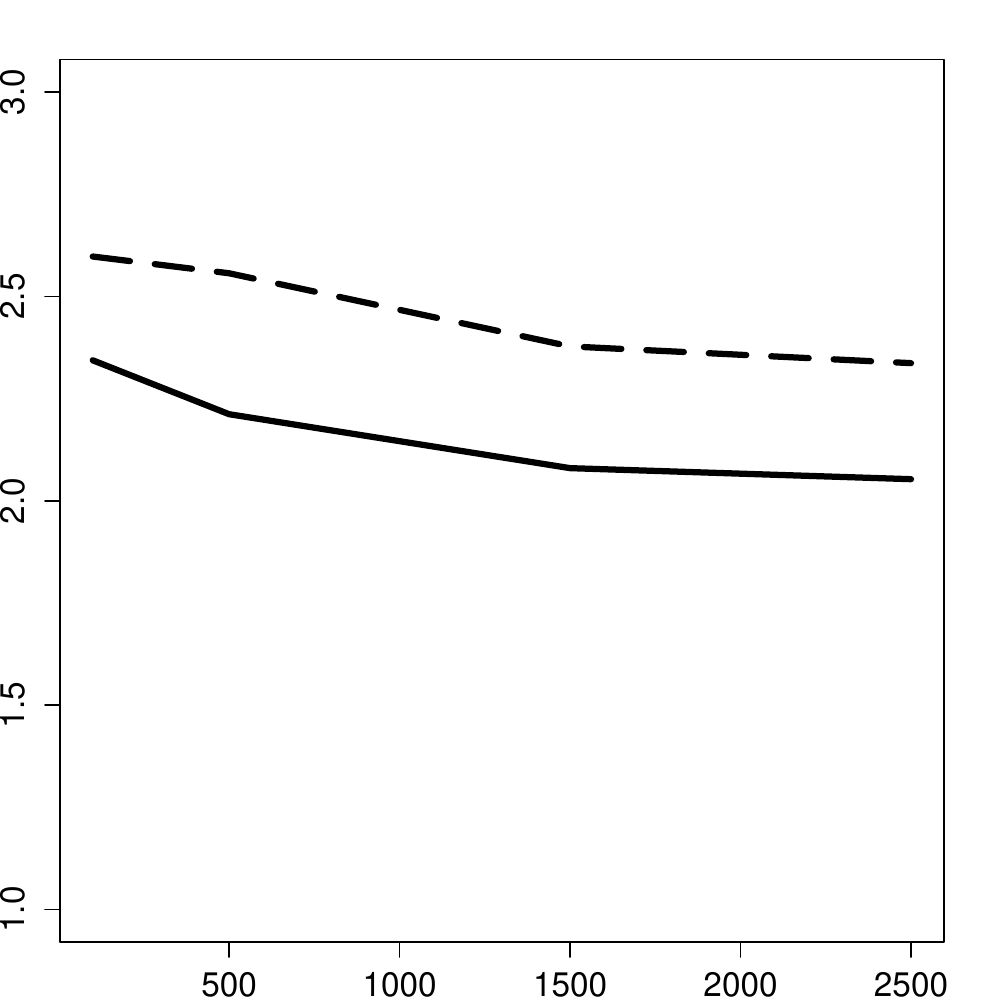} &
\includegraphics[width=0.30\textwidth]{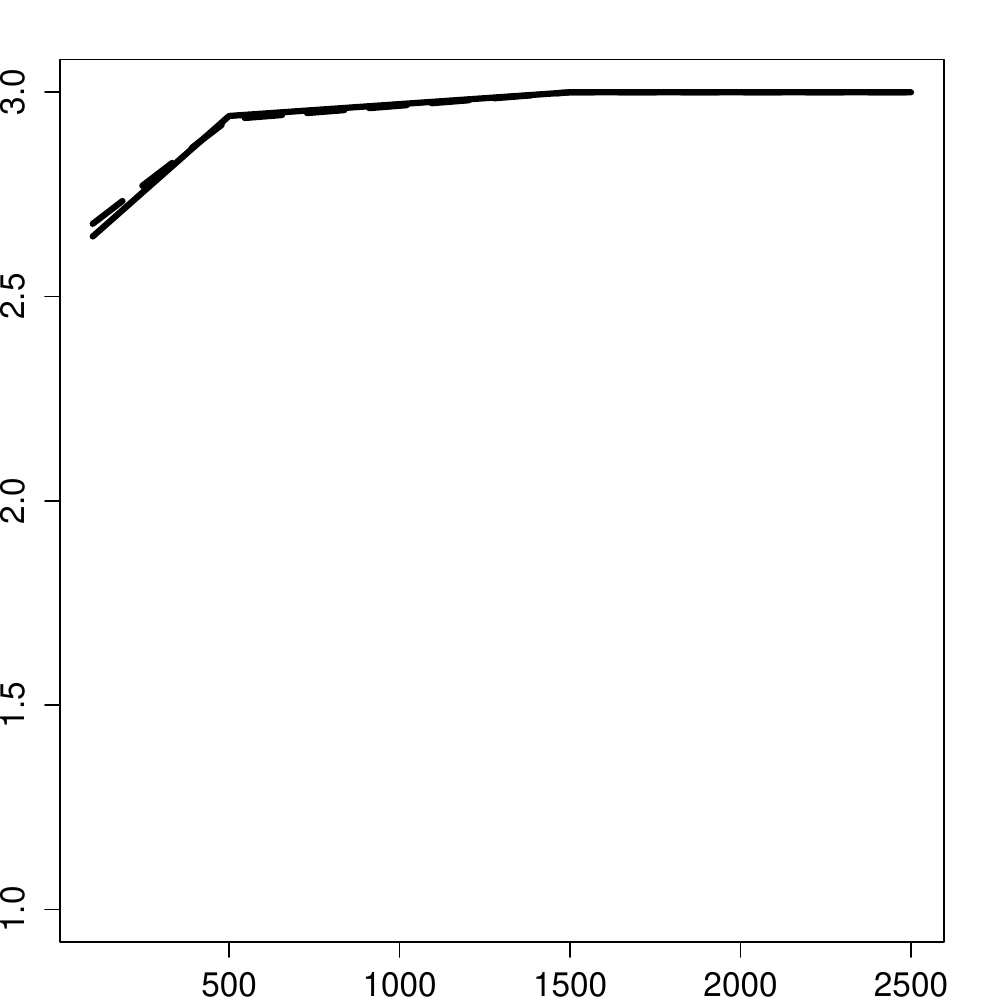} \\
Case 5 ($k^{*}=1$, p=2, q=16.5) & Case 6 ($k^{*}=2$, p=2, q=16.5) \\
\includegraphics[width=0.30\textwidth]{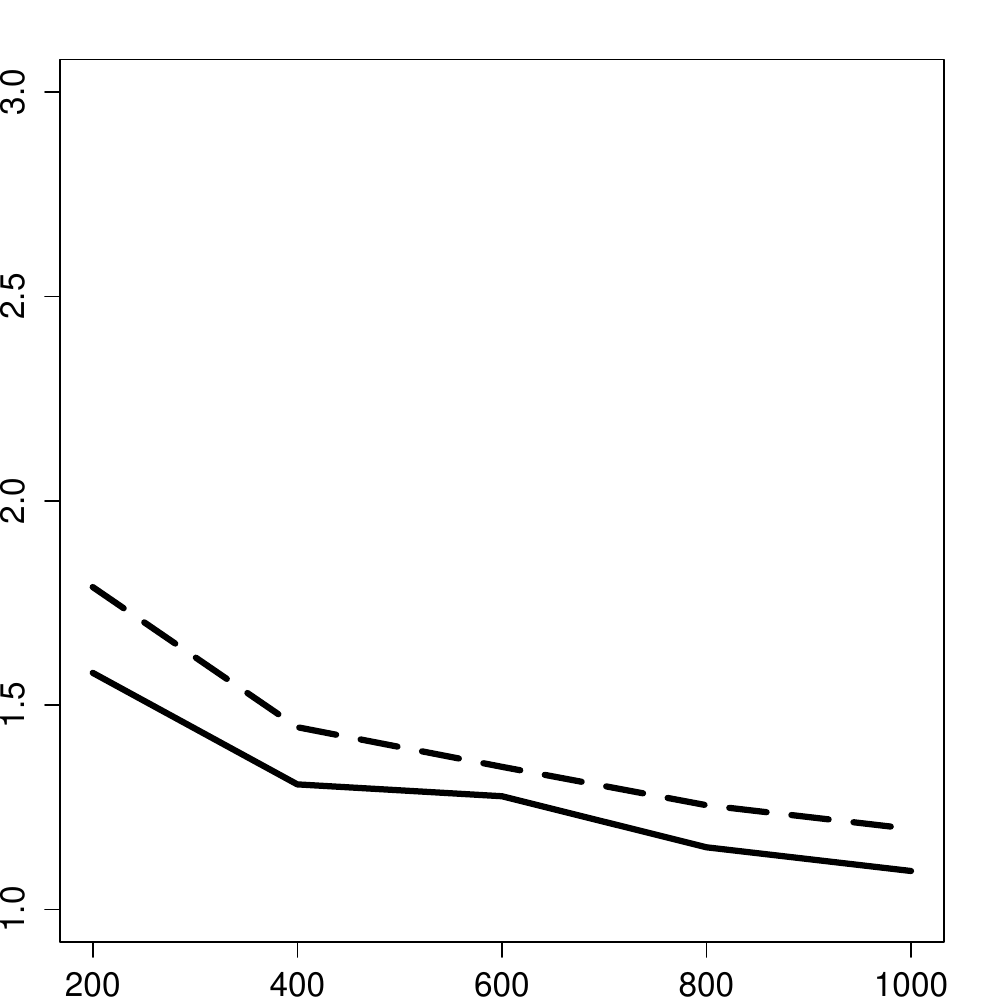} &
\includegraphics[width=0.30\textwidth]{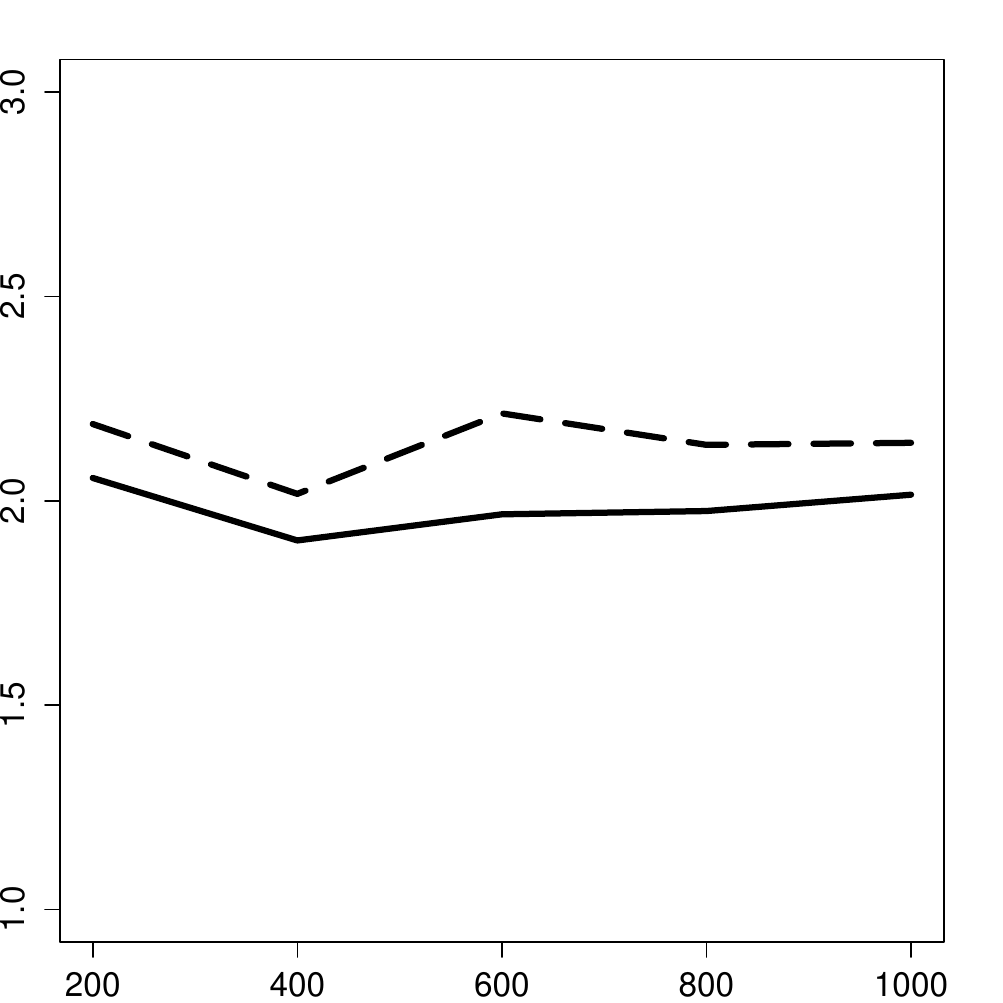} \\
Case 7 ($k^{*}=2$, p=2, q=16.5) & Case 8 ($k^{*}=3$, p=2, q=16.5) \\
\includegraphics[width=0.30\textwidth]{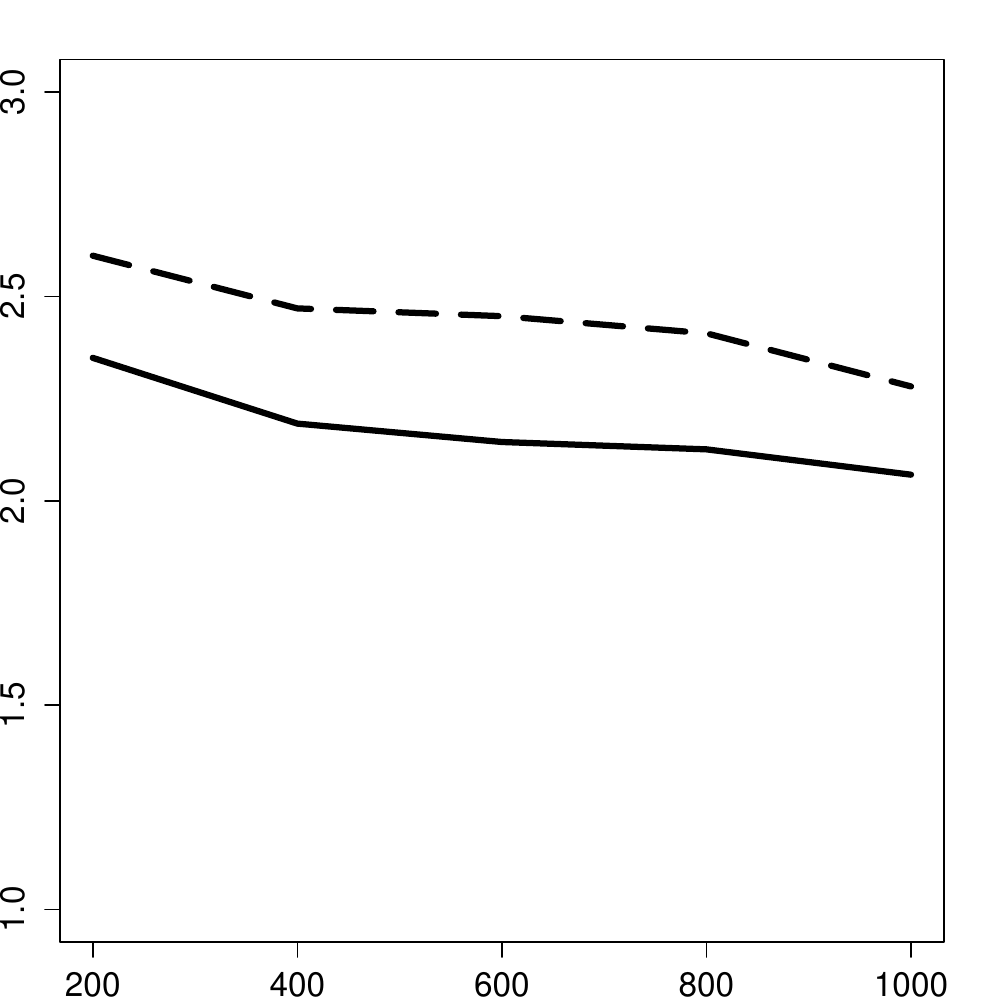} &
\includegraphics[width=0.30\textwidth]{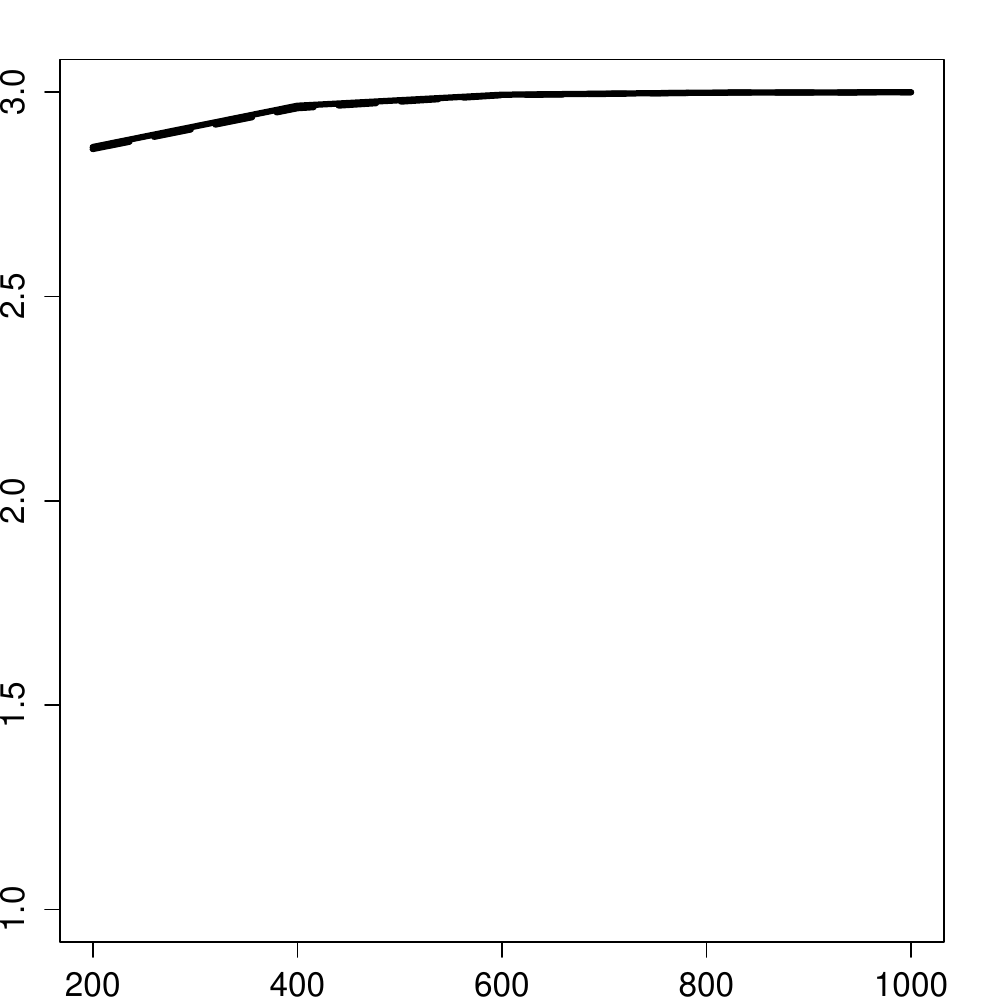} \\
\end{tabular}
\end{center}
\caption{Simulation study. Posterior expected model size $E(k \mid \by)$  versus $n$ for $q=4$
and $q=16.5$ for univariate and bivariate Normal mixtures as recommended by \cite{book_silvia} for the MOM-IW-Dir (solid line) and Normal-IW-Dir (dotted line).}
\label{supfig:synthetic_allmodels2}
\end{figure}

Regarding the univariate Normal mixtures in Cases 1-4,
the four top panels in Figure \ref{supfig:synthetic_allmodels} show the posterior expected number of components given by
$E(k \mid \by)= P(\mathcal{M}_{1} \mid \by) + 2P(\mathcal{M}_{2} \mid \by) + 3P(\mathcal{M}_{3} \mid \by)$
for the alternative prior specification $q=2$ and $P(\kappa < 4)=0.05$.
The four top panels in  Figure \ref{supfig:synthetic_allmodels2} show analogous results
for $q=4$ and $P(\kappa < 4)=0.05$, showing that the findings are fairly robust to mild deviations from our default $q$.

Regarding the bivariate Normal mixtures in Cases 5-8,
the four bottom panels in Figure \ref{supfig:synthetic_allmodels} shows $E(k \mid \by)$
for $q=3$ and $P(\kappa < 4)=0.05$. The four bottom panels in Figure \ref{supfig:synthetic_allmodels2} show the same results
for $q=16.5$ (a value recommended in \cite{book_silvia} and \cite{robert4}, Chapter 10) and $P(\kappa < 4)=0.05$,
showing again that the findings are fairly robust to mild deviations from our recommended prior setting.

\begin{figure}[ht]
\begin{center}
\begin{tabular}{cc}
Case 1 ($k^{*}=1$, p=1, q=2) & Case 2 ($k^{*}=2$, p=1, q=2) \\
\includegraphics[width=0.30\textwidth]{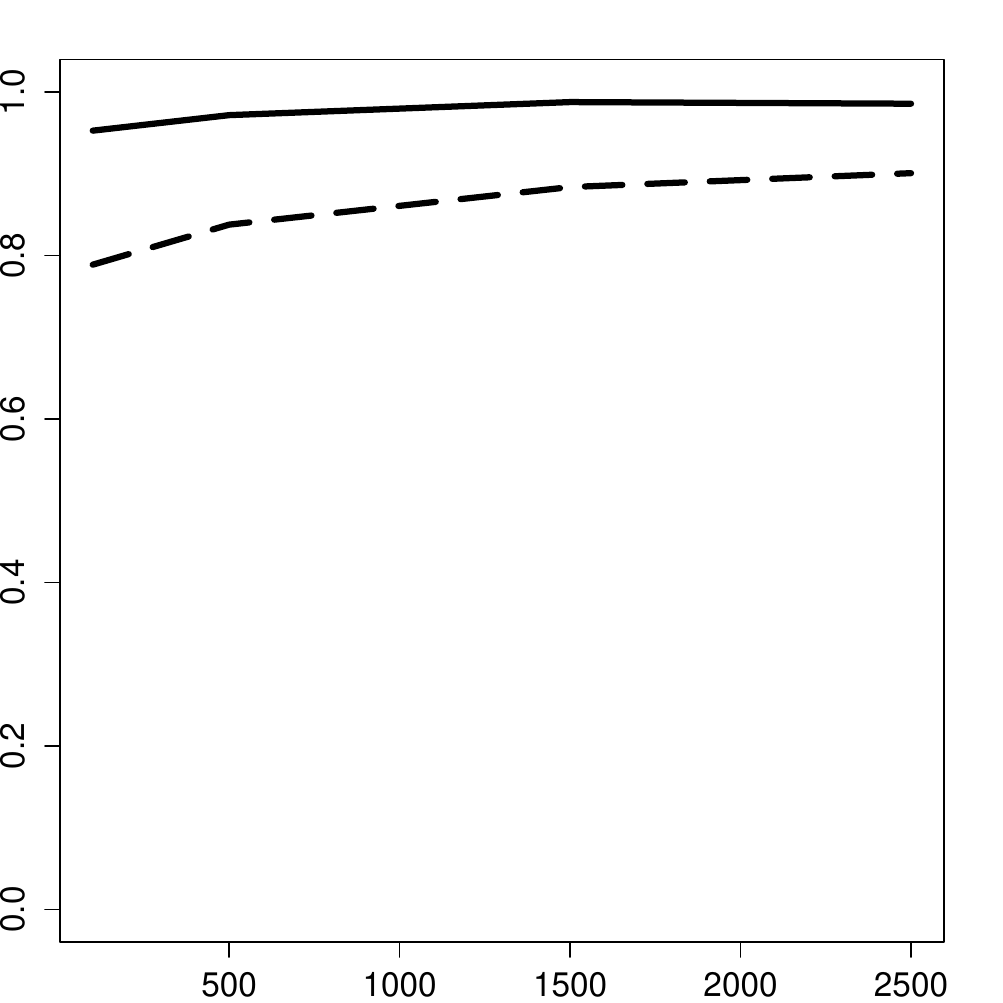} &
\includegraphics[width=0.30\textwidth]{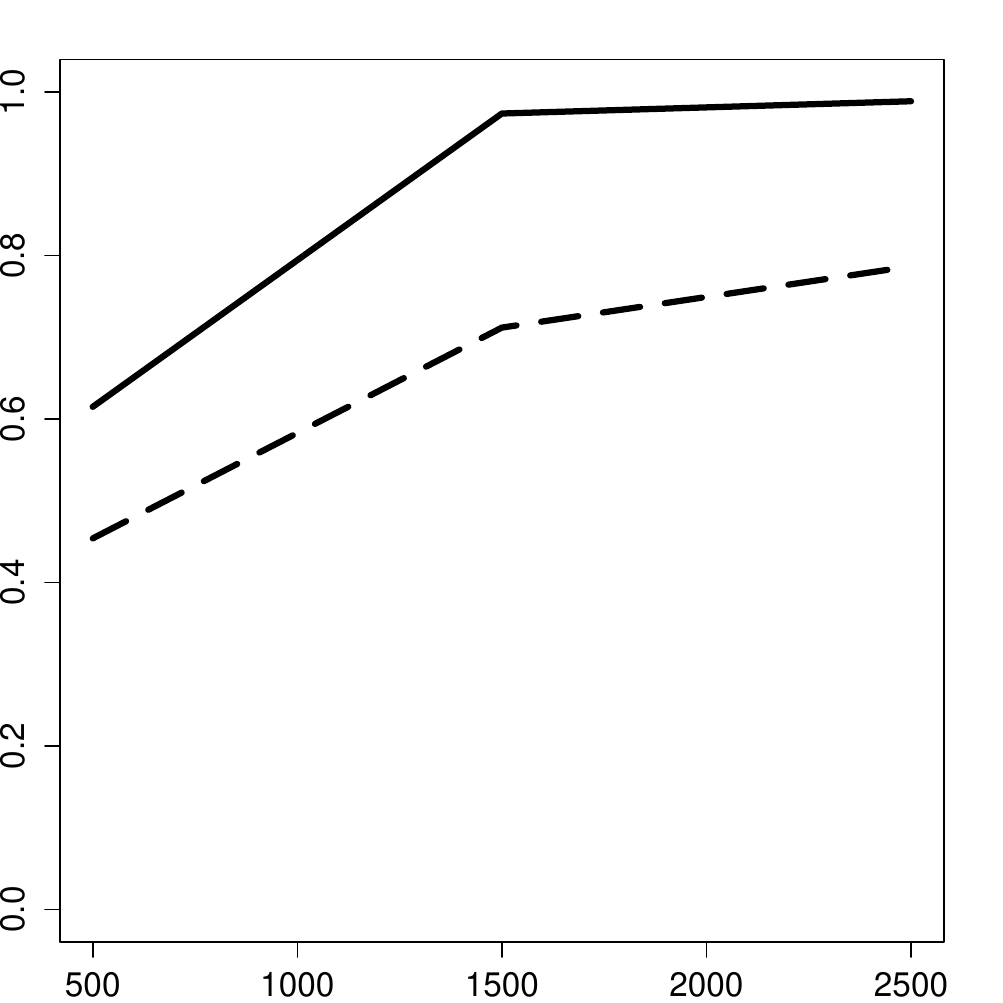} \\
Case 3 ($k^{*}=2$, p=1, q=2) & Case 4 ($k^{*}=3$, p=1, q=2) \\
\includegraphics[width=0.30\textwidth]{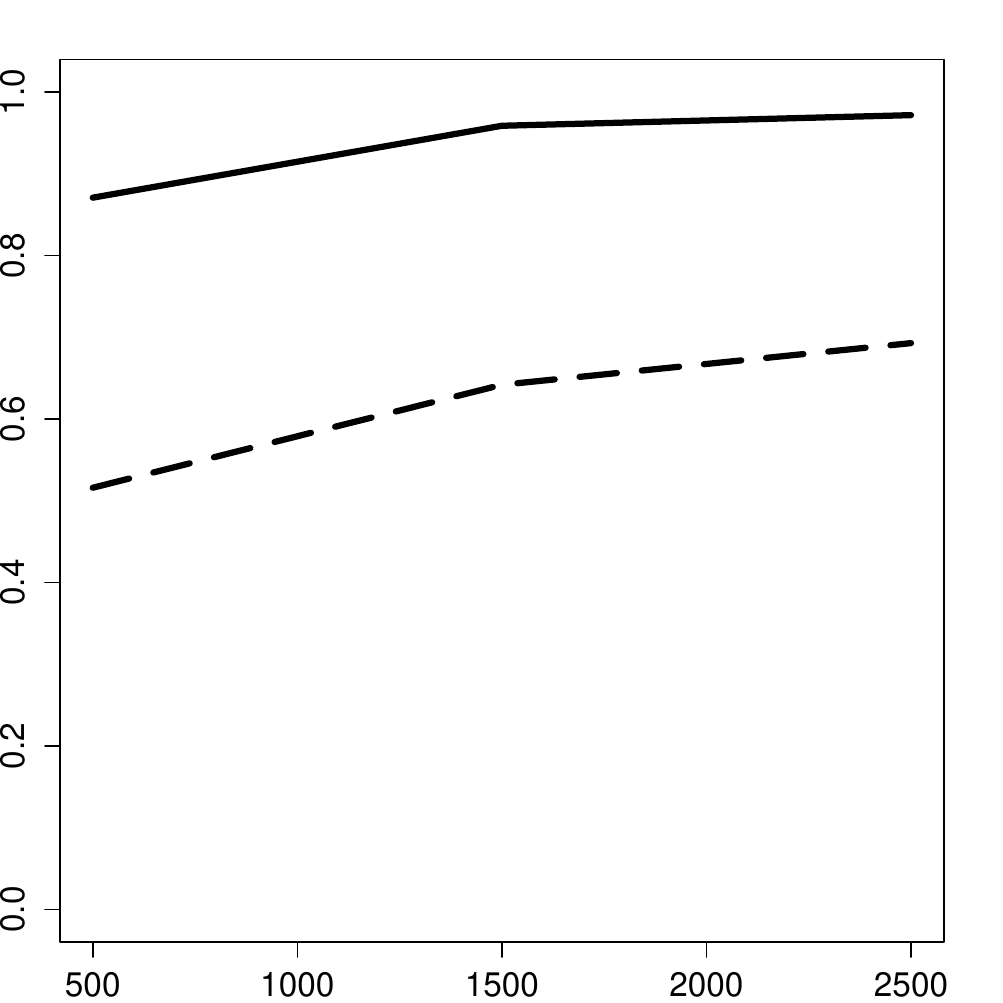} &
\includegraphics[width=0.30\textwidth]{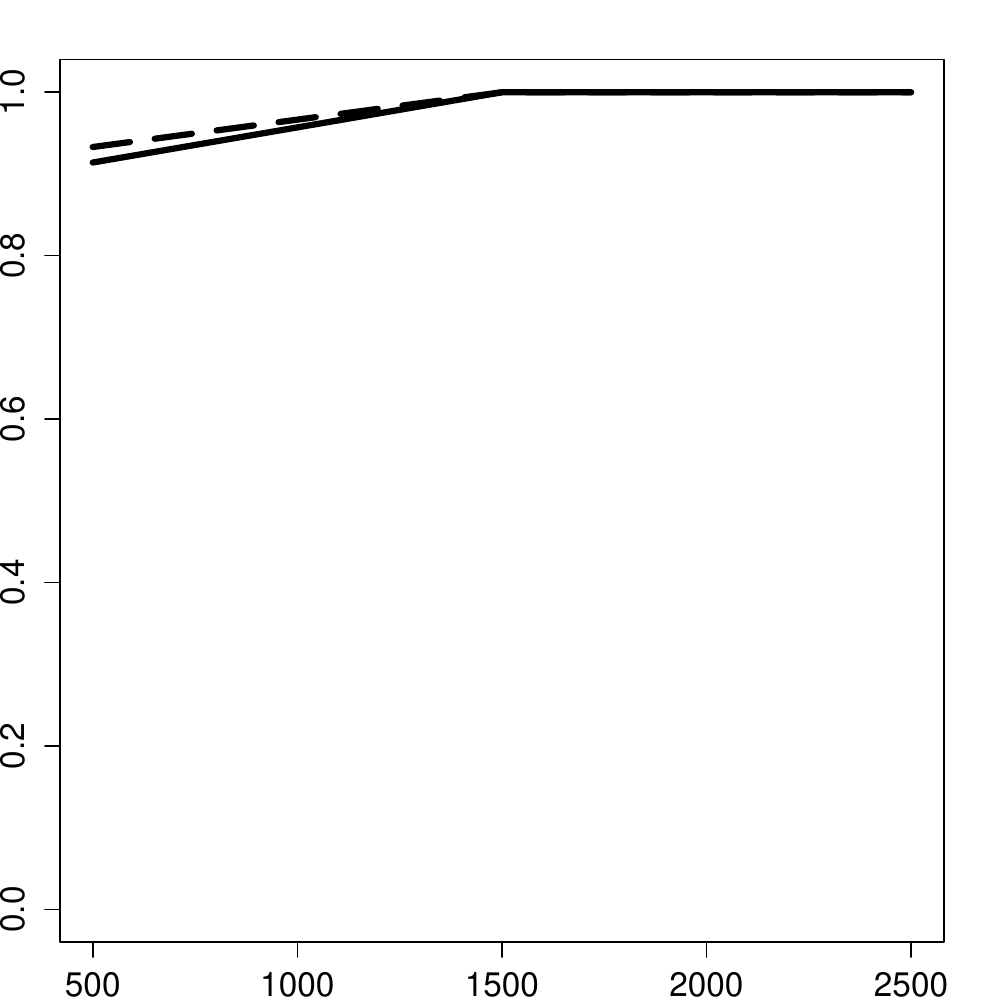} \\
Case 5 ($k^{*}=1$, p=2, q=3) & Case 6 ($k^{*}=2$, p=2, q=3) \\
\includegraphics[width=0.30\textwidth]{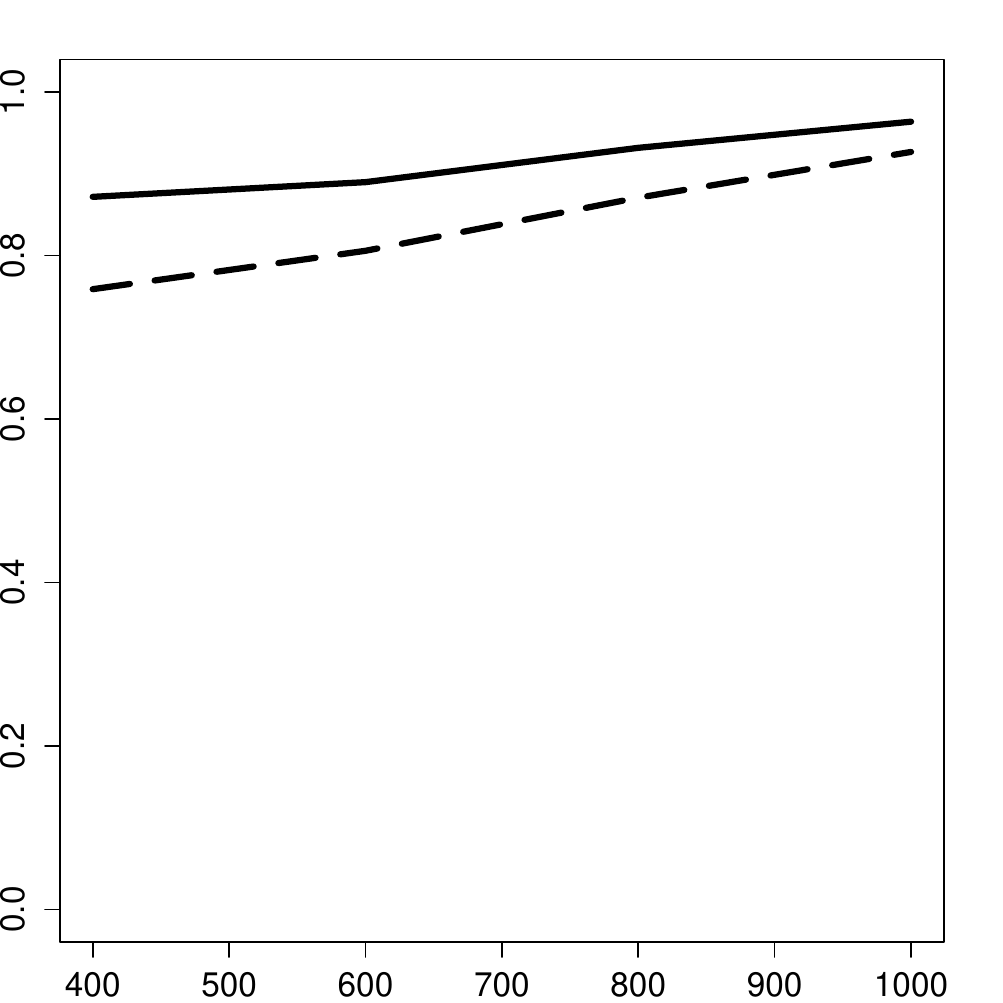} &
\includegraphics[width=0.30\textwidth]{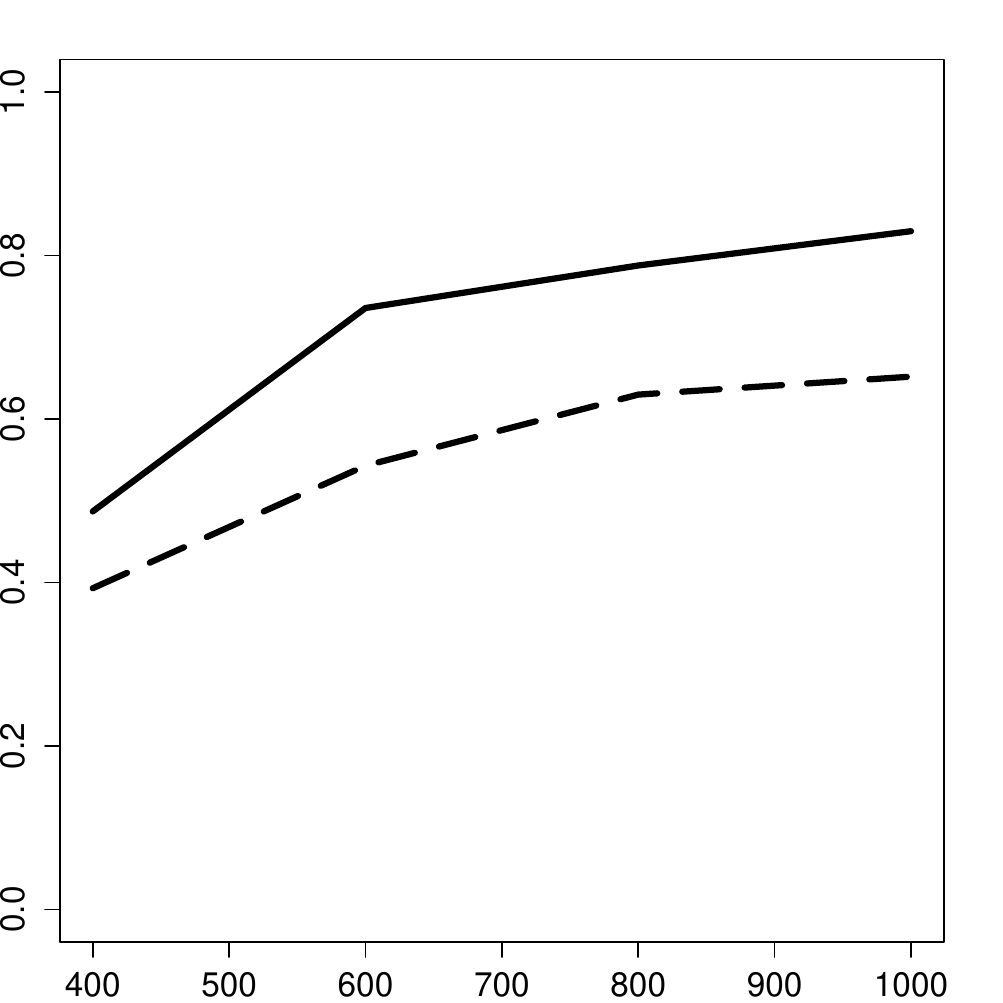} \\
Case 7 ($k^{*}=2$, p=2, q=3) & Case 8 ($k^{*}=3$, p=2, q=3) \\
\includegraphics[width=0.30\textwidth]{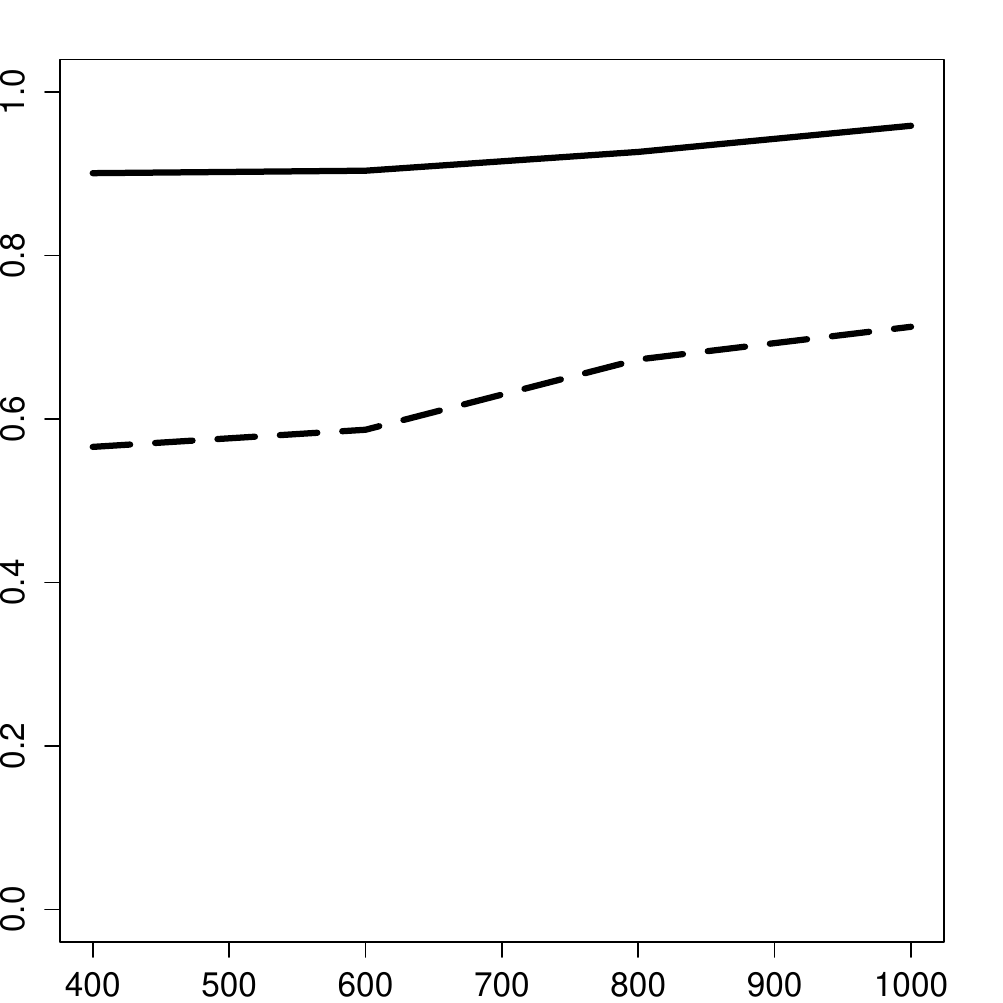} &
\includegraphics[width=0.30\textwidth]{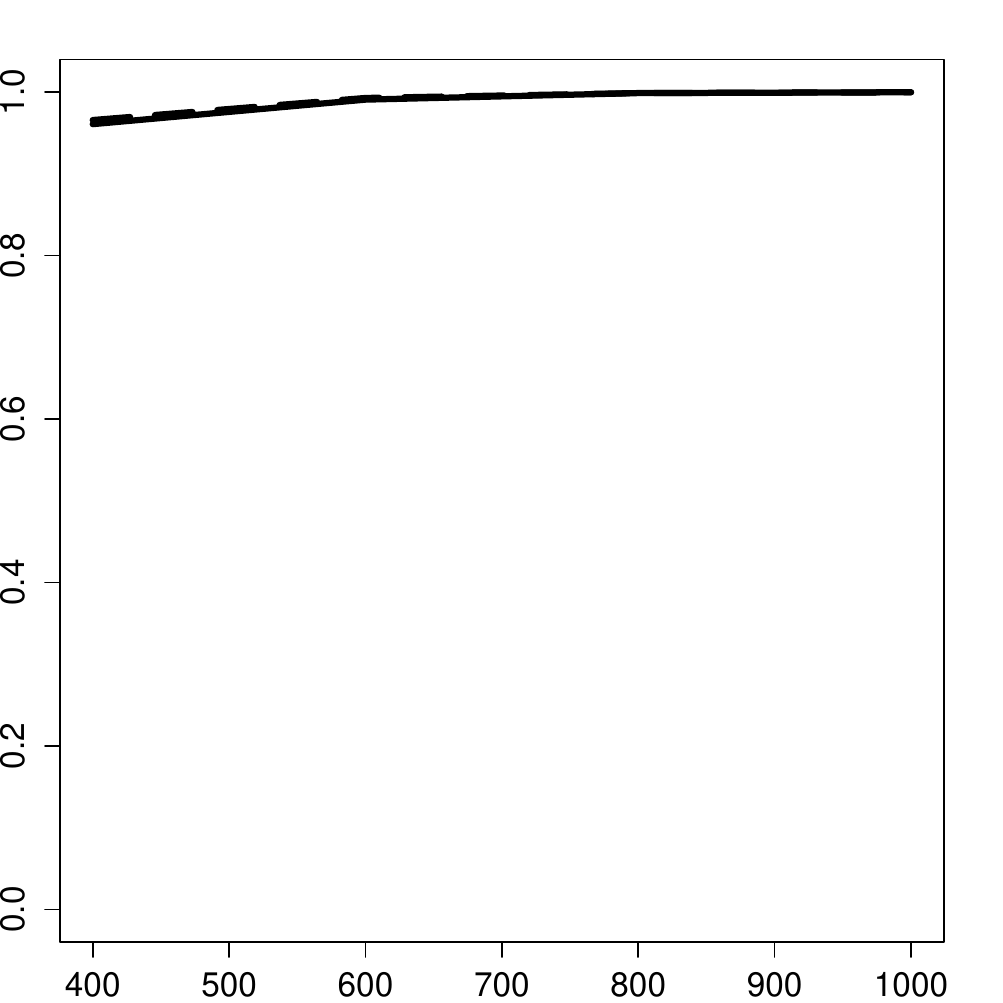} \\
\end{tabular}
\end{center}
\caption{Simulation study. $P(\mathcal{M}_{k^*} \mid \by)$ versus $n$ under $P(\kappa < 4 \mid \Mk)=0.1$ for the MOM-IW-Dir (solid line) and Normal-IW-Dir (dotted line).}
\label{supfig:synthetic_truth2}
\end{figure}

Finally, to assess sensitivity to the prior elicitation of $g$,
Figure \ref{supfig:synthetic_truth2} shows the average posterior probability $P(\mathcal{M}_{k^*} \mid \by)$ for Cases 1-8
with $P(\kappa < 4)=0.1$ and $q$ set as in Figure \ref{supfig:synthetic_truth}.
Although the results are largely similar to those in Figure \ref{supfig:synthetic_truth},
the benefits in parsimony enforcement are somewhat reduced in some situations ({\it e.g.} Case 5),
indicating that $P(\kappa<4 \mid g,\mathcal{M}_{K})=0.05$ may be slightly preferable to 0.1
to achieve a better balance between parsimony and detection power.

\begin{figure}[ht]
\begin{center}
\begin{tabular}{cc}
Case 1 ($k^{*}=1$, q=2) & Case 2 ($k^{*}=2$, p=1, q=2) \\
\includegraphics[width=0.30\textwidth]{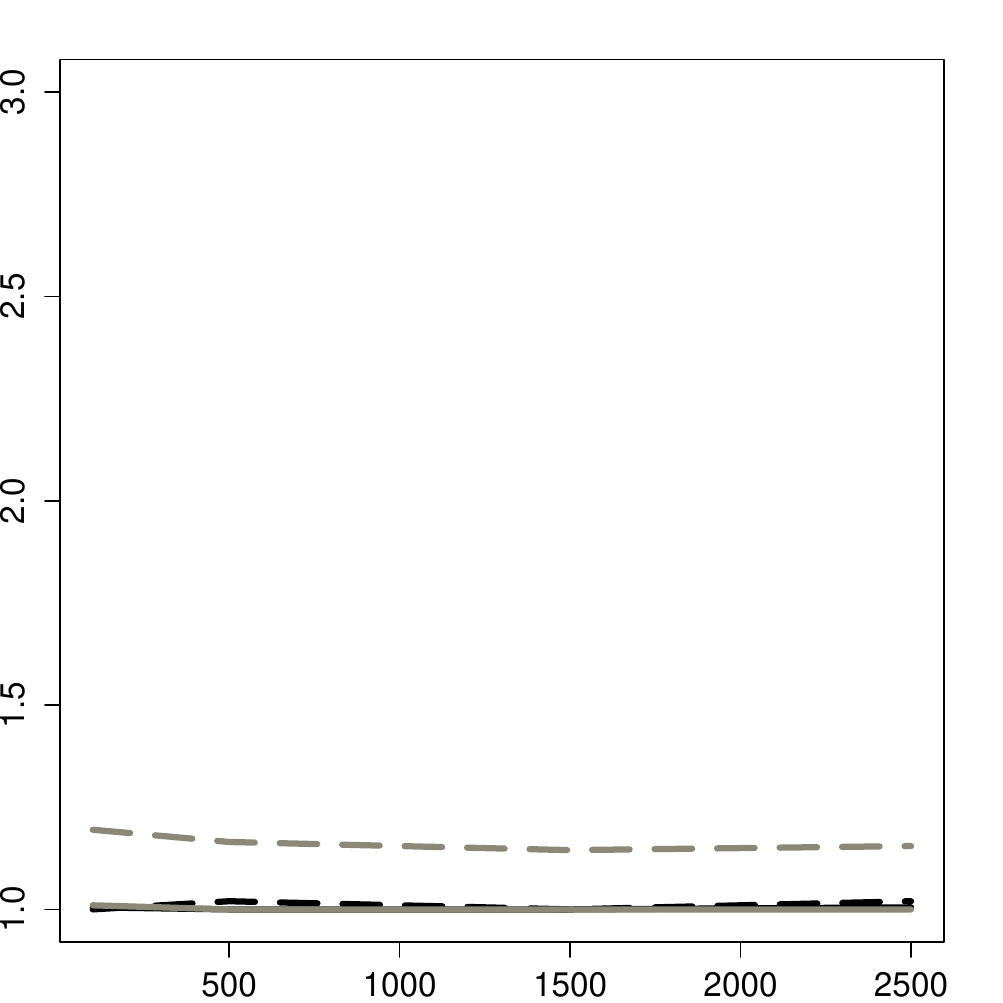} &
\includegraphics[width=0.30\textwidth]{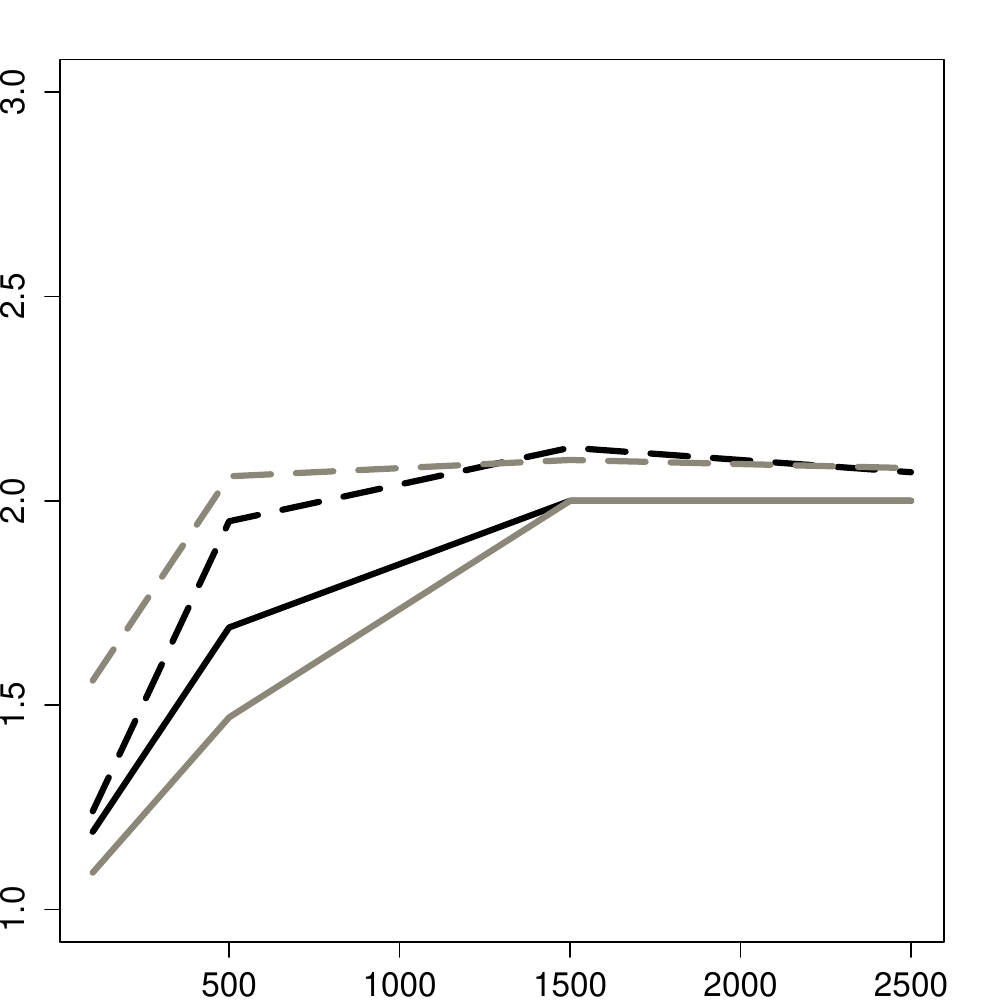} \\
Case 3 ($k^{*}=2$, p=1, q=2) & Case 4 ($k^{*}=3$, p=1, q=2) \\
\includegraphics[width=0.30\textwidth]{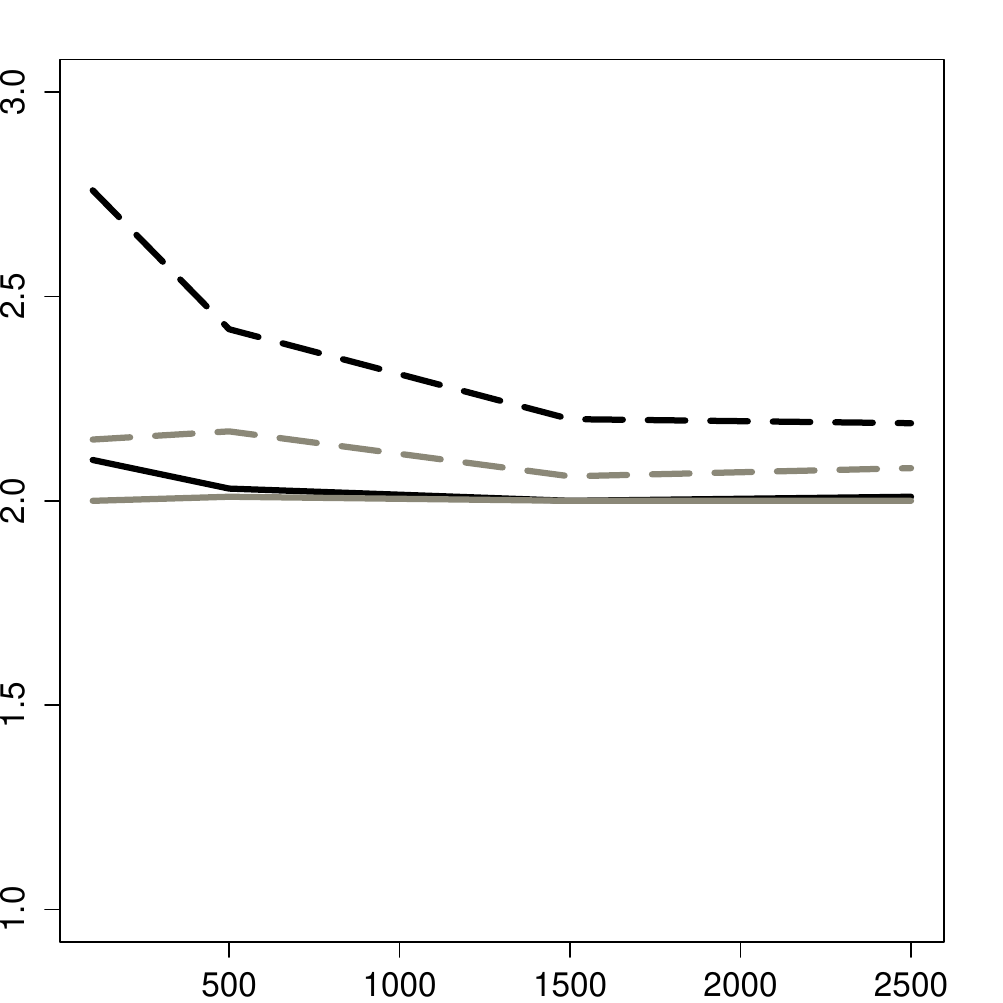} &
\includegraphics[width=0.30\textwidth]{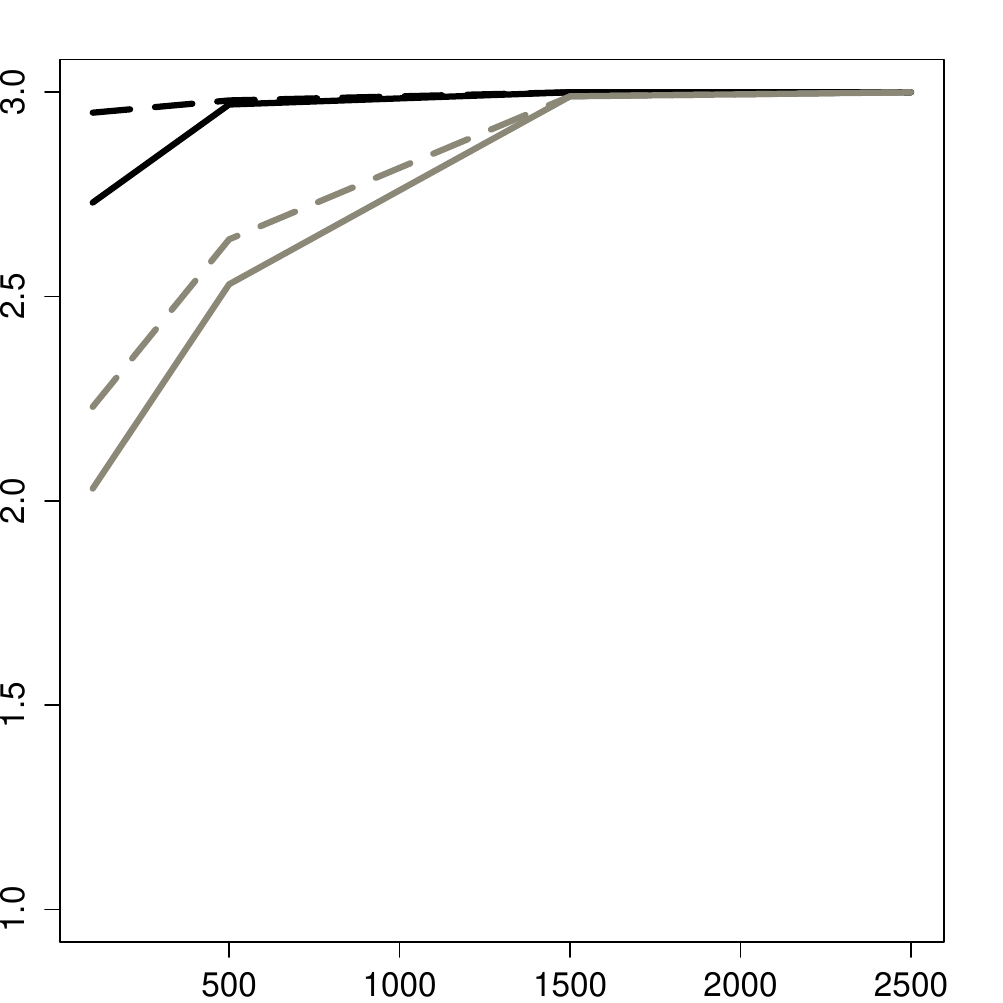} \\
Case 5 ($k^{*}=1$, p=2, q=3) & Case 6 ($k^{*}=2$, p=2, q=3) \\
\includegraphics[width=0.30\textwidth]{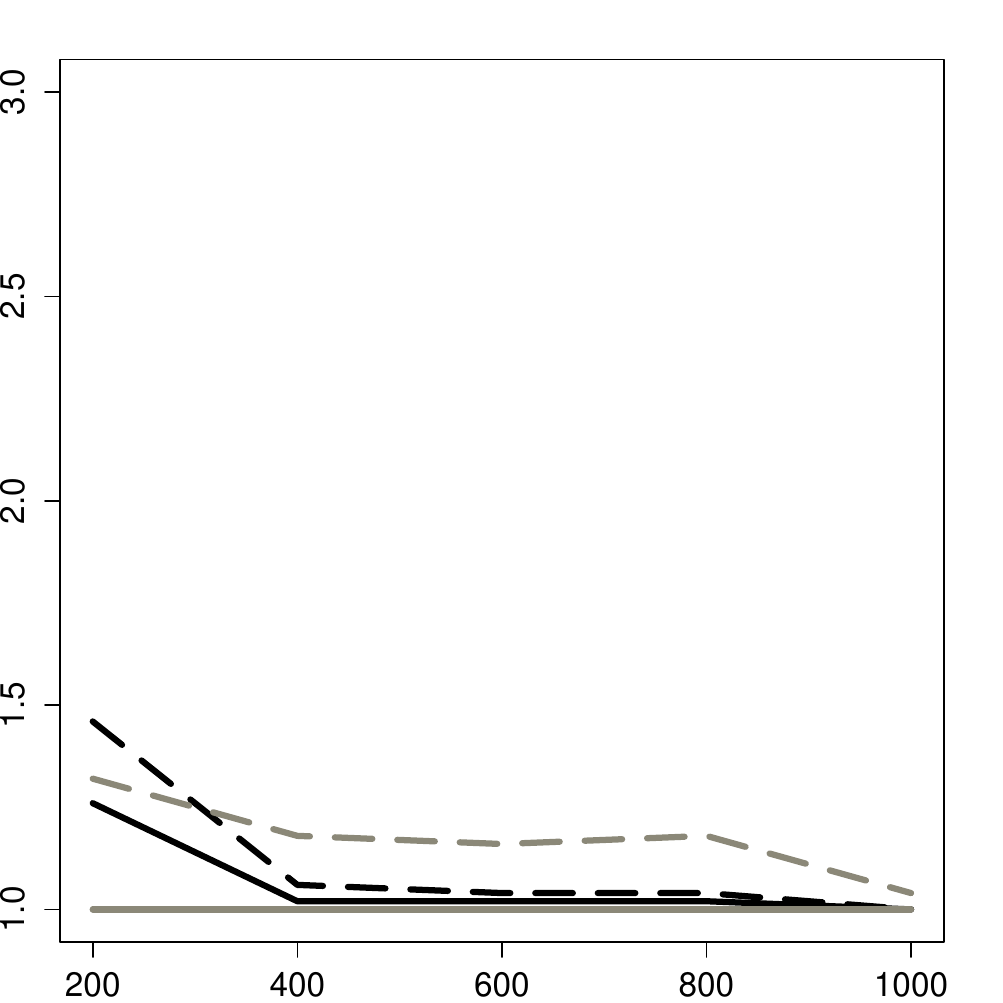} &
\includegraphics[width=0.30\textwidth]{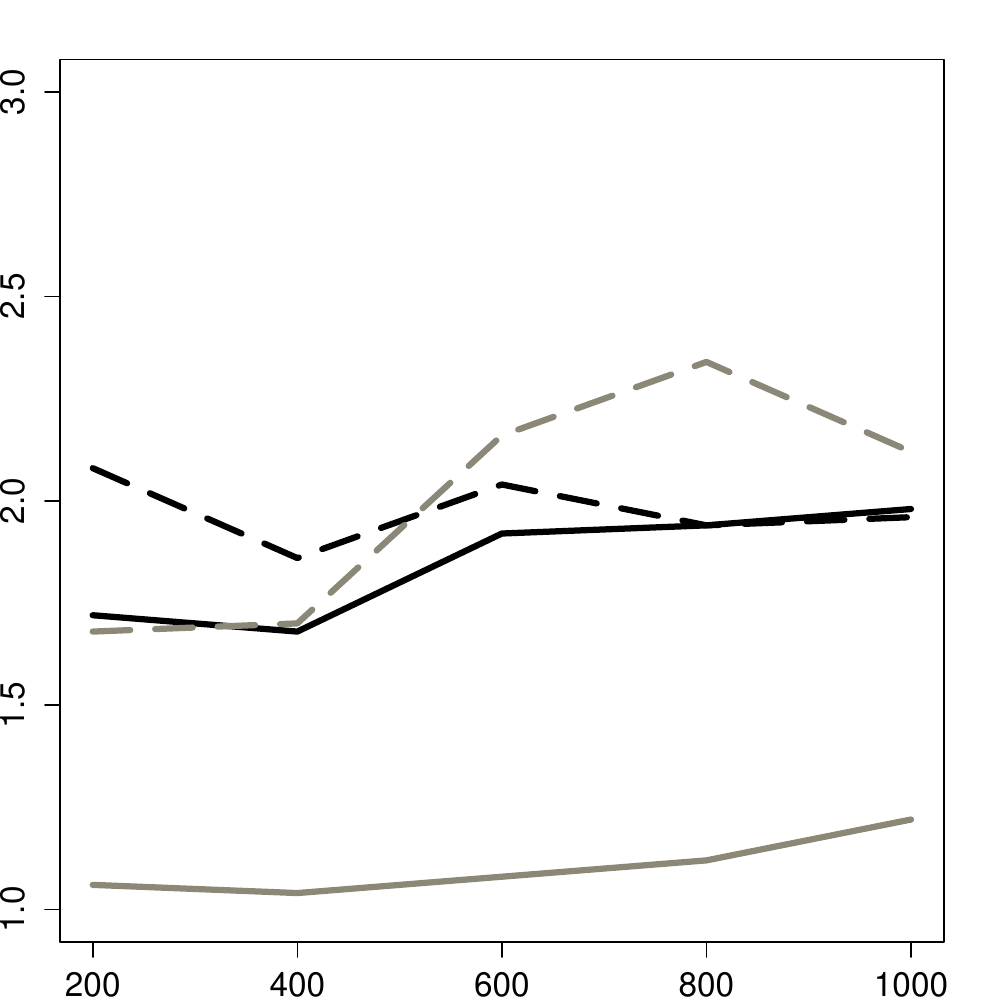} \\
Case 7 ($k^{*}=2$, p=2, q=3) & Case 8 ($k^{*}=3$, p=2, q=3) \\
\includegraphics[width=0.30\textwidth]{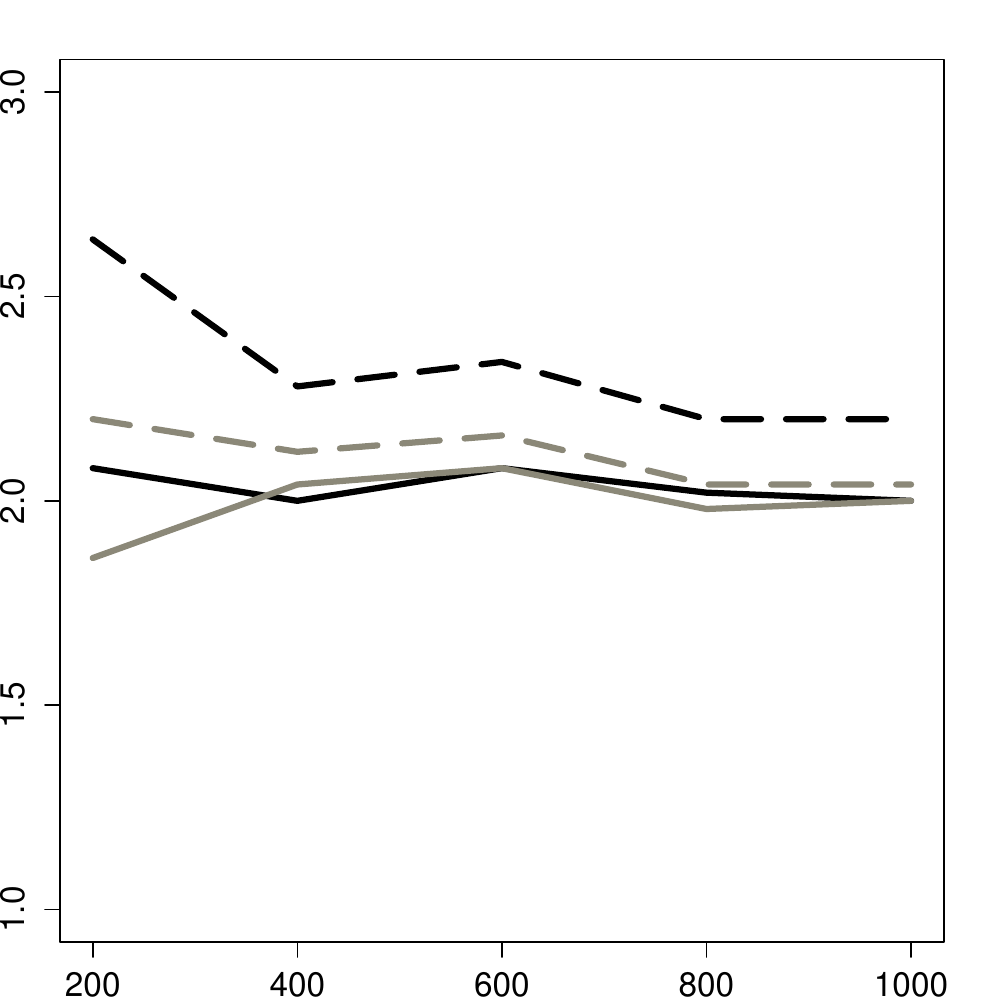} &
\includegraphics[width=0.30\textwidth]{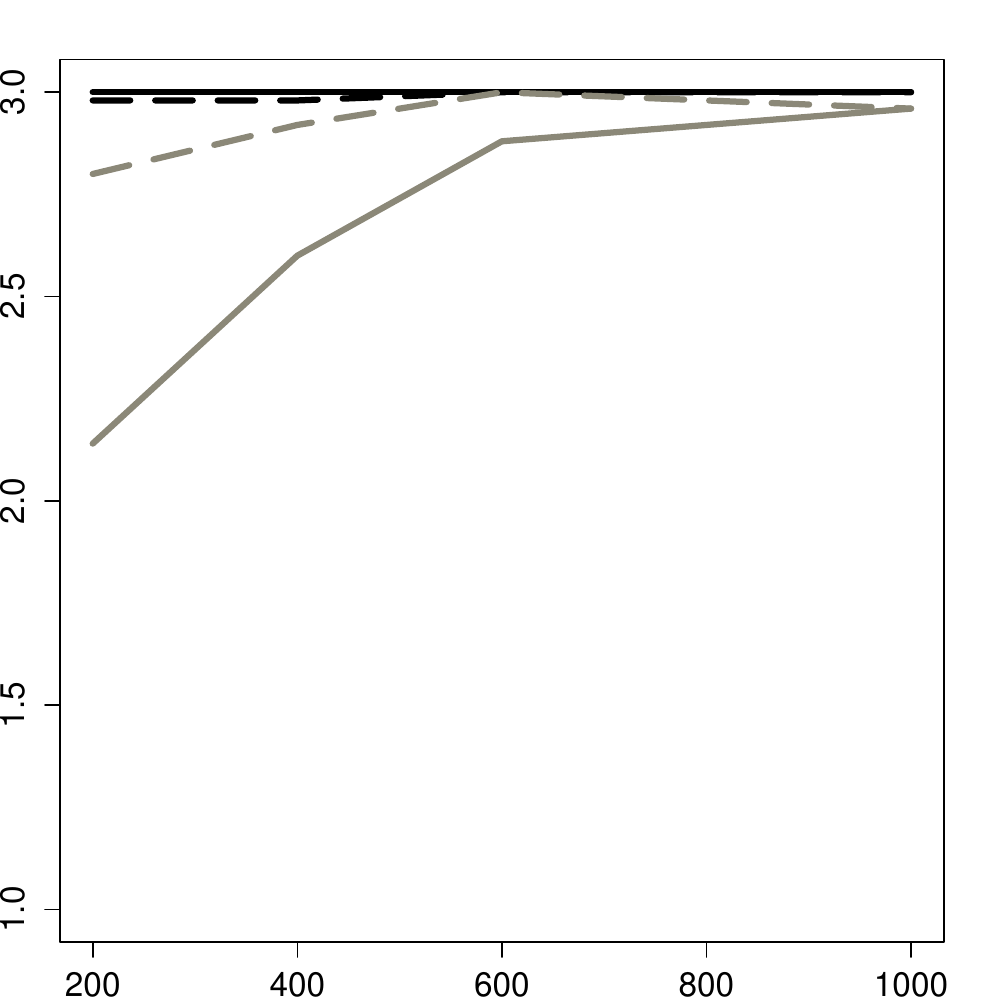} \\
\end{tabular}
\end{center}
\caption{Simulation study. Average $\hat{k}$ versus $n$ for MOM-IW-Dir (solid black), Normal-IW-Dir (dotted black), AIC (dotted gray) and BIC (solid gray).}
\label{supfig:synthetic_allmodels3}
\end{figure}

\section{Supplementary results for the applications}
\label{supplsec:results2}

\begin{figure}[!ht]
\includegraphics[width=0.80\textwidth]{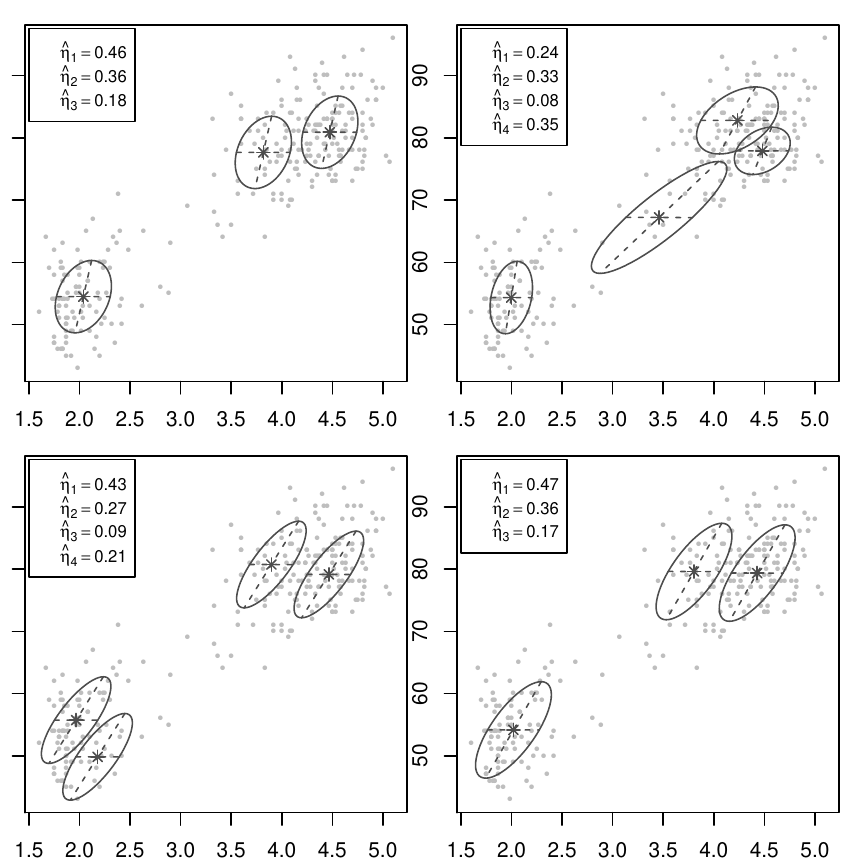}
\caption{Faithful dataset. The x-axis portrays the eruption time (minutes) and the y-axis the waiting time until the next eruption (minutes).
  Contours for the model chosen by BIC/sBIC and AIC (top),
Normal-IW and MOM-IW (bottom), from left to right and the points indicate the data.}
\label{supfig:faithful}
\end{figure}

\begin{figure}[ht]
\includegraphics[width=0.65\textwidth]{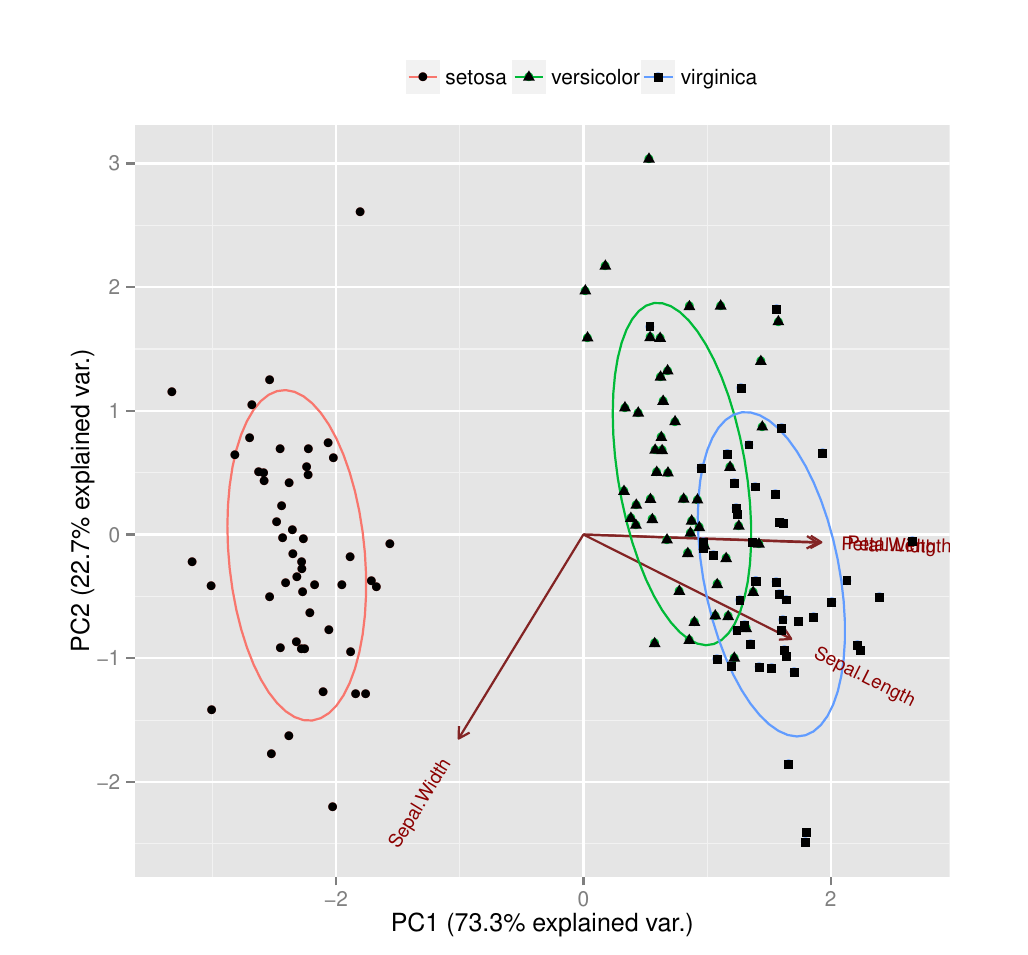}
\caption{Principal components for the Fisher's Iris data-set, classification of observations
and contours using EM algorithm under MOM-IW.}
\label{supfig:iris}
\end{figure}

Table  \ref{tab:resultsmiss} provides more detailed results for the misspecified  Normal model
(Section \ref{ssec:mispec}). It indicates the posterior probability
of 11 models with $k=1,...,6$ components, for each $k$, considering either
homogeneous ($\Sigma_{j}=\Sigma$) or heterogeneous ($\Sigma_{i}\neq\Sigma_{j}$)
covariance matrices.
The model with highest posterior, BIC and AIC is indicated in bold face.
Table  \ref{tab:Faithful} shows analogous results for the Faithful data (Section \ref{ssec:oldfaithful}),
Table \ref{tab:Iris} for the Iris data (Section \ref{ssec:iris}) and Table \ref{tab:Cytometry} for the Cytometry data (Section
\ref{ssec:Cytometry}).

As an alternative to formal Bayesian model selection suppose one fits
a model with a large number of components ($k=6$ in our examples)
to successively discard those deemed unnecessary.
One strategy to discard components is to set a threshold on the estimated $\hat{\bm{\eta}}$,
which results in the addition of spurious components.
An alternative illustrated in Table \ref{tab:many13} and  Table \ref{tab:many23}
is to describe the number $m=\sum_{j=1}^k \mbox{I}(n_j>0)$ of non-empty components
(no allocated observations) at each MCMC iteration when obtaining posterior draws from
$p^{L}(\bz, \bvartheta\mid \by,\mathcal{M}_{6})$ and
$p(\bz, \bvartheta\mid \by,\mathcal{M}_{6})$ (respectively).
For instance, for the misspecified model roughly 95\% of the MCMC iterations
had 6 components with some allocated observations, and similarly
for other data sets, which naively suggest that at least $k=6$ components are needed.
This  is in stark contrast with posterior model probabilities $P(\mathcal{M}_{k}\mid \by)$
in Tables \ref{tab:resultsmiss}-\ref{tab:Cytometry},
which suggest more parsimonious models.
This difference is explained by the fact that $P(m \mid \by,\mathcal{M}_6)$
reported in Table \ref{tab:many13} and Table \ref{tab:many23}
conditions on the larger model
whereas $P(\Mk \mid \by)$ is a formal measure of uncertainty for each of the models under consideration conditional on the observed data.

\begin{table}[ht]
\caption{Misspecified model.
  $P(\Mk \mid \by)$ under Normal-IW-Dir and MOM-IW-Dir priors, BIC, AIC and sBIC
  for $k \in \{1,\ldots,6\}$ and homogeneous ($\Sigma_{j}=\Sigma$) or heterogeneous ($\Sigma_{i}\neq\Sigma_{j}$)}
\small
\centering
\begin{tabular}{rrrrrrrrr}
  \hline
 & &  Normal-IW-Dir  & MOM-IW-Dir & BIC & AIC & sBIC &
\\  \hline
& $k$  & $P(\mathcal{M}_{k}\mid \by)$ & $P(\mathcal{M}_{k}\mid \by)$ &  &  &  &  & \\ \hline
 \hline
 & 1 & 0.000 & 0.000 & -2992.820 & -2981.828 &  &  & \\
$\Sigma_{j}=\Sigma$   & 2 & 0.000 & 0.000 & -2549.767 & -2532.179 &  & & \\
   & 3 & 0.003 & \textbf{1.000} & -2548.774 & -2524.591 &  &  & \\
   & 4 & 0.062 & 0.000 & -2556.581 & -2525.803 &  &  & \\
   & 5 & \textbf{0.469} & 0.000 & -2566.122 & -2528.748 &  &  & \\
   & 6 & 0.465 & 0.000 & -2574.371 & -2530.402  &  &  &  \\ \hline
$\Sigma_{i}\neq\Sigma_{j}$ & 2 & 0.000 & 0.000 & -2545.129 & -2520.946 & -2548.942  &  & \\
    & 3 & 0.000 & 0.000 & -2529.037 & -2491.663 & -2534.729   &  & \\
   &  4 & 0.000 & 0.000 & \textbf{-2522.954} & -2472.389 & \textbf{-2527.448}
 &  & \\
   & 5 & 0.000 & 0.000 & -2535.703 & -2471.948 & -2528.207  &  & \\
   & 6 & 0.000 & 0.000 & -2546.878 & \textbf{-2469.931} & -2529.068 &  &  \\ \hline
   \hline
\end{tabular}
\label{tab:resultsmiss}
\end{table}

\begin{table}[ht]
\caption{Cytometry data.
  $P(\Mk \mid \by)$ under Normal-IW-Dir and MOM-IW-Dir priors, BIC, AIC and sBIC
  for $k \in \{1,\ldots,6\}$ and homogeneous ($\Sigma_{j}=\Sigma$) or heterogeneous ($\Sigma_{i}\neq\Sigma_{j}$)}
\small
\centering
\begin{tabular}{rrrrrrrrr}
  \hline
 & &  Normal-IW-Dir  & MOM-IW-Dir & BIC & AIC & sBIC &
\\  \hline
& $k$  & $P(\mathcal{M}_{k}\mid \by)$ & $P(\mathcal{M}_{k}\mid \by)$ &  &  &  &  & \\ \hline
  \hline
 & 1 & 0.000 & 0.000 & -28337.23 & -28295.02 &  &  & \\
$\Sigma_{j}=\Sigma$   & 2 & 0.000 & 0.000 & -27720.64 & -27665.86 &  & & \\
   & 3 & 0.000 & 0.000 & -27541.73 & -27474.39 &  &  & \\
   & 4 & 0.000 & 0.000 & -27443.22 & -27363.31 &  &  & \\
   & 5 & 0.000 & 0.000 & -27271.67 & -27179.19 &  &  & \\
   & 6 & 0.000 & 0.000 & -27226.41 & -27121.36  &  &  &  \\ \hline
$\Sigma_{i}\neq\Sigma_{j}$ & 2 & 0.072 & 0.005 & -27357.56 & -27277.65 & -37869.06
 &  & \\
    & 3 & \textbf{0.928} & \textbf{0.995} & \textbf{-27015.35} &-26897.74  & -36478.20 &  & \\
   &  4 & 0.000 & 0.000 & -27048.60 & -26893.29 & -35247.11 &  & \\
   & 5 & 0.000 & 0.000 & -27041.50 & -26848.50 & -34415.96 &  & \\
   & 6 & 0.000 & 0.000 & -27075.18 & \textbf{-26844.48} & \textbf{-33888.20} &  &  \\ \hline
   \hline
\end{tabular}
\label{tab:Cytometry}
\end{table}

\begin{table}[ht]
\caption{Faithful data.
  $P(\Mk \mid \by)$ under Normal-IW-Dir and MOM-IW-Dir priors, BIC, AIC and sBIC
  for $k \in \{1,\ldots,6\}$ and homogeneous ($\Sigma_{j}=\Sigma$) or heterogeneous ($\Sigma_{i}\neq\Sigma_{j}$)}
\small
\centering
\begin{tabular}{rrrrrrrrr}
  \hline
 & &  Normal-IW-Dir  & MOM-IW-Dir & BIC & AIC & sBIC &
\\  \hline
& $k$  & $P(\mathcal{M}_{k}\mid \by)$ & $P(\mathcal{M}_{k}\mid \by)$ &  &  &  &  & \\ \hline
  \hline
 & 1 & 0.000 & 0.000 & -558.006 & -548.992 &  &  & \\
$\Sigma_{j}=\Sigma$   & 2 & 0.000 & 0.000 & -416.805 & -402.382 &  & & \\
   & 3 & 0.132 & \textbf{0.967} &
\textbf{-411.356} & -391.524 &  &  & \\
   & 4 & \textbf{0.473} & 0.000 &
-419.748 & -394.507 &  &  & \\
   & 5 & 0.353 & 0.000 & -418.019 & -387.369 &  &  & \\
   & 6 & 0.042 & 0.000 & -427.821 & -391.763    &  &  &  \\ \hline
$\Sigma_{i}\neq\Sigma_{j}$ & 2 & 0.000 & 0.000 & -415.291 & -395.459 & -419.103   &  & \\
    & 3 & 0.000 & 0.000 & -422.609 & -391.960  & \textbf{-415.938}   &  & \\
   &  4 & 0.000 & 0.000 & -425.370 & \textbf{-383.903}  & -417.278  &  & \\
   & 5 & 0.000 & 0.000 & -439.754 & -387.470  & -420.569   &  & \\
   & 6 & 0.000 & 0.000 & -448.896 & -385.795 & -422.231 &  &  \\ \hline
   \hline
\end{tabular}
\label{tab:Faithful}
\end{table}

\begin{table}[ht]
\caption{Iris data.
    $P(\Mk \mid \by)$ under Normal-IW-Dir and MOM-IW-Dir priors, BIC, AIC and sBIC
  for $k \in \{1,\ldots,6\}$ and homogeneous ($\Sigma_{j}=\Sigma$) or heterogeneous ($\Sigma_{i}\neq\Sigma_{j}$)}
\small
\centering
\begin{tabular}{rrrrrrrrr}
  \hline
 & &  Normal-IW-Dir  & MOM-IW-Dir & BIC & AIC & sBIC &
\\  \hline
& $k$  & $P(\mathcal{M}_{k}\mid \by)$ & $P(\mathcal{M}_{k}\mid \by)$ &  &  &  &  & \\ \hline
  \hline
 & 1 & 0.000 & 0.000 & -414.989 & -393.915 &  &  & \\
$\Sigma_{j}=\Sigma$   & 2 & 0.000 & 0.000 & -344.049 & -315.448 &  & & \\
   & 3 & \textbf{0.809} & \textbf{1.000} &
-316.483 & -280.355 &  &  & \\
   & 4 & 0.029 & 0.000 &
-295.705 & -252.051 &  &  & \\
   & 5 & 0.132 & 0.000 & -302.465 & -251.284 &  &  & \\
   & 6 & 0.030 & 0.000 & -310.909 & -252.201    &  &  &  \\ \hline
$\Sigma_{i}\neq\Sigma_{j}$ & 2 & 0.000 & 0.000 & \textbf{-287.009} & -243.355 & -415.449  &  & \\
    & 3 & 0.000 & 0.000 & -290.420 & -224.186  & -410.122   &  & \\
   &  4 & 0.000 & 0.000 & -314.483 & -225.669  & \textbf{-408.839}   &  & \\
   & 5 & 0.000 & 0.000 & -341.910 & -230.517  & -414.190   &  & \\
   & 6 & 0.000 & 0.000 & -355.786 & \textbf{-221.813} & -422.209 &  &  \\ \hline
   \hline
\end{tabular}
\label{tab:Iris}
\end{table}

\begin{table}[ht]
\caption{Posterior distribution on non-empty components $m=\sum_{j=1}^k \mbox{I}(n_j>0)$ in repulsive overfitted mixtures under  $\Sigma_{j}=\Sigma$. The Misspecified, Faithful, Iris and cytometry data considered in Section \ref{sec:results}.}
\centering
\begin{tabular}{rrrrrrrrr}\\  \hline
& \multicolumn{6}{c}{$\hat{P}(m\mid \by,\mathcal{M}_{6})$} \\
& $m=1$ & $m=2$ & $m=3$ & $m=4$ & $m=5$ & $m=6$  \\ \hline \hline
&  &  & $q=1$ &  &  &   \\ \hline \hline
Misspecified & 0.00 & 0.00 & 0.00 & 0.00 & 0.02 & 0.98 \\
Faithful & 0.00 & 0.00 & 0.00 & 0.00 & 0.26 & 0.74 \\
Iris  & 0.00 & 0.99 & 0.00 & 0.01 & 0.00 & 0.00 \\
Cytometry & 0.00 & 0.00 & 0.00 & 0.00 & 0.00 & 1.00 \\ \hline \hline
&  &  & $q=0.01$ &  &  &   \\ \hline \hline
Misspecified & 0.00 & 0.00 & 0.00 & 0.35 & 0.63 & 0.02 \\
Faithful & 0.00 & 0.00 & 0.76 & 0.23 & 0.01 & 0.00 \\
Iris  & 0.00 & 1.00 & 0.00 & 0.00 & 0.00 & 0.00 \\
Cytometry & 0.00 & 0.00 & 0.00 & 0.00 & 0.00 & 1.00 \\ \hline \hline
&  &  & $q=3.10^{-8}$ &  &  &   \\ \hline \hline
Misspecified & 0.00 & 0.00 & 0.83 & 0.00 & 0.00 & 0.17 \\
Faithful & 0.00 & 0.00 & 0.99 & 0.01 & 0.00 & 0.00 \\
Iris & 0.00 & 1.00 & 0.00 & 0.00 & 0.00 & 0.00 \\
Cytometry & 0.00 & 0.00 & 0.00 & 0.00 & 0.00 & 1.00 \\ \hline
\end{tabular}
\label{tab:many23}
\end{table}

\begin{table}[ht]
\caption{Combined words in the political blogs data}
\normalsize
\centering
\begin{tabular}{rrrrrrrrr}
  \hline \hline
clinton & clintons &  \\
obama & obamas & barack \\
america & american & americans \\
candidate & candidates &  \\
democratic & democrats &  \\
new & news &  \\
president & presidential &  \\
senate & senator &  \\
year & years &  \\
vote & voters &  \\
thing & things &  \\
   \hline \hline
\end{tabular}
\label{tab:blogs2}
\end{table}

\begin{table}[ht]
\caption{Political blogs data. $P(\Mk \mid \by)$ under a MOM-Beta and Beta priors and $k \in \{1,\ldots,6\}$, BIC and BIC.}
\small
\centering
\begin{tabular}{rrrrrrrrr}
  \hline
 & &  MOM-Beta & Beta & BIC & AIC & &\\  \hline
& $k$  & $P(\mathcal{M}_{k}\mid \by)$ & $P(\mathcal{M}_{k}\mid \by)$ &  &  &  &  & \\ \hline
  \hline
 & 1 & 0.000 & 0.000 & -257405.2 & -256317.0 &  &  & \\
  & 2 & \textbf{1.000} & 0.000 & -255488.8 & -253307.8 &  & & \\
   & 3 & 0.000 & 0.000 &   \textbf{-255329.7} & -252055.9 &  & \\
   & 4 & 0.000 & \textbf{1.000} &   -255358.8  & -250992.2 &  & \\
   & 5 & 0.000 & 0.000 &   -255712.2 & -250252.7 &  & \\
   & 6 & 0.000 & 0.000 &   -256366.0  & \textbf{-249813.7}  &  &  \\ \hline
\end{tabular}
\label{tab:blogs}
\end{table}

\begin{figure}[ht]
\begin{center}
\begin{tabular}{cc}
BIC  & AIC \\
\includegraphics[width=0.5\textwidth]{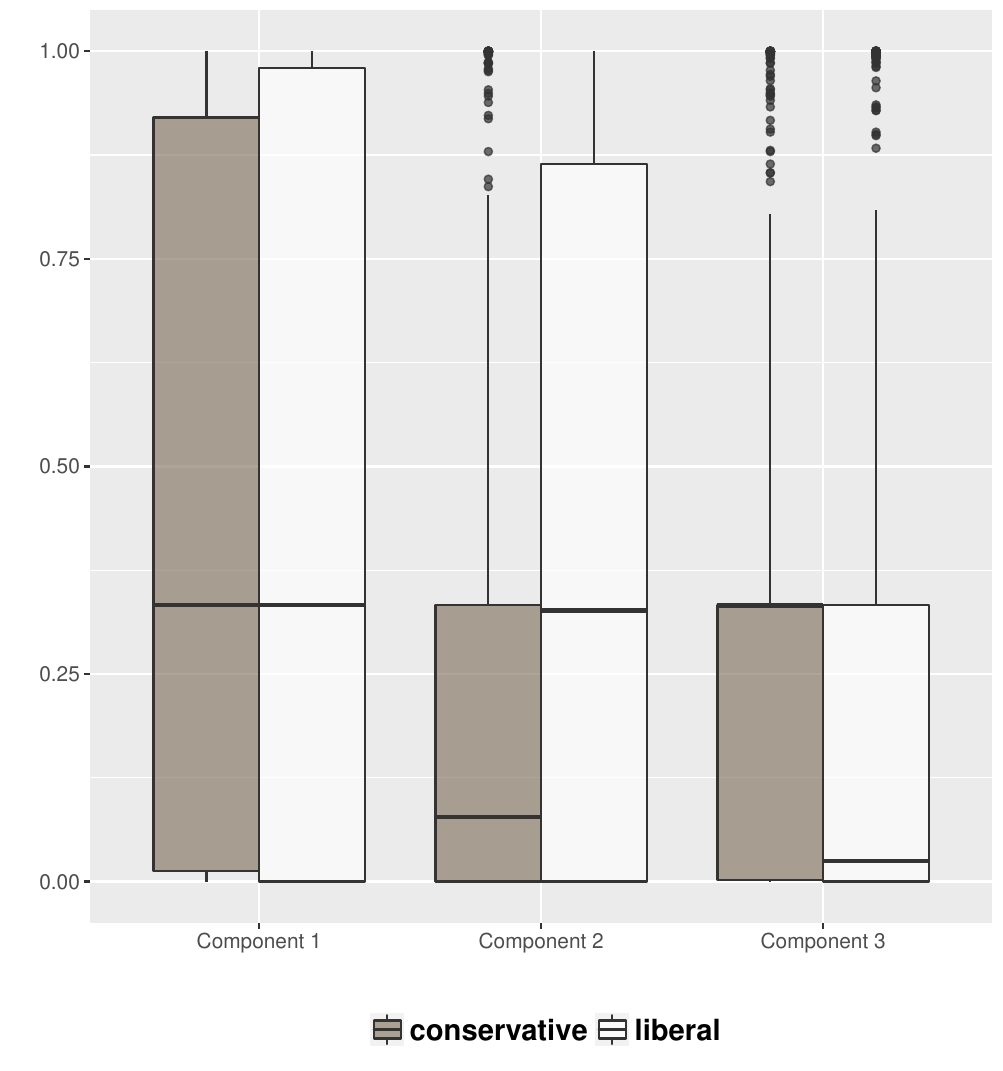} &
\includegraphics[width=0.5\textwidth]{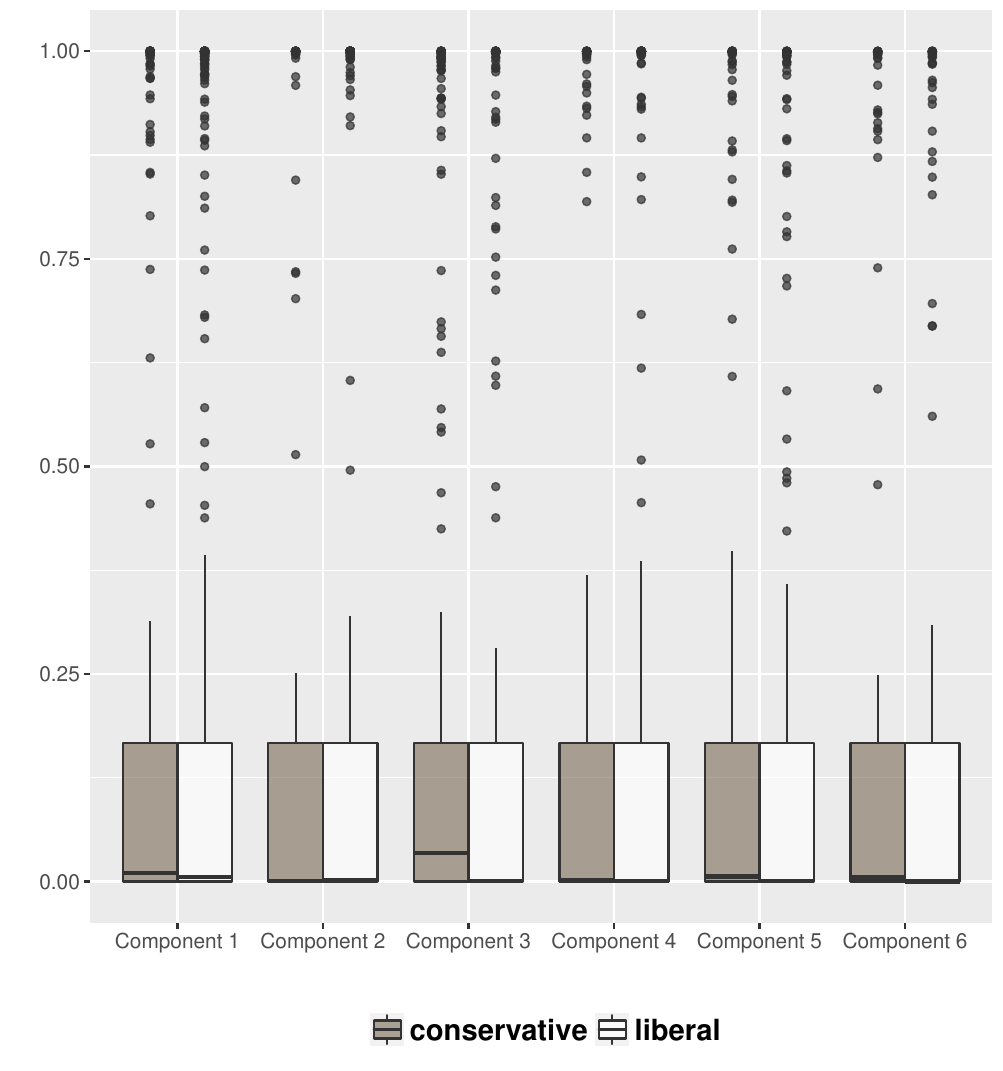} \\
Beta  & MOM-Beta \\
\includegraphics[width=0.5\textwidth]{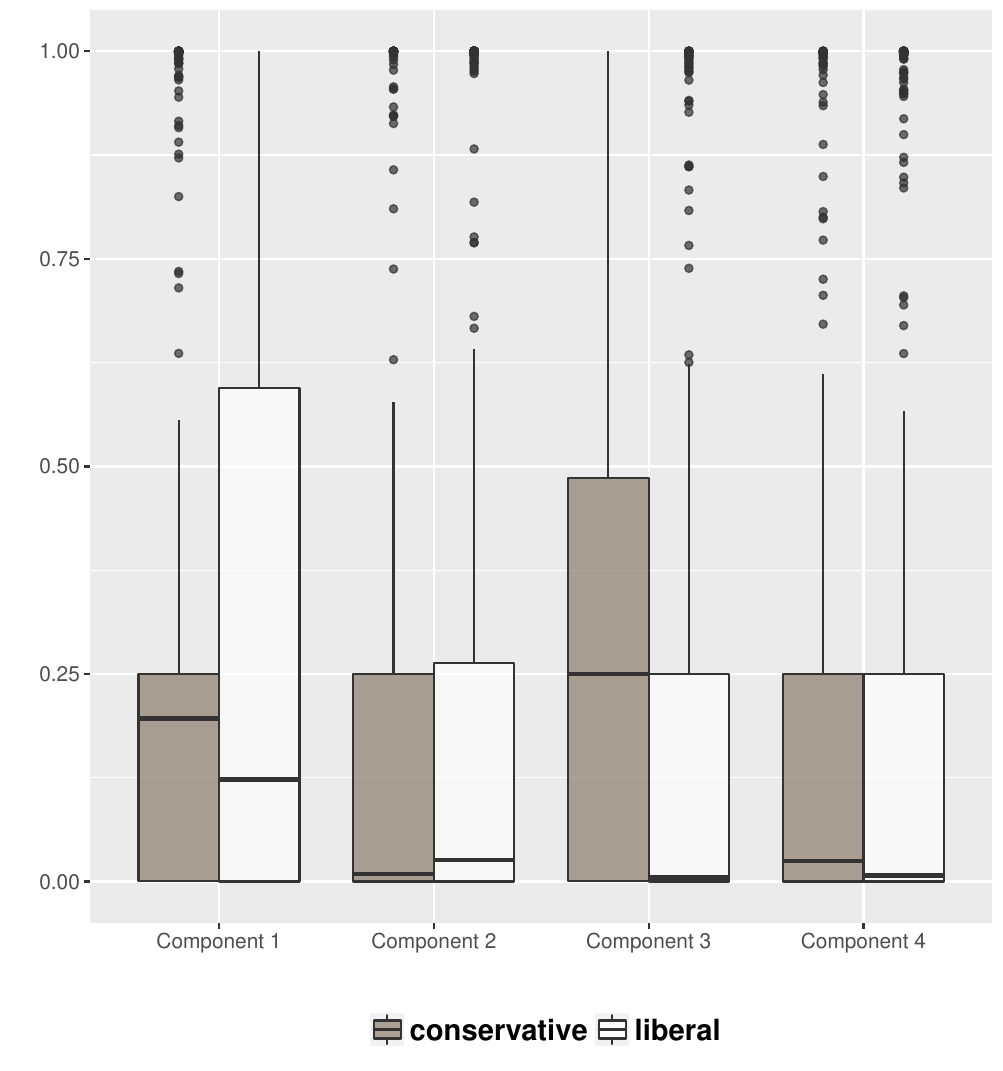} &
\includegraphics[width=0.5\textwidth]{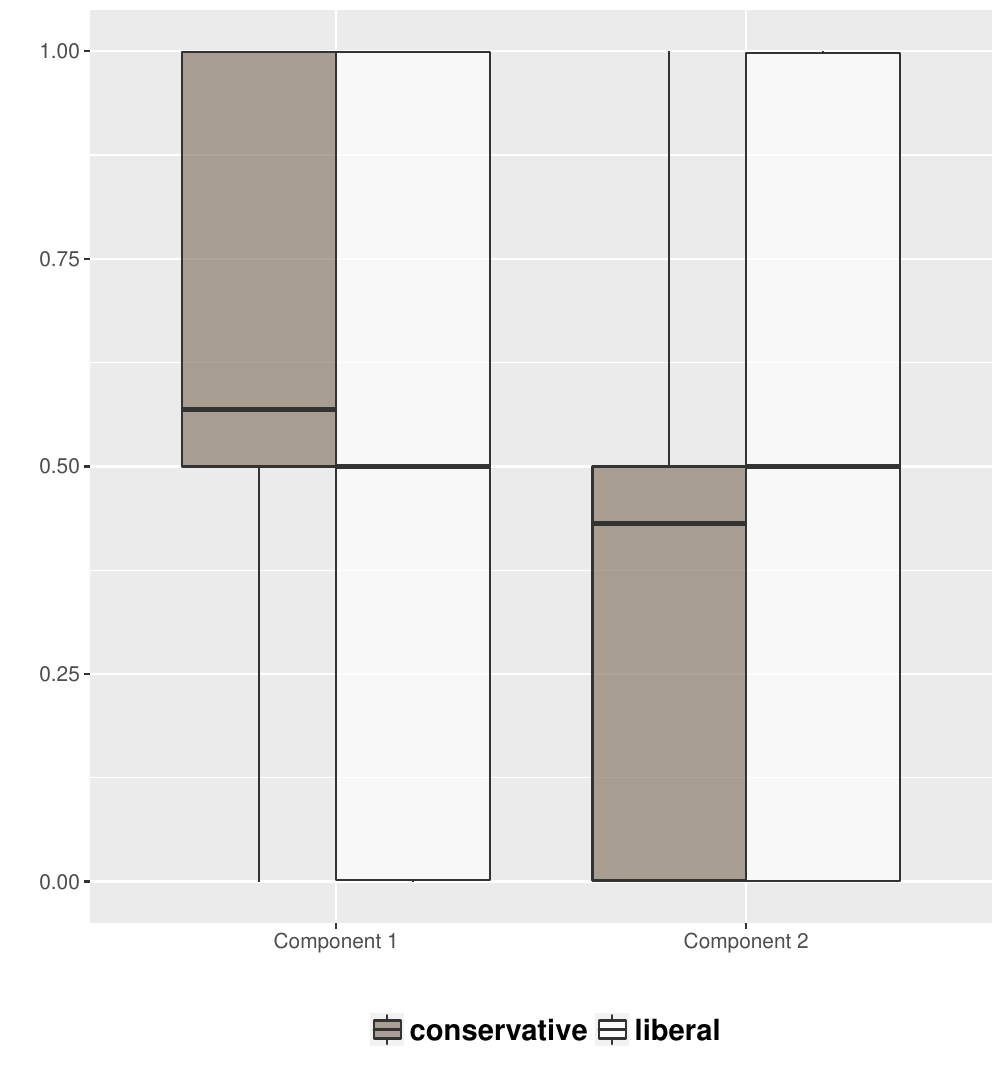}\\
\end{tabular}
\end{center}
\caption{Estimated cluster probabilities $p(z_{i}=j |\by,\mathcal{M}_{j})$ under BIC, AIC, Beta and MOM-Beta for
documents labelled as conservative or liberal}
\label{supfig:postcNLP}
\end{figure}

\section{Sample R code}
\label{supplsec:rcode}

The R code below generates bivariate data from a single-component Normal mixture
and uses \texttt{bfnormmix} from R package \texttt{mombf} to obtain posterior probabilities for 1-3 components,
both under MOM-IW-Dir and Normal-IW-Dir priors, under default prior parameters.
The obtained estimates are $P(\mathcal{M}_1 \mid \by)=0.889$ and $0.771$ for MOM-IW-Dir and Normal-IW-Dir priors, respectively.

\scriptsize
\begin{verbatim}
> library(mombf)
> set.seed(1)
>      x <- matrix(rnorm(100*2),ncol=2)
>      bfnormmix(x=x,k=1:3)
mixturebf object with 2 variables

Use draw() to obtain posterior samples. postProb() returns posterior probabilities as given below

  k   pp.momiw     pp.niw logprobempty logbf.momiw     logpen logbf.niw
1 1 0.88864310 0.77117451         -Inf    0.000000  0.0000000  0.000000
2 2 0.09992687 0.18573876    -3.191547   -2.185257 -0.7616836 -1.423573
3 3 0.01143003 0.04308673    -2.470700   -4.353451 -1.4687516 -2.884700
        model
1 Normal, VVV
2 Normal, VVV
3 Normal, VVV
\end{verbatim}
\normalsize

\section{An illustration for the computations for the product of Binomial mixture under MOM-Beta priors}
\label{supplsec:EMMCMC}

We illustrate some computational issues and diagnostics related to posterior multi-modality, the EM and MCMC algorithms
in product Binomial mixtures.
We considered a simulation with $k^{*}=4$ components, $n=500$, $p=8$ variables and equal component weights $\eta_1^*=\eta_2^*=\eta_3^*=\eta_4^*=1/4$.
Each component had two large success probabilities $\theta_{jf}^*=0.32$ whereas the remaining probabilities were small (0.04 and 0.08),
specifically
\begin{align}\label{truesim}
\btheta&=
\begin{pmatrix}
0.32 & 0.04 & 0.04 & 0.04 \\
0.32 & 0.08 & 0.08 & 0.08 \\
0.04 & 0.32 & 0.04 & 0.04 \\
0.08 & 0.32 & 0.08 & 0.08 \\
0.04 & 0.04 & 0.32 & 0.04 \\
0.08 & 0.08 & 0.32 & 0.08 \\
0.04 & 0.04 & 0.04 & 0.32 \\
0.08 & 0.08 & 0.08 & 0.32 \\
\end{pmatrix}.
\end{align}
The default MOM-Beta prior parameters are $g=2.6$ and $q=2$ (Section \ref{ssec:priorelicitation}).
Although our EM algorithm is guaranteed to increase the log-posterior at each iteration,
in practice there are potential issues with slow convergence or reaching local maxima/saddlepoints.
To address this in our implementation we run the EM algorithm (Algorithm \ref{alg:em1})
from 30 different random starting values and keep the
estimate achieving the highest log-posterior value.
The obtained estimates were fairly close to the simulation truth, specifically
\begin{align}
\hat{\bm{\eta}}&=(0.28, 0.26, 0.24, 0.22); &
\hat{\btheta}&=
\begin{pmatrix}
0.34 & 0.05 & 0.04 & 0.04\\
0.28 & 0.07 & 0.08 & 0.07\\
0.04 & 0.31 & 0.04 & 0.05\\
0.08 & 0.31 & 0.08 & 0.08\\
0.05 & 0.05 & 0.35 & 0.06\\
0.09 & 0.08 & 0.31 & 0.08\\
0.03 & 0.04 & 0.04 & 0.33\\
0.09 & 0.09 & 0.07 & 0.30\\
\end{pmatrix}
\end{align}

    We also studied the ability of the BIC, AIC, and Beta and MOM-Beta priors to recover $k^{*}=4$,
    finding that all except for the AIC returned the correct value (Table \ref{tab:simulaLCA}).
    Recall that the posterior probabilities require estimating the integrated likelihood,
    for which in turn we run an MCMC algorithm.
    To assess practical MCMC convergence we used
    trace plots for 2,000 iterations targetting $p(\bvartheta_4 \mid \by, \mathcal{M}_4)$ after a burn period of 1,000.
    The plots did not reveal any issues with the chain's mixing.

\begin{table}[ht]
\caption{Product Binomial simulation.
$P(\Mk \mid \by)$ for $k \in \{1,\ldots,6\}$ under Beta and MOM-Beta priors, BIC and AIC.}
\small
\centering
\begin{tabular}{rrrrrrrrr}
  \hline
 &  Beta & MOM-Beta & BIC & AIC & &
\\  \hline
 $k$  & $P(\mathcal{M}_{k}\mid \by)$ & $P(\mathcal{M}_{k}\mid \by)$ &  &  &  &  & \\ \hline
  \hline
1 & 0.000 & 0.000 & -22702.00 & -22668.29 &  &   \\
  2 & 0.000 & 0.000 & -21569.65 & -21498.00 &  &  \\
  3 & 0.000 & 0.000 & -20782.58 & -20673.00 &  &  \\
  4 & \textbf{1.000} & \textbf{1.000} & \textbf{-20051.63} & -19904.11 &  &   \\
  5 & 0.000 & 0.000 & -20074.65 & -19889.21 &  &   \\
  6 & 0.000 & 0.000 & -20099.17 & \textbf{-19875.80} &  &  \\ \hline
\end{tabular}
\label{tab:simulaLCA}
\end{table}


\end{document}